\documentclass[aps,twocolumn,prd,nofootinbib,showpacs]{revtex4}

\usepackage{amsmath}
\usepackage{graphicx}
\usepackage{dcolumn}
\usepackage{bm}
\usepackage{amssymb}
\usepackage{latexsym}
\usepackage{sparticles}
\usepackage{feynmf}
\usepackage{rotating}
\usepackage{hyperref}

\hypersetup{
    bookmarks=true,         
    unicode=false,          
    pdftoolbar=true,        
    pdfmenubar=true,        
    pdffitwindow=false,     
    pdfstartview={FitH},    
    pdftitle={My title},    
    pdfauthor={Author},     
    pdfsubject={Subject},   
    pdfcreator={Creator},   
    pdfproducer={Producer}, 
    pdfkeywords={keyword1} {key2} {key3}, 
    pdfnewwindow=true,      
    colorlinks=true,       
    linkcolor=red,          
    citecolor=cyan,        
    filecolor=magenta,      
    urlcolor=green,           
    linktocpage=true
}

\newcommand{\sss}[1]{{\scriptscriptstyle{#1}}}

\newcommand{\lta}{\lesssim}
\newcommand{\gta}{\gtrsim}
\newcommand{\uPl}{\mathrm{Pl}}

\newcommand{\usssPl}{\sss{\uPl}}

\newcommand{\ie}{\textsl{i.e.~}}

\def\spose#1{\hbox to 0pt{#1\hss}}
\def\lta{\mathrel{\spose{\lower 3pt\hbox{$\mathchar"218$}}
     \raise 2.0pt\hbox{$\mathchar"13C$}}}
\def\gta{\mathrel{\spose{\lower 3pt\hbox{$\mathchar"218$}}
     \raise 2.0pt\hbox{$\mathchar"13E$}}}

\newcommand{\de}[2]{\kern - #1 em \mathrm{d} #2}

\newcommand{\mpl}{m_\usssPl}
\newcommand{\Mp}{M_\usssPl}

\begin{document}

\title{Everything You Always Wanted To Know About The Cosmological
  Constant Problem (But Were Afraid To Ask)}

\author{J\'er\^ome Martin} \email{jmartin@iap.fr}
\affiliation{Institut d'Astrophysique de Paris, UMR7095-CNRS,
  Universit\'e Pierre et Marie Curie, 98bis boulevard Arago, 75014
  Paris, France}

\date{\today}

\begin{abstract}
  This article aims at discussing the cosmological constant problem at
  a pedagogical but fully technical level. We review how the vacuum
  energy can be regularized in flat and curved space-time and how it
  can be understood in terms of Feynman bubble diagrams. In
  particular, we show that the properly renormalized value of the
  zero-point energy density today (for a free theory) is in fact far
  from being $122$ orders of magnitude larger than the critical energy
  density, as often quoted in the literature. We mainly consider the
  case of scalar fields but also treat the cases of fermions and gauge
  bosons which allows us to discuss the question of vacuum energy in
  super-symmetry. Then, we discuss how the cosmological constant can
  be measured in cosmology and constrained with experiments such as
  measurements of planet orbits in our solar system or atomic
  spectra. We also review why the Lamb shift and the Casimir effect
  seem to indicate that the quantum zero-point fluctuations are not an
  artifact of the quantum field theory formalism. We investigate how
  experiments on the universality of free fall can constrain the
  gravitational properties of vacuum energy and we discuss the status
  of the weak equivalence principle in quantum mechanics, in
  particular the Collela, Overhausser and Werner experiment and the
  quantum Galileo experiment performed with a Salecker-Wigner-Peres
  clock. Finally, we briefly conclude with a discussion on the
  solutions to the cosmological constant problem that have been
  proposed so far.
\end{abstract}

\maketitle
\addtocontents{toc}{\protect\sloppy}
\tableofcontents

\section{Introduction}
\label{sec:intro}

This review article aims at discussing the cosmological constant
problem~\cite{Weinberg:1988cp,Carroll:1991mt,Dolgov:1997za,
  Sahni:1999gb,Straumann:1999ia,Weinberg:2000yb,
  Carroll:2000fy,Rugh:2000ji,Padmanabhan:2002ji,Yokoyama:2003ii,
  Sarkar:2005bc,Polchinski:2006gy,Copeland:2006wr,Brax:2009ae,Sapone:2010iz}. This
question is central in modern physics because its resolution would
certainly mean a very important step forward in our attempts to
understand physics beyond the current standard model.

\par

The history of the cosmological constant problem is a long and rich
one~\cite{Straumann:1999ia}. Its premises were in fact already
present immediately after the birth of quantum field theory. Indeed,
from the Heisenberg uncertainty principle, we know that the ground
state energy of the quantum mechanical oscillator cannot be zero
because the potential and the kinetic energies cannot vanish at the
same time. Since a (free) quantum field can be viewed as an infinite
collection of harmonic oscillators, it immediately comes that its
ground state energy must be infinite. Of course, this is not yet the
cosmological constant problem because gravity does not enter the
stage. But, clearly, potentially, a very severe problem is already
present.

\par

The infinity mentioned above is the first infinity that one encounters
in quantum field
theory~\cite{Peskin:1995ev,Greiner:1996zu,Itzykson:1980rh,
  Bailin:1994qt,LeBellac:1991cq,Ryder:1985wq,Mandl:1985bg}. However,
the presence of this infinity did not prevent the founding fathers of
quantum field theory to develop the theory since, in absence of
gravity (a working assumption of quantum field theory justified by the
weakness of gravity), only differences in energies are
observable. Therefore, in this context, this infinity can be merely
ignored and the rest of the formalism can be worked out without any
problem. As is well-known, other types of infinities appear and, this
time, there is no way to get around them. Treating and taming them is
the goal of renormalization. The very impressive agreement between
high precision measurements in accelerators and the predictions of
quantum field theory in presence of radiative corrections is the proof
that renormalization is able to correctly regulate these infinities.

\par

Therefore, we see that these two types of infinities are treated very
differently. The problem of the zero-point energy is just avoided
while the problem of the radiative corrections is directly and
explicitly addressed. Clearly, one cannot help thinking that the first
problem is in fact swept under the carpet. And, indeed, as soon as
gravity is turned on, it badly strikes back.

\par

At this point, it is worth noticing the following. One should not get
the impression that the zero-point energy cannot be renormalized. As a
matter of fact, as will be discussed in this review, it can be made
perfectly finite. However, this finite, renormalized, value of the
zero-point energy seems to be too large to be compatible with the
observations. Again, this is very different from the usual case of
quantum field theory where the finite part extracted from a divergent
expression always leads to a good agreement with the experiments. At
first sight, the cosmological constant problem is therefore neither
the presence of a new infinity nor our inability to regularize it but
rather the apparent failure of the renormalization scheme to produce,
at the quantitative level, a finite vacuum energy compatible with the
observational data.

\par

In fact, a failure of renormalization is not the only logical
possibility. The vacuum fluctuations could also be a mathematical
artifact of the quantum field theory framework and/or their
gravitational properties could be abnormal. However, when one tries to
test (theoretically or experimentally) these ideas, it seems that one
never encounters a problem of the kind mentioned above. This fact must
also be considered as a part of the cosmological constant problem
which, therefore, appears to have many different ramifications. We see
that this problem is in fact a very deep problem which lies at the
cross roads between different branches of physics (gravitational
physics, quantum field theory, cosmology etc \dots). In brief, it has
to do with the gravitational properties of the quantum vacuum.

\par

The importance of this question has recently been reinforced by the
discovery that the expansion of our Universe is
accelerated~\cite{Perlmutter:1998np,Riess:1998cb}. According to the
standard lore, this could be the first observational evidence that the
quantum vacuum is able to curve space-time. On the other hand, the
effect of a cosmological constant is visible only on large scales and,
therefore, it seems to be problematic to check this result elsewhere
than in cosmology. But, if true, the cosmological constant problem
represents a unique situation where some aspects of quantum gravity
are at play and where, at the same time, corresponding observational
signatures are not hopelessly beyond our technical capabilities. This
makes it a valuable opportunity to go beyond our current understanding
of theoretical physics. The only other situation where the three above
mentioned aspects (quantum mechanics, gravity and the possibility to
detect sizable observational effects) are mixed is the theory of
cosmological
inflation~\cite{Starobinsky:1980te,Guth:1980zm,Mukhanov:1981xt,Linde:1981mu,Mukhanov:1982nu,Starobinsky:1982ee,Guth:1982ec,Hawking:1982cz,Albrecht:1982wi,Linde:1983gd}
(for a review, see
Refs.~\cite{Martin:2003bt,Martin:2004um,Martin:2007bw}) where the
Cosmic Microwave Background (CMB) anisotropies, first detected by the
COBE (COsmic Background Explorer) satellite~\cite{Smoot:1992td}, are
supposed to originate from the quantum fluctuations of the inflation
and gravitational fields (for the observational status of inflation,
see Ref.~\cite{Martin:2006rs}).

\par

In this review article, we go through all the issues mentioned
before. Our prejudice is to present the question at the technical
level, in full details, in order for the paper to be reasonably
self-contained. The price to pay for this approach is that, sometimes,
we are led to consider problems that are well treated in standard
textbooks and/or are not, strictly speaking, directly concerned with
the cosmological constant problem. On the other hand, this is also an
opportunity to review many different subjects belonging to different
types of physics.

\par

The outline of the article is the following one. In
Sec.~\ref{sec:introcc}, we discuss how the cosmological constant
$\Lambda$ is introduced in the Einstein field equations and we explain
why the vacuum energy also participates in the value of $\Lambda$. In
Sec.~\ref{sec:classicalcc}, we treat the classical cosmological
constant problem, \ie we show that, in the presence of gravity, it is
no longer true that only differences of energy are observable. We
illustrate this discussion by means of the electro-weak phase
transition. In Sec.~\ref{sec:calculatecc}, we demonstrate that, even
if the classical ground state is tuned to zero, the quantum vacuum
fluctuations still give a contribution to the cosmological
constant. We discuss how the corresponding quantum vacuum energy can
be calculated and argue that the method often used in the literature
and which consists in introducing a cut-off, is not appropriate. We
also investigate how the quantum vacuum energy can be expressed in
terms of Feynman diagrams (bubble diagrams) and investigate the case
where interactions are present. In Sec.~\ref{sec:bubble}, we focus on
the so-called bubble diagrams. We discuss them in the context of
quantum field theory but also in ordinary quantum mechanics. We then
show that they have very different properties than the other loop
diagrams. In Sec.~\ref{sec:GEP}, we use the Gaussian effective
potential approach to calculate the vacuum energy in a situation where
non perturbative effects are present. Since all the calculations are
carried out in the case of a scalar field, we treat in
Sec.~\ref{sec:otherfield} the case of other types of fields (spinor
and vector fields). This also gives us the opportunity to discuss the
idea of super-symmetry. In Sec.~\ref{sec:vaccurve}, we present the
cosmological constant problem in a more rigorous way. In particular,
we show that a flat space-time calculation is a good approximation
although a fully consistent approach should be formulated in curved
space-time. One concludes the first part of this review article with
Sec.~\ref{sec:valuelambda} where we estimate the vacuum energy and
find a value very far from the often quoted ``$122$ orders of
magnitude''. In Sec.~\ref{sec:measuringlambda}, we explain, from a
theoretical point of view, how the value of the cosmological constant
was recently measured in cosmology. In particular, we discuss the
hypothesis that are implicitly assumed in order to obtain this
result. In Sec.~\ref{sec:lambdaelsewhere}, we discuss whether the
cosmological constant can be measured elsewhere than in cosmology and
argue that no other experimental context can compete with
cosmology. However, we also show that other experiments can put
constraints on the vacuum energy that are interesting from the point
of view of the cosmological constant problem. In
Sec.~\ref{sec:existence}, we review the experiments (Lamb shift,
Casimir effect) that seem to prove the existence of the vacuum
fluctuations. In Sec.~\ref{sec:vacuumweight}, we investigate whether
there are observations that can probe their gravitational
properties. In Sec.~\ref{sec:wepqm}, we discuss whether the weak
equivalence principle still holds in quantum mechanics since this has
obvious implications for the question of the vacuum weight. Finally,
in Sec.~\ref{sec:conclusions}, we present our conclusions and very
briefly review the solutions to the cosmological constant problem that
have been proposed so far. Let us mention that this article is part of
a more general review on the dark energy question, see
Refs.~\cite{Astier:2012ba,Kunz:2012aw,Clarkson:2012bg,deRham:2012az}.

\par

Before concluding this introduction, we would like to make the
following remark. This article aims at discussing what the
cosmological constant problem is. This article does not aim at
discussing what solutions to this problem have been proposed even if,
as already mentioned, in Sec.~\ref{sec:conclusions}, we say a few
words on this topic (and choose, in a totally arbitrary way which only
reflects the author prejudices, to focus more on some of them). The
justification for the above mentioned choice is that the question of
the gravitational properties of vacuum zero-point fluctuations is
already a highly non-trivial one as the variety of the subjects listed
above demonstrate. Therefore, it seems to us better to deeply
understand what is at play before trying to propose a way out. Of
course, it is because this problem is very difficult and has many
different ramifications in many different branches of physics that it
is so interesting.

\section[The CC]{The Cosmological Constant}
\label{sec:introcc}

In this first section, we introduce the cosmological constant as 
a free parameter in the classical action of the gravitational field.
This action, together with the action describing matter, can be expressed as 
\begin{eqnarray}
\label{eq:totalaction}
S&=&\frac{1}{2\kappa}\int {\rm d}^4x \sqrt{-g}
\left(R-2\Lambda_{_{\rm B}}\right)+S_{\rm matter}\left[g_{\mu \nu},\Psi\right],
\end{eqnarray}
where $\kappa \equiv 8\pi G/c^4\equiv 8\pi/\mpl^2\equiv 1/\Mp^2$,
$\mpl$ and $\Mp$ being the Planck mass and the reduced Planck mass
respectively. The first term is the standard Einstein-Hilbert action
and represents the gravitational part of the total action given by
Eq.~(\ref{eq:totalaction}). The cosmological constant appears in the
second term of the above expression. As announced, at this level, it
is merely a new parameter of the total action and for this reason we
write it as $\Lambda _{_{\rm B}}$ for ``bare cosmological
constant''. It has the dimension of the inverse of a square length. It
is compatible with general covariance and is of course compatible with
a conserved energy momentum tensor since $\nabla ^{\mu}g_{\mu
  \nu}=0$. Therefore, this term appears to be totally natural from the
relativistic point of view and there is a priori no reason to discard
it. Since, according to the standard lore of field theory, everything
which is not forbidden should be considered, the cosmological constant
should clearly be included in our description of the gravitational
field. Finally, the third term in the above equation denotes the
matter action where $\Psi$ represents a generic matter field that we
do not need to specify at this stage. Variation of the total action
with respect to the metric tensor leads to the Einstein equations of
motion which read
\begin{equation}
R_{\mu \nu}-\frac{1}{2}Rg_{\mu \nu}+\Lambda_{_{\rm B}} g_{\mu \nu}
=\kappa T_{\mu \nu}\, ,
\end{equation}
where the stress-energy tensor is defined by
\begin{equation}
\label{eq:defstresstensor}
T_{\mu \nu}=-\frac{2}{\sqrt{-g }}
\frac{\delta S_{\rm matter}}{\delta g
^{\mu \nu}}\, .
\end{equation}
As already mentioned, $\Lambda_{_{\rm B}}$ appears to be just a
parameter of the model. At this level, the only thing that can be done
is to try to constrain it using various observations. But, clearly, in
the present framework, there is no way to calculate its value from
more fundamental considerations.

\par

However, as originally shown by Sakharov~\cite{Sakharov:1967pk}, the
nature of the discussion is crucially changed when one takes into
account quantum field theory. The point is that the stress energy
tensor of a field placed in the vacuum state must be given by
\begin{equation}
  \langle 0\vert T_{\mu \nu}\vert 0\rangle 
=-\rho_{_{\rm vac}}g_{\mu \nu},
\end{equation}
where $\rho_{\rm vac}$ is the constant energy density of the
vacuum. This equation is valid for all the fields present in the
Universe, as will be shown below. There are several ways to prove the
above result. Firstly, one can use the fact that, in flat space-time
(in Minkowski space-time), the only invariant tensor is $\eta_{\mu
  \nu}$. Since the vacuum state must be the same for all observers,
one necessarily has $\langle T_{\mu \nu}\rangle \propto \eta _{\mu
  \nu}$. In curved space-time, this means that
\begin{equation}
\label{eq:qmvac}
\langle T_{\mu \nu}\rangle =-\rho_{_{\rm vac}}(t,{\bf x})g_{\mu \nu},
\end{equation}
and from the fact that the stress-energy tensor must be conserved, we
reach the conclusion that $\rho _{_{\rm vac}}$ must be a
constant. Therefore, one obtains the formula shown above. 

\par

Another way to obtain the same result is to consider a specific
example, for instance a scalar field $\Phi$. The corresponding action
reads
\begin{equation}
\label{eq:actionsf}
S_{\Phi}=-\int {\rm d}^4x\sqrt{-g }
\biggl[\frac{1}{2}g^{\mu \nu}\partial
_{\mu} \Phi \partial _{\nu}\Phi +V(\Phi )\biggr].
\end{equation}
where $V(\Phi)$ is the potential. Using the
definition~(\ref{eq:defstresstensor}), the corresponding stress-energy
tensor can be written explicitly as
\begin{equation}
\label{eq:tmunuscalarfield}
T_{\mu \nu}=\partial _{\mu }\Phi \partial _{\nu }\Phi 
-g_{\mu \nu}\biggl[\frac{1}{2}g^{\alpha \beta }
\partial _{\alpha }\Phi \partial _{\beta }\Phi 
+V(\Phi )\biggr]\, .
\end{equation}
From this expression, one sees that the scalar field is in fact a
perfect fluid. Now, the vacuum state is the minimum energy
state. Clearly, in order to minimize the energy we have to consider a
situation where the kinetic energy vanishes and where the field sits
at the minimum of its potential. In this case, the stress energy
tensor reduces to
\begin{equation}
\label{eq:classicalvac}
\langle T_{\mu \nu}\rangle =-V(\Phi_{_{\rm min}})g_{\mu \nu},
\end{equation} 
which has exactly the expected form with $\rho_{_{\rm
    vac}}=V(\Phi_{_{\rm min}})$. 

\par

In fact what was shown before is that there are at least two sources
for the vacuum energy. There is a ``classical'' contribution given by
Eq.~(\ref{eq:classicalvac}) which originates from the value of the
potential at its minimum. The corresponding vacuum energy will be
calculated in Sec.~\ref{sec:classicalcc}. There is also a
``quantum-mechanical'' source given by Eq.~(\ref{eq:qmvac}) which
originates from the zero point fluctuations of the ground state. This
problem will be treated in Sec.~\ref{sec:calculatecc}.

\par

Having established the form of the stress energy momentum tensor in
the vacuum, one can now proceed with the Sakharov argument. The next
step consists in assuming that the equivalence principle applies to
the zero-point fluctuations (here, and from now on, we mean to the
``classical'' and ``quantum-mechanical'' contributions as discussed in
the last paragraph), that is to say that the zero-point fluctuations
gravitate. Clearly, this is not a trivial step but, after all, the
vacuum fluctuations are just a specific type of energy and, in general
relativity, all forms of energy gravitate. Therefore, a consistent way
of writing the Einstein equations when quantum field theory is taken
into account seems to be
\begin{equation}
R_{\mu \nu}-\frac{1}{2}Rg_{\mu \nu}+\Lambda_{_{\rm B}} g_{\mu \nu}
=\kappa T_{\mu \nu}^{\rm matter}+\kappa \langle T_{\mu \nu}\rangle\, ,
\end{equation}
where on the right hand side the first contribution comes from
ordinary matter while the second one represents the contribution
originating from the vacuum (of course, it is to be understood that we
sum up the contributions coming form all the fields present in the
universe). Using the form of $\langle T_{\mu \nu}\rangle$ established
before, one arrives at
\begin{equation}
\label{eq:einsteineqs}
R_{\mu \nu}-\frac{1}{2}Rg_{\mu \nu}+\Lambda_{_{\rm eff}} g_{\mu \nu}
=\kappa T_{\mu \nu}^{\rm matter}\, ,
\end{equation}
where 
\begin{equation}
\label{eq:deflambdaeff}
\Lambda_{_{\rm eff}}=\Lambda_{_{\rm B}}+\kappa \rho_{_{\rm vac}}.
\end{equation}
Therefore, we conclude that the effective cosmological constant is the
sum of the bare cosmological constant and of a contribution
originating from the vacuum fluctuations. The effective cosmological
constant $\Lambda_{_{\rm eff}}$ is the quantity that one can observe
and constrain when tests of the Einstein equations are carried out.

\par

As we now discuss, the problem is that $\kappa \rho_{_{\rm vac}}$ is
made of several terms which are all huge in comparison with the
observed value of $\Lambda_{_{\rm eff}}$.

\section{The Classical Cosmological Constant Problem}
\label{sec:classicalcc}

\begin{figure*}
\begin{center}
\includegraphics[width=14.5cm]{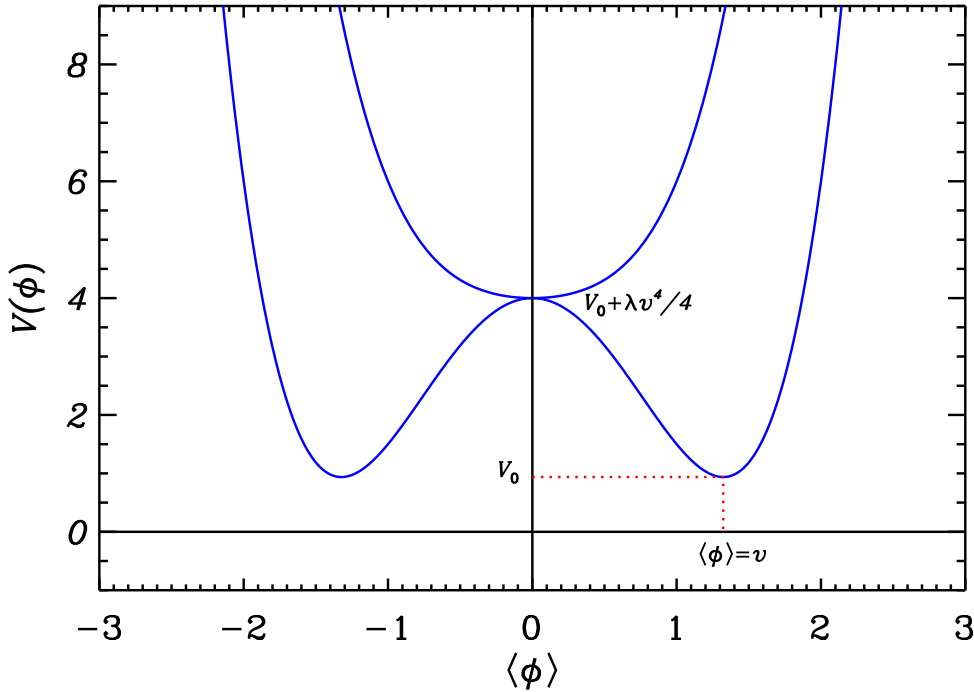}
\caption{The effective potential given by
  Eq.~(\ref{eq:effectivepotT}). Before the transition, for $T>T_{\rm
    cri}$, the minimum of the potential is located at the origin and
  the vacuum energy is given by $V_0+\lambda v^4/4$. After the
  transition, for $T<T_{\rm cri}$, the minimum is located at $\Phi=v$
  and the corresponding vacuum energy has changed and now equals
  $V_0$. It is clear that $V_0$ can always be chosen such that the
  vacuum energy vanishes either before or after the transition. It is
  equally clear that one cannot choose the parameters of the potential
  such that $\rho_{_{\rm vac}}$ is zero before and after the phase
  transition.}
\label{fig:ssbzero}
\end{center}
\end{figure*}

The cosmological constant problem is in fact a multi facets question
as the quantity $\Lambda _{_{\rm eff}}$ receives contributions from
different origins. As was demonstrated in the last section on the
example of a scalar field, a classical contribution to $\Lambda
_{_{\rm eff}}$ comes from the value of the potential at its
minimum. If this is not zero, then it affects the value of the
effective cosmological constant. This point is discussed in this
section and is exemplified with the electroweak transition. Another
contribution comes from the quantum zero point fluctuations but this
will be discussed in the next sections.

\subsection{Phase Transition}
\label{subsec:transition}

It is possible to study the dynamics of a phase transition with the
help of the following simple model. Let us consider a scalar field
$\Phi$ in interaction with another scalar field $\Psi$ such that
\begin{equation}
V(\Phi,\Psi)=V(\Phi)+\frac{\bar{g}}{2}\Phi^2\Psi^2,
\end{equation}
where $\bar{g}$ is a dimensionless coupling constant and where the 
self-interacting potential is given by
\begin{equation}
V(\Phi)=V_0+\frac{\lambda}{4}\left(\Phi^2-v^2\right)^2,
\end{equation}
$\lambda$ being a coupling constant describing the self-interaction
and $v$ the value of the scalar field at the minimum of its potential
in absence of an interaction with $\Psi$. The quantity $V_0$ denotes
the classical off-set. If the field $\Psi$ is in thermal equilibrium,
then one is entitled to replace $\Psi^2$ with $\langle \Psi ^2
\rangle_{T}$ where the average is taken in a thermal state with
temperature $T$. Since $\langle \Psi ^2 \rangle_{T}\propto
T^2$~\cite{Kapusta:2006pm}, the effective potential becomes
\begin{equation}
\label{eq:effectivepotT}
V_{\rm eff}(\Phi)=V_0+\frac{\lambda}{4}\left(\Phi^2-v^2\right)^2
+\frac{\bar{g}}{2}T^2\Phi^2,
\end{equation}
where we have slightly redefined the coupling constant $\bar{g}$ in
order to take into account the proportionality constant between the
thermal average and the temperature. This potential can also be
expressed as
\begin{equation}
\label{eq:pottemperature}
V_{\rm eff}(\Phi)=V_0+\frac{\lambda v^4}{4}+\frac{\lambda v^2}{2}
\left(\frac{T^2}{T_{\rm cri}^2}-1\right)\Phi^2+\frac{\lambda}{4}\Phi^4,
\end{equation}
where we have defined $T_{\rm cri}=v\sqrt{\lambda/\bar{g}}$. One sees
on the above equation that the interaction with the field $\Psi$ gives
a temperature dependence to the effective mass of the $\Phi$ field,
namely
\begin{equation}
m^2_{\rm eff}\left(T\right)\equiv\lambda v^2
\left(\frac{T^2}{T_{\rm cri}^2}-1\right).
\end{equation}
As a consequence, when $T>T_{\rm cri}$, that is to say before the
transition, the square of the effective mass is positive while, after
the transition, when $T<T_{\rm cri}$, it becomes negative. In the
first situation, the minimum is located at $\Phi=0$ and the
corresponding value of the vacuum energy is $V_0+\lambda v^4/4$, see
Fig.~\ref{fig:ssbzero}. In the second situation, the minimum is
located at $\Phi=v$ and the off-set is simply given by $V_0$, see
Fig.~\ref{fig:ssbzero}. We now clearly see the problem. If we require
the vacuum energy to vanish before the transition, we must choose
$V_0=-\lambda v^4/4$. But, then, after the transition, the vacuum
energy is no longer zero and is given by the negative value
$\rho_{_{\rm vac}}=-\lambda v^4/4$. On the other hand, one can choose
$\rho_{_{\rm vac}}$ to vanish after the transition. In this case, one
must choose $V_0=0$. But this means that it was not zero before the
transition, $\rho_{_{\rm vac}}=\lambda v^4/4$. This last option can
maybe viewed as the preferred solution since we do not have direct
observational constraints on the matter content of the universe prior
to Big Bang Nucleosynthesis (BBN). Anyhow, we see that the choice of a
parameter in the potential allows us to change the vacuum energy at
the classical level, hence the name ``classical cosmological constant
problem''. But, clearly, the off-set cannot be zero before and after
the transition. Of course, in order to really establish that there is
a problem, the question remains to estimate this vacuum energy for
realistic phase transition and to compare it with the observational
constraints. In the next subsection, we turn to the first issue and
consider the case of the electroweak phase transition.

\subsection{The Electro-weak Phase Transition}
\label{subsec:ewtransition}

In this section, we calculate the vacuum energy ``induced'' by the
electroweak phase transition. In order to achieve this goal, we first
must recall some basics facts about the standard model of particle
physics, see Refs.~\cite{Peskin:1995ev,Greiner:1996zu,Itzykson:1980rh,
  Bailin:1994qt,LeBellac:1991cq,Ryder:1985wq,Mandl:1985bg}.

\par

\begin{figure*}
\begin{center}
\includegraphics[width=8.5cm]{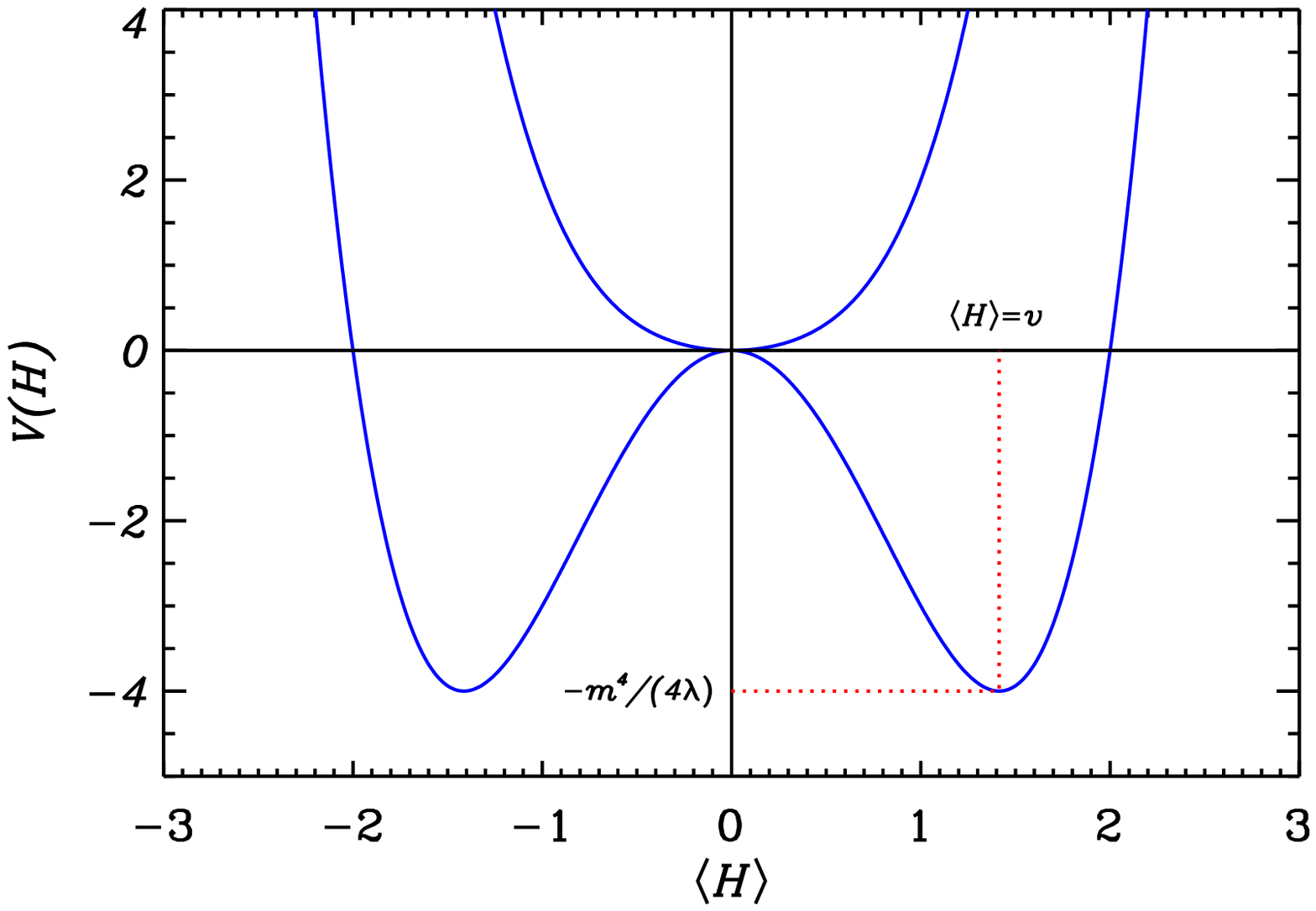}
\includegraphics[width=8.5cm]{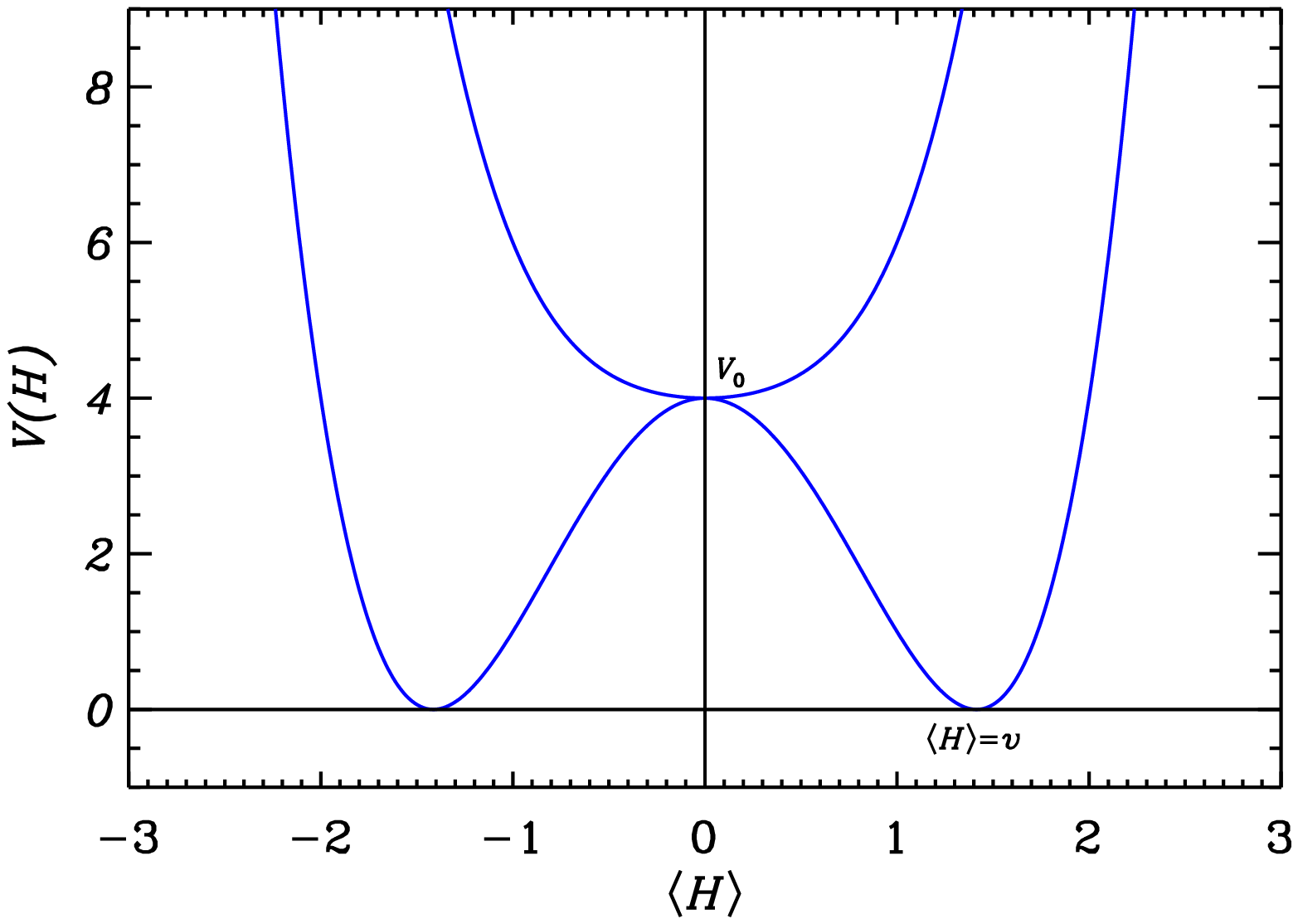}
\caption{Effective potential of the Higgs boson before and after the
  electroweak phase transition. The left panel corresponds to a
  situation where the vacuum energy vanishes at high temperature. As a
  consequence $\rho_{_{\rm vac}}$ is negative at temperature smaller
  than the critical temperature. This is the situation treated in the
  text where the quantity $-m^4/(4\lambda)$ is explicitly
  calculated. On the right panel, the off-set parameter $V_0$ is
  chosen such that the vacuum energy is zero after the transition. As
  a consequence, it does not vanish at high temperatures.}
\label{fig:ssb}
\end{center}
\end{figure*}

The Higgs field is a doublet of complex scalar fields. It is charged
under the group $\mbox{U}(1)_{_{\rm Y}}\times \mbox{SU}(2)_{_{\rm L}}$
characterized by two coupling constant $g'$ and $g$. The Higgs field
Lagrangian reads
\begin{equation} 
\label{eq:lagrangianhiggs}
{\cal L}_{\rm Higgs}=-\left(D_{\mu}\Sigma \right)^{\dagger}D^{\mu}\Sigma
-V\left(\Sigma,\Sigma^{\dagger}\right),
\end{equation}
where the covariant derivative can be expressed as
\begin{equation}
\label{eq:covderive}
D_\mu \Sigma=\partial _{\mu}\Sigma+i\frac{g'}{2}Y_{_{\rm H}}B_{\mu}\Sigma
+ig T_aW^a_{\mu}\Sigma,
\end{equation}
where $a$ runs from one to three. The quantities $B_{\mu}$ and
$W_{\mu}^a$ are the gauge bosons, $Y_{_{\rm H}}$ is the Higgs weak
hyper-charge and $T_a$ are the generators of the $\mbox{SU}(2)$ group
(see below for more details about these quantities). The potential is
chosen to be
\begin{equation}
\label{eq:selfpotHiggs}
V\left(\Sigma,\Sigma^{\dagger}\right)=\frac{m^2}{2}\Sigma ^{\dagger}\Sigma
+\frac{\lambda}{4}\left(\Sigma ^{\dagger}\Sigma\right)^2,
\end{equation}
and clearly resembles Eqs.~(\ref{eq:effectivepotT})
and~(\ref{eq:pottemperature}) for $T<T_{\rm cri}$. Indeed, in this
expression, the quantity $m^2$ is negative (as explained before, $m^2$
is in fact the square of the effective mass). This also means that the
minimum of the potential is not located at $\langle \Sigma\rangle =0$
but at
\begin{equation}
\langle \Sigma \rangle=v=\sqrt{-\frac{m^2}{\lambda}},
\end{equation}
obtained from the condition $\partial V/\partial \left(\Sigma
  ^{\dagger}\Sigma\right)=0$. The quantity $v$ is in fact the new
vacuum expectation value of the Higgs after spontaneous symmetry
breaking. In particle physics, one studies the theory after the
transition, when the Higgs field has stabilized at its new
minimum. But, in reality, as explained before, we must consider that
the transition is a dynamical process. Before the transition we had
$m^2>0$ and the minimum was located at $\langle \Sigma \rangle =0$ and
after the transition, $m^2$ has become negative and, consequently, the
new vacuum is given by $v$. Returning to the minimum after the
transition, it is easy to show that the value of the potential at
$\langle \Sigma \rangle =v$ is given by
\begin{equation}
\label{eq:rhovacew}
V\left(\langle \Sigma \rangle=v\right)=-\frac{m^4}{4\lambda},
\end{equation}
and is negative. Clearly, this is because we have chosen $V_0=0$ in
Eq.~(\ref{eq:selfpotHiggs}). As already explained in
Sec.~\ref{subsec:transition}, this means that the vacuum energy
vanishes before the transition and becomes negative after, see also
Fig.~\ref{fig:ssbzero}. This means that we have considered the
situation corresponding to the left panel in Fig.~\ref{fig:ssb}. Of
course, this is arbitrary and we could also have considered the
situation corresponding to the right panel in
Fig.~\ref{fig:ssb}. Again, as explained in the previous subsection, in
this case, the vacuum energy would vanish after the transition but not
before. 

The numerical value of the vacuum energy given by
Eq.~(\ref{eq:rhovacew}) is fixed by the electroweak physics in
particular by the parameters $m$ and $\lambda$. Therefore, in order to
compute $\rho_{_{\rm vac}}$ explicitly, we must explain how $m$ and
$\lambda$ are evaluated. For this purpose, we now briefly recall the
main features of the standard model of particle physics. Let us come
back to the Higgs Lagrangian given at the beginning of this section,
see Eq.~(\ref{eq:lagrangianhiggs}). In the expression of ${\cal
  L}_{\rm Higgs}$, the quantities $T_a$ are the generator of
$\mbox{SU}(2)_{\rm L}$ and are given by $T_a=\sigma _a/2$ where
$\sigma _a$ are the Pauli matrices
\begin{equation}
\sigma _1=\begin{pmatrix} 0 & 1 \cr 1 & 0\end{pmatrix}, \quad 
\sigma _2=\begin{pmatrix} 0 & -i \cr i & 0\end{pmatrix}, \quad 
\sigma _3=\begin{pmatrix} 1 & 0 \cr 0 & -1\end{pmatrix}.
\end{equation}
Using the
gauge invariance, one can always rewrite the complex doublet as
\begin{equation}
\Sigma =\left(v+\frac{H}{\sqrt{2}}\right)
\begin{pmatrix} 0 \cr 1
\end{pmatrix},
\end{equation}
where $H$ is a real scalar field, the ``Higgs boson''.  Therefore, the
covariant derivative~(\ref{eq:covderive}) can be expressed as
\begin{widetext}
\begin{equation}
D_{\mu}\Sigma =
\begin{pmatrix}
\displaystyle
i\frac{g}{2}\left(v+\frac{H}{\sqrt{2}}\right)
\left(W_{\mu}^1-iW_{\mu}^2\right) \cr \cr
\displaystyle
\frac{1}{\sqrt{2}}\partial _{\mu}H+i\left(v+\frac{H}{\sqrt{2}}\right)
\left(\frac{g'}{2}Y_{_{\rm H}}B_{\mu}-\frac{g}{2}W_{\mu}^3\right)
\end{pmatrix},
\end{equation}
and the Higgs Lagrangian~(\ref{eq:lagrangianhiggs}) becomes
\begin{eqnarray}
\label{eq:higgslagrangianinter}
{\cal L}_{\rm Higgs}&=&-\frac{1}{2}\partial _{\mu}H\partial ^{\mu}H-
V\left(\Sigma,\Sigma ^{\dagger}\right)
-\left(v^2+\sqrt{2}vH+\frac{1}{2}H^2\right)
\biggl(\frac{g^2}{4}W_{\mu}^1W^{1 \mu}
+\frac{g^2}{4}W_{\mu}^2W^{2 \mu}+\frac{g^2}{4}W_{\mu}^3W^{3 \mu}
\nonumber \\ & &
+\frac{g'^2}{4}Y_{_{\rm H}}^2B_{\mu}B^{\mu}
-\frac{gg'}{2}Y_{_{\rm H}}B_{\mu}W^{3 \mu}\biggr),
\end{eqnarray}
where the scalar potential can be written as
\begin{equation}
\label{eq:higgspot}
V\left(\Sigma ,\Sigma ^{\dagger}\right)=-\frac{\lambda v^4}{4}
+\frac{1}{2}\lambda v^2H^2+\frac{\lambda}{2}\frac{v}{\sqrt{2}}H^3
+\frac{\lambda }{16}H^4.
\end{equation}
We now introduce the physical gauge bosons. As is well-known, they are
defined by
\begin{eqnarray}
\label{eq:defgaugebosons}
B_{\mu}=-\sin \theta Z_{\mu}+\cos \theta A_{\mu}\, ,\quad 
W^3_{\mu}=\cos \theta Z_{\mu}+\sin \theta A_{\mu}\, ,\quad 
W^2_{\mu}=\frac{W_{\mu}^+-W_{\mu}^-}{i\sqrt{2}}\, ,\quad 
W^1_{\mu}=\frac{W_{\mu}^++W_{\mu}^-}{\sqrt{2}}\, ,
\end{eqnarray}
where $\theta $ is the Weinberg angle, $A_{\mu}$ represents the photon
and $Z_{\mu}$, $W_{\mu}^{\pm}$ the massive gauge bosons. Expressing
the Higgs Lagrangian~(\ref{eq:higgslagrangianinter}) in terms of these
new fields, one obtains the following expression
\begin{eqnarray}
\label{eq:higgslagrangianexplicit}
{\cal L}_{\rm Higgs}&=&-\frac{1}{2}\partial _{\mu}H\partial ^{\mu}H-
V\left(\Sigma,\Sigma ^{\dagger}\right)
-\left(v^2+\sqrt{2}vH+\frac{1}{2}H^2\right)
\biggl[\frac{g^2}{4}W_{\mu}^+W^{-\mu}
+\frac{g^2}{4}W_{\mu}^-W^{+ \mu}
\nonumber \\ & &
+Z_{\mu}Z^{\mu}\left(\frac{g^2}{4}\cos ^2\theta +\frac{g'^2}{4}
Y_{_{\rm H}}^2\sin ^2\theta 
+\frac{gg'}{2}Y_{_{\rm H}}\sin \theta \cos \theta\right)
+A_{\mu}A^{\mu}\left(\frac{g^2}{4}\sin ^2\theta 
+\frac{g'^2}{4}Y_{_{\rm H}}^2\cos ^2\theta 
-\frac{gg'}{2}Y_{_{\rm H}}\sin \theta \cos \theta\right)
\nonumber \\ & & 
+Z_{\mu}A^{\mu}\left(\frac{g^2}{2}\cos \theta \sin \theta 
-\frac{g'^2}{2}Y_{_{\rm H}}^2\cos \theta
\sin \theta  
-\frac{gg'}{2}Y_{_{\rm H}}\cos ^2\theta 
+\frac{gg'}{2}Y_{_{\rm H}}\sin ^2\theta \right)
\biggr].
\end{eqnarray}
Of course ${\cal L}_{\rm Higgs}$ is not the only term in the standard
model Lagrangian. There are also the terms describing the kinetic terms
of the gauge fields
\begin{equation}
{\cal L}_{\rm K-Gauge}=-\frac{1}{4}B_{\mu \nu}B^{\mu \nu}
-\frac{1}{4}W_{a \mu \nu}W^{a \mu \nu}=-\frac{1}{4}A_{\mu \nu}A^{\mu \nu}
-\frac{1}{4}Z_{\mu \nu}Z^{\mu \nu}
-\frac{1}{4}W_{+ \mu \nu}W^{+\mu \nu}
-\frac{1}{4}W_{-\mu \nu}W^{-\mu \nu}+\cdots ,
\end{equation}
\end{widetext}
where $B_{\mu \nu}$ is the $B$ field strength, namely $B_{\mu
  \nu}\equiv \partial_{\mu}B_{\nu}-\partial_{\nu}B_{\mu}$ and where a
similar definition applies to the other gauge fields.

\par

Finally, one must also specify how leptons are described. In the
standard model of particle physics, a fermion is represented by a
four-component Dirac spinor $\Psi$, see also Sec.~\ref{subsec:fermion}
where the zero-point fluctuations of spinors are calculated. Let us
now recall what chirality is. For this purpose, we introduce the
matrix $\gamma ^5$ such that $\left(\gamma ^5\right)^2=\mathbb{I}$ and
$\left(\gamma ^5\right)^{\dagger }=\gamma ^5$. This allows us to
define two chiral projectors by
\begin{equation}
P_{_{\rm R}}\equiv \frac{\mathbb{I}_4+\gamma _5}{2},
\quad P_{_{\rm L}}=\frac{\mathbb{I}_4-\gamma _5}{2},
\end{equation}
where $\mathbb{I}_4$ denotes the identity four by four matrix. We now
define the Dirac matrices, see Sec.~\ref{subsec:fermion} for a more
general approach. In the chiral representation, they are given by
\begin{eqnarray}
\label{eq:chiralrepresentation}
\gamma ^0 &=& \begin{pmatrix} 0 
& -\mathbb{I}_2 \cr -\mathbb{I}_2 & 0\end{pmatrix},
\quad 
\gamma ^i=\begin{pmatrix} 0 & \sigma ^i\cr -\sigma ^i & 0\end{pmatrix},
\nonumber \\  
\gamma ^5 &=& \begin{pmatrix} \mathbb{I}_2 & 0 \cr 0 
& -\mathbb{I}_2 \end{pmatrix}, 
\end{eqnarray}
and the two projectors can be written as
\begin{eqnarray}
  P_{_{\rm R}}&=& \begin{pmatrix} \mathbb{I}_2 & 0 \cr 0 & 0\end{pmatrix},
  \quad 
  P_{_{\rm L}}= \begin{pmatrix} 0  & 0 \cr 0 & \mathbb{I}_2 \end{pmatrix}.
\end{eqnarray}
Given a Dirac spinor $\Psi$ (we denote the four-component spinors by a
capital Greek letter and the two-component spinors by an ordinary
Greek letter)
\begin{equation}
\label{eq:spinorrl}
\Psi = \begin{pmatrix} \psi _{_{\rm R}}\cr \psi _{_{\rm L}} \end{pmatrix}, 
\end{equation}
one can thus define the two states of chirality (right and left
handed) by means of the two following expressions
\begin{equation}
\Psi_{_{\rm R}}\equiv P_{_{\rm R}}\Psi=
\begin{pmatrix} \psi _{_{\rm R}}\cr 0 \end{pmatrix}, 
\quad \Psi_{_{\rm L}}\equiv 
P_{_{\rm L}}\Psi
\begin{pmatrix} 0 \cr \psi _{_{\rm L}} \end{pmatrix}.
\end{equation}
The quantities $\psi _{_{\rm R,L}}$ are two-component spinors known as
Weyl spinors. They play a very important role in super-symmetry, see
Sec.~\ref{subsec:susy}. Notice that we also have $\overline\Psi\equiv
\Psi ^{\dagger}\gamma ^0=-\left(\psi _{_{\rm L}}^{\dagger},\psi_{_{\rm
      R}}^{\dagger}\right)$. This implies that the (mass-less) Dirac
Lagrangian can be expressed as
\begin{equation}
\label{eq:diracmasslessL}
{\cal L}_{\rm Dirac}=-i\overline \Psi \gamma ^{\mu}\partial_{\mu}\Psi
=-i\psi _{_{\rm L}}^{\dagger}
\overline{\sigma}^{\mu}\partial _{\mu}\psi_{_{\rm L}}-
i\psi _{_{\rm R}}^{\dagger}
{\sigma}^{\mu}\partial _{\mu}\psi_{_{\rm R}},
\end{equation}
where we have defined the covariant Pauli matrices $\sigma
^{\mu}=\left(\mathbb{I}_2,\sigma ^i\right)$ and $\overline \sigma
^{\mu}=\left(\mathbb{I}_2,-\sigma ^i\right)$.

\par

After these brief reminders, we now consider the first family of the
standard model leptons (the other two families can be treated in the
same fashion and the quarks must be considered separately), \ie the
electron and the neutrino. Following the usual convention, we denote
by $e_{_{\rm R}}$ the right-handed electron Weyl spinor and by
$e_{_{\rm L}}$ the left-handed one. With regards to the neutrino, only
the left-handed particle is present in the standard model and will be
denoted as $\nu _{_{\rm L}}$. Then, we have the following
properties. The particles $e_{_{\rm L}}$ and $\nu _{_{\rm L}}$ are
charged under ${\rm U}(1)_{_{\rm Y}}$ but also under ${\rm
  SU}(2)_{_{\rm L}}$, \ie they are considered as a complex doublet of
two-component spinors
\begin{equation}
L_e = \begin{pmatrix} \nu _{_{\rm L}}\cr e_{_{\rm L}} \end{pmatrix}, 
\end{equation}
On the other hand, $e_{_{\rm R}}$ is charged under ${\rm U}(1)_{_{\rm
    Y}}$ but is postulated to be a ${\rm SU}(2)_{_{\rm L}}$
singlet. Therefore, concretely, the Lagrangian can be written as
\begin{eqnarray}
{\cal L}_{\rm Leptons}&=&-iL_e^{\dagger}\overline{\sigma}^{\mu }
\left(\partial _{\mu}+i\frac{g'}{2}Y_{_{\rm L}}B_{\mu}+igT_aW_{\mu}^a\right)L_e
\nonumber \\ & &
-ie_{_{\rm R}}^{\dagger}\bar{\sigma}^{\mu }
\left(\partial _{\mu}+i\frac{g'}{2}Y_{_{\rm R}}B_{\mu}\right)e_{_{\rm R}},
\end{eqnarray}
where $L_e^{\dagger}$ means $\left(\nu_{_{\rm L}}^{\dagger}, e_{_{\rm
      L}}^{\dagger}\right)$. Notice the factor $1/2$ in front of the
gauge boson $B_{\mu}$ which is the usual convention [this factor was
already present in Eq.~(\ref{eq:covderive})]. The quantities $Y_{_{\rm
    L}}$ and $Y_{_{\rm R}}$ are the leptonic weak hyper-charges. Let us
focus on the first term in the above expression. The term between the
parenthesis is a two by two matrix that can be written in terms of the
physical gauge bosons, using the
definitions~(\ref{eq:defgaugebosons}). Similarly, the second term can
be expressed in terms of the fields $A_{\mu}$ and $Z_{\mu}$.
Then, the Lagrangian for the leptons can be rewritten as
\begin{widetext}
\begin{eqnarray}
\label{eq:lagrangianleptons}
{\cal L}_{\rm Leptons} &=& -i\nu_{_{\rm L}}^{\dagger}\overline{\sigma }_{\mu}
\left[\partial _\mu -i\left(\frac{g'}{2}Y_{_{\rm L}}\sin \theta -\frac{g}{2}\cos
\theta\right)Z_{\mu} +i\left(\frac{g'}{2}Y_{_{\rm L}}\cos \theta 
+\frac{g}{2}\sin \theta \right)A_{\mu }\right]\nu_{_{\rm L}}
\nonumber \\ & & 
-ie_{_{\rm L}}^{\dagger}\overline{\sigma }_{\mu}
\left[\partial _\mu -i\left(\frac{g'}{2}Y_{_{\rm L}}\sin \theta +\frac{g}{2}\cos
\theta\right)Z_{\mu} +i\left(\frac{g'}{2}Y_{_{\rm L}}\cos \theta 
-\frac{g}{2}\sin \theta \right)A_{\mu }\right]e_{_{\rm L}}
\nonumber \\ & &
+\frac{g}{\sqrt{2}}\nu_{_{\rm L}}^{\dagger}\overline{\sigma}^{\mu}W_{\mu}^-
e_{_{\rm L}}
+\frac{g}{\sqrt{2}}e_{_{\rm L}}^{\dagger}\overline{\sigma}^{\mu}W_{\mu}^+
\nu_{_{\rm L}}
-ie_{_{\rm R}}^{\dagger}\overline{\sigma}^{\mu }
\left(\partial _{\mu}-i\frac{g'}{2}Y_{_{\rm R}}\sin \theta Z_{\mu}
+i\frac{g'}{2}Y_{_{\rm R}}\cos \theta A_{\mu}\right)e_{_{\rm R}}.
\end{eqnarray}
\end{widetext}
Therefore, the total Lagrangian is now given by ${\cal L}={\cal
  L}_{\rm Higgs} +{\cal L}_{\rm K-Gauge}+{\cal L}_{\rm Leptons}$. This
Lagrangian depends on various free parameters that we now discuss and
choose. Firstly, the neutrino is not electromagnetically charged and
from Eq.~(\ref{eq:lagrangianleptons}) one has that
\begin{equation}
\label{eq:Yrel1}
g'Y_{_{\rm L}}\cos \theta +g\sin \theta =0.
\end{equation}
Secondly, the electron carries the charge $e$. This means that one
must recover the following Lagrangian,
\begin{equation}
{\cal L}_{_{\rm EM}}=-i\overline \Psi \gamma ^{\mu}
\left(\partial _{\mu}+iQA_{\mu }\right)\Psi,
\end{equation}
where $Q$ is the electromagnetic charge, $Q=e$ in our case. Working
out the interaction part of the above Lagrangian, one obtains
\begin{equation}
{\cal L}_{_{\rm EM}}^{\rm int}=
-Qe_{_{\rm L}}^{\dagger}\overline \sigma_{\mu}A_{\mu }e_{_{\rm L}}
-Qe_{_{\rm R}}^{\dagger}\sigma_{\mu}A_{\mu }e_{_{\rm R}},
\end{equation}
and, looking at Eq.~(\ref{eq:lagrangianleptons}), one deduces that
\begin{eqnarray}
\label{eq:Yrel2}
\frac{g'}{2}Y_{_{\rm L}}\cos \theta -\frac{g}{2}\sin \theta 
&=&\frac{g'}{2}Y_{_{\rm R}}\cos \theta=e.
\end{eqnarray}
As a consequence, comparing Eq.~(\ref{eq:Yrel1}) and~(\ref{eq:Yrel2}),
one can establish that $ 2Y_{_{\rm L}} = Y_{_{\rm R}}$. It is
conventional to choose $Y_{_{\rm L}}=-1$ and therefore $Y_{_{\rm
    R}}=-2$. Then, Eq.~(\ref{eq:Yrel1}) leads to $ g' = g\tan \theta
$, from which one can write $\cos \theta =g/\sqrt{g^2+g'^2}$ and $\sin
\theta =g'/\sqrt{g^2+g'^2}$. Finally, Eq.~(\ref{eq:Yrel2}) implies
that $g\sin \theta =e$. Everything can be summarized by the Gell-Mann
formula
\begin{equation}
Q=e\left(I_3+\frac{Y}{2}\right),
\end{equation}
if the leptonic isospin doublet is assigned a ``weak isospin'' $I=1/2$
such that the neutrino has a third component $I_3=1/2$ while the
electron is such that $I_3=-1/2$. Of course the right handed particles
are such that $I=0$.  Returning to
Eq.~(\ref{eq:higgslagrangianexplicit}) and using the above equations,
we see that the photons remains mass-less if $Y_{_{\rm H}}=1$. This
also allows us to determine the mass of the gauge bosons and one
arrives at
\begin{eqnarray}
m_{W^{\pm}}^2&=&\frac{g^2v^2}{2},\\
m_Z^2 &=&  2v^2\left(\frac{g}{2}\cos \theta +\frac{g'}{2}\sin \theta \right)^2
=\frac{m_{W^{\pm}}^2}{\cos ^2 \theta}.
\end{eqnarray}
One also easily proves that the coefficient of the cross-term
$Z_{\mu}A^{\mu}$ vanishes. Finally, looking at
Eq.~(\ref{eq:higgspot}), one establishes that the Higgs mass is
given by
\begin{equation}
m_{_{\rm H}}^2=\lambda v^2,
\end{equation}
and, from the expression of the potential ~(\ref{eq:higgspot}),
one finds the expression of its value at the minimum
\begin{equation}
  V(\langle \Sigma \rangle=v)=-\lambda\frac{v^4}{4}
=-\frac{m_{_{\rm H}}^4}{4\lambda},
\end{equation}
in agreement with what we had before. Using the expression of the
Higgs mass, one can re-write this expression as
\begin{equation}
\label{eq:vacew}
  \rho_{_{\rm vac}}=V(\langle \Sigma \rangle=v)=-\frac14 m_{_{\rm H}}^2v^2.
\end{equation}
We see that in order to calculate the vacuum energy, we need to know
$v$ and the Higgs mass. Let us first discuss how the vacuum
expectation value of the Higgs can be obtained. For this purpose, we
focus on the term in the electroweak Lagrangian describing the
interaction between leptons and charged gauge bosons (the so-called
charged currents). From the above considerations, see
Eq.~(\ref{eq:lagrangianleptons}), they are given by
\begin{equation}
{\cal L}_{\rm int L-W}=\frac{g}{\sqrt{2}}\nu_{_{\rm L}}^{\dagger}
\overline{\sigma}^{\mu}W_{\mu}^-
e_{_{\rm L}}
+\frac{g}{\sqrt{2}}e_{_{\rm L}}^{\dagger}\overline{\sigma}^{\mu}W_{\mu}^+
\nu_{_{\rm L}}.
\end{equation}
It is easy to verify by an explicit calculation that it can be
re-expressed in terms of Dirac spinors\footnote{Let us recall that
  they are defined by the following expression
\begin{equation}
\Psi_e=
\begin{pmatrix} e_{_{\rm L}}\cr e_{_{\rm R}} \end{pmatrix},
\end{equation}
and a similar expression for $\Psi_{\nu}$.}as
\begin{eqnarray}
{\cal L}_{\rm int L-W}&=& \frac{g}{\sqrt{2}}\overline{\Psi}_{\nu}
\gamma^{\mu}\frac{1-\gamma_5}{2}W_{\mu}^-
\Psi_{e} \nonumber \\
&+&\frac{g}{\sqrt{2}}\overline{\Psi}_{e}
\gamma^{\mu}\frac{1-\gamma_5}{2}W_{\mu}^+
\Psi_{\nu},
\end{eqnarray}
which is similar to the Fermi theory for the weak interaction described 
by the following Lagrangian
\begin{eqnarray}
{\cal L}_{\rm Fermi}=\frac{G_{\rm F}}{\sqrt{2}}J_{\lambda}J^{\lambda \dagger},
\end{eqnarray}
with $J_{\lambda}=J_{\lambda}^{(e)}+J_{\lambda}^{(\mu)}$ (\ie an
electronic and a muonic part) where the current is given by
$J_{\lambda}^{(e)}=\overline{\Psi}_e\gamma_{\lambda}(1-\gamma_5)\Psi_{\nu_e}$
and a similar definition for $J_{\lambda}^{(\nu)}$. If, for instance
we consider the decay of muon into an electron, a muonic neutrino and
an anti electronic neutrino, described by the graph
\begin{equation}
\vspace{0.3cm}
\begin{fmffile}{muon}
\parbox{35mm}{
\begin{fmfgraph*}(60,30)
\fmfleft{i}
\fmfright{o1,o2,o3}
\fmf{fermion}{i,v1}
\fmf{fermion}{v1,o1}
\fmf{boson,tension=1.5}{v1,v2}
\fmf{fermion,tension=1.5}{v2,o2}
\fmf{fermion,tension=1.5}{v2,o3}
\fmflabel{$\mu^-$}{i}
\fmflabel{$\nu_{\mu}$}{o3}
\fmflabel{$\overline{\nu}_e$}{o1}
\fmflabel{$e^-$}{o2}
\end{fmfgraph*}
}
\end{fmffile}
\vspace{0.3cm}
\end{equation}
the two theories will lead to the same leading contribution at small
energies if the following identification is made
\begin{equation}
\frac{g^2}{8m_W^2}=\frac{G_{\rm F}}{\sqrt{2}}.
\end{equation}
Indeed, the appearance of the gauge boson mass comes from the
propagators of $W_{\mu}$. At small momenta compared to the mass, only
$m_W$ remains and the above graph is in fact equivalent to the graph
corresponding to the Fermi theory. Then, using the expression of the
$W$-bosons mass established before, $m_W^2=g^2v^2/2$, one arrives at
\begin{equation}
v^2=\frac{\sqrt{2}}{4G_{\rm F}^2}.
\end{equation}
Notice that there is a factor $\sqrt{2}$ difference with respect to
the textbook~\cite{LeBellac:1991cq} because we have defined the Higgs
as $v+H/\sqrt{2}$ instead of $(v+H)/\sqrt{2}$. The Fermi constant is
given by $G_{\rm F}\simeq 1.16\times 10^{-5}\,
\left(\mbox{GeV}\right)^{-2}$. Therefore,
\begin{equation}
v\simeq 175 \, \mbox{GeV}.
\end{equation}
or $\sqrt{2}v\simeq 246\, \mbox{GeV}$. We see that we have been able
to calculate the Higgs vacuum expectation value $v$. As a consequence,
Eq.~(\ref{eq:vacew}) reads
\begin{equation}
\rho_{_{\rm vac}}=-\frac{\sqrt{2}}{16}\frac{m_{_{\rm H}}^2}{G_{\rm F}^2}.
\end{equation}
The mass of the Higgs boson is not known although, at the time of
writing, there are reasons to believe that $m_{_{\rm H}}<129 \,
\mbox{GeV}$ at $95\%$ CL~\cite{Chatrchyan:2012tx} (there are even
reasons to believe that $m_{_{\rm H}}\simeq 125 \, \mbox{GeV}$ but
this is not yet established at a sufficient statistical level at the
time of writing). Therefore, we find that
\begin{equation}
\rho_{_{\rm vac}}^{_{\rm EW}}\simeq -1.2 \times 10^8 \, \mbox{GeV}^4
\simeq -10^{55}\rho_{\rm cri},
\end{equation}
where $\rho_{\rm cri}\simeq 10^{-47}\, \mbox{GeV}^4$ is the critical
energy density today. As will be discussed in the following, this
result is in fact in contradiction with various observations.

\par

To conclude this section, let us recall that one can always adjust the
vacuum energy today to zero by tuning the parameter $V_0$. This is
clearly not a very satisfactory method and this implies that the
vacuum energy density was huge prior to the electroweak phase
transition. It is also worth noticing that this problem is not
specific to the electroweak transition. Exactly the same discussion
could be presented for the Quantum Chromo Dynamics (QCD)
transition~\cite{Dolgov:1997za,Rugh:2000ji} (for instance) and would
lead to the same problem since
\begin{equation}
\rho_{_{\rm vac}}^{_{\rm QCD}}\simeq 10^{-2}\, \mbox{GeV}^4
\simeq 10^{45}\rho_{\rm cri}.
\end{equation}
Moreover, as we are now going to discuss in detail, even if we
``solve'' the classical cosmological constant by tuning the vacuum
energy to zero, see Eq.~(\ref{eq:classicalvac}), the problem reappears
at the quantum level. Indeed, we have seen that the zero-point quantum
fluctuations also give a contribution to the cosmological constant,
see Eq.~(\ref{eq:qmvac}). The goal of the next section is to estimate
the corresponding energy density.

\section{The Quantum-Mechanical Cosmological 
Constant Problem}
\label{sec:calculatecc}

We have seen that, even if we solve the classical cosmological
constant problem and find a convincing reason to put the minimum of
the potential to zero, $\rho_{_{\rm vac}}$ is a quantity which also
receives contributions from the zero-point fluctuations of all the
quantum fields present in the Universe. We now discuss this
``quantum-mechanical'' cosmological constant problem. In principle,
the corresponding vacuum energy density can be estimated from first
principles. The goal of this section (and of Sec.~\ref{sec:GEP}) is to
carry out this task. We will first discuss how the Lorentz invariance
affects the calculation of the vacuum energy, see the next
subsection~\ref{subsec:zeroenergy}. Then, in
sub-Sec.~\ref{subsec:feynman}, we will show how this question can be
formulated in terms of Feynman diagrams. In
sub-Sec.~\ref{subsec:interactingvacuum}, we will also evaluate the
vacuum energy not only for a free theory as usually done but also in
the more realistic case where interactions are present. Finally, in
order to understand better the origin of the vacuum energy density, in
sub-Secs.~\ref{subsec:qmqft} and~\ref{subsec:pertqm}, we will address
the same question but in the context of ordinary quantum
mechanics. This will allow us in sub-Sec.~\ref{subsec:bubbleqm}
and~\ref{subsec:vacqm} to discuss in more detail the properties of the
so-called bubble diagrams from which the cosmological constant
originates. As mentioned above, all the fields present in the universe
participate in the vacuum energy. Therefore, we need to do the
calculation for scalar, fermion and vector fields. Here, in a first
step, we carry out the calculation for a real scalar field. The other
types will be treated later on in Sec.~\ref{sec:otherfield}. For
simplicity, we will also consider that the metric is flat and that the
fields live in Minkowski space-time. A sloppy justification is that
the main contribution to the vacuum energy density comes from modes
with very high momenta, corresponding to scales at which the curvature
of space-time is negligible. In fact, we will come back to this very
important question in more detail in what follows and will try to
discuss this point in a more rigorous way, see
Sec.~\ref{sec:vaccurve}.

\subsection{The Zero-Point Energy Density}
\label{subsec:zeroenergy}

Let us consider a simple real free scalar field with the potential
$V(\Phi)=m^2\Phi^2/2$ where $m$ is the mass of the scalar particle. In
flat space-time, the equation of motion is nothing but the
Klein-Gordon equation, namely
\begin{equation}
\label{eq:kgequation}
-\ddot{\Phi}+\delta ^{ij}\partial _i\partial _j\Phi-m^2\Phi=0.
\end{equation}
Since the scalar field is free, the equation of motion is linear and,
as a consequence, one can Fourier expand $\Phi(t,{\bm x})$ as
\begin{eqnarray}
\label{eq:fourierfield}
\Phi\left(t,{\bm x}\right)&=&\frac{1}{\left(2\pi\right)^{3/2}}
\int \frac{{\rm d}^3{\bm k}}{\sqrt{2\omega(k)}}\biggl(c_{\bm k}
{\rm e}^{-i\omega t +i{\bm k}\cdot {\bm x}}
\nonumber \\ 
& & +
c_{\bm k}^{\dagger}{\rm e}^{i\omega t -i{\bm k}\cdot {\bm x}}
\biggr),
\end{eqnarray}
with 
\begin{equation}
\label{eq:defomegasf}
  \omega (k)\equiv \sqrt{k^2+m^2}.
\end{equation}
The four-dimensional momentum has been written as
$k^{\mu}=\left(k^0,{\bm k}\right)$ and the integration is performed
over its spatial part only. We have used the notation $\vert {\bm
  k}\vert =k$. Its time component is given by $k^0=\omega$. As usual,
the field has been quantized by considering $c_{\bm k}$ and $c_{\bm
  k}^{\dagger}$ (the so-called annihilation and creation operators) as
quantum operators satisfying the following commutation relation
\begin{equation}
\label{eq:comsf}
\left[c_{\bm k},c^{\dagger}_{\bm k'}\right]=\delta ^{(3)}\left(
{\bm k}-{\bm k'}\right).
\end{equation}
This equation is in fact equivalent to the standard commutation rule
between the field operator and its conjugate momentum.

\par

We are now in a position where one can compute $\langle 0\vert T_{\mu
  \nu}\vert 0\rangle$. From Eq.~(\ref{eq:tmunuscalarfield}), we see
that we must first evaluate the mean values of various quantities
depending on the field operator and its derivatives. Straightforward
calculations lead to
\begin{eqnarray}
\label{eq:phidot2}
\langle 0\vert \dot{\Phi}^2\vert 0\rangle
&=&\frac{1}{(2\pi)^3}\int 
\frac{{\rm d}^3{\bm k}}{2\omega(k)}\, \omega ^2(k),
\\
\label{eq:partialphi2}
\langle 0\vert \delta ^{ij}\partial _i\Phi\partial _j\Phi \vert 0\rangle
&=&\frac{1}{(2\pi)^3}\int \frac{{\rm d}^3{\bm k}}{2\omega(k)}\, {\bm k}^2,
\\
\label{eq:phi2}
\langle 0\vert \Phi^2\vert 0\rangle
&=&\frac{1}{(2\pi)^3}\int \frac{{\rm d}^3{\bm k}}{2\omega(k)}.
\end{eqnarray}
These equations can then be used to determine the energy density and
pressure of the vacuum. From the above
expression~(\ref{eq:tmunuscalarfield}) of the stress-energy tensor,
one has
\begin{equation}
\label{eq:hamiltonscalr}
T_{00}={\cal H}=\frac12 \dot{\Phi}^2+\frac12 
\delta ^{ij}\partial _i\Phi\partial _j\Phi
+\frac12 m^2\Phi^2,
\end{equation}
where ${\cal H}$ denotes the Hamiltonian density. This leads to the
following expression for the energy density $\rho=u^{\mu}u^{\nu}T_{\mu
  \nu}$ measured by a fundamental observer characterized by its
four-velocity vector $u^{\mu}=(1,{\mathbf 0})$ 
\begin{equation}
\label{eq:rhovac}
\langle \rho\rangle =\langle 0\vert u^0u^0 T_{0 0}\vert 0\rangle
= \frac{1}{\left(2\pi\right)^3}
\frac12 \int {\rm d}^3{\bm k}\, \omega (k).
\end{equation}
In the same manner, one can easily estimate the pressure. One obtains
\begin{eqnarray}
\langle p\rangle &=& \left \langle 0\left\vert 
\frac13 \perp ^{\mu \nu}T_{\mu \nu} \right \vert 0\right \rangle
\nonumber \\
&=& \left \langle 0\left\vert 
\frac12 \dot{\Phi}^2-\frac16 \delta ^{ij}\partial _i\Phi\partial _j\Phi
-\frac12 m^2\Phi^2\right \vert 0\right \rangle,
\end{eqnarray}
where $\perp^{\mu \nu}\equiv g^{\mu \nu}+u^{\mu }u^{\nu}$ is a
projector. Upon using the expressions~(\ref{eq:phidot2}),
(\ref{eq:partialphi2}) and~(\ref{eq:phi2}), one arrives at the
following expression
\begin{equation}
\label{eq:pressurevac}
\langle p\rangle = \frac{1}{\left(2\pi\right)^3}
\frac16 \int {\rm d}^3{\bm k}\, \frac{k^2}{\omega (k)}.
\end{equation}
Notice that we also have 
\begin{equation}
\langle 0\vert T_{0 i}\vert 0\rangle= -\frac{1}{\left(2\pi\right)^3}
\frac12 \int {\rm d}^3{\bm k}\, k_i=0,
\end{equation}
and, hence, we do not need to consider to off-diagonal elements. This
completes our calculation of the vacuum stress-energy tensor.

\par

From the above considerations, using Eq.~(\ref{eq:deflambdaeff}), we
deduce that
\begin{equation}
\Lambda_{_{\rm eff}}=\Lambda_{_{\rm B}}+\frac{\kappa}{\left(2\pi\right)^3}
\int {\rm d}^3{\bm k}\, \frac12\omega (k).
\end{equation}
The obvious problem with the previous calculation is that the
integrals expressing the energy density $\langle \rho \rangle$ and the
pressure $\langle p\rangle$ blow up in the ultra-violet regime (and,
therefore, strictly speaking, the effective cosmological constant is
in fact infinite). This is of course a well-known problem in quantum
field theory. Usually, this type of divergences is ignored (or
sometimes removed by mean of the normal ordering product) because, in
absence of gravity, only differences of energy can be
detected~\cite{Peskin:1995ev,Greiner:1996zu}. However, in the present
context, this is clearly no longer the case since the absolute value
of the vacuum energy weighs and, therefore, can be measured. This is
why regularizing these infinities becomes a crucial issue that we now
discuss in some details.

\par

The common method found in the literature is to introduce a cut-off at
$k=M$, where the physical interpretation of $M$ is that this is the
scale at which the effective theory used before breaks
down~\cite{Carroll:1991mt,
  Sahni:1999gb,Straumann:1999ia,Carroll:2000fy,Padmanabhan:2002ji}. The
quantity $M$ is not known: it could be the Planck scale, the string
scale or even the super-symmetric breaking scale. As we will see later
on, this is in fact not important because the cosmological constant
problem is present regardless of the precise value of the cut-off. In
this approach, the energy density becomes
\begin{eqnarray}
\langle \rho \rangle &=& \frac{1}{4\pi ^2}\int _0^{M} {\rm d}k 
k^2 \sqrt{k^2+m^2}
\\
&=& \frac{M^4}{16 \pi^2}\Biggl[\sqrt{1+\frac{m^2}{M^2}}
\left(1+\frac12\frac{m^2}{M^2}\right)
\nonumber \\ & &
-\frac12\frac{m^4}{M^4}\ln \left(
\frac{M}{m}+
\frac{M}{m}\sqrt{1+\frac{m^2}{M^2}}\right)
\Biggr]
\\
\label{eq:rhocut}
&=& \frac{M^4}{16 \pi^2}\left(1+\frac{m^2}{M^2}
+\cdots \right),
\end{eqnarray}
where, in the last expression, we have expanded the exact expression
in terms of the small parameter $m/M$. We see that the divergence is
quartic in the cut-off scale. Let us now perform the same calculation
for the pressure. Using Eq.~(\ref{eq:pressurevac}), one obtains
\begin{eqnarray}
\langle p \rangle &=& \frac13 \frac{1}{4\pi ^2}\int _0^{M} {\rm d}k 
\frac{k^4}{\sqrt{k^2+m^2}}
\\
&=& \frac13\frac{M^4}{16 \pi^2}
\Biggl[\sqrt{1+\frac{m^2}{M^2}}
\left(1-\frac32\frac{m^2}{M^2}\right)
\nonumber \\ & &
+\frac32\frac{m^4}{M^4}\ln \left(
\frac{M}{m}+
\frac{M}{m}\sqrt{1+\frac{m^2}{M^2}}\right)
\Biggr]
\\
\label{eq:pcut}
&=& \frac13 \frac{M^4}{16 \pi^2}\left(1-\frac{m^2}{M^2}
+\cdots \right).
\end{eqnarray}
It is clear from the previous expressions that $\langle p \rangle
/\langle \rho \rangle\neq -1$ which indicates that the stress energy
tensor is not of the form $\propto -\rho g_{\mu \nu}$. In the limit
$m\rightarrow 0$, as can be easily shown from Eqs.~(\ref{eq:rhocut})
and~(\ref{eq:pcut}), the equation of state is in fact $\langle p
\rangle /\langle \rho \rangle=1/3$. This would mean that the zero
point fluctuations do not behave like a cosmological constant but
rather like radiation. On the other hand, this result seems to be
strange since we know from fundamental considerations that the
stress-energy tensor of a system in its ground state must be
proportional to the metric tensor. Clearly, something has gone wrong
is our calculation. 

\par

Interestingly enough, let us also note in passing that if one only
focuses on the logarithmic terms, namely
\begin{eqnarray}
\label{eq:rholog}
\langle \rho \rangle ^{\rm log}&=& -\frac{m^4}{32\pi ^2}\ln \left(
\frac{M}{m}+
\frac{M}{m}\sqrt{1+\frac{m^2}{M^2}}\right),
\\
\label{eq:plog}
\langle p \rangle ^{\rm log}&=& \frac{m^4}{32\pi ^2}\ln \left(
\frac{M}{m}+
\frac{M}{m}\sqrt{1+\frac{m^2}{M^2}}\right),
\end{eqnarray}
then the equation of state of the vacuum is correctly reproduced (even
if the energy density can be negative). In fact, as we explain below,
this is not a coincidence~\cite{Akhmedov:2002ts}.

\par

It is easy to check that the previous considerations are not an
artifact of the fact that we have introduced a sharp cut-off. Indeed,
if rather than a sharp cut-off in the upper limit of the integrals,
one introduces a smooth exponential cut-off, one obtains (for
simplicity, one only considers the case $m=0$),
\begin{eqnarray}
\langle \rho \rangle &=& \frac{1}{4\pi ^2}\int _0^{\infty} {\rm d}k 
\, k^3\, {\rm e}^{-\alpha k}=\frac{1}{4\pi ^2}\frac{\Gamma(4)}{\alpha ^4}
\\ &=&
\label{eq:rhoexpcut}
\frac{1}{4\pi ^2}6 M^4,
\end{eqnarray}
where we have written $\alpha =1/M$ and where $\Gamma(z)$ is the
Euler's integral of the second
kind~\cite{Abramovitz:1970aa,Gradshteyn:1965aa}. One recovers our
familiar quartic divergence. Doing the same manipulation for the
pressure, one obtains again that $\langle p \rangle /\langle \rho
\rangle=1/3$.

\par

In fact, what has gone wrong is that we have used schemes of
regularization that do not respect the symmetry of our underlying
theory~\cite{Akhmedov:2002ts,Ossola:2003ku,Koksma:2011cq}. Indeed, the
Lorentz invariance is broken when we impose a cut-off on the spatial
momentum only. As exemplified by Eq.~(\ref{eq:rhoexpcut}), it is easy
to show that this is a feature of any regularization scheme breaking
Lorentz invariance and not a specific property of the particular
cut-off scheme used above (and usually used in the literature). In
fact, this property is well-known in the context of gauge theories. If
one regulates divergent graphs with a method that is not gauge
invariant, then one obtains incorrect
results~\cite{Rosenberg:1962pp}. Here, the situation is very similar.

\par

Therefore, in order to meaningfully evaluate the zero-point energy
density, one must use a regularization scheme that does not break
Lorentz
invariance~\cite{Akhmedov:2002ts,Ossola:2003ku,Koksma:2011cq}. An
obvious choice is dimensional regularization and we now explore this
route. In order to use this method, we first reformulate our problem
in a $d$ dimensional Minkowski space-time. Obviously, the Klein-Gordon
equation~(\ref{eq:kgequation}) remains unchanged except that the
indices of the Kr\"onecker symbol now run from $0$ to $d-1$. The
Fourier expansion of the field becomes [compare with
Eq.~(\ref{eq:fourierfield})]
\begin{eqnarray}
\Phi\left(t,{\bm x}\right)&=&\frac{1}{\left(2\pi\right)^{(d-1)/2}}
\int \frac{{\rm d}^{d-1}{\bm k}}{\sqrt{2\omega(k)}}\biggl(c_{\bm k}
{\rm e}^{-i\omega t +i{\bm k}\cdot {\bm x}}
\nonumber \\ 
& & +
c_{\bm k}^{\dagger}{\rm e}^{i\omega t -i{\bm k}\cdot {\bm x}}
\biggr),
\end{eqnarray}
and, as a consequence, the energy density can be expressed as
\begin{eqnarray}
\label{eq:rhodim}
\langle \rho\rangle 
&=& \frac{\mu^{4-d}}{\left(2\pi\right)^{(d-1)}}
\frac12 \int {\rm d}^{d-1}{\bm k}\, \omega (k)
\\
&=&
\frac{\mu^{4-d}}{\left(2\pi\right)^{(d-1)}}
\frac12 \int _0^{\infty}
{\rm d}k k^{d-2}{\rm d}^{d-2}\Omega\, \omega(k),
\end{eqnarray}
where we have introduced a scale $\mu$ in order for the equation to be
dimensionally correct. This equation should be compared with
Eq.~(\ref{eq:rhovac}). Then, using the fact that the angular integrals
can be easily performed, namely $\int {\rm
  d}^{d-2}\Omega=2\pi^{(d-1)/2}/\Gamma(d/2-1/2)$, where, again,
$\Gamma(z)$ is the Euler's integral of the second
kind~\cite{Abramovitz:1970aa,Gradshteyn:1965aa}, one obtains
\begin{eqnarray}
\label{eq:rhodimresult}
\langle \rho\rangle 
&=& \frac{\mu^4}{2\left(4\pi\right)^{(d-1)/2}}
\frac{\Gamma(-d/2)}{\Gamma(-1/2)}\left(\frac{m}{\mu}\right)^d.
\end{eqnarray}
Performing the same calculation for the pressure, one arrives at a
similar expression
\begin{eqnarray}
\label{eq:pdim}
\langle p \rangle 
&=& 
\frac{\mu^{4-d}}{\left(2\pi\right)^{(d-1)}}
\frac{1}{2(d-1)} \int {\rm d}^{d-1}{\bm k}\, \frac{k^2}{\omega (k)}\\
&=& \frac{\mu^4}{4\left(4\pi\right)^{(d-1)/2}}
\frac{\Gamma(-d/2)}{\Gamma(1/2)}\left(\frac{m}{\mu}\right)^d.
\end{eqnarray}
But $\Gamma(-1/2)=-2\Gamma(1/2)$ and, as a consequence,
\begin{equation}
\langle p \rangle= -\langle \rho\rangle.
\end{equation}
This time, as expected, we have obtained the equation of state of the
vacuum. The fact that the regularization method used satisfies the
Lorentz symmetry has led us to the correct result. 

\par

The consistency of the above result can be checked in a different
manner~\cite{Akhmedov:2002ts}. Using Eqs.~(\ref{eq:rhodim})
and~(\ref{eq:pdim}) for the energy density and the pressure, one sees
that they satisfy the following differential equation
\begin{equation}
\langle \rho\rangle -2m^2 \frac{{\rm d}\langle \rho\rangle}{{\rm d}m^2}
=(d-1)\langle p\rangle.
\end{equation}
Using the fact that $\langle p\rangle =-\langle \rho \rangle$ one obtains
\begin{equation}
m^2 \frac{{\rm d}\langle \rho\rangle}{{\rm d}m^2}
=\frac{d}{2}\langle \rho \rangle,
\end{equation}
which can be integrated to give $\langle \rho \rangle \propto m^d$ in
full agreement with Eq.~(\ref{eq:rhodimresult}). Therefore, we
conclude that the above considerations are all consistent.

\par

Let us now analyze in more details the structure of the
divergences. For this purpose one writes $d=4-\epsilon $, where
$\epsilon$ is supposed to be a small parameter, and we expand $\langle
\rho\rangle$ in terms of $\epsilon $. The Euler function is such that
\begin{equation}
\Gamma\left(-2+\frac{\epsilon}{2}\right)=\frac{1}{-2+\epsilon/2}
\frac{1}{-1+\epsilon/2}\frac{1}{\epsilon/2}\Gamma\left(1+\frac{\epsilon}{2}
\right),
\end{equation}
with $\Gamma(1+\epsilon/2)\simeq \Gamma(1)+\epsilon \Gamma'(1)/2$,
$\Gamma'(1)=\Psi(1)\Gamma(1)$ where $\Psi$ is the Polygamma
function~\cite{Abramovitz:1970aa,Gradshteyn:1965aa} and
$\Psi(1)=-\gamma\simeq 0.57772$ the Euler-Mascheroni
constant~\cite{Abramovitz:1970aa,Gradshteyn:1965aa}.  If, in addition,
we use the two following expansions
\begin{eqnarray}
\left(4\pi\right)^{-3/2+\epsilon/2}& \simeq &\frac{1}{(4\pi)^{3/2}}
\left[1+\frac{\epsilon}{2}\ln \left(4 \pi\right)\right],\\
\left(\frac{m}{\mu}\right)^{4-\epsilon}&\simeq &
\left(\frac{m}{\mu}\right)^4\left(1-\epsilon \frac{m}{\mu}\right),
\end{eqnarray}
then one arrives at the following expression for the energy density of
the vacuum
\begin{equation}
\label{eq:rhovaceps}
\langle \rho\rangle \simeq -\frac{m^4}{64 \pi^2}
\left[\frac{2}{\epsilon}+\frac32-\gamma -\ln 
\left(\frac{m^2}{4\pi \mu^2}\right)\right]+\cdots .
\end{equation}
Using a $\overline{\mbox{MS}}$ renormalization scheme convention
[$\overline{\mbox{MS}}$ means ``modified minimal subtraction'' and
simply consists in subtracting the pole $\sim 1/\epsilon$ together
with the accompanying terms $\gamma $ and $\ln (4\pi)$, see chapter~11
of Ref.~\cite{Peskin:1995ev}; in fact, here, we slightly modify the
convention such that the term $3/2$ is subtracted as well], the
regularized, finite, energy density of the vacuum now reads
\begin{equation}
\label{eq:rhovacrenorm}
\langle \rho\rangle =\frac{m^4}{64 \pi^2}
\ln \left(\frac{m^2}{\mu^2}\right).
\end{equation}
This result is very different from the result obtained by imposing a
sharp cut-off. In fact, one recovers the structures of the logarithm
terms found in Eqs.~(\ref{eq:rholog}) and~(\ref{eq:plog}) since it is
well-known that the dimensional regularization scheme removes the
power law terms. As discussed around Eqs.~(\ref{eq:rholog})
and~(\ref{eq:plog}), this is the reason why those terms lead to the
correct vacuum equation of state. We see that the result is not
proportional to the cut-off scale to the power four, as usually
claimed on dimensional grounds, but to the mass of the particle to the
power four, which is also dimensionally correct. Physically, for
instance, this means that the photon does not contribute to the vacuum
energy density contrary to what a sharp cut-off calculation would
predict. Therefore, the order of magnitude of the final result can a
priori be very different. We also notice that the sign of $\langle
\rho\rangle$ can change according to $m>\mu$ or $m<\mu$. Of course, it
remains to be seen whether this could solve the cosmological constant
problem and we will argue below that, unfortunately, the answer to
this question is no. It should also be kept in mind that the above
calculation is valid for free fields. It is clear that a more
realistic version should take into account the interactions and we
will also consider this issue in the following. However, before
addressing these various points and in order to gain further physical
insights, we would like to investigate again the above issues, this
time from the Feynman diagrams point of view. We turn to this question
in the next sub-section.

\subsection{The Vacuum and the Feynman Diagrams}
\label{subsec:feynman}

We have seen in the last sub-section that the vacuum contribution to
the cosmological constant is given in terms of divergent integrals. A
priori, there is no reason to be surprised since divergent integrals
are routinely found in quantum field theory. We have the method of
renormalization at our disposal to get rid of these infinities and the
so accurate predictions of particle physics are here to remind us that
this is a very convincing way of taming these divergences. However, we
will see that the divergent integrals that we encounter when we
calculate the vacuum energy are in fact of a different nature than the
divergent integrals met when we calculate, say, a cross section. In
order to discuss this point in more detail and to understand better
the structure of these divergences, we now briefly return to some
textbook
considerations~\cite{Peskin:1995ev,Greiner:1996zu,Itzykson:1980rh,
  Bailin:1994qt,LeBellac:1991cq,Ryder:1985wq,Mandl:1985bg}.

\par

In quantum field theory, the basic object is the generating functional
defined by the following path integral (we still restrict our
considerations to a simple scalar field)
\begin{equation}
\label{eq:defZ}
Z\left[J\right]={\cal N}\int {\cal D}\Phi \exp\left\{i\int {\rm d}^4x
\left[{\cal L}+J(x)\Phi(x)\right]\right\},
\end{equation}
where ${\cal N}$ is a number chosen such that $Z[J=0]=1$. The quantity
${\cal L}$ is the Lagrangian density while $J$ represents a source
coupled to the scalar field. From the generating functional, one can
compute the various correlation functions
\begin{eqnarray}
\label{eq:defcorrelation}
G^{(n)}(x_1 \cdots x_n) &=& \left \langle \Omega\left \vert 
T\left[\Phi(x_1)\cdots \Phi(x_n)\right]\right\vert \Omega \right \rangle \\
&=& \left(\frac{1}{i}\right)^n
\frac{\delta^n Z\left[J\right]}{\delta J(x_1)
\cdots \delta J(x_n)}\biggl \vert _{J=0},\nonumber \\
\end{eqnarray}
where $\vert \Omega\rangle $ denotes the vacuum state of the
theory. We distinguish it from $\vert 0\rangle $ which denotes the
vacuum state of the theory without interaction. In this last case, the
generating functional can be computed exactly. It reads (notice that
we have $Z_0[0]=1$ and that the ``$0$'' subscript indicates that we
refer to the free theory)
\begin{equation}
\label{eq:freeZ}
Z_0[J]=\exp\left[-\frac12\int {\rm d}^4x{\rm d}^4y J(x)D_{_{\rm F}}(x-y)
J(y)\right],
\end{equation}
where $D_{_{\rm F}}$ is the Feynman propagator defined by the
following expression
\begin{equation}
\label{eq:propagator}
D_{_{\rm F}}\left(x_1-x_2\right)=\left \langle \Omega\left \vert 
T\left[\Phi(x_1)\Phi(x_2)\right]\right\vert \Omega \right \rangle,
\end{equation}
$T$ denoting the time ordered product. As is well-known, the
propagator can be conveniently expressed in Fourier space as
\begin{equation}
\label{eq:propafourier}
D_{_{\rm F}}\left(x_1-x_2\right)=\frac{i}{(2\pi)^4}
\int \frac{{\rm d}^4k}{k^2+m^2}{\rm e}^{ik_{\mu}\cdot \left(x_1^{\mu}
-x_2^{\mu}\right)}.
\end{equation}
In presence of interactions, one can write ${\cal L}={\cal L}_0+{\cal
  L}_{\rm int}$ where ${\cal L}_{\rm int}$ represents the
self-interacting part of the Lagrangian density. In this case the
generating functional can be written as
\begin{equation}
\label{eq:geneint}
Z[J]={\cal N}\exp\left \{i\int {\rm d}^4x \,{\cal L}_{\rm int}
\left[\frac{1}{i}\frac{\delta }{\delta J(x)}\right]\right\}Z_0[J],
\end{equation}
where, in order to have the correct normalization, the factor ${\cal
  N}$ must be given by
\begin{equation}
\label{eq:normgeneint}
{\cal N}^{-1}\equiv \exp\left \{i\int {\rm d}^4x \, {\cal L}_{\rm int}
\left[\frac{1}{i}\frac{\delta }{\delta J(x)}\right]\right\}
Z_0[J]\biggl\vert _{J=0}.
\end{equation}
As explained in the standard quantum field theory textbooks, the above
expressions~(\ref{eq:geneint}) and~(\ref{eq:normgeneint}) are
particularly well suited for a perturbative expansion. For instance,
if we want to calculate the generating functional for the case of a
quartic theory ${\cal L}_{\rm int}=\lambda \Phi^4/4!$, where $\lambda
$ is the coupling constant, we are led to the following expansion in
$\lambda$
\begin{eqnarray}
\label{eq:expansionZ}
Z[J]={\cal N}\left[1-i\frac{\lambda}{4!}\int {\rm d}^4x
\frac{1}{i^4}\frac{\delta ^4Z_0[J]}{\delta J(x)^4}+\cdots \right].
\end{eqnarray}
As usual, the various terms that appear in this expansion can be
graphically represented by Feynman diagrams. The Feynman rules of the
theory in real space are given by
\begin{eqnarray}
\label{eq:fruleone}
D(x_1-x_2) &\equiv & \qquad 
\begin{fmffile}{ruleone}
\parbox{20mm}{
\begin{fmfgraph*}(40,30)
\fmfleft{i}
\fmfright{o}
\fmf{plain}{i,o}
\fmfdot{i}
\fmfdot{o}
\fmflabel{$x_1$}{i}
\fmflabel{$x_2$}{o}
\end{fmfgraph*}} 
\end{fmffile},
\\
\label{eq:fruletwo}
-i\lambda \int {\rm d}^4x &\equiv & 
\begin{fmffile}{ruletwo}
\parbox{20mm}{
\begin{fmfgraph*}(40,30)
\fmfleft{i1,i2}
\fmfright{o1,o2}
\fmf{plain}{i1,v,o2}
\fmf{plain}{i2,v,o1}
\fmfdot{v}
\end{fmfgraph*}}
\end{fmffile},
\\
i\int {\rm d}^4x J(x)&\equiv & 
\begin{fmffile}{rulethree}
\parbox{20mm}{
\begin{fmfgraph*}(40,30)
\fmfleft{i}
\fmfright{o}
\fmf{plain}{i,o}
\fmfv{decor.shape=circle,decor.filled=0,decor.size=0.1w}{i}
\end{fmfgraph*}}
\end{fmffile},
\end{eqnarray}
and a loop for a propagator evaluated at the same point in
space-time. This completes our quick reminder of the basics of field
theory that we need in the following. 

\par

Now, we compute the energy density of a free scalar field (the
interacting case is treated in the next subsection) using the tools
that we have just presented. In fact, it is interesting to calculate
the trace of the stress-energy tensor. Upon using
Eq.~(\ref{eq:tmunuscalarfield}), one obtains
\begin{equation}
\label{eq:traceoft}
\langle T\rangle =\langle \eta^{\mu \nu}T_{\mu \nu}\rangle=-\langle \rho\rangle
+(d-1)\langle p\rangle.
\end{equation}
Then, since the energy density and the pressure are given by
Eqs.~(\ref{eq:rhodim}) and~(\ref{eq:pdim}), the above relation can be
re-expressed as
\begin{equation}
\langle T\rangle =-\mu^{4-d}\int \frac{{\rm d}^{d-1}{\bm k}}{(2\pi)^{d-1}}
\frac{m^2}{2\omega (k)}
\end{equation} 
But, from Eqs.~(\ref{eq:propagator}) and~(\ref{eq:propafourier}), and
now specifying to the case $d=4$, one has
\begin{equation}
D_{_{\rm F}}(0)=\frac{i}{(2\pi)^4}\int \frac{{\rm d}^4k}{k^2+m^2}
=\frac{i}{(2\pi)^4}\int \frac{{\rm d}k^0{\rm d}^{3}{\bm k}}
{-(k^0)^2+\omega^2},
\end{equation}
where we have distinguished the integration over the time and space
components of the momentum. Using the identity $\int {\rm
  d}k^0/[-(k^0)^2+\omega^2]=-\pi/(i\omega)$, see also
Eq.~(\ref{eq:propagatoronedim}), this leads to
\begin{equation}
D_{_{\rm F}}(0)=-\int \frac{{\rm d}^3{\bm k}}{(2\pi)^32\omega}.
\end{equation}
or, in other words,
\begin{equation}
\label{eq:linkTpropa}
\langle T\rangle =m^2 D_{_{\rm F}}(0).
\end{equation}
But, if we now use the fact that $\langle p\rangle
=-\langle\rho\rangle$, Eq.~(\ref{eq:traceoft}) can also be written as
$\langle T\rangle =-d\langle \rho\rangle$. As a consequence,
Eq.~(\ref{eq:linkTpropa}) implies that
\begin{equation}
\label{eq:rhopropa}
\langle \rho\rangle =-\frac{m^2}{4}D_{_{\rm F}}(0).
\end{equation}
This equation can also be obtained by a direct calculation. Indeed one has 
\begin{eqnarray}
D_{_{\rm F}}(0)
&=& \frac{i\mu^{4-d}}{(2\pi)^d}\int \frac{{\rm d}k^0{\rm d}^{d-1}{\bm k}}
{-(k^0)^2+\omega ^2} \\
&=&  -\frac{d}{m^2}\frac{\mu^4}{2\left(4\pi\right)^{(d-1)/2}}
\frac{\Gamma(-d/2)}{\Gamma(-1/2)}\left(\frac{m}{\mu}\right)^d
\\
&=& -\frac{d}{m^2}\langle \rho \rangle,
\end{eqnarray}
where, in the second step of the calculation, we have used dimensional
regularization. Now, using Eq.~(\ref{eq:deflambdaeff}) and the Feynman
rule stipulating that a propagator evaluated at the same point is
represented by a loop, one can re-write the cosmological constant as
\begin{equation}
\Lambda_{_{\rm eff}}=\Lambda_{_{\rm B}}-\kappa \frac{m^2}{4}
\begin{fmffile}{lambdabubble}
\quad \, \, 
\parbox{20mm}{
\begin{fmfgraph*}(40,30)
\fmfleft{i}
\fmfright{o}
\fmfdot{i}
\fmfv{decor.shape=circle,decor.filled=0,decor.size=0.5w}{i}
\end{fmfgraph*}}
\end{fmffile}\hspace{-1.3cm}.
\end{equation}
In other words, the cosmological constant is given by a ``bubble
diagram'', \ie a diagram that has no external leg.

\par

We have just established that the vacuum energy density, or
equivalently the cosmological constant, can be expressed in term of a
bubble diagram. This type of diagram plays a special role in quantum
field theory and we will discuss this point in more detail in the
following next sub-sections. In order to develop our intuition, let
us first calculate explicitly the generating function in the case of
a self-interacting scalar field. The calculation is straightforward
and upon using Eq.~(\ref{eq:freeZ}), this leads to
\begin{widetext}
\begin{eqnarray}
-i\frac{\lambda}{4!}\int {\rm d}^4x
\frac{1}{i^4}\frac{\delta ^4Z_0[J]}{\delta J(x)^4}
&=&-i\frac{\lambda}{4!}
\Biggl\{3\int {\rm d}^4x D^2(0)-6\int {\rm d}^4xD(0)
\left[\int {\rm d}^4yD(x-y)J(y)\right]^2
\nonumber \\ & &
+\int {\rm d}^4x\left[\int {\rm d}^4yD(x-y)J(y)\right]^4\Biggr\}
Z_0[J]+\cdots \\ 
\label{eq:expansionderZ}
&=& \frac{1}{4!}
\left(
3 \quad \quad \quad
\begin{fmffile}{numZ}
\raisebox{-0.95cm}{
\begin{rotate}{90}
\parbox{25mm}{
\begin{fmfgraph*}(61,30)
\fmfleft{i}
\fmfright{o}
\fmf{phantom}{i,v}
\fmf{plain,left}{v,v}
\fmf{plain,right}{v,v}
\fmf{phantom}{v,o}
\fmfdot{v}
\end{fmfgraph*}}
\end{rotate}}
\quad \quad
+6
\quad
\parbox{20mm}{
\begin{fmfgraph*}(40,30)
\fmfleft{i}
\fmfright{o}
\fmf{plain}{i,v,v,o}
\fmfv{decor.shape=circle,decor.filled=0,decor.size=0.1w}{i}
\fmfdot{v}
\fmfv{decor.shape=circle,decor.filled=0,decor.size=0.1w}{o}
\end{fmfgraph*}}
+\parbox{20mm}{
\begin{fmfgraph*}(40,30)
\fmfleft{i1,i2}
\fmfright{o1,o2}
\fmf{plain}{i1,v,o2}
\fmf{plain}{i2,v,o1}
\fmfdot{v}
\fmfv{decor.shape=circle,decor.filled=0,decor.size=0.1w}{i1}
\fmfv{decor.shape=circle,decor.filled=0,decor.size=0.1w}{i2}
\fmfv{decor.shape=circle,decor.filled=0,decor.size=0.1w}{o1}
\fmfv{decor.shape=circle,decor.filled=0,decor.size=0.1w}{o2}
\end{fmfgraph*}}
\end{fmffile}
+\cdots 
\right)Z_0[J].
\end{eqnarray}
One notices the appearance of a ``bubble diagram'' (the first one),
similar (\ie with no external leg) but not identical to the one
encountered in the calculation of the cosmological constant. We also
have a diagram with a loop (the second one) but with external legs and
a diagram with no loop (the last one). All the diagram with loops
(that is to say the two first ones) are divergent in quantum field
theory. Nevertheless, the bubble diagram and the loop diagram play in
fact a very different role as we are going to see. Let us now
calculate the normalization ${\cal N}$, see
Eq.~(\ref{eq:normgeneint}). It uses the same expression as before but
with $J=0$ which means that all the diagram containing a source
disappear. As a consequence, we have
\begin{equation}
\label{eq:expansionnorm}
{\cal N}^{-1}=
1+
\frac{3}{4!}
\quad \quad \quad
\begin{fmffile}{norm}
\raisebox{-0.95cm}{
\begin{rotate}{90}
\parbox{25mm}{
\begin{fmfgraph*}(61,30)
\fmfleft{i}
\fmfright{o}
\fmf{phantom}{i,v}
\fmf{plain,left}{v,v}
\fmf{plain,right}{v,v}
\fmf{phantom}{v,o}
\fmfdot{v}
\end{fmfgraph*}}
\end{rotate}}
\end{fmffile}
\end{equation}
We can now gather everything and compute the generating
functional. Upon using the expression~(\ref{eq:expansionZ}) and the
results~(\ref{eq:expansionderZ}) and~(\ref{eq:expansionnorm}), one
arrives at
\begin{eqnarray}
Z[J] &=& \left[1+\frac{1}{4!}
\left(
\begin{fmffile}{finalZ}
6
\quad
\parbox{20mm}{
\begin{fmfgraph*}(40,30)
\fmfleft{i}
\fmfright{o}
\fmf{plain}{i,v,v,o}
\fmfv{decor.shape=circle,decor.filled=0,decor.size=0.1w}{i}
\fmfdot{v}
\fmfv{decor.shape=circle,decor.filled=0,decor.size=0.1w}{o}
\end{fmfgraph*}}
+\parbox{20mm}{
\begin{fmfgraph*}(40,30)
\fmfleft{i1,i2}
\fmfright{o1,o2}
\fmf{plain}{i1,v,o2}
\fmf{plain}{i2,v,o1}
\fmfdot{v}
\fmfv{decor.shape=circle,decor.filled=0,decor.size=0.1w}{i1}
\fmfv{decor.shape=circle,decor.filled=0,decor.size=0.1w}{i2}
\fmfv{decor.shape=circle,decor.filled=0,decor.size=0.1w}{o1}
\fmfv{decor.shape=circle,decor.filled=0,decor.size=0.1w}{o2}
\end{fmfgraph*}}
\end{fmffile}
\right)+\cdots\right]
{\rm e}^{\frac12 \, 
\begin{fmffile}{exp}
\parbox{20mm}{
\begin{fmfgraph*}(15,15)
\fmfleft{i}
\fmfright{o}
\fmf{plain}{i,o}
\fmfv{decor.shape=circle,decor.filled=0,decor.size=0.1w}{i}
\fmfv{decor.shape=circle,decor.filled=0,decor.size=0.1w}{o}
\end{fmfgraph*}}
\end{fmffile}
}\hspace{-1cm}.
\end{eqnarray}
\end{widetext}
We see that the bubble diagram has canceled in the final
expression. Since all the predictions (cross-sections etc ...) follow
from the calculation of the generating functional, we see that the
bubble diagrams never contribute, in absence of gravity, to any
observables. On the contrary, when the gravitational field is turned
on, the bubble diagrams become important and can affect the predictions
of the theory. In summary, the cosmological constant problem can be
viewed as the appearance of a new type of divergent graphs. As
demonstrated with Eq.~(\ref{eq:rhovacrenorm}), these graphs can be
regularized but, as will be discussed below, the regularized value
seems to be incompatible with some astrophysical observations.

\subsection{The Vacuum in Presence of Interactions}
\label{subsec:interactingvacuum}

So far we have calculated the vacuum energy density for a free
field. It is natural to ask what happens when one considers
interaction~\cite{Ossola:2003ku} since we know that a realistic model
necessarily contains such terms. For this purpose, let us therefore
consider again a simple self-interacting scalar field with ${\cal
  L}_{\rm int}=\lambda\Phi^4/4!$. As a consequence, this adds the
following contribution to the energy density of the vacuum
\begin{eqnarray}
\hspace{0.4cm}
\label{eq:Deltarhointer}
\Delta \rho &=& \frac{\lambda}{4!}\left\langle \Phi^4\right\rangle 
=\frac{3\lambda}{4!}\left\langle \Phi^2\right \rangle^2
=\frac{\lambda}{8}D_{_{\rm F}}^2(0)
\\
\label{eq:vacuumdiagram}
&=&\frac{i}{8\int {\rm d}^4x}
\quad \quad \quad
\begin{fmffile}{norm}
\raisebox{-0.95cm}{
\begin{rotate}{90}
\parbox{25mm}{
\begin{fmfgraph*}(61,30)
\fmfleft{i}
\fmfright{o}
\fmf{phantom}{i,v}
\fmf{plain,left}{v,v}
\fmf{plain,right}{v,v}
\fmf{phantom}{v,o}
\fmfdot{v}
\end{fmfgraph*}}
\end{rotate}}
\end{fmffile},
\end{eqnarray}
where, on the first line in the above expression, we have used the
fact that the vacuum wave-functional is a Gaussian. As expected, this
new term is also given by a bubble diagram. However, one should not
forget that, when interaction is present, one must also renormalize
the mass. As is well-known, at one loop, the renormalization of the
mass comes from the tadpole diagram. More precisely, one has to
calculate the correction to the propagator. It is given by the
following expression
\begin{widetext}
\begin{eqnarray}
\label{eq:2pointgraph}
\left \langle \Omega\left \vert 
T\left[\Phi(x_1)\Phi(x_2)\right]\right\vert \Omega \right \rangle
&=&\qquad 
\begin{fmffile}{ruleone}
\parbox{20mm}{
\begin{fmfgraph*}(40,30)
\fmfleft{i}
\fmfright{o}
\fmf{plain}{i,o}
\fmfdot{i}
\fmfdot{o}
\fmflabel{$x_1$}{i}
\fmflabel{$x_2$}{o}
\end{fmfgraph*}} 
\end{fmffile}
+\frac12 \qquad
\begin{fmffile}{tadpole}
\parbox{20mm}{
\begin{fmfgraph*}(40,30)
\fmfleft{i}
\fmfright{o}
\fmf{plain}{i,v,v,o}
\fmfdot{i}
\fmfdot{o}
\fmfdot{v}
\fmflabel{$x_1$}{i}
\fmflabel{$x_2$}{o}
\end{fmfgraph*}}
\end{fmffile}
+\cdots ,
\end{eqnarray}
or, in terms of explicit expressions [see Eq.~(\ref{eq:propafourier})]
\begin{eqnarray}
\label{eq:propaand2point}
\left \langle \Omega\left \vert 
T\left[\Phi(x_1)\Phi(x_2)\right]\right\vert \Omega \right \rangle
&=& 
D_{_{\rm F}}(x_1-x_2)
-\frac{i\lambda}{2} \int{\rm d}^4xD_{_{\rm F}}(x_1-x)
D_{_{\rm F}}(0)D_{_{\rm F}}(x-x_2)+\cdots
\\
&=&
\frac{i}{(2\pi)^4}
\int \frac{{\rm d}^4p}{p^2+m^2}{\rm e}^{ip_{\mu}\cdot \left(x_1^{\mu}
-x_2^{\mu}\right)}
+\frac{i\lambda}{2}D_{_{\rm F}}(0)\frac{1}{(2\pi)^4}
\int {\rm d}^4p\frac{{\rm e}^{ip_{\mu}\cdot (x_1^{\mu}-x_2^{\mu})}}{(p^2+m^2)^2}
+\cdots
\\
&=& \frac{i}{(2\pi)^4}
\int \frac{{\rm d}^4p}{p^2+m^2}{\rm e}^{ip_{\mu}\cdot \left(x_1^{\mu}
-x_2^{\mu}\right)}\left[1+\frac{\lambda}{2}D_{_{\rm F}}(0)\frac{1}{p^2+m^2}
\right]+\cdots
\\
&=&
\frac{i}{(2\pi)^4}
\int \frac{{\rm d}^4p}{p^2+m^2-\lambda D_{_{\rm F}}(0)/2}
{\rm e}^{ip_{\mu}\cdot \left(x_1^{\mu}
-x_2^{\mu}\right)}+\cdots ,
\end{eqnarray}
\end{widetext}
where the last equation is equivalent to the previous one at leading
order in the coupling constant. We see that this implies the following
redefinition (or renormalization) of the mass
\begin{equation}
\label{eq:renormmassinter}
m_{\rm ren}^2=m^2-\frac{\lambda}{2}D_{_{\rm F}}(0)
\end{equation}
As a consequence, the vacuum energy which is now the sum of the free
contribution~(\ref{eq:rhopropa}) and of the
contribution~(\ref{eq:Deltarhointer}), which originates from the
self-interaction of the field, can be re-expressed as
\begin{eqnarray}
\langle \rho\rangle &=&-\frac{m^2}{4}D_{_{\rm F}}(0)
+\frac{\lambda}{8}D_{_{\rm F}}^2(0)
\nonumber \\
&=&-\frac{m^2_{\rm ren}}{4}D_{_{\rm F}}(0)
-\frac14\left[\frac{\lambda}{2}D_{_{\rm F}}(0)\right]D_{_{\rm F}}(0)
+\frac{\lambda}{8}D_{_{\rm F}}^2(0)\nonumber
\\
&=&-\frac{m^2_{\rm ren}}{4}D_{_{\rm F}}(0),
\end{eqnarray}
where, in the second line, we have used
Eq.~(\ref{eq:renormmassinter}). The extra contribution coming from the
interacting term is exactly canceled by the renormalization of the
mass~\cite{Birrell:1982ix,Ossola:2003ku}. In other words, one sees
that the presence of the interaction has not modified the expression
of the vacuum energy density~(\ref{eq:rhopropa}), at least at one
loop, provided the final result is expressed in terms of the
renormalized mass rather than the bare one.

\section{The Bubble Diagrams}
\label{sec:bubble}

The previous sections have shown the central role played by the bubble
diagrams in the vacuum energy problem. These diagrams do not affect
the physical predictions when the gravitational field is turned off
and this is why, in standard quantum field theory, there are usually
ignored. In this section, we wish to discuss in more detail this type
of diagram. For this purpose, we will analyze them in ordinary quantum
mechanics where the technical aspects are less complicated and where
the interpretation is easier~\cite{Bender:1969si,abbott,minahan}. 

\subsection{Quantum Mechanics as a Field Theory}
\label{subsec:qmqft}

Let us start with the simplest quantum-mechanical system, namely the
one-dimensional harmonic oscillator. Let us denote by $x(t)$ the
position of the particle evolving into a parabolic potential. Then,
the corresponding Lagrangian is given by
\begin{equation}
L(\dot{x},x)=\frac{1}{2}m\dot{x}^2-\frac{1}{2}m\omega ^2x^2\, ,
\end{equation}
where $m$ is the mass of the particle and $\omega$ is the effective
frequency. Then, let us define the ``scalar field'' $\Phi(t)$ by
\begin{equation}
\label{eq:defphiqm}
\Phi(t)\equiv \sqrt{m}x(t)\, ,
\end{equation}
which implies that $\Phi $ has dimension $-1/2$. As a consequence, the
Lagrangian now reads
\begin{equation}
L=\frac{\dot{\Phi}^2}{2}-\frac{\omega ^2}{2}\Phi^2\, .
\end{equation}
This Lagrangian is similar to the one one encounters in field theory
for a free scalar field. This means that all the techniques of quantum
field theory can in fact be used in ordinary quantum mechanics. Of
course, this does not lead to genuine new results since both
approaches are equivalent but this can shed a new light of some
physical results. This is the strategy that we use here applied to the
vacuum energy problem. We now quickly remind how the formalism of
quantum field theory can be implemented in quantum mechanics. We will
also compare this approach to the standard approach of quantum
mechanics. The first step is to define the conjugate momentum. It is
given by
\begin{equation}
\Pi(t) =\frac{\partial L}{\partial \dot \Phi}=\dot{\Phi}(t)\, .
\end{equation}
and this leads to the following Hamiltonian
\begin{equation}
H=\frac{\Pi^2}{2}+\frac{\omega ^2}{2}\Phi^2\, .
\end{equation}
The next step is to introduce the creation and annihilation
$c^{\dagger}$ and $c$ operators. They are defined by the expressions
\begin{eqnarray}
\Phi &=& \frac{1}{\sqrt{2\omega }}\left(c^{\dagger}+c\right)\, ,\\
\Pi &=& i\sqrt{\frac{\omega }{2}}\left(c^{\dagger}-c\right)\, .
\end{eqnarray}
As usual one can check that $\left[\Phi,\Pi\right]=i$ implies
$\left[c,c^{\dagger}\right]=1$. As it is standard in field theory, we
now work in the Heisenberg picture where the operators are time
dependent. A straightforward calculation shows that the Hamiltonian
can be expressed as
\begin{equation}
\label{eq:hamiltonian}
H=\frac{\omega }{2}\left(cc^{\dagger}+c^{\dagger}c\right)\, .
\end{equation}
This allows us to calculate the time evolution of the operator
$c$. Indeed, the Heisenberg equation reads
\begin{equation}
i\frac{{\rm d}c}{{\rm d}t}=\left[c,H\right]=\omega c\, ,
\end{equation}
from which we deduce that
\begin{equation}
c(t)=c(0){\rm e}^{-i\omega t}=c_0{\rm e}^{-i\omega t}\,.
\end{equation}
As a consequence, the field operator can be written in terms of the
creation and annihilation operators in a way which is very similar to
Eq.~(\ref{eq:fourierfield}), namely
\begin{equation}
\Phi(t)=\frac{1}{\sqrt{2\omega }}\left(c_0{\rm e}^{-i\omega t}
+c_0^{\dagger}{\rm e}^{i\omega t}\right)\, .
\end{equation}
Of course we do not have an integral over the Fourier modes because,
in quantum mechanics, we deal with a single harmonic oscillator as
opposed to an infinite collection of harmonic oscillators in field
theory.

\par

\begin{figure*}
\begin{center}
\includegraphics[width=10cm]{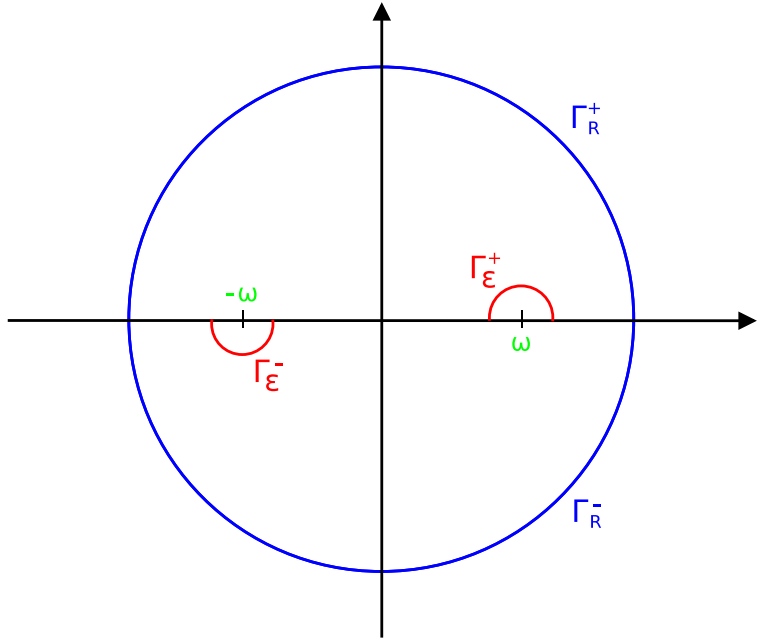}
\caption{Contours in the complex time plane used to calculate the
  Feynman propagators of the one-dimensional harmonic oscillator. As
  usual, there are two poles on the real axis that can be avoided by
  going along the two small red circles. The final contour is closed
  by a large blue circle of radius $R$.}
\label{fig:complexplane}
\end{center}
\end{figure*}

\par

At this level, the theory is known exactly. It is clear that the most
interesting part is in fact the treatment of the interactions which,
in field theory, is based on the calculation of the Feynman
diagrams. Since this calculation is itself based on the calculation of
the propagator, one must now evaluate it for our simple system. It is
defined by an equation similar to Eq.~(\ref{eq:propagator}), namely
\begin{equation}
D(t_1,t_2)\equiv \langle 0\vert T\left[\Phi(t_1)
\Phi(t_2)\right]\vert 0\rangle\, ,
\end{equation}
where we recall that the time ordered product is defined by
\[
T\left[\Phi(t_1)\Phi(t_2)\right]=
\begin{cases}
\Phi(t_1)\Phi(t_2), & t_1>t_2\\
\Phi(t_2)\Phi(t_1), & t_2>t_1.
\end{cases}
\]
Then, the calculation proceeds straightforwardly and one obtains for
$t_1>t_2$
\begin{eqnarray}
D(t_1,t_2)&=& \biggl\langle 0\biggl \vert \frac{1}{\sqrt{2\omega }}
\left(c_0{\rm e}^{-i\omega t_1}
+c_0^{\dagger}{\rm e}^{i\omega t_1}\right) \nonumber \\ & &
\frac{1}{\sqrt{2\omega }}
\left(c_0{\rm e}^{-i\omega t_2}
+c_0^{\dagger}{\rm e}^{i\omega t_2}\right)\biggr\vert 0 \biggr \rangle \\
&=& \frac{1}{2\omega}{\rm e}^{-i\omega (t_1-t_2)}\, ,
\end{eqnarray}
and, in a similar way if $t_2>t_1$,  
\begin{eqnarray}
D(t_1,t_2)&=& \biggl\langle 0\biggl \vert \frac{1}{\sqrt{2\omega }}
\left(c_0{\rm e}^{-i\omega t_2}
+c_0^{\dagger}{\rm e}^{i\omega t_2}\right) \nonumber \\ & &
\frac{1}{\sqrt{2\omega }}
\left(c_0{\rm e}^{-i\omega t_1}
+c_0^{\dagger}{\rm e}^{i\omega t_1}\right)\biggr\vert 0 \biggr \rangle \\
&=& \frac{1}{2\omega}{\rm e}^{i\omega (t_1-t_2)}\, .
\end{eqnarray}
The two previous equations can be summarized into a single one
\begin{equation}
\label{eq:propaqm}
D(t_1,t_2)=\frac{1}{2\omega }{\rm e}^{-i\omega \vert t_2-t_1\vert }\, .
\end{equation}
We see that the system is so simple that we obtain an explicit
expression of the propagator in real space (in fact, in field theory,
there is also an expression of the propagator in real space but it is
not so often used because of its complexity -- it is given in terms of
ordinary and modified Bessel functions --). Nevertheless, in order to
compare with the standard approach and Eq.~(\ref{eq:propafourier}), it
is interesting to obtain the expression of the propagator in Fourier
space. The Fourier transform of a function $f(t)$ is defined by
\begin{equation}
f(t)=\frac{1}{\sqrt{2\pi}}\int _{-\infty}^{+\infty}f(E)\, {\rm e}^{iEt}
\, {\rm d}E.
\end{equation}
The calculation is easy and well-known. Here, for convenience, we
quickly remind how it proceeds. For that purpose, let us consider the
function
\begin{equation}
f(z)\equiv \frac{{\rm e}^{-izt}}{z^2-\omega ^2}\, .
\end{equation}
It has two poles at $z=\pm \omega $ on the real axis, see
Fig.~\ref{fig:complexplane}. We assume that $t>0$ and consider the
following contour: along the real axis with two small circles $\Gamma
_{\epsilon}^{\pm}$ around the two poles (the circle around $z=-\omega$
goes into the lower part of the complex plane while the circle around
$z=+\omega$ goes into the upper part); and we complete by a big circle
$\Gamma_{R}^{-}$ going into the lower part of the complex plane. In
this case, only the pole $z=+\omega$ is inside the contour $\Gamma=
\mathbb{R}\cup \Gamma_{\epsilon}^{\pm}\cup\Gamma_R^{-}$, see
Fig.~\ref{fig:complexplane}. We have
\begin{equation}
  \lim _{\vert z\vert \rightarrow \infty}\vert z f(z)\vert =
  \lim _{\vert z\vert \rightarrow \infty} \frac{\vert z\vert 
{\rm e}^{\Im(z)t}}{z^2-\omega ^2}=0\, ,
\end{equation} 
since $\Im(z)<0$ in the lower part of the complex plan. According to
the first Jordan lemma, this means that the integral on the big circle
vanishes. In the same way, we have $\lim _{ z\rightarrow 0}\vert z
f(z)\vert =0$ and the second Jordan lemma tells that the integral on
$\Gamma _{\epsilon}^{\pm}$ is also zero. Then, using the fact that
$f(z)\equiv P(z)/Q(z)$ and that, in this case, the residue can be
written as $\mbox{Res}(f,z)=P(z)/Q'(z)$, one obtains
\begin{equation}
\mbox{Res}(f,z=+\omega)=\frac{{\rm e}^{-i\omega t}}{2\omega}.
\end{equation}
As a consequence
\begin{eqnarray}
\lim_{R\rightarrow +\infty}
\left[\frac{i}{2\pi}\int _{\Gamma}f(z){\rm d}z\right] &=&
\frac{i}{2\pi}\int _{-\infty}^{\infty}f(z){\rm d}z
\nonumber \\
&=&\frac{i}{2\pi}\times -
2i\pi \mbox{Res}(f,z=+\omega) \nonumber \\
&=& \frac{{\rm e}^{-i\omega t}}{2\omega},
\end{eqnarray}
where the minus sign in the second line comes from the fact that the
contour is clockwise (\ie opposite to the direct direction).

\par

Now if $t<0$, then the big circle is taken in the upper plane. As a
consequence, the pole inside the contour is now the one located at
$z=-\omega$. Therefore, 
\begin{eqnarray}
\frac{i}{2\pi}\int _{\Gamma}f(z){\rm d}z &=&
\frac{i}{2\pi}\int _{-\infty}^{\infty}f(z){\rm d}z
\nonumber \\
&=&\frac{i}{2\pi}\times +
2i\pi \mbox{Res}(f,z=-\omega) \nonumber \\
&=& \frac{{\rm e}^{i\omega t}}{2\omega}.
\end{eqnarray}
The previous considerations allow us to conclude that
\begin{equation}
\label{eq:propagatoronedim}
\frac{i}{2\pi}
\int _{-\infty}^{+\infty} \frac{{\rm e}^{-iEt}}{E^2-\omega ^2}
{\rm d}E=\frac{{\rm e}^{-i\omega \vert t\vert }}{2\omega}
=D(t),
\end{equation}
that is to say we have established the Fourier transform of the
propagator. It should be obvious that the above equation is the
counterpart of Eq.~(\ref{eq:propafourier}). As in field theory, we
will show that it can be used to develop a perturbative method which
allows us to take into account interactions. 

\subsection{Perturbations Theory in Quantum Mechanics}
\label{subsec:pertqm}

Before discussing our main subject, we need to explain how the Feynman
diagrams can be used in quantum
mechanics~\cite{abbott,minahan}. Usually, in quantum mechanics, the
perturbation theory is not based on the Feynman
diagrams~\cite{tannoudji2006}. This gives the impression that the
techniques of field theory are very different. In fact, as we now
demonstrate, these techniques can also be utilized in ordinary quantum
mechanics~\cite{Bender:1969si} and, obviously (as could have been
guessed from the very beginning), they lead to the same physical
predictions. In the following, we will discuss this question and
compare the two approaches in some details. We believe that this can
improve our physical understanding of the vacuum energy problem.

\par

In order to achieve the above mentioned task, we need to recall how
interactions are treated in quantum mechanics. For this purpose, let
us consider a system obeying the following Schr\"odinger equation
\begin{equation}
i\frac{{\rm d}\vert \Psi (t)\rangle_{_{\rm S}}}{{\rm d}t}
=H(p_{_{\rm S}},x_{_{\rm S}})\vert \Psi(t) \rangle_{_{\rm S}},
\end{equation}
where $H(p_{_{\rm S}},x_{_{\rm S}})$ is the Hamiltonian operator. The
above equation is written in the Schr\"odinger representation, hence
the subscript ``S''. Then, one can define the evolution operator
$U(t,t_{\rm i})$ which relates the state vector at the time $t$ to the
state vector at some initial time $t_{\rm i}<t$ by the following
expression
\begin{equation}
\vert \Psi (t)\rangle_{_{\rm S}}=U(t,t_{\rm i})
\vert \Psi (t_{\rm i})\rangle_{_{\rm S}}.
\end{equation}
This operator obeys
\begin{equation}
i\frac{\partial U(t,t_{\rm i})}{{\partial }t}=HU(t,t_{\rm i})\, ,
\end{equation}
and, provided the Hamiltonian is not explicitly time dependent, it can
also be written as
\begin{equation}
U(t,t_{\rm i})={\rm e}^{-iH(t-t_{\rm i})}.
\end{equation}
Of course, in this representation, the operator $p_{_{\rm S}}$ and
$x_{_{\rm S}}$ are time independent since only the state can change
with time.

\par

One the contrary, in the Heisenberg representation, the state vector
does not evolve and can be expressed in terms of the Schr\"odinger
state vector evaluated at some reference time $t=t_0$,
\begin{equation}
\vert \Psi \rangle_{_{\rm H}}\equiv \vert \Psi (t_0)\rangle_{_{\rm S}}.
\end{equation}
Notice that the reference time is arbitrary and is not necessarily
$t_{\rm i}$. In this picture, the operators are time-dependent and,
for instance, $x_{_{\rm H}}(t)$ is given by
\begin{equation}
x_{_{\rm H}}(t)\equiv U^{\dagger}(t,t_0)x_{_{\rm S}}U(t,t_0),
\end{equation}
such that $x_{_{\rm H}}(t_0)=x_{_{\rm S}}$. In this picture, the time
evolution of the operators is computed from the Heisenberg equation
which reads
\begin{eqnarray}
i\frac{{\rm d}x_{_{\rm H}}}{{\rm d}t}&=&
i\frac{\partial U^{\dagger}(t,t_0)}{\partial t}
x_{_{\rm S}}U(t,t_0)+iU^{\dagger}(t,t_0)x_{_{\rm S}}
\frac{\partial U(t,t_0)}{\partial t}
\nonumber \\
\\
&=& -U^{\dagger}Hx_{_{\rm S}}U+U^{\dagger}x_{_{\rm S}}HU \\
&=& \left[x_{_{\rm S}},H\right],
\end{eqnarray}
where in the second line we have used that $-i\partial
U^{\dagger}/\partial t=U^{\dagger}H$.

\par

Let us now consider the case where interactions are explicitly
present. We assume that the corresponding Hamiltonian can be split
into two parts, one corresponding to the free theory and one
corresponding to the interactions
\begin{equation}
H=H_0+H_{\rm int}.
\end{equation}
We also assume that $H_{\rm int}$ is a small perturbation. The free
evolution, due to the free Hamiltonian $H_0$, is described by the
evolution operator $U_0$. This operator is of course different from
$U$ which describes the full evolution with the interactions taken
into account. Clearly, $U_0$ can be expressed as
\begin{equation}
U_0(t,t_{\rm i})={\rm e}^{-iH_0(t-t_{\rm i})}\, .
\end{equation}
Then the idea is to factor out the free evolution (which is supposed
to be known) and to focus on the time evolution due to the
perturbation on top of the free evolution. For this purpose, we define
the interaction representation by (again, the representations are
supposed to coincide when $t=t_0$)
\begin{eqnarray}
\label{eq:defintpicture}
x_{_{\rm I}}(t)&=& U_0^{\dagger}(t,t_0)x_{_{\rm S}}U_0(t,t_0)\\
&=& {\rm e}^{iH_0(t-t_0)}x_{_{\rm S}}{\rm e }^{-iH_0(t-t_0)}.
\end{eqnarray}
Clearly $x_{_{\rm I}}(t)$ is different from $x_{_{\rm H}}(t)$ because
its time evolution is ``generated'' only by the free Hamiltonian and
not by the total one. Let us now calculate the equation of motion of
$x_{_{\rm I}}(t)$. One has
\begin{eqnarray}
\frac{{\rm d}x_{_{\rm I}}}{{\rm d}t}&=&iH_0x_{_{\rm I}}
+{\rm e}^{iH_0(t-t_0)}x_{_{\rm S}}\times -iH_0{\rm e}^{-iH_0(t-t_0)}\\
&=& i\left[H_0,x_{_{\rm I}}\right].
\end{eqnarray}
Therefore, the time evolution of $x_{_{\rm I}}(t)$ is controlled by
the free Hamiltonian. This evolution is supposed to be known. In the
same manner, we define the state vector in the interaction picture by
\begin{equation}
\label{eq:psiintshrodinger}
\vert \Psi(t)\rangle _{_{\rm I}}=U_0^{\dagger}(t,t_0)
\vert \Psi(t)\rangle _{_{\rm S}}.
\end{equation}
As a consequence, in the interaction picture, both the state vectors
and the operators are time dependent. The equation obeyed by the state
vector is given by
\begin{eqnarray}
  i\frac{{\rm d}\vert \Psi \rangle _{_{\rm I}}}{{\rm d}t}
  &=&-H_0\vert \Psi\rangle _{_{\rm I}}+{\rm e}^{iH_0(t-t_0)}i
  \frac{{\rm d}\vert \Psi \rangle _{_{\rm S}}}{{\rm d}t} \\
  &=&-H_0{\rm e}^{iH_0(t-t_0)}\vert \Psi\rangle _{_{\rm S}}
\nonumber \\ & &
+{\rm e}^{iH_0(t-t_0)}(H_0+H_{\rm int})
  \vert \Psi\rangle _{_{\rm S}}\\
  &=& {\rm e}^{iH_0(t-t_0)}H_{\rm int}\vert \Psi\rangle _{_{\rm S}}\\
  &=& {\rm e}^{iH_0(t-t_0)}H_{\rm int}{\rm e}^{-iH_0(t-t_0)}
  \vert \Psi \rangle _{_{\rm I}}\\
\label{eq:schrodingerint}
  &=& H_{_{\rm I}}\vert \Psi \rangle _{_{\rm I}},
\end{eqnarray}
where $H_{_{\rm I}}=U_0^{\dagger}(t,t_0)H_{\rm int}U_0(t,t_0)$ in
agreement with the definition~(\ref{eq:defintpicture}). The time
evolution of $\vert \Psi\rangle _{_{\rm I}}$ is therefore generated
only by the interacting Hamiltonian. Then, it is natural to define the
evolution operator in the interaction picture. This operator is
associated with the Schr\"odinger
equation~(\ref{eq:schrodingerint}). Its definition reads
\begin{equation}
\label{eq:defintstate}
\vert \Psi(t)\rangle _{_{\rm I}}=U_{_{\rm I}}(t,t_{\rm i})\vert \Psi (t_{\rm i})
\rangle _{_{\rm I}}.
\end{equation}
Of course $U_{_{\rm I}}$ is different from $U_0$ (and from $U$). A
straightforward calculation shows that it can also be expressed as
\begin{equation}
U_{_{\rm I}}(t,t_{\rm i})={\rm T}\exp\left[-i\int _{t_{\rm i}}^{t}{\rm d}\tau 
H_{_{\rm I}}(\tau)\right].
\end{equation}
Another expression for $U_{_{\rm I}}$ is based on the following
considerations. From Eqs.~(\ref{eq:defintstate})
and~(\ref{eq:psiintshrodinger}), we have
\begin{eqnarray}
\vert \Psi (t)\rangle_{_{\rm I}}&=& U_{_{\rm I}}(t,t_{\rm i})\vert 
\Psi(t_{\rm i})\rangle _{_{\rm I}}\\
&=& U_{_{\rm I}}(t,t_{\rm i})U_0^{\dagger}(t_{\rm i},t_0)\vert 
\Psi(t_{\rm i})\rangle _{_{\rm S}}\\
&=& U_0^{\dagger}(t,t_0)\vert 
\Psi(t)\rangle _{_{\rm S}}.
\end{eqnarray}
Therefore, one obtains
\begin{eqnarray}
\vert \Psi(t)\rangle _{_{\rm S}} &=& U_0(t,t_0)U_{_{\rm I}}(t,t_{\rm i})
U_0^{\dagger}(t_{\rm i},t_0)\vert \Psi(t_{\rm i})\rangle _{_{\rm S}}\\
&=& U(t,t_{\rm i})
\vert \Psi (t_{\rm i})\rangle_{_{\rm S}},
\end{eqnarray}
from which we deduce that
\begin{equation}
U(t,t_{\rm i})=U_0(t,t_0)U_{_{\rm I}}(t,t_{\rm i})U_0^{\dagger}(t_{\rm i},t_0),
\end{equation}
or
\begin{eqnarray}
  U_{_{\rm I}}(t,t_{\rm i})&=&U_0^{\dagger}(t,t_0)U(t,t_{\rm i})
  U_0^{\dagger}(t_{\rm i},t_0),\\
  &=& {\rm e}^{iH_0(t-t_0)}{\rm e}^{-iH(t-t_{\rm i})}
{\rm e}^{-iH_0(t_{\rm i}-t_0)}.
\end{eqnarray}
This expression is very natural since it expresses nothing but the
fact that the full evolution of the system is just the combination of
the free evolution and of the evolution due to the interacting
term. Let us also remark that one can write
\begin{eqnarray}
x_{_{\rm I}}(t)&=&U_0^{\dagger}(t,t_0)x_{_{\rm S}}U_0(t,t_0)\\
&=& U_0^{\dagger}(t,t_0)U(t,t_0)x_{_{\rm H}}U^{\dagger}(t,t_0)U_0(t,t_0)\\
&=& {\rm e}^{iH_0(t-t_0)}{\rm e}^{-iH(t-t_0)}x_{_{\rm H}}(t)
{\rm e}^{iH(t-t_0)}{\rm e}^{-iH_0(t-t_0)} \nonumber \\ \\
&=& U_{_{\rm I}}(t,t_0)x_{_{\rm H}}(t)U_{_{\rm I}}^{\dagger}(t,t_0).
\end{eqnarray}
Again, this result is very intuitive and expresses the fact that, in
the interacting picture, the free evolution of the operators is
factorized out.

\par

After these preliminaries, one can now derive the Gell-Man Law
equation which is at the heart of the perturbative treatment presented
in what follows. Let $\vert n_{_{\rm I}}\rangle $ be the energy
eigenstates of the full Hamiltonian, \ie the free Hamiltonian plus the
Hamiltonian describing the interactions. We denote the true vacuum
state by $\vert 0_{_{\rm I}}\rangle \equiv \vert \Omega\rangle$, in
agreement with the conventions adopted in the previous
subsections. Then, one can always expand the vacuum state of the free
theory according to
\begin{eqnarray}
\vert 0\rangle &=& \sum _{n=0}^{\infty}\vert n_{_{\rm I}}\rangle \langle 
n_{_{\rm I}}\vert 0\rangle, \\
&=& \vert \Omega \rangle \langle \Omega \vert 0\rangle 
+\sum _{n\neq 0}^{\infty}\vert n_{_{\rm I}}\rangle \langle 
n_{_{\rm I}}\vert 0\rangle.
\end{eqnarray}
As a consequence, applying the operator ${\rm e}^{-iH(t+t_0)}$ to the
above relation, one obtains
\begin{eqnarray}
\label{eq:eHintervac}
{\rm e}^{-iH(t+t_0)}\vert 0\rangle 
&=& {\rm e}^{-iE_0(t+t_0)}\vert \Omega \rangle \langle \Omega \vert 0\rangle 
\nonumber \\ 
&+&\sum _{n\neq 0}^{\infty}{\rm e}^{-iE_n(t+t_0)}\vert n_{_{\rm I}}\rangle \langle 
n_{_{\rm I}}\vert 0\rangle.
\end{eqnarray}
Then, we take the limit $t\rightarrow \infty (1-i\epsilon)$. This
kills all the terms on the right hand side but the first one. Indeed,
we have $E_n>E_0$ ($n\neq 0$) and this means that ${\rm e}^{-iE_0t}$
goes to zero less rapidly than all the other exponential
terms. Therefore, in this limit, Eq.~(\ref{eq:eHintervac}) reads ${\rm
  e}^{-iH(t+t_0)}\vert 0\rangle \simeq {\rm e}^{-iE_0(t+t_0)}\vert
\Omega \rangle \langle \Omega \vert 0\rangle $ and, by inverting this
formula, this allows us to express the interacting vacuum as
\begin{eqnarray}
\vert \Omega \rangle &=&\lim_{t\rightarrow \infty (1-i\epsilon)}
\frac{{\rm e}^{iE_0(t+t_0)}}{\langle \Omega \vert 0\rangle}
{\rm e}^{-iH(t+t_0)}\vert 0\rangle\\
&=&\lim_{t\rightarrow \infty (1-i\epsilon)}
\frac{{\rm e}^{iE_0(t+t_0)}}{\langle \Omega \vert 0\rangle}
{\rm e}^{-iH(t+t_0)}{\rm e}^{iH_0(t+t_0)}\vert 0\rangle \nonumber\\
&=& \lim_{t\rightarrow \infty (1-i\epsilon)}
\frac{{\rm e}^{iE_0(t+t_0)}}{\langle \Omega \vert 0\rangle}
U_{_{\rm I}}(t_0,-t)\vert 0\rangle.
\end{eqnarray}
In the second line, we have used the fact that $H_0\vert 0\rangle=0$
since we can always define the ground state as the state associated to
the zero of energy. This means that ${\rm e}^{iH_0(t+t_0)}\vert
0\rangle =\vert 0\rangle$ and explains the appearance of the new
factor ${\rm e}^{iH_0(t+t_0)}$. In the same fashion, using ${\rm
  e}^{iH(t-t_0)}\vert \Omega \rangle={\rm e}^{iE_0(t-t_0)}\vert \Omega
\rangle $, one has
\begin{eqnarray}
\langle 0\vert {\rm e}^{-iH(t-t_0)}
&=& \lim_{t\rightarrow \infty (1-i\epsilon)} \langle 0\vert \Omega \rangle \langle 
\Omega \vert {\rm e}^{-iE_0(t-t_0)},
\end{eqnarray}
from which one deduces that
\begin{eqnarray}
\langle \Omega \vert &=&  \lim_{t\rightarrow \infty (1-i\epsilon)}
\frac{{\rm e}^{iE_0(t-t_0)}}{\langle 0\vert \Omega \rangle}\langle 0\vert 
{\rm e}^{-iH(t-t_0)}\\
 &=&  \lim_{t\rightarrow \infty (1-i\epsilon)}
\frac{{\rm e}^{iE_0(t-t_0)}}{\langle 0\vert \Omega \rangle}\langle 0\vert 
{\rm e}^{iH_0(t-t_0)}{\rm e}^{-iH(t-t_0)}\nonumber \\
&=& \lim_{t\rightarrow \infty (1-i\epsilon)}
\frac{{\rm e}^{iE_0(t-t_0)}}{\langle 0\vert \Omega \rangle}\langle 0\vert 
U_{_{\rm I}}(t,t_0).
\end{eqnarray}
Therefore, the norm of the vacuum state is given by
\begin{equation}
\label{eq:normomega}
\langle \Omega\vert \Omega \rangle=\lim_{t\rightarrow \infty (1-i\epsilon)}
\frac{{\rm e}^{2iE_0t}}{\vert \langle 0\vert \Omega \rangle \vert^2}
\langle 0\vert U_{_{\rm I}}(t,t_0)U_{_{\rm I}}(t_0,-t)\vert 0\rangle.
\end{equation}
In the following, we take $\langle \Omega\vert \Omega \rangle=1$
although this choice is not mandatory (but, then, when we compute a
correlation function, it would be necessary to divide the
corresponding expression by $\langle \Omega\vert \Omega \rangle$).

\par

We are now in a position where one can calculate the two-point
correlation function of the field operator [recall that $x(t)$ is
directly proportional to the field operator, see
Eq.~(\ref{eq:defphiqm})]. Using the previous results, one obtains
\begin{widetext}
\begin{eqnarray}
\langle \Omega \vert x_{_{\rm H}}(t_2)x_{_{\rm H}}(t_1)\vert \Omega 
\rangle &=& 
\lim_{t\rightarrow \infty (1-i\epsilon)}
\frac{{\rm e}^{2iE_0t}}{\vert \langle 0\vert \Omega \rangle \vert^2}
\langle 0\vert U_{_{\rm I}}(t,t_0)U_{_{\rm I}}^{\dagger}(t_2,t_0)
x_{_{\rm I}}(t_2)U_{_{\rm I}}(t_2,t_0)U_{_{\rm I}}^{\dagger}(t_1,t_0)
x_{_{\rm I}}(t_1)U_{_{\rm I}}(t_1,t_0)U_{_{\rm I}}(t_0,-t)\vert 0\rangle
\nonumber \\
\\
&=& 
\lim_{t\rightarrow \infty (1-i\epsilon)}
\frac{
\langle 0\vert U_{_{\rm I}}(t,t_0)U_{_{\rm I}}(t_0,t_2)
x_{_{\rm I}}(t_2)U_{_{\rm I}}(t_2,t_0)U_{_{\rm I}}(t_0,t_1)
x_{_{\rm I}}(t_1)U_{_{\rm I}}(t_1,t_0)U_{_{\rm I}}(t_0,-t)\vert 0\rangle}
{\langle 0\vert U_{_{\rm I}}(t,t_0)U_{_{\rm I}}(t_0,-t)\vert 0\rangle }\\
&=&
\lim_{t\rightarrow \infty (1-i\epsilon)}
\frac{
\langle 0\vert U_{_{\rm I}}(t,t_2)
x_{_{\rm I}}(t_2)U_{_{\rm I}}(t_2,t_1)
x_{_{\rm I}}(t_1)U_{_{\rm I}}(t_1,-t)\vert 0\rangle}
{\langle 0\vert U_{_{\rm I}}(t,-t)\vert 0\rangle }\\
&=&
\lim_{t\rightarrow \infty (1-i\epsilon)}
\frac{
\langle 0\vert {\rm T}\left[ 
x_{_{\rm I}}(t_2)x_{_{\rm I}}(t_1)U_{_{\rm I}}(t,-t)\right]\vert 0\rangle}
{\langle 0\vert U_{_{\rm I}}(t,-t)\vert 0\rangle }\\
&=&
\label{eq:gellmann}
\lim_{t\rightarrow \infty (1-i\epsilon)}
\frac{
\langle 0\vert {\rm T}\left[ 
x_{_{\rm I}}(t_2)x_{_{\rm I}}(t_1)
\exp\left(-i\int_{-t}^t{\rm d}\tau H_{_{\rm I}}\right)\right]\vert 0\rangle}
{\langle 0\vert \exp\left(-i\int_{-t}^t{\rm d}\tau H_{_{\rm I}}\right)
\vert 0\rangle },
\end{eqnarray}
\end{widetext}
where, in order to go from the first to the second line, we have used
Eq.~(\ref{eq:normomega}) and the normalization $\langle \Omega\vert
\Omega \rangle=1$ to express the term $\vert \langle 0\vert \Omega
\rangle \vert^2$. Eq.~(\ref{eq:gellmann}) is the Gell-Mann Law
equation~\cite{Peskin:1995ev}. It will be the basis for our
perturbative treatment of quantum mechanics. It plays the same role as
Eq.~(\ref{eq:defcorrelation}) since it allows us to perturbatively
evaluate the various correlation functions in terms of Feynman
diagrams (the Gell-Mann Law equation can be generalized to higher
correlation function if needed).

\subsection{Quantum Mechanics and Bubble Diagrams}
\label{subsec:bubbleqm}

In order to mimic the self-interacting field theory studied before, we
now consider a quartic oscillator in quantum
mechanics~\cite{abbott,minahan}, \ie a system where the potential is
given by $V(x)\sim x^2+x^4$. More precisely, this means that the
interacting Hamiltonian can be expressed as
\begin{equation}
H_{\rm int}=\frac{\lambda}{4!}\Phi^4,
\end{equation}
where we recall that $\Phi$ is given by Eq.~(\ref{eq:defphiqm}). Let
us then compute the two-point correlation function. For this purpose,
as explained before, we use the Gell-Mann Law equation. The numerator
reads
\begin{equation}
\left\langle 0\left\vert {\rm T}\left[ 
\Phi(t_2)\Phi(t_1)
\exp\left(-i\int {\rm d}\tau H_{_{\rm I}}\right)\right]\right\vert 0
\right\rangle,
\end{equation}
where, for simplicity, we do not write the limits (but there will be
reestablished when needed). It is also important to remember that the
field are written in the interacting picture (and, again for
simplicity we do not write the subscript ``${\rm I}$'', they can be
considered as free fields). Obviously, one cannot calculate exactly
this quantity so we do it perturbatively by expanding the
exponential. At first order in the coupling constant $\lambda $, we
obtain
\begin{widetext}
\begin{eqnarray}
\left\langle 0\left\vert {\rm T}\left\{
\Phi(t_2)\Phi(t_1)\left[1-i\frac{\lambda }{4!}
\int {\rm d}\tau \Phi^4(\tau)+\cdots \right]\right\}
\right\vert 0\right \rangle
&=&\left\langle 0\left\vert {\rm T}\left[
\Phi(t_2)\Phi(t_1)\right]\right\vert 0\right \rangle
\nonumber \\ & &
-i\frac{\lambda}{4!}\int {\rm d}\tau 
\left\langle 0\left\vert {\rm T}\left[
\Phi(t_2)\Phi(t_1)\Phi^4(\tau)\right]\right\vert 0\right \rangle
+\cdots ,\\
&=&D(t_1,t_2)
-i\frac{\lambda}{4!}\int {\rm d}\tau 
\left\langle 0\left\vert {\rm T}\left[
\Phi(t_2)\Phi(t_1)\Phi^4(\tau)\right]\right\vert 0\right \rangle
+\cdots .\nonumber \\
\end{eqnarray}
It should now be clear that the same expression can also be written in
terms on Feynman diagrams. One arrives at the following expression
\begin{equation}
\label{eq:gmlnume}
\left\langle 0\left\vert {\rm T}\left[ 
\Phi(t_2)\Phi(t_1)
\exp\left(-i\int {\rm d}\tau H_{_{\rm I}}\right)\right]\right\vert 0
\right\rangle= \qquad 
\begin{fmffile}{three}
\parbox{20mm}{
\begin{fmfgraph*}(40,30)
\fmfleft{i}
\fmfright{o}
\fmf{plain}{i,o}
\fmfdot{i}
\fmfdot{o}
\fmflabel{$t_1$}{i}
\fmflabel{$t_2$}{o}
\end{fmfgraph*}}
+\frac{3}{4!}
\qquad
\parbox{20mm}{
\begin{fmfgraph*}(40,30)
\fmfleft{i}
\fmfright{o}
\fmf{plain}{i,o}
\fmfdot{i}
\fmfdot{o}
\fmflabel{$t_1$}{i}
\fmflabel{$t_2$}{o}
\end{fmfgraph*}}
\hspace{-1.35cm}
\raisebox{-0.40cm}{
\begin{rotate}{90}
\parbox{25mm}{
\begin{fmfgraph*}(60,30)
\fmfleft{i}
\fmfright{o}
\fmf{phantom}{i,v}
\fmf{plain,left}{v,v}
\fmf{plain,right}{v,v}
\fmf{phantom}{v,o}
\fmfdot{v}
\end{fmfgraph*}}
\end{rotate}}
\qquad \qquad
+\frac{12}{4!}
\qquad
\parbox{20mm}{
\begin{fmfgraph*}(40,30)
\fmfleft{i}
\fmfright{o}
\fmf{plain}{i,v,v,o}
\fmfdot{i}
\fmfdot{v}
\fmfdot{o}
\fmflabel{$t_1$}{i}
\fmflabel{$t_2$}{o}
\end{fmfgraph*}}
\end{fmffile}
+{\cal O}\left(\lambda^2\right),
\end{equation}
\end{widetext}
where we have defined the following Feynman rules~\cite{abbott,minahan}
\begin{eqnarray}
\label{eq:fruleoneqm}
D(t_1,t_2) &\equiv & \qquad 
\begin{fmffile}{one}
\parbox{20mm}{
\begin{fmfgraph*}(40,30)
\fmfleft{i}
\fmfright{o}
\fmf{plain}{i,o}
\fmfdot{i}
\fmfdot{o}
\fmflabel{$t_1$}{i}
\fmflabel{$t_2$}{o}
\end{fmfgraph*}} 
\end{fmffile},
\\
\label{eq:fruletwoqm}
-i\lambda \int {\rm d}\tau &\equiv & 
\begin{fmffile}{two}
\parbox{20mm}{
\begin{fmfgraph*}(40,30)
\fmfleft{i1,i2}
\fmfright{o1,o2}
\fmf{plain}{i1,v,o2}
\fmf{plain}{i2,v,o1}
\fmfdot{v}
\end{fmfgraph*}}
\end{fmffile}.
\end{eqnarray}
Clearly, these Feynman rules are nothing but the one-dimensional
version of the standard Feynman rules~(\ref{eq:fruleone})
and~(\ref{eq:fruletwo}) of quantum field theory. The only difference
is that four dimensional space-time integrations are simply replaced
with one-dimensional time integration. In fact, in terms of diagrams,
the expressions in field theory or in quantum mechanics are exactly
similar. In particular, we see the appearance of diagrams with loops
and external legs (the tadpole) and also of diagrams with loops but no
external legs, our bubble diagrams. The advantage of our approach now
becomes clear. We are in a position where not only one can compute
these diagrams in quantum mechanics but we can also investigate their
deep meaning by comparing the result with the standard perturbative
theory of quantum mechanics where everything is well under control.

\par

Let us start with the tadpole diagram. Using
Eqs.~(\ref{eq:fruleoneqm}) and~(\ref{eq:fruletwoqm}), the
corresponding expression can be written as
\begin{eqnarray}
\begin{fmffile}{four}
\parbox{20mm}{
\begin{fmfgraph*}(40,30)
\fmfleft{i}
\fmfright{o}
\fmf{plain}{i,v,v,o}
\fmfdot{i}
\fmfdot{o}
\fmfdot{v}
\fmflabel{$t_1$}{i}
\fmflabel{$t_2$}{o}
\end{fmfgraph*}}
\end{fmffile}
&=&-i\lambda\int {\rm d}\tau D(t_1,\tau)D(\tau,\tau)D(\tau,t_2)
\nonumber \\
\\
&=& -\frac{i\lambda}{8\omega ^3}\int {\rm d}\tau {\rm e}^{-i\omega 
\vert t_1-\tau\vert -i\omega \vert \tau -t_2\vert}\\
&=& -\frac{i\lambda}{4\omega ^2}D(t_2,t_1)\left(\vert t_2-t_1\vert 
-\frac{i}{\omega}\right).\nonumber  \\
\end{eqnarray}
The most striking difference with the case of quantum field theory is
that the tadpole diagram is now finite. In some sense, this is the
signal that ordinary quantum mechanics does not need renormalization:
the perturbative approach leads to finite result. Technically, this is
due to the fact that we deal with one-dimensional integrals that have
better convergence properties. Physically, this is due to the fact
that a single quantum-mechanical oscillator leads to finite
predictions while an infinite collections of such systems apparently
lead to infinite physical quantities.

\par

Now let us calculate the ``bubble diagram'' present in the second term
of the above expansion~(\ref{eq:gmlnume}). Having established that the
tadpole diagram is now finite, it is particularly interesting to see
what happens for this type of diagram. Using again the Feynman
rules~(\ref{eq:fruleoneqm}) and~(\ref{eq:fruletwoqm}), it is
straightforward to establish that
\begin{eqnarray}
\begin{fmffile}{fifth}
\hspace{-2.2cm}
\raisebox{-0.95cm}{
\begin{rotate}{90}
\parbox{25mm}{
\begin{fmfgraph*}(60,30)
\fmfleft{i}
\fmfright{o}
\fmf{phantom}{i,v}
\fmf{plain,left}{v,v}
\fmf{plain,right}{v,v}
\fmf{phantom}{v,o}
\fmfdot{v}
\end{fmfgraph*}}
\end{rotate}}
\end{fmffile}
\qquad &=&-i\lambda\int D^2(\tau,\tau){\rm d}\tau\\
\qquad &=& -i\frac{\lambda}{4\omega ^2}\int {\rm d}\tau=\infty.
\end{eqnarray}
Therefore, even in quantum mechanics, the bubble diagram remains
divergent. This shows that the divergent nature of the loop and bubble
diagrams is, in some sense, different. They are both divergent in
quantum field theory but one (loop diagrams with external legs)
becomes finite in the limit of quantum mechanics while the other
(bubble diagram) remains infinite.

\par

But this now leads to the following question. We know from the
standard perturbation theory of quantum mechanics that all the
physical predictions are finite, even without
renormalization. However, using the other approach based on the
Feynman diagrams, we have just seen that it leads to terms that can be
divergent. How these two facts can be consistent with each other?  The
answer is of course that, as in field theory, the bubble diagrams
exactly cancel out in the calculation of the correlation functions.
This can easily be seen if one notices that
\begin{widetext}
\begin{align}
\begin{fmffile}{six}
\parbox{20mm}{
\begin{fmfgraph*}(40,30)
\fmfleft{i}
\fmfright{o}
\fmf{plain}{i,o}
\fmfdot{i}
\fmfdot{o}
\fmflabel{$t_1$}{i}
\fmflabel{$t_2$}{o}
\end{fmfgraph*}}
+\frac{3}{4!}
\qquad
\parbox{20mm}{
\begin{fmfgraph*}(40,30)
\fmfleft{i}
\fmfright{o}
\fmf{plain}{i,o}
\fmfdot{i}
\fmfdot{o}
\fmflabel{$t_1$}{i}
\fmflabel{$t_2$}{o}
\end{fmfgraph*}}
\hspace{-1.35cm}
\raisebox{-0.40cm}{
\begin{rotate}{90}
\parbox{25mm}{
\begin{fmfgraph*}(60,30)
\fmfleft{i}
\fmfright{o}
\fmf{phantom}{i,v}
\fmf{plain,left}{v,v}
\fmf{plain,right}{v,v}
\fmf{phantom}{v,o}
\fmfdot{v}
\end{fmfgraph*}}
\end{rotate}}
\qquad \qquad
+\frac{12}{4!}
\qquad
\parbox{20mm}{
\begin{fmfgraph*}(40,30)
\fmfleft{i}
\fmfright{o}
\fmf{plain}{i,v,v,o}
\fmfdot{i}
\fmfdot{v}
\fmfdot{o}
\fmflabel{$t_1$}{i}
\fmflabel{$t_2$}{o}
\end{fmfgraph*}}
\end{fmffile}
&=\left(\qquad
\begin{fmffile}{seven} 
\parbox{20mm}{
\begin{fmfgraph*}(40,30)
\fmfleft{i}
\fmfright{o}
\fmf{plain}{i,o}
\fmfdot{i}
\fmfdot{o}
\fmflabel{$t_1$}{i}
\fmflabel{$t_2$}{o}
\end{fmfgraph*}}
+
\frac{12}{4!}
\qquad
\parbox{20mm}{
\begin{fmfgraph*}(40,30)
\fmfleft{i}
\fmfright{o}
\fmf{plain}{i,v,v,o}
\fmfdot{i}
\fmfdot{o}
\fmfdot{v}
\fmflabel{$t_1$}{i}
\fmflabel{$t_2$}{o}
\end{fmfgraph*}}
\end{fmffile}+\cdots
\right)\\
& \times 
\biggl(1+\frac{3}{4!}
\qquad \quad 
\begin{fmffile}{height}
\raisebox{-0.95cm}{
\begin{rotate}{90}
\parbox{25mm}{
\begin{fmfgraph*}(60,30)
\fmfleft{i}
\fmfright{o}
\fmf{phantom}{i,v}
\fmf{plain,left}{v,v}
\fmf{plain,right}{v,v}
\fmf{phantom}{v,o}
\fmfdot{v}
\end{fmfgraph*}}
\end{rotate}}
\end{fmffile}
\qquad \biggr)+{\cal O}\left(\lambda^2\right),
\end{align}
and that 
\begin{equation}
\left \langle 0\left \vert \exp\left(-i\int {\rm d}\tau H_{_{\rm I}}\right)
\right \vert 0\right \rangle=
1+\frac{3}{4!}
\qquad \quad 
\begin{fmffile}{nine}
\raisebox{-0.95cm}{
\begin{rotate}{90}
\parbox{25mm}{
\begin{fmfgraph*}(60,30)
\fmfleft{i}
\fmfright{o}
\fmf{phantom}{i,v}
\fmf{plain,left}{v,v}
\fmf{plain,right}{v,v}
\fmf{phantom}{v,o}
\fmfdot{v}
\end{fmfgraph*}}
\end{rotate}}
\end{fmffile}
\qquad  +{\cal O}\left(\lambda^2\right).
\end{equation}
Therefore (as expected) the infinities of the bubble diagram never
play a role in ordinary quantum mechanics. As a consequence, the
two-point correlation function is given by
\begin{equation}
\langle \Omega \vert \Phi(t_2)\Phi(t_1)\vert \Omega 
\rangle =
\qquad 
\begin{fmffile}{ten}
\parbox{20mm}{
\begin{fmfgraph*}(40,30)
\fmfleft{i}
\fmfright{o}
\fmf{plain}{i,o}
\fmfdot{i}
\fmfdot{o}
\fmflabel{$t_1$}{i}
\fmflabel{$t_2$}{o}
\end{fmfgraph*}}
+\frac{12}{4!}
\qquad
\parbox{20mm}{
\begin{fmfgraph*}(40,30)
\fmfleft{i}
\fmfright{o}
\fmf{plain}{i,v,v,o}
\fmfdot{i}
\fmfdot{v}
\fmfdot{o}
\fmflabel{$t_1$}{i}
\fmflabel{$t_2$}{o}
\end{fmfgraph*}}
\end{fmffile}
+{\cal O}\left(\lambda^2\right),
\end{equation}
\end{widetext}
which is perfectly finite. Again, in terms of Feynman diagrams the
above equation is strictly equivalent to the corresponding one in
quantum field theory, see Eq.~(\ref{eq:2pointgraph}).

\subsection{Quantum Mechanics and Vacuum Energy}
\label{subsec:vacqm}

Let us now discuss how the vacuum energy is evaluated in ordinary
quantum mechanics. The calculation is straightforward when there is no
interaction and, from Eq.~(\ref{eq:hamiltonian}), one obtains
\begin{equation}
\langle H\rangle =\frac{\omega }{2}.
\end{equation}
This is of course nothing but the ground state energy of an harmonic
oscillator. It cannot be zero because, due the Heisenberg uncertainty
principle, the kinetic and potential energy cannot vanish at the same
time. It can also be rewritten as
\begin{equation}
\label{eq:relaHdzero}
\langle H\rangle =\omega ^2D(0),
\end{equation}
where we have used Eq.~(\ref{eq:propaqm}). This exactly corresponds to
Eq.~(\ref{eq:rhopropa}) with $m=\omega$ and $4\rightarrow 1$ (since we
now work in one dimension rather than in four), except that there is a
sign difference which comes from the fact that, in field theory, we
have used the signature (- + + +). Of course, the big difference is
that $D(0)$ is a finite quantity in quantum mechanics. In some sense,
the vacuum energy problem appears to be very simple, almost
trivial. The ground state of a quantum-mechanical oscillator is finite
but a quantum field is a system which can be described by an infinite
collection of harmonic oscillators and, as a consequence, its ground
state energy is infinite.

\par

Let us now consider a case where an interaction is present as we did
in Sec.~\ref{subsec:interactingvacuum}. For simplicity, we consider
again the case of a quartic oscillator with $V_{\rm
  int}=\lambda\Phi^4/4!$. Upon using the standard perturbation theory
in quantum mechanics, one can calculate the displacement of the energy
levels. They are given by the standard formula~\cite{tannoudji2006}
\begin{equation}
E_n=E_n^0+\langle n\vert V_{\rm int}\vert n\rangle ,
\end{equation}
where $\vert n\rangle $ are the unperturbed eigenvectors. It is easy
to show that they can be expressed as
\begin{equation}
\label{eq:spectrum}
\Delta E_n\equiv E_n-E_n^0=\frac{\lambda}{32\omega ^2}\left(2n^2+2n+1\right),
\end{equation}
which means that, for the first two levels, the shift in energy is
given by
\begin{eqnarray}
\label{eq:shiftenergy}
\Delta E_0 &=& \frac{\lambda}{32\omega ^2}, \quad
\Delta E_1 = \frac{5\lambda}{32\omega ^2}.
\end{eqnarray}
We can now use the other method, based on the Feynman diagrams. It is easy 
to obtain that
\begin{eqnarray}
\Delta E_0&=&
\frac{\lambda}{4!}\left\langle \Phi^4\right\rangle 
=\frac{3\lambda}{4!}\left\langle \Phi^2\right \rangle^2
=\frac{\lambda}{8}D^2(0)
\nonumber \\
&=&\frac{i}{8\int {\rm d}t}
\quad \quad \quad
\begin{fmffile}{norm}
\raisebox{-0.95cm}{
\begin{rotate}{90}
\parbox{25mm}{
\begin{fmfgraph*}(61,30)
\fmfleft{i}
\fmfright{o}
\fmf{phantom}{i,v}
\fmf{plain,left}{v,v}
\fmf{plain,right}{v,v}
\fmf{phantom}{v,o}
\fmfdot{v}
\end{fmfgraph*}}
\end{rotate}}
\end{fmffile}.
\end{eqnarray}
Clearly, we have just obtained the strict equivalent of
Eq.~(\ref{eq:vacuumdiagram}), the term $\int {\rm d}^4x$ being
replaced by $\int {\rm d}t$ as could have been guessed in the context
of quantum mechanics. Then, using the fact that $D(0)=1/(2\omega)$,
one obtains that
\begin{equation}
\Delta E_0=\frac{\lambda}{32\omega ^2},
\end{equation}
that is to say we recover the result established before, see
Eq.~(\ref{eq:shiftenergy}). Therefore, we have shown explicitly that
the two approaches lead to the same result. It is also interesting to
notice that $\Delta E_0$ is given by a bubble diagram but that this
divergent graph is canceled by the infinite term $\int {\rm d}t$ so
that a finite quantity is left.

\par

The previous considerations also lead to the following question:
before we had to deal with the renormalization of the mass. How does
this aspect appear in the present context? In this case, we deal with
the same diagrams, namely
\begin{widetext}
\begin{eqnarray}
\left \langle \Omega\left \vert 
T\left[\Phi(t_1)\Phi(t_2)\right]\right\vert \Omega \right \rangle
&=&\qquad 
\begin{fmffile}{ruleone}
\parbox{20mm}{
\begin{fmfgraph*}(40,30)
\fmfleft{i}
\fmfright{o}
\fmf{plain}{i,o}
\fmfdot{i}
\fmfdot{o}
\fmflabel{$x_1$}{i}
\fmflabel{$x_2$}{o}
\end{fmfgraph*}} 
\end{fmffile}
+\frac12 \qquad
\begin{fmffile}{tadpole}
\parbox{20mm}{
\begin{fmfgraph*}(40,30)
\fmfleft{i}
\fmfright{o}
\fmf{plain}{i,v,v,o}
\fmfdot{i}
\fmfdot{o}
\fmfdot{v}
\fmflabel{$x_1$}{i}
\fmflabel{$x_2$}{o}
\end{fmfgraph*}}
\end{fmffile}
+\cdots ,
\end{eqnarray}
or, in terms of explicit expressions [compare with
Eqs.~(\ref{eq:2pointgraph}) and~(\ref{eq:propaand2point})]
\begin{eqnarray}
\left \langle \Omega\left \vert 
T\left[\phi(t_1)\phi(t_2)\right]\right\vert \Omega \right \rangle
&=& 
D_{_{\rm F}}(t_1-t_2)
-\frac{i\lambda}{2} \int{\rm d}\tau D(t_1-\tau)D_{_{\rm F}}(0)D_{_{\rm F}}(\tau-t_2)
\\
&=&D_{_{\rm F}}(t_1-t_2)
+\frac{i\lambda}{2}D_{_{\rm F}}(0)\frac{1}{2\pi}
\int {\rm d}E\frac{{\rm e}^{-iE(t_1-t_2)}}{(E^2-\omega^2)^2},
\end{eqnarray}
\end{widetext}
and this leads to 
\begin{equation}
\omega^2_{\rm ren}=\omega^2+\frac{\lambda}{2}D(0),
\end{equation}
which is exactly Eq.~(\ref{eq:renormmassinter}) (again, there is a
sign difference which originates from the sign difference in the
propagator, see above). In quantum mechanics, the propagator is finite
and therefore
\begin{equation}
\omega^2_{\rm ren}=\omega^2+\frac{\lambda}{4\omega},
\end{equation}
or
\begin{equation}
\omega_{\rm ren}=\omega+\frac{\lambda}{8\omega^2}.
\end{equation}
Then, one can express the spectrum in terms of the renormalized
``mass'' $\omega_{\rm ren}$ as we have done in quantum field theory
although this is not necessary here since the result is anyway
finite. Straightforward calculations lead to
\begin{equation}
E_n=\omega_{\rm ren}\left(n+\frac12\right)
-\frac{\lambda}{32\omega^2_{\rm ren}}\left(2n^2-2n-1\right).
\end{equation}
It does not come as a surprised that, as a function of the
renormalized mass, the spectrum is a different function of $n$, see
Eq.~(\ref{eq:spectrum}). However, from the above expression one also
obtains
\begin{equation}
E_1-E_0=\omega_{\rm ren}.
\end{equation}
In presence of an interaction, the difference between the first
excited state and the fundamental level has been ``modified'' (or,
rather, is expressed differently). But, observationally, we define or
measure the spectrum with respect to $E_1-E_0$. In other words what we
define as $\omega$ is in fact $E_1-E_0$. Therefore, renormalizing
$\omega $ consists in preserving this definition even in presence of
an interaction. Obviously, $E_1-E_0$ is not the bare $\omega$ (the one
which appears in the Lagrangian) when the an-harmonic term is
present. What we do here is to define $\omega $ as the observed one,
namely as $E_1-E_0$, and then calculate the spectrum in terms of that
observational quantity. What is done in quantum field theory is
clearly exactly similar to the above described procedure except of
course that the bare quantities are in fact infinite because $D_{_{\rm
    F}}(0)$ is infinite. Therefore, this exercise illustrates nicely
the deep meaning of renormalization.

\par

This concludes our discussion of the vacuum energy density in terms of
Feynman diagrams. We have shown that the cosmological constant is in
fact given in terms of very peculiar graphs, the so-called bubble
diagrams. These diagrams are, in a sense, more divergent than the
usual loop diagrams because they remain infinite even in the limit of
quantum mechanics contrary to the last ones. In non gravitational
physics, this is not a problem because the bubble diagrams always
cancel out in the equations describing observable quantities. When the
gravitational field is turned on, we face the tasks of renormalizing
these diagrams which is more difficult than in the standard situation
because of their bad behavior mentioned before. Given that
astrophysical observations seem to indicate that the vacuum energy is
non vanishing (see below), the details of the renormalizing procedure
become a crucial issue. Not only we have to extract a finite quantity
from a divergent graph but this finite quantity must be in agreement
with the observations. As argued before, the technique which consists
in imposing a spatial cut-off is certainly not consistent (and, as
well-known, gives a very large contribution to the vacuum energy). A
more reasonable approach seems to be dimensional regularization since
this satisfies Lorentz invariance and is therefore consistent with the
vacuum equation of state. As shown before, this typically leads to
Eq.~(\ref{eq:rhovacrenorm}). In the next section, we present a new
argument, based on the Gaussian effective potential, which seems to
support this conclusion.

\section{The Gaussian Effective Potential}
\label{sec:GEP}

The Gaussian effective potential is a non perturbative approach to
quantum field
theory~\cite{Stevenson:1984rt,Stevenson:1985zy,Stevenson:1985kr,Stevenson:1986na,Stevenson:1986nb,Stevenson:1986bq,Hajj:1987gk,Stevenson:1986sb,IbanezMeier:1992fp}. Originally
developed in the context of quantum mechanics, it is has been
generalized to field theory. Much less known is the fact that the
Gaussian effective potential method has something to say about the
vacuum energy problem and the aim of this section is to present these
considerations~\cite{Stevenson:1985zy}. In the next subsection, we
quickly recall the main idea and how this approach can be applied to
ordinary quantum mechanics. Then, we explain how it can be implemented
in quantum field theory and, finally, we discuss its implication for
the cosmological constant problem.

\subsection{The Gaussian Effective Potential in Quantum Mechanics}
\label{subsec:GEPQM}

The main idea underlying the Gaussian effective potential approach is
the following one~\cite{Stevenson:1984rt}. At the classical level, a
system is described by a potential. Once this one is specified, one
can compute the evolution of the system. When quantum effects are
taken into account, the behavior of the system will be modified and,
by means of the formalism of quantum mechanics, one can work out the
corresponding physical predictions. The idea of the Gaussian effective
potential is to find an effective potential which, at the same time,
can be used as if it were a ``classical'' potential (that is to say
can be used in the framework of Newtonian dynamics) and takes into
account the quantum effects. At first sight, this approach seems
equivalent to what is known in the literature as the effective
potential~\cite{Coleman:1973jx,Weinberg:1987vp}. In this method, one
requires the wave-function mean value to be centered at a fixed point
$\Phi_0$. Then, the effective potential gives the exact ground state
energy of the system in presence of quantum corrections. However, in
some situation, the result can be quite artificial. Consider for
instance a situation where the wave-function possesses two large peaks
on each side of $\Phi_0$. In such a case, the effective potential does
not necessarily lead to a correct description of what happens in the
vicinity of $\Phi_0$ because its shape could just be determined by
some average procedure between the two distant peaks. Typically, one
would obtain a convex potential centered at $\Phi_0$ while our
physical intuition rather indicates that the effective potential
should be a double-well potential, the two wells corresponding to the
two peaks of the wave-function.  Another problem is that the standard
effective potential is usually computed perturbatively in
$\hbar$. Clearly, when the quantum corrections become important, this
approach breaks down. Since the Gaussian effective potential does not
suffer from the two above mentioned drawbacks, we will use it in this
review. When possible, we will discuss the difference between the two
approaches in order to stress their differences and
similarities~\cite{Curtright:1983cf}.

\par

We now turn to the precise definition of the Gaussian effective
potential. This definition is based on a variational principle. Here
we consider a one-dimensional system as we have already done in
Sec.~\ref{subsec:qmqft}.  In particular, this means that the scalar
field is given by Eq.~(\ref{eq:defphiqm}). Then, the Gaussian
effective potential is defined by~\cite{Stevenson:1984rt}
\begin{equation}
\label{eq:defgepqm}
V_{_{\rm G}}\left(\Phi_0\right)=\min _{\Omega}\left \langle \Psi\left \vert H
\right \vert \Psi\right \rangle, 
\end{equation}
where $H$ is the Hamiltonian of the system and where the wave-function
is taken to be a simple Gaussian (hence the name Gaussian effective
potential)
\begin{equation}
\label{eq:trialwave}
\Psi\left(\Phi\right)=\left(\frac{\Omega}{\hbar \pi}\right)^{1/4}
{\rm e}^{-\Omega \left(\Phi-\Phi_0\right)^2/(2\hbar)}.
\end{equation}
The parameter $\Omega $ is related to the width of the
wave-function. Here the idea is to find the value of $\Omega $ which
minimizes the energy of the system. In practice, the mean value of the
Hamiltonian that appears in Eq.~(\ref{eq:defgepqm}) can be expressed
as
\begin{equation}
\label{eq:meanH}
\left \langle H\right \rangle 
=\int _{-\infty}^{\infty}{\rm d}\Phi \Psi^*\left(\Phi\right)
\left[-\frac{\hbar^2}{2}\frac{{\rm d}^2}{{\rm d}\Phi^2}+V\left(\Phi\right)
\right]\Psi\left(\Phi\right).
\end{equation}
To illustrate how the calculation of the Gaussian effective potential
works on a concrete example, we choose the following potential
\begin{equation}
V\left(\Phi\right)=\frac12m^2\Phi^2+\lambda \Phi^4,
\end{equation} 
that is to say the potential of an an-harmonic oscillator. Another
motivation for this choice is that the comparison with field theory is
easy since the prototypical model possesses the same potential. Upon
using Eq.~(\ref{eq:trialwave}) in Eq.~(\ref{eq:meanH}), one arrives at
\begin{eqnarray}
\label{eq:meanHquartic}
\left \langle H\right \rangle \left(\Phi_0,\Omega\right)
&=&\frac{\hbar \Omega }{4}+\frac12 m^2\left(\Phi_0^2+\frac{\hbar}{2\Omega}
\right)
\nonumber \\ & &
+\lambda\left[\Phi_0^4+6\Phi_0^2\frac{\hbar}{2\Omega}
+\frac{3\hbar^2}{(2\Omega)^2}\right].
\end{eqnarray}
We must now determine the optimal $\Omega$. As explained above, this
is achieved by requiring $\partial \left \langle H\right \rangle
/\partial \Omega=0$, which leads to the equation
\begin{equation}
\label{eq:gapqm}
\Omega^3-\left(m^2+12\lambda \Phi_0^2\right)\Omega-6\hbar \lambda =0.
\end{equation}
This equation allows us to determine $\Omega$ in terms of $\Phi_0$. In
the following, we write this result as $\Omega_0\equiv
\Omega(\Phi_0)$. Upon using the above relation in the expression of
$\langle H\rangle$, one obtains the following expression for the
Gaussian effective potential
\begin{equation}
\label{eq:gepaho}
V_{_{\rm G}}\left(\Phi_0\right)
=\frac12m^2\Phi_0^2+\lambda \Phi_0^4+\frac{\hbar\Omega _0}{2}
-\frac{3\hbar^2\lambda}{4\Omega_0}.
\end{equation}
Of course, as already emphasized before, one should not forget that
$\Omega_0$ is a function of $\Phi_0$.

\par

A complete study of the above potential has been provided in
Ref.~\cite{Stevenson:1984rt} and we will not repeat this
analysis. Here, we just want to illustrate that the Gaussian effective
potential is an accurate method to estimate the ground state of a
system (even in the strong coupling limit) since this is directly
relevant for the cosmological constant problem. Let us consider the
case where $m^2>0$ (the case of the double-well potential, reminiscent
of symmetry breaking in field theory has been studied in
Ref.~\cite{Stevenson:1984rt}). The ground state of the an-harmonic
oscillator is given by Eq.~(\ref{eq:gepaho})
\begin{equation}
\label{eq:groundaho}
V_{_{\rm G}}\left(\Phi_0=0\right)=\frac{\hbar\Omega_0}{2}
\left(1-\frac{3}{2}\frac{\hbar \lambda}{\Omega _0^3}\right),
\end{equation}
where $\Omega _0$ is solution of 
\begin{equation}
\label{eq:cubicground}
\Omega^3-m^2\Omega-6\hbar \lambda=0.
\end{equation} 
The discriminant of this cubic equation can be expressed as
\begin{equation}
\label{eq:discri}
\Delta \left(\Phi_0=0\right)=4m^6\left(1-\frac{972}{16}\frac{1}{\xi^6}
\right),
\end{equation}
where the quantity $\xi $ in the above relation is given by
\begin{equation}
\xi^2\equiv \frac{m^2}{(2\lambda)^{2/3}}.
\end{equation}
The awkward coefficient $972/16$ in Eq.~(\ref{eq:discri}) originates
from our will to work with the definition of $\xi$ used in
Ref.~\cite{Stevenson:1984rt}. The cubic
equation~(\ref{eq:cubicground}) can be explicitly solved. We deal with
different branches according to the sign of the discriminant. This
sign is determined by $\xi_{\rm lim}=(972/16)^{1/6}\simeq 1.9827$. If
$\xi <\xi_{\rm lim}$ ($\xi=0$ represent the quartic oscillator) we are
in the strong coupling regime which is of particular interest in order
to demonstrate the usefulness of the Gaussian effective
potential. Therefore, let us focus on this situation. In this case,
the exact solution of Eq.~(\ref{eq:cubicground}) can be written as
\begin{eqnarray}
\Omega_0&=& m\left[\frac{3}{2\xi^3}-\frac{1}{27}
\sqrt{27\left(\frac{972}{16}\frac{1}{\xi^6}-1\right)}\right]^{1/3}
\nonumber \\ & & 
+m\left[\frac{3}{2\xi^3}+\frac{1}{27}
\sqrt{27\left(\frac{972}{16}\frac{1}{\xi^6}-1\right)}\right]^{1/3}.
\end{eqnarray}
Together with Eq.~(\ref{eq:groundaho}), the above equation gives an
excellent approximation for the ground state. For instance, if $m=1$
and $\lambda =10$, then $\xi^2\simeq (1/20)^{2/3}\simeq 0.1357$ and
$\Omega_0=4$. As a consequence, this gives $V_{_{\rm
    G}}\left(\Phi_0=0\right)\simeq 1.53$, a result that is accurate at
$\simeq 1.75\%$ according to Ref.~\cite{Stevenson:1984rt}.

\par

This simple example has shown that the Gaussian effective potential
can be an efficient tool to calculate the ground state of a quantum
system even in a non perturbative regime. Therefore, it appears as an
interesting approach for the vacuum energy problem. However, before
turning to the calculation of the Gaussian effective potential in
field theory, it is interesting to compare it with the standard
effective potential in more detail. This is the purpose of the next
subsection.

\subsection{Comparison with the One Loop Effective Potential}

In order to explain how the effective potential is obtained, we need
to quickly return to the basics of quantum field theory. In
Eq.~(\ref{eq:defZ}), we have defined the generating functional
$Z[J]$. Let us now define another generating functional $W[J]$
by~\cite{Peskin:1995ev,Greiner:1996zu}
\begin{equation}
\label{eq:defW}
Z[J]={\rm e}^{iW[J]}.
\end{equation}
Following the procedure in Eq.~(\ref{eq:defcorrelation}), one can
define new correlation functions $G^{(n)}_{\rm c}$ according to
\begin{eqnarray}
G^{(n)}_{\rm c} &=& \left(\frac{1}{i}\right)^{n-1}
\frac{\delta^n W\left[J\right]}{\delta J(x_1)
\cdots \delta J(x_n)}\biggl \vert _{J=0}.
\end{eqnarray}
As is well-known, these functions represent in fact the connected
$n$-points functions. Then, one introduces the classical field
$\Phi_{\rm c}$
\begin{equation}
\label{eq:defphic}
\Phi_{\rm c}\equiv \frac{\delta W[J]}{\delta J(x)}\biggl \vert_{J=0},
\end{equation}
where we remind that $J(x)$ is a source. This definition can be easily
justified. Indeed, from Eqs.~(\ref{eq:defZ}) and~(\ref{eq:defW}), one
sees that
\begin{eqnarray}
\Phi_{\rm c}&=&\frac{1}{iZ[J]}\frac{\delta Z[J]}{\delta J(x)}
\\ 
&=&\frac{{\cal N}\int {\cal D}\Phi \exp\left\{i\int {\rm d}^4x
\left[{\cal L}+J(x)\Phi(x)\right]\right\}\Phi(x)}
{{\cal N}\int {\cal D}\Phi \exp\left\{i\int {\rm d}^4x
\left[{\cal L}+J(x)\Phi(x)\right]\right\}} \nonumber \\
&=& \frac{\left\langle \Omega \left\vert \Phi(x)\right \vert
\Omega \right\rangle }
{\left \langle \Omega \vert \Omega \right \rangle}
\,.
\end{eqnarray}
Therefore, we see that the classical field is in fact the vacuum
expectation value of the field operator in the vacuum state of the
theory. When the source vanishes, the classical field does not
necessarily goes to zero, a typical and well-known example being of
course symmetry breaking.

\par

The next step consists in introducing a new generating function, the
so-called effective action, defined by
\begin{equation}
\label{eq:defeffectiveaction}
\Gamma [\Phi_{\rm c}]=W[J]-\int {\rm d}^4xJ(x)\Phi_{\rm c}(x).
\end{equation}
Clearly, $\Gamma [\Phi_{\rm c}]$ does not depend on the source since
$\delta \Gamma/\delta J=0$. Let us now calculate the functional
derivative of $\Gamma $ with respect with the classical field. This
gives
\begin{eqnarray}
\frac{\delta \Gamma [\Phi_{\rm c}]}{\delta \Phi_{\rm c}(x)}
&=& \frac{\delta W}{\delta \Phi_{\rm c}(x)}
-\int {\rm d}^4y\frac{\delta J(y)}{\delta \Phi_{\rm c}(x)}
\Phi_{\rm c}(y)
\nonumber \\ & & -
\int {\rm d}^4yJ(y)\frac{\delta \Phi_{\rm c}(y)}{\delta \Phi_{\rm c}(x)}
\\
&=&\int {\rm d}^4y \frac{\delta W}{\delta J(y)}
\frac{\delta J(y)}{\delta \Phi_{\rm c}(x)}
-\int {\rm d}^4y\frac{\delta J(y)}{\delta \Phi_{\rm c}(x)}
\Phi_{\rm c}(y)
\nonumber \\ & &
-J(x)= -J(x).
\end{eqnarray}
In order to interpret these equations and to better understand their
meaning, it is interesting to apply them to the case of a free theory
where all the calculations can be carried out explicitly. From
Eq.~(\ref{eq:freeZ}), we immediately see that (the subscript zero
indicates that the calculation is performed for the free theory)
\begin{equation}
\label{eq:freeW}
iW_0[J]=-\frac12\int {\rm d}^4x\, {\rm d}^4y J(x)D_{_{\rm F}}(x-y)J(y).
\end{equation}
Upon using Eq.~(\ref{eq:defphic}), one obtains
\begin{equation}
\label{eq:freephic}
\Phi_{\rm c}(z)=-\frac{1}{i}\int {\rm d}^4x\, D_{_{\rm F}}(z-x)J(x).
\end{equation}
But $iD_{_{\rm F}}$ is the Green function of the Klein-Gordon
operator. As a consequence, the above equation implies that
\begin{equation}
\label{eq:eomphic}
\left(\eta ^{\mu \nu}\partial _{\mu }\partial _{\nu}-m^2\right)
\Phi_{\rm c}(x)=J(x),
\end{equation}
\ie the classical field obeys the classical equation of motion (hence
its name). Then, using Eqs.~(\ref{eq:freeW}) and~(\ref{eq:freephic})
into the definition~(\ref{eq:defeffectiveaction}), the effective
action can be expressed as
\begin{equation}
\Gamma_0[\Phi_{\rm c}]=\frac{1}{2i}
\int {\rm d}^4x{\rm d}^4y J(x)D_{_{\rm F}}(x-y)J(y).
\end{equation}
The next step is to use the equation of motion of the classical
field~(\ref{eq:eomphic}) to rewrite the effective action as
\begin{eqnarray}
\Gamma_0[\Phi_{\rm c}]&=&\frac{1}{2i}
\int{\rm d}^4x\, {\rm d}^4y D_{_{\rm F}}(x-y)
\left(\eta ^{\mu \nu}\partial _{\mu }^x\partial _{\nu}^x-m^2\right)
\Phi_{\rm c}(x)\nonumber \\ & & \times 
\left(\eta ^{\mu \nu}\partial _{\mu }^y\partial _{\nu}^y-m^2\right)
\Phi_{\rm c}(y),
\end{eqnarray}
where $\partial_{\mu}^{x,y}$ means a partial derivative with respect
to the coordinates at point $x$ and $y$. Then, integrating by part the
$x$ dependent terms in the above integral leads to
\begin{eqnarray}
\Gamma_0[\Phi_{\rm c}]&=&\frac{1}{2i}
\int{\rm d}^4x{\rm d}^4y \Phi_{\rm c}(x)
\left(\eta ^{\mu \nu}\partial _{\mu }^x\partial _{\nu}^x-m^2\right)
D_{_{\rm F}}(x-y)
\nonumber \\ & & \times 
\left(\eta ^{\mu \nu}\partial _{\mu }^y\partial _{\nu}^y-m^2\right)
\Phi_{\rm c}(y).
\end{eqnarray}
Finally, using the fact that $D_{_{\rm F}}/i$ is the Green function
satisfying $-\eta ^{\mu \nu}\partial _{\mu }\partial _{\nu}D_{_{\rm
    F}}+m^2D_{_{\rm F}}=\delta$, one obtains
\begin{eqnarray}
\Gamma_0[\Phi_{\rm c}]&=&\frac{1}{2}
\int{\rm d}^4x \Phi_{\rm c}(x)
\left(\eta ^{\mu \nu}\partial _{\mu }\partial _{\nu}-m^2\right)
\Phi_{\rm c}(x) \nonumber \\
&=& -\frac{1}{2}
\int{\rm d}^4x \left(\eta ^{\mu \nu}\partial _{\mu }\Phi_{\rm c}
\partial _{\nu }\Phi_{\rm c}+m^2\Phi_{\rm c}\right),
\end{eqnarray}
and one recovers the action of a free scalar field theory. In presence
of interactions, $\Gamma [\Phi_{\rm c}]$ will no longer agree with the
classical action. The quantum corrections will transform it into a
complicated non local functional. In this situation, one typically
expects an expression that can be written as
\begin{eqnarray}
\label{eq:effectiveaction}
  \Gamma [\Phi_{\rm c}]
  &=& -
  \int{\rm d}^4x \biggl[\frac12 A(\Phi_{\rm c})
  \eta ^{\mu \nu}\partial _{\mu }\Phi_{\rm c}
\partial _{\nu }\Phi_{\rm c}
\nonumber \\ &+& 
V_{\rm eff}\left(\Phi_{\rm c}\right)
+B(\Phi_{\rm c})\left(\eta ^{\mu \nu}\partial _{\mu }\Phi_{\rm c}
\partial _{\nu }\Phi_{\rm c}\right)^4
+\cdots \biggr],
\nonumber \\
\end{eqnarray}
where $A(\Phi_{\rm c})$ and $B(\Phi_{\rm c})$ are functions (not
functionals) of $\Phi_{\rm c}$ and the dots represent higher
derivative terms. Clearly, the previous considerations justify the
name ``effective action'' for $\Gamma [\Phi_{\rm c}]$.

\par

Another interesting aspect is that the effective action is also the
generating functional of the irreducible correlation functions $\Gamma
^{(n)}$ (also known as proper vertices). In other words, one has
\begin{eqnarray}
  \Gamma ^{(n)}(x_1, \cdots, x_n) &=& 
  \frac{\delta^n \Gamma\left[\Phi_{\rm c}\right]}{\delta \Phi_{\rm c}(x_1)
    \cdots \delta \Phi_{\rm c}(x_n)}\biggl \vert _{\Phi_{\rm c}=0}.
\end{eqnarray}
Equivalently, one can express the effective action as a Volterra
expansion, namely
\begin{eqnarray}
\Gamma[\Phi_{\rm c}]&=&\sum_{n=0}^{\infty}\frac{1}{n!}
\int {\rm d}^4x_1 \cdots {\rm d}^4x_n \Gamma ^{(n)}(x_1, \cdots, x_n)
\nonumber \\ & & \times 
\Phi_{\rm c}(x_1) \cdots \Phi_{\rm c}(x_n),
\end{eqnarray}
which is nothing but another way to write the
functional~(\ref{eq:effectiveaction}).

\par

Now, suppose that we are interested in determining the effective
action concretely. It is clear that this is a complicated
calculation. As a first step, it is worth calculating the effective
potential in Eq.~(\ref{eq:effectiveaction}). It is true that it does
not represent all the information contained in the effective action
but this would allow us to determine the location of the minimum of
the system in presence of quantum corrections. Therefore, this is an
interesting quantity. For this purpose, it is sufficient to consider a
case where $\Phi_{\rm c}$ is constant. Indeed, in this situation, the
effective action reduces to $\Gamma =-V V_{\rm eff}(\Phi_{\rm c})$,
where $V$ is the space-time volume. On the other hand, the Volterra
expansion can be re-expressed as
\begin{eqnarray}
\label{eq:volterraaction}
\Gamma[\Phi_{\rm c}]&=&\sum_{n=0}^{\infty}\frac{1}{n!}
\int {\rm d}^4x_1 \cdots {\rm d}^4x_n \Gamma ^{(n)}(x_1, \cdots, x_n)
\Phi_{\rm c}^n .
\nonumber \\
\end{eqnarray}
This expression can be further simplified if one takes the Fourier
transform of the proper vertex,
\begin{eqnarray}
\label{eq:fourierproper}
\Gamma ^{(n)}(x_1, \cdots, x_n)&=& \int \frac{{\rm d}^4k_1}{(2\pi)^4}
\cdots \frac{{\rm d}^4k_n}{(2\pi)^4}\Gamma^{(n)}(k_1, \cdots, k_n)
\nonumber \\ & &  \times
{\rm e}^{-ik_1\cdot x_1} \cdots {\rm e}^{-ik_n\cdot x_n}.
\end{eqnarray}
In fact, for convenience, we introduce the coefficients
$\overline{\Gamma}^{(n)}(k_1,\cdots ,k_n)$ that are easier to
manipulate. They are defined by the following expression
\begin{equation}
\Gamma^{(n)}\left(k_1,\cdots ,k_n\right)\equiv (2\pi)^4
\delta\left(k_1+\cdots k_n\right)
\overline{\Gamma}^{(n)}\left(k_1, \cdots k_n\right),
\end{equation}
Then, we insert the Fourier expansion~(\ref{eq:fourierproper}) into
the expression~(\ref{eq:volterraaction}) of the effective action. The
integrals over space of the exponentials lead to Dirac functions in
momentum. As a consequence, one arrives at the following expression
\begin{equation}
\Gamma[\Phi_{\rm c}]
=V\sum_{n=0}^{\infty}\frac{1}{n!}
\overline{\Gamma}^{(n)}(\bm{k}_i=0)\Phi_{\rm c}^n,
\end{equation}
where, as already mentioned, $V=(2\pi)^4 \delta (0)$ is the space-time
volume. One finally reaches the result that
\begin{equation}
\label{eq:veff}
V_{\rm eff}(\phi_{\rm c})=-\sum_{n=0}^{\infty}\frac{1}{n!}
\overline{\Gamma}^{(n)}(\bm{k}_i=0)\Phi_{\rm c}^n.
\end{equation}
Therefore, in order to calculate the effective potential, one just has
to evaluate the proper vertices with vanishing momenta. At one loop,
this gives rise to the following sum
\begin{equation}
\label{eq:veffdiagrams}
V_{\rm eff}= \qquad 
\begin{fmffile}{poteff}
\parbox{20mm}{
\begin{fmfgraph*}(40,30)
\fmfleft{v1}
\fmfv{decor.shape=circle,decor.filled=0,decor.size=0.7w}{v1}
\fmffreeze
\fmfforce{(0.3w,0.8h)}{v2}
\fmfdot{v2}
\fmfforce{(.6w,.95h)}{d1}
\fmfforce{(.6w,.75h)}{d2}
\fmf{plain}{v2,d1}
\fmf{plain}{v2,d2}
\fmffreeze
\end{fmfgraph*}}
\end{fmffile}
\hspace{-0.85cm}
+
\hspace{1cm}
\begin{fmffile}{potefftwo}
\parbox{20mm}{
\begin{fmfgraph}(40,30)
\fmfleft{v1}
\fmfv{decor.shape=circle,decor.filled=0,decor.size=0.7w}{v1}
\fmffreeze
\fmfforce{(0.3w,0.8h)}{v2}
\fmfdot{v2}
\fmfforce{0.3w,0.2h}{v3}
\fmfdot{v3}
\fmfforce{(0.6w,0.95h)}{d1}
\fmfforce{(0.6w,0.75h)}{d2}
\fmfforce{(0.6w,0.25h)}{d3}
\fmfforce{(0.6w,.05h)}{d4}
\fmf{plain}{v2,d1}
\fmf{plain}{v2,d2}
\fmf{plain}{v3,d3}
\fmf{plain}{v3,d4}
\end{fmfgraph}}
\end{fmffile}
\hspace{-0.9cm}
+\cdots.
\end{equation}
It turns out that this series can be summed up as we are now going to
show.

\par

In order to see how it works, let us calculate the second diagram in
the above expansion. For this purpose, we start with the following
diagram
\begin{eqnarray}
\nonumber \\
{\cal D}=
\begin{fmffile}{fish}
\parbox{30mm}{
\begin{fmfgraph*}(50,30)
\fmfleft{i1,i2}
\fmfright{o1,o2}
\fmf{plain}{i1,v1}
\fmf{plain}{i2,v1}
\fmf{plain,left=0.8,tension=0.2}{v1,v2}
\fmf{plain,right=0.8,tension=0.2}{v1,v2}
\fmf{plain}{v2,o1}
\fmf{plain}{v2,o2}
\fmfdot{v1,v2}
\fmflabel{$t_1$}{i1}
\fmflabel{$t_2$}{i2}
\fmflabel{$t_3$}{o1}
\fmflabel{$t_4$}{o2}
\end{fmfgraph*}}
\end{fmffile}
\\
\nonumber
\end{eqnarray}
According to the Feynman rules~(\ref{eq:fruleone})
and~(\ref{eq:fruletwo}), this diagram is equivalent to the expression
\begin{eqnarray}
{\cal D}&=&\left(-i\lambda\right)^2\int {\rm d}u{\rm d}vD(t_1-u)
D(t_2-u)D^2(u-v) \nonumber \\
& & \times D(v-t_3)D(v-t_4),
\end{eqnarray}
where $u$ and $v$ denotes the two internal vertices. Upon using
Eq.~(\ref{eq:propagatoronedim}), this expression takes the form [here,
we use the quantum-mechanical theory introduced in
Sec.~\ref{subsec:qmqft} where the ``mass'' is denoted $\omega$]
\begin{widetext}
\begin{eqnarray}
{\cal D}&=&\left(-i\lambda\right)^2\int {\rm d}u{\rm d}v
\int \frac{{\rm d}E_1}{2\pi}
\frac{{\rm d}E_2}{2\pi}
\frac{{\rm d}E_3}{2\pi}
\frac{{\rm d}E_4}{2\pi}
\frac{{\rm d}E}{2\pi}
\frac{{\rm d}\tilde{E}}{2\pi}
\frac{i{\rm e}^{-iE_1(t_1-u)}}{E_1^2-\omega^2}
\frac{i{\rm e}^{-iE_2(t_2-u)}}{E_2^2-\omega^2}
\frac{i{\rm e}^{-iE_3(v-t_3)}}{E_3^2-\omega^2}
\nonumber \\ & & \times
\frac{i{\rm e}^{-iE_4(v-t_4)}}{E_4^2-\omega^2}
\frac{i{\rm e}^{-iE(u-v)}}{E ^2-\omega^2}
\frac{i{\rm e}^{-i\tilde{E}(u-v)}}{\tilde{E}^2-\omega^2}.
\end{eqnarray}
The integrals over $u$ and $v$ are easily performed and leads to two
Dirac functions. Then, a further integration over $\tilde{E}$ can be
performed and we are left with the expression
\begin{eqnarray}
\label{eq:calDtransitory}
{\cal D}&=&\left(-i\lambda\right)^2(2\pi)^2\delta \left(E_1+E_2-E_3-E_4\right)
\int \frac{{\rm d}E_1}{2\pi}
\frac{{\rm d}E_2}{2\pi}
\frac{{\rm d}E_3}{2\pi}
\frac{{\rm d}E_4}{2\pi}
\frac{{\rm d}E}{2\pi}
\frac{i{\rm e}^{-iE_1t_1}}{E_1^2-\omega^2}
\frac{i{\rm e}^{-iE_2t_2}}{E_2^2-\omega^2}
\frac{i{\rm e}^{-iE_3t_3}}{E_3^2-\omega^2}
\frac{i{\rm e}^{-iE_4t_4}}{E_4^2-\omega^2}
\nonumber \\ & & \times
\frac{i}{E^2-\omega^2}
\frac{i}{\left(E_3+E_4-E\right)^2-\omega^2}.
\end{eqnarray}
In order to compute the expression~(\ref{eq:veffdiagrams}) (or rather
the contribution of the second diagram to this expression), we have
seen that we must amputate the diagram by appending the external leg
propagators. This amounts to ignore the factors $i/(E_i^2-\omega^2)$
in the above expression. Moreover, since the graph is given by the
Fourier transform of ${\cal D}$, this means that only the expression
inside the above quadruple integrals over $E_i$ should be
considered. Finally, the ``momentum'' (in the present context the
energy) of the external legs should be put to zero (\ie we require all
the $E_i$'s left in the expression of ${\cal D}$ to vanish) and we
must factorize out a Dirac function expressing energy conservation
[times $(2\pi)^2$ since we are considering a $0+1$ theory]. This means
that Eq.~(\ref{eq:calDtransitory}) takes the following form
\begin{equation}
\begin{fmffile}{potefftwo}
\parbox{20mm}{
\begin{fmfgraph}(40,30)
\fmfleft{v1}
\fmfv{decor.shape=circle,decor.filled=0,decor.size=0.7w}{v1}
\fmffreeze
\fmfforce{(0.3w,0.8h)}{v2}
\fmfdot{v2}
\fmfforce{0.3w,0.2h}{v3}
\fmfdot{v3}
\fmfforce{(0.6w,0.95h)}{d1}
\fmfforce{(0.6w,0.75h)}{d2}
\fmfforce{(0.6w,0.25h)}{d3}
\fmfforce{(0.6w,.05h)}{d4}
\fmf{plain}{v2,d1}
\fmf{plain}{v2,d2}
\fmf{plain}{v3,d3}
\fmf{plain}{v3,d4}
\end{fmfgraph}}
\end{fmffile}
=\left(-i\lambda\right)^2
\int
\frac{{\rm d}E}{2\pi}
\frac{i}{E^2-\omega^2}
\frac{i}{E^2-\omega^2}.
\end{equation}
\end{widetext}
The fact that the propagator appears squared is of course not
fortuitous. It comes from the fact that we have two ``double legs''
attached to the loop. For $n$ ``double legs'', it is clear that we
would have the propagator to the power $n$ and an overall factor
$(-i\lambda)^n$. As a consequence, at one loop, Eq.~(\ref{eq:veff})
can be expressed as
\begin{equation}
V_{\rm eff}\left(\Phi_{\rm c}\right)
=-\sum_{p=1}^{\infty}\frac{\Phi_{\rm c}^{2p}}{2p}
\left(\frac{\lambda 4!}{2}\right)^p
\int \frac{{\rm d}E}{2\pi}
\frac{1}{(E^2-\omega^2)^p},
\end{equation}
where we have taken into account a symmetry factor and where we have
noted $n=2p$, $p$ being the number of external points. The sum can now
be performed and we obtain
\begin{eqnarray}
V_{\rm eff}\left(\Phi_{\rm c}\right)
&=&-\frac12 
\int _{-\infty}^{\infty}\frac{{\rm d}E}{2\pi}
\ln \left(1+\frac{12\lambda \Phi_{\rm c}^2}{E^2-\omega^2}\right)
\\ 
&=& \frac12\left(\omega -\sqrt{12\lambda \Phi_{\rm c}^2
-\omega ^2}\right)\\
\label{eq:vefqm}
&=& \frac{\hbar}{2}
\left(-m+\sqrt{m^2+12\lambda \Phi_{\rm c}^2}
\right),
\end{eqnarray}
where we have re-established $\hbar$ and where we have taken into
account the sign difference for $\omega$ (or $m$) due to the fact that
we work with a $(-+++)$ signature, see the remark after
Eq.~(\ref{eq:relaHdzero}). This expression should be compared to
Eq.~(\ref{eq:gepaho}). In fact the first term just comes from the fact
that the method of the effective potential assumes that $V_{\rm
  eff}(0)=0$. Therefore, in order to compare with the Gaussian
effective potential, one must ignore this rescaling. As a consequence, 
the total potential (including the leading order in $\hbar^0$) reads
\begin{equation}
\label{eq:vefffinalqm}
V_{\rm eff}\left(\Phi_{\rm c}\right)=\frac12 m^2\Phi_{\rm c}^2
+\frac{\hbar}{2}\sqrt{m^2+12\lambda \Phi_{\rm c}^2}.
\end{equation}
One sees that the vacuum energy is predicted to be $V_{\rm
  eff}(0)=\hbar m/2$ a result which is not realistic in the non
perturbative regime.

\par

In fact the comparison can be made in an even more explicit manner if
one remembers that the one loop Gaussian effective potential consists
in retaining the dominant term is a systematic expansion in
$\hbar$. As a consequence, one should ignore the $\hbar ^2$ term in
Eq.~(\ref{eq:meanHquartic}) and the $\hbar $ term in
Eq.~(\ref{eq:gapqm}). In particular, this last approximation leads to
the following explicit solution for the gap equation
\begin{equation}
\label{eq:oneloopgapqm}
\overline{\Omega}^2=m^2+12 \lambda \Phi_0^2
=\frac{{\rm d}^2V}{{\rm d}\Phi_0^2}.
\end{equation}
If we then use Eq.~(\ref{eq:gepaho}) and neglects the $\hbar^2$ term in 
that equation, one obtains
\begin{eqnarray}
  V_{_{\rm G}}(\Phi_0) &=& V(\Phi_0)
+\frac{\hbar}{2}\sqrt{\frac{{\rm d}^2V}{{\rm d}\Phi_0^2}}
+{\cal O}(\hbar^2)\\
&=&  \frac12 m^2\Phi_0^2
+\frac{\hbar}{2}\sqrt{m^2+12\lambda \Phi_0^2}+\cdots ,
\end{eqnarray}
that is to say exactly Eq.~(\ref{eq:vefffinalqm}). We conclude that
the effective potential is in fact contained in the Gaussian effective
potential approach. It just consists in neglecting higher power of
$\hbar$ since this is a one loop approach. As a consequence, the
effective potential cannot be reliable in the non perturbative regime
where the quantum effects are strong.

\par

Having demonstrated that the Gaussian effective potential approach is
an efficient tool, we now need to study how it works in quantum field
theory. Then, we will able to use it in order to calculate the energy
density of the vacuum state.

\subsection{The Gaussian Effective Potential in Field Theory 
and the Vacuum Energy}
\label{subsec:gepqft}

We now turn to the main question of this section, namely the
calculation of the Gaussian effective potential in quantum field
theory~\cite{Stevenson:1985zy,Stevenson:1985kr,Stevenson:1986bq,Stevenson:1986sb}. For
this purpose, we now consider a model similar to the one considered in
the previous subsections, namely a free model plus a quartic self
interaction
\begin{equation}
{\cal H}=\frac12\dot{\Phi}^2+\frac{1}{2}\delta ^{ij}\partial _i\Phi 
\partial_j\Phi+\frac{1}{2}m_{_{\rm B}}^2\Phi^2+\lambda_{_{\rm B}}\Phi^4,
\end{equation}
where $m_{_{\rm B}}$ and $\lambda _{_{\rm B}}$ are the bare mass and
coupling constant, respectively. Then, we write the field operator as
(we work in $d$ space-time dimensions)
\begin{eqnarray}
\Phi\left(t,{\bm x}\right)&=&\Phi_0+\frac{1}{\left(2\pi\right)^{(d-1)/2}}
\int \frac{{\rm d}^{d-1}{\bm k}}{\sqrt{2\omega_{\Omega}(k)}}\biggl(c_{\bm k}
{\rm e}^{-i\omega t +i{\bm k}\cdot {\bm x}}
\nonumber \\ 
& & +
c_{\bm k}^{\dagger}{\rm e}^{i\omega t -i{\bm k}\cdot {\bf x}}
\biggr),
\end{eqnarray}
with
\begin{equation}
\omega_{\Omega}(k)\equiv \sqrt{k^2+\Omega^2}.
\end{equation}
The quantity $\Phi_0$ represents a classical and constant field. The
parameter $\Omega$ denotes the mass of the excitations around this
classical value. The idea at the basis of the calculation is the same
as before, namely minimizing the energy with respect to the parameter
$\Omega$. Using the properties of the creation and annihilation
operators, it is straightforward to calculate the mean value of the
Hamiltonian in the vacuum state. We find
\begin{eqnarray}
\label{eq:Hmean}
\left \langle {\cal H}\right \rangle\left(\Omega,\Phi_0\right) 
&=& I_1(\Omega)+\frac12
\left(m_{_{\rm B}}^2-\Omega^2\right)I_0(\Omega)+\frac12m_{_{\rm B}}^2\Phi_0^2
\nonumber \\ &+&
\lambda_{_{\rm B}}\Phi_0^4
+6\lambda_{_{\rm B}}\Phi_0^2I_0(\Omega)
+3\lambda_{_{\rm B}}I_0^2(\Omega),
\end{eqnarray}
where we have defined the integrals $I_p(\Omega)$ by
\begin{equation}
\label{eq:defI}
I_p(\Omega)= \frac{1}{\left(2\pi\right)^{d-1}} 
\int \frac{{\rm d}^{d-1}{\bm k}}
{2\omega_{\Omega}(k)}\omega_{\Omega}^{2p}(k).
\end{equation}
Of course, the integrals $I_p(\Omega)$ can be divergent and this
signals, as usual in field theory, the need for renormalization. In
fact, these integrals can be performed exactly and one is led to
\begin{equation}
\label{eq:Ieuler}
I_p(\Omega)=\frac{1}{2(4\pi)^{(d-1)/2}}
\frac{\Gamma[(2-d-2p)/2]}{\Gamma[(1-2p)/2]}\Omega^{d+2p-2}.
\end{equation}
It is clear that, for some values of $d$ and $p$, the above expression
does not exist and is only formal.

\par

As already mentioned, in order to calculate the Gaussian effective
potential, one must vary $ \left \langle {\cal H}\right \rangle$ with
respect to $\Omega$. Upon using Eq.~(\ref{eq:Hmean}) and ${\rm
  d}I_p/{\rm d}\Omega=(2p-1)\Omega I_{p-1}$, one arrives at the
following ``gap'' equation, the field theory counterpart of
Eq.~(\ref{eq:gapqm})
\begin{equation}
\label{eq:gapqft}
\Omega^2=m_{_{\rm B}}^2+12\lambda_{_{\rm B}}\left[\Phi_0^2
+I_0\left(\Omega\right)\right].
\end{equation}
This is an algebraic equation for $\Omega$. Its solution can be
written as $\overline{\Omega}=\overline{\Omega}\left(\Phi_0;m_{_{\rm
      B}},\lambda_{_{\rm B}}\right)$, that is to say it depends on
$\Phi_0$ and on the bare parameters (in the following we no longer
write the dependence on the bare parameters). Then, the Gaussian
effective potential is defined by inserting this solution into
Eq.~(\ref{eq:Hmean}). This leads to
\begin{equation}
\label{eq:defgep}
V_{_{\rm G}}(\Phi_0)\equiv\left \langle {\cal H}\right \rangle
\left[\overline{\Omega}(\Phi_0),\Phi_0\right]. 
\end{equation}
Upon using the gap equation, one can also find a more compact
expression, namely
\begin{eqnarray}
\label{eq:gepqft}
V_{_{\rm G}}(\Phi_0)&=&I_1\left[\overline{\Omega}\left(\Phi_0\right)\right]
-3\lambda_{_{\rm B}}I_0^2\left[\overline{\Omega}\left(\Phi_0\right)\right]
+\frac12m_{_{\rm B}}^2\Phi_0^2
\nonumber \\ & &
+\lambda_{_{\rm B}}\Phi_0^4.
\end{eqnarray}
At this level, this expression is in fact still formal because of the
divergences of the integrals $I_p$. Therefore, we must first tame
those divergences. This is done by means of a renormalization of the
parameters appearing in the Lagrangian of the model.

\par

Renormalization consists in parametrizing the physical quantities in
terms of new, observable parameters, different from the bare
parameters that appear in the Lagrangian. We now define the
renormalized mass by
\begin{eqnarray}
m_{_{\rm R}}^2 &=& \frac{{\rm d}^2V_{_{\rm G}}}{{\rm d}\Phi_0^2}\biggl \vert 
_{\Phi_0=0} .
\end{eqnarray}
Upon using Eq.~(\ref{eq:gepqft}) and the following relation [which
comes from differentiating Eq.~(\ref{eq:gapqft})]
\begin{equation}
\frac{{\rm d}\overline{\Omega}}{{\rm d}\Phi_0}=
\frac{\Phi_0}{\overline{\Omega}}\frac{12\lambda_{_{\rm B}}}
{1+6\lambda_{_{\rm B}}
I_{-1}(\overline{\Omega})},
\end{equation}
one finds that
\begin{eqnarray}
\frac{{\rm d}^2V_{_{\rm G}}}{{\rm d}\Phi_0^2}
&=&m_{_{\rm B}}^2+12\lambda_{_{\rm B}}\left[\Phi_0^2
+I_0(\overline{\Omega})\right]
\nonumber \\ & &
-\frac{(12\lambda_{_{\rm B}}\Phi_0)^2I_{-1}(\overline{\Omega})}
{1+6\lambda_{_{\rm B}}I_{-1}(\overline{\Omega})}.
\end{eqnarray}
As a consequence, putting $\Phi_0=0$ in the above equation, the
renormalized mass can be expressed as
\begin{equation}
m_{_{\rm R}}^2=m_{_{\rm B}}^2+12\lambda_{_{\rm B}}I_0(\overline{\Omega}_0),
\end{equation}
where $\overline{\Omega}_0$ denotes the solution of the gap
equation~(\ref{eq:gapqft}) when $\Phi_0=0$, \ie $\Omega^2=m_{_{\rm
    B}}^2+12\lambda_{_{\rm B}}I_0\left(\Omega\right)$. It is obvious
that $\overline{\Omega}_0$ is not a function of $\Phi_0$ but depends
only on $m_{_{\rm B}}$ and $\lambda_{_{\rm B}}$. In fact, this can
also been seen from Eq.~(\ref{eq:gapqft}) since it tells us that
$m_{_{\rm R}}^2=\overline{\Omega}_0^2$. As a consequence, the above
equation can be re-written as
\begin{equation}
\label{eq:renormmass}
m_{_{\rm B}}^2=m_{_{\rm R}}^2-12\lambda_{_{\rm B}}I_0\left(m_{_{\rm R}}\right).
\end{equation}
We are now in a position where we can calculate the Gaussian effective
potential in terms of the renormalized mass. For this purpose, we
restart from Eq.~(\ref{eq:defgep}) and substitute the
expression~(\ref{eq:renormmass}). One obtains
\begin{eqnarray}
\label{eq:gepinter}
V_{_{\rm G}}(\Phi_0)&=&I_{1}\left(\overline{\Omega}\right)
+\frac12\left(m_{_{\rm R}}^2-\overline{\Omega}^2\right)
I_0(\overline{\Omega})
+\frac12m_{_{\rm R}}^2\Phi_0^2
\nonumber \\ & & 
+\lambda_{_{\rm B}}\Phi_0^4
+6\lambda_{_{\rm B}}\left[I_{0}\left(\overline{\Omega}\right)
-I_0\left(m_{_{\rm R}}\right)\right]\Phi_0^2
\nonumber \\ & & 
+3\lambda_{_{\rm B}}I_{0}^2\left(\overline{\Omega}\right)
-6\lambda_{_{\rm B}}I_{0}\left(\overline{\Omega}\right)
I_0\left(m_{_{\rm R}}\right).
\end{eqnarray}
However, at this stage, the expression of the Gaussian effective
potential is not yet a function of the renormalized mass only as one
still needs to express $I_0(\overline{\Omega})$ and
$I_1(\overline{\Omega})$ in terms of $I_0(m_{_{\rm R}})$ and
$I_1(m_{_{\rm R}})$. There are several ways to do it, see
Ref.~\cite{Stevenson:1985zy}. Here we choose to use a method where a
cut-off is temporarily introduced. We work in $d=4$ dimensions to
simplify the calculations. Let us exemplify the method with $I_0$. We
have
\begin{eqnarray}
I_0(\Omega)&=&\frac{1}{2(2\pi)^3}\int {\rm d}^3\bm{k}\, \omega^{-1}_{\Omega}
\\
&=&\frac{1}{4\pi^2}\int_0^M{\rm d}kk^2\left(k^2+\Omega^2\right)^{-1/2},
\end{eqnarray}
where we have performed the angular integrals and where we have
introduced a cut-off $M$ to regulate the otherwise divergent
integral. This gives
\begin{eqnarray}
I_0(\Omega)&=&\frac{1}{4\pi^2}\Biggl[\frac{M}{2}\sqrt{\Omega^2+M^2}
\nonumber \\ & &
-\frac{\Omega^2}{2}\ln \left(M+\sqrt{\Omega^2+M^2}\right)
+\frac{\Omega^2}{2}
\ln \Omega\Biggr].\nonumber \\
\end{eqnarray}
As a consequence, one can write
\begin{widetext}
\begin{eqnarray}
\label{eq:diffI0}
I_0(\Omega)-I_0(m)&=& \frac12\left(m^2-\Omega^2\right)I_{-1}(m)+\frac{1}{4\pi^2}
\Biggl[\frac{M}{2}\sqrt{\Omega^2+M^2}-\frac{M}{2}\sqrt{m^2+M^2}
+\frac{\Omega^2}{2}\ln \left(\frac{M+\sqrt{m^2+M^2}}{M+\sqrt{\Omega^2+M^2}}
\right)
\nonumber \\ & &
+\frac{\Omega^2}{2}\ln \Omega -\frac{m^2}{2}\ln m+
\frac12\left(m^2-\Omega^2\right)\left(\frac{M}{\sqrt{m^2+M^2}}+\ln m\right)
\Biggr],
\end{eqnarray}
\end{widetext}
where we have used
\begin{eqnarray}
\label{eq:Iminus1}
I_{-1}(\Omega)&=&\frac{1}{4\pi^2}\Biggl[\frac{-M}{\sqrt{\Omega^2+M^2}}
+\ln \left(M+\sqrt{\Omega^2+M^2}\right)\nonumber \\ & & 
-\ln \Omega\Biggr],
\end{eqnarray}
an expression which can be obtained with the same method as described
above. Then we take the limit $M\rightarrow +\infty$ and
Eq.~(\ref{eq:diffI0}) becomes
\begin{eqnarray}
\label{eq:diffI0final}
I_0(\Omega)-I_0(m)&=& -\frac12\left(\Omega^2-m^2\right)I_{-1}(m)
\nonumber \\ & &
+\frac{m^2}{16\pi^2}
\left[x\ln x-(x-1)\right],
\end{eqnarray}
where $x\equiv \Omega^2/m^2$. Notice that, at this stage of the
calculation, there is no need to write $x$ in terms of
$\overline{\Omega}$ and/or $m_{_{\rm R}}$. The above expression is
valid for any $\Omega$ and $m$. Following
Ref.~\cite{Stevenson:1985zy}, we call the function between the square
brackets in the above equation $L_2(x)$. Exactly in the same manner,
one can also show that
\begin{align}
\label{eq:diffI1final}
I_1(\Omega)& -I_1(m) = \frac12\left(\Omega^2-m^2\right)I_{0}(m)
\nonumber \\ &-
\frac18\left(\Omega^2-m^2\right)^2I_{-1}(m)+\frac{m^4}{32\pi^2}L_3(x),
\end{align}
with $L_3(x)=[2x^2\ln x-2(x-1)-3(x-1)^2]/4$. Then, upon using
Eqs.~(\ref{eq:diffI0final}) and~(\ref{eq:diffI1final}) in
Eq.~(\ref{eq:gepinter}), one obtains
\begin{widetext}
\begin{eqnarray}
\label{eq:gepfinal}
V_{_{\rm G}}(\Phi_0) &=& I_1\left(m_{_{\rm R}}\right)
-3\lambda_{_{\rm B}}I_0^2\left(m_{_{\rm R}}\right)
+\frac12m_{_{\rm R}}^2\Phi_0^2+\lambda_{_{\rm B}}\Phi_0^4
+\frac{m_{_{\rm R}}^4}{32\pi^2}L_3(\overline{x})
+\frac34\lambda_{_{\rm B}}m_{_{\rm R}}^4
\left(\overline{x}-1\right)^2I_{-1}^2\left(m_{_{\rm R}}\right)
\nonumber \\ & & 
+\frac18I_{-1}\left(m_{_{\rm R}}\right)m_{_{\rm R}}^4(\overline{x}-1)
\left\{(\overline{x}-1)
-\frac{3\lambda_{_{\rm B}}}{2\pi^2}\left[L_2(\overline{x})+
\frac{16\pi^2}{m_{_{\rm R}}^2}\Phi_0^2\right]\right\}
\nonumber \\ & &
-\frac{m_{_{\rm R}}^4}{32\pi^2}L_2(\overline{x})
\left\{\left(\overline{x}-1\right)-
\frac{3\lambda_{_{\rm B}}}{8\pi^2}\left[L_2(\overline{x})
+32\pi^2\frac{\Phi_0^2}{m_{_{\rm R}}^2}\right]\right\},
\end{eqnarray}
\end{widetext}
with $\overline{x}\equiv \overline{\Omega}^2/m_{_{\rm R}}^2$. Finally,
the gap equation~(\ref{eq:gapqft}) can also be re-written in terms of
the renormalized mass. The corresponding expression reads
\begin{eqnarray}
\label{eq:newgap}
(\overline{x}-1)\left[1+6\lambda_{_{\rm B}}I_{-1}\left(m_{_{\rm R}}\right)\right]
&=&\frac{3\lambda_{_{\rm B}}}{4\pi^2}
\biggl[L_2(\overline{x})
\nonumber \\ & &
+16\pi^2 \frac{\Phi_0^2}{m_{_{\rm R}}^2}\biggr].
\end{eqnarray}
We see that the dependence on $\Phi_0$ of the Gaussian effective
potential is quite complicated. Indeed, solving the gap equation
provides us with $\overline{x}=\overline{x}(\Phi_0)$ that we must then
insert in Eq.~(\ref{eq:gepfinal}).

\par

We shall now discuss the renormalization of the coupling constant
$\lambda_{_{\rm B}}$. It is defined by
\begin{eqnarray}
\lambda _{_{\rm R}} &=& \frac{1}{4!}
\frac{{\rm d}^4V_{_{\rm G}}}{{\rm d}\Phi_0^4}\biggl \vert 
_{\Phi_0=0} .
\end{eqnarray}
Proceeding in the same manner as before, one obtains the following
equation for the quantity $\lambda _{_{\rm R}}$
\begin{equation}
\label{eq:renormcoupling}
\lambda_{_{\rm R}}=\lambda_{_{\rm B}}\frac{1-12\lambda_{_{\rm B}}I_{-1}
\left(m_{_{\rm R}}\right)}
{1+6\lambda_{_{\rm B}}I_{-1}
\left(m_{_{\rm R}}\right)}.
\end{equation}
The renormalization of the coupling constant has been discussed at
length in Refs.~\cite{Stevenson:1985zy, Stevenson:1986bq} and we will
not repeat all the details since, here, our main concern is the vacuum
energy. In fact we will focus on the case named ``precarious'' in
Ref.~\cite{Stevenson:1985zy}. Let us return to
Eq.~(\ref{eq:renormcoupling}); one can invert this expression and
express the bare coupling constant in terms of the renormalized
one. One obtains
\begin{eqnarray}
\label{eq:lambdabr}
\lambda_{_{\rm B}}&=& \frac{1-
6\lambda_{_{\rm R}}I_{-1}\left(m_{_{\rm R}}\right)}{24I_{-1}\left(m_{_{\rm R}}\right)}
\nonumber \\ & & \times
\left\{1\pm \sqrt{1-\frac{48\lambda_{_{\rm R}}I_{-1}\left(m_{_{\rm R}}\right)}
{\left[1-6\lambda_{_{\rm R}}I_{-1}\left(m_{_{\rm R}}\right)\right]^2}}\right\}.
\end{eqnarray}
Then, from Eq.~(\ref{eq:Iminus1}), we see that $I_{-1}\left(m_{_{\rm
      R}}\right)$ is in fact a logarithmically divergent quantity when
the cut-off $M$ is sent to infinity. Therefore, we must treat
$I_{-1}\left(m_{_{\rm R}}\right)$ as an arbitrarily large quantity. In
this limit, if one chooses the plus sign in Eq.~(\ref{eq:lambdabr}),
one obtains
\begin{equation}
\label{eq:sillylambda}
\lambda_{_{\rm B}}=-\frac12 \lambda_{_{\rm R}}+{\cal O}
\left[\frac{1}{I_{-1}\left(m_{_{\rm
      R}}\right)}\right].
\end{equation}
As shown in Ref.~\cite{Stevenson:1985zy}, this case is not
viable. Intuitively, it is clear that a large negative coupling
constant leads to a potential which is not bounded from
below. Therefore, we are left with the minus sign for which one finds
that
\begin{eqnarray}
\label{eq:precariouslambda}
\lambda_{_{\rm B}}&=&-\frac{1}{6I_{-1}\left(m_{_{\rm R}}\right)}
\Biggl\{1+\frac{1}{2\lambda_{_{\rm R}}I_{-1}\left(m_{_{\rm
      R}}\right)}
\nonumber \\ & & 
+{\cal O}
\left[\frac{1}{I_{-1}^2\left(m_{_{\rm
      R}}\right)}\right]\Biggr\}.
\end{eqnarray}
In passing, it is worth signaling that this equation can be
re-expressed in a different manner as follows. Let us first introduce
the new characteristic scale $M_{_{\rm C}} $ (not to be confused with
the cut-off $M$ discussed before) defined by
\begin{equation}
\frac{1}{\lambda_{_{\rm R}}}\equiv -\frac{1}{4\pi^2}
\ln \left(\frac{m_{_{\rm R}}^2}{M_{_{\rm C}}^2}\right).
\end{equation}
Then, if we use the fact that
\begin{equation}
I_{-1}(M_{_{\rm C}})-I_{-1}(m_{_{\rm R}})=-\frac{1}{8\pi^2}L_1(x),
\end{equation}
where $L_1(x)=\ln x$ [an expression which can be derived with the same
method that has led to Eqs.~(\ref{eq:diffI0})
and~(\ref{eq:diffI1final})], the bare coupling constant can be
re-written as
\begin{equation}
\label{eq:defmc}
\lambda_{_{\rm B}}=-\frac{1}{6I_{-1}(M_{_{\rm C}})}.
\end{equation}
Therefore, it has a negative but infinitesimal value which implies
that the stability properties of the corresponding scenario are much
better compared to the case~(\ref{eq:sillylambda}). As argued in
Ref.~\cite{Stevenson:1985zy}, besides simplifying the expression of
the renormalized coupling constant, the scale $M_{_{\rm C}}$ plays an
important role in a complete study of the precarious case (even if, in
the present context, this is not apparent since we focus on the vacuum
energy only). Inserting the expression~(\ref{eq:precariouslambda}) [or
Eq.~(\ref{eq:defmc})] into Eq.~(\ref{eq:gepfinal}), one can re-write
the effective potential as
\begin{eqnarray}
\label{eq:geprenormfinal}
V_{_{\rm G}}(\Phi_0) &=& I_1\left(m_{_{\rm R}}\right)
-3\lambda_{_{\rm B}}I_0^2\left(m_{_{\rm R}}\right)
+\frac12m_{_{\rm R}}^2\overline{x}\Phi_0^2
\nonumber \\ & &
+\frac{m_{_{\rm R}}^4}{32\pi^2}L_3(\overline{x})
-\frac{m_{_{\rm R}}^4}{16\lambda_{_{\rm R}}}
\left(\overline{x}-1\right)^2
\nonumber \\ & &
+{\cal O}\left[\frac{1}{I_{-1}(m_{_{\rm R}})}\right],
\end{eqnarray}
while the gap equation~(\ref{eq:newgap}) becomes
\begin{eqnarray}
\label{eq:gaprenormfinal}
\overline{x}-1=\frac{\lambda_{_{\rm R}}}{4\pi^2}
\left[L_2\left(\overline{x}\right)
+16\pi^2\frac{\Phi_0^2}{m_{_{\rm R}}^2}\right].
\end{eqnarray}
Let us notice that, in Eq.~(\ref{eq:geprenormfinal}), we have left the
quantity $\lambda_{_{\rm B}}$ into the field independent term
$I_1\left(m_{_{\rm R}}\right) -3\lambda_{_{\rm B}}I_0^2\left(m_{_{\rm
      R}}\right)$ since it represents the zero point energy that will
be treated in detail in the next
sub-section~\ref{subsec:vacuumgep}. Finally, if one uses the explicit
form of the functions $L_2$ and $L_3$, then the Gaussian effective
potential takes the form
\begin{eqnarray}
\label{eq:vgfinalfinal}
V_{_{\rm G}}(\Phi_0) &=& I_1\left(m_{_{\rm R}}\right)
-3\lambda_{_{\rm B}}I_0^2\left(m_{_{\rm R}}\right)
+\frac12m_{_{\rm R}}^2\overline{x}\Phi_0^2
\nonumber \\
&+&\frac{m_{_{\rm R}}^4}{128\pi^2}
\Biggl[2\overline{x}^2\ln \overline{x}-2(\overline{x}-1)
-3(\overline{x}-1)^2
\nonumber \\ & &
-2 \frac{4\pi^2}{\lambda _{_{\rm R}}}(\overline{x}-1)^2\biggr], 
\end{eqnarray}
and the gap equation can be re-expressed as
\begin{equation}
\label{eq:gapfinalfinal}
(\overline{x}-1)\left(1+\frac{4\pi^2}{\lambda _{_{\rm R}}}\right)
-16\pi^2\frac{\Phi_0^2}{m_{_{\rm R}}^2}
=\overline{x}\ln \overline{x}.
\end{equation}
The two equations~(\ref{eq:vgfinalfinal}) and~(\ref{eq:gapfinalfinal})
give a complete expression of the Gaussian effective potential for the
renormalized precarious ``$\lambda \Phi^4$''
theory~\cite{Stevenson:1985zy, Stevenson:1986bq}.

\par

As we have done for the quantum mechanics case, it is interesting to
compare the above result to the standard effective potential
approach. We know that this potential can be obtained from the
Gaussian effective potential method by neglecting the terms
proportional to $\hbar^2$. Therefore, we restart from
Eq.~(\ref{eq:Hmean}) and neglect the term $3\lambda _{_{\rm B}}I_0^2$
since this corresponds to the standard one loop method (\ie it is of
order $\hbar^2$). One obtains
\begin{eqnarray}
\label{eq:meanHeff}
\left \langle {\cal H}\right \rangle\left(\Omega,\Phi_0\right) 
&=& I_1(\Omega)+\frac12
\left(m_{_{\rm B}}^2-\Omega^2\right)I_0(\Omega)+\frac12m_{_{\rm B}}^2\Phi_0^2
\nonumber \\ &+&
\lambda_{_{\rm B}}\Phi_0^4
+6\lambda_{_{\rm B}}\Phi_0^2I_0(\Omega)+{\cal O}(\hbar^2).
\end{eqnarray}
As a consequence, this leads to the new following gap equation
\begin{equation}
\label{eq:newgapqft}
\Omega^2=m_{_{\rm B}}^2+12\lambda_{_{\rm B}}\Phi_0^2.
\end{equation}
Clearly, this relation is the exact counter-part of
Eq.~(\ref{eq:oneloopgapqm}). Then, one has to calculate the
renormalized mass and coupling constant. one arrives at
\begin{eqnarray}
m_{_{\rm R}}^2 &=& m_{_{\rm B}}^2+12\lambda_{_{\rm B}}I_0(m_{_{\rm B}}), \\
\lambda_{_{\rm R}} &=&\lambda_{_{\rm B}}\left[1-18\lambda_{_{\rm B}}
I_{-1}(m_{_{\rm B}})\right].
\end{eqnarray} 
These equations should be compared to Eqs.~(\ref{eq:renormmass})
and~(\ref{eq:renormcoupling}). We notice that it is now $m_{_{\rm B}}$
that appears in the argument of the integrals rather than $m_{_{\rm
    R}}$ as in Eqs.~(\ref{eq:renormmass})
and~(\ref{eq:renormcoupling}). This is because
Eq.~(\ref{eq:newgapqft}) implies that
$\overline{\Omega}(\Phi_0=0)=m_{_{\rm B}}$ and not
$\overline{\Omega}(\Phi_0=0)=m_{_{\rm R}}$ as before. The final step
consists in using the two previous equation in
Eq.~(\ref{eq:meanHeff}). One obtains
\begin{widetext}
\begin{eqnarray}
\label{eq:potcw}
V_{_{\rm G}}(\Phi_0)&=& I_1\left(m_{_{\rm B}}\right)+\frac12 m_{_{\rm R}}^2\Phi_0^2
+\lambda_{_{\rm R}}\Phi_0^4+\frac{m_{_{\rm B}}^4}{64\pi^2}
\Biggl[\Biggl(1+12\lambda_{_{\rm B}}\frac{\Phi_0^2}{m_{_{\rm B}}^2}
\Biggr)^2\ln \Biggl(1+12\lambda_{_{\rm B}}\frac{\Phi_0^2}{m_{_{\rm B}}^2}
\Biggr)-12\lambda_{_{\rm B}}\frac{\Phi_0^2}{m_{_{\rm B}}^2}
\nonumber \\ & &
-216\lambda_{_{\rm B}}^2\frac{\Phi_0^4}{m_{_{\rm B}}^4}\Biggr]+\cdots ,
\end{eqnarray}
\end{widetext}
the dots corresponding to the higher order terms in $\hbar$. If one
neglects those terms, then $V_{_{\rm G}}$ reduces to $V_{\rm eff}$.
The corresponding expression is nothing but the Coleman-Weinberg
potential. The previous analysis confirms that the effective potential
is indeed a particular case of the Gaussian effective potential
approach. An even more detailed comparisons of the two methods can be
found in Refs.~\cite{Stevenson:1985zy, Stevenson:1986bq}. Here, we do
not pursue this issue and now come to the question of the vacuum
energy problem.

\subsection{Vacuum Energy and Gaussian Effective Potential}
\label{subsec:vacuumgep}

We are now in a position where the main problem of this section can be
addressed. From Eqs.~(\ref{eq:gepfinal})
and~(\ref{eq:geprenormfinal}), we see that the vacuum energy density
predicted by the Gaussian effective potential method (\ie the term
left after having taken $\Phi_0=0$ in those expressions) can be
expressed as
\begin{equation}
\rho_{_{\rm vac}}=I_1(m_{_{\rm R}})-3\lambda _{_{\rm B}}I_0(m_{_{\rm R}}).
\end{equation}
As we have already stressed many times, it is important to remember
that this result is a priori non perturbative. Using the expression of
the precarious coupling constant~(\ref{eq:precariouslambda}), one
obtains
\begin{eqnarray}
\label{eq:rhovacgep}
\rho_{_{\rm vac}}&=&I_1(m_{_{\rm R}})
\nonumber \\ &+ &
\frac12\frac{I_0(m_{_{\rm R}})}{I_{-1}(m_{_{\rm R}})}
\left[1+\frac{1}{2\lambda I_{-1}(m_{_{\rm R}})}\right]I_0(m_{_{\rm R}}).
\end{eqnarray}
The expressions of the integrals $I_p$ have been given in
Eqs.~(\ref{eq:defI}) and~(\ref{eq:Ieuler}). The three integrals that
appear in the above expression are all divergent. Nevertheless, the
ratio $I_0/I_{-1}$ is finite and reads
\begin{equation}
\frac{I_0(m_{_{\rm R}})}{I_{-1}(m_{_{\rm R}})}=\frac{m_{_{\rm R}}^2}{2-d}.
\end{equation}
Nevertheless, even after having inserted the above ratio into
Eq.~(\ref{eq:rhovacgep}), this expression still contains singular
terms. Explicitly, one has
\begin{eqnarray}
\rho_{_{\rm vac}}&=&\frac{m_{_{\rm R}}^4}{4(2-d)}
\Biggl[-m_{_{\rm R}}^{d-4} \frac{4-d}{2\sqrt{\pi}(4\pi)^{(d-1)/2}}
\Gamma \left(-\frac{d}{2}\right)
\nonumber \\ & &
+\frac{1}{\lambda_{_{\rm R}}(2-d)}
\Biggr].
\end{eqnarray}
However, writing this expression for $d=4+\epsilon$ and taking the
limit $\epsilon\rightarrow 0$, we have the remarkable result that a
second cancellation occurs and that the final expression is in fact
finite. It can be expressed as
\begin{equation}
\label{eq:rhovacgegrenorm}
\rho_{_{\rm vac}}=\frac{m_{_{\rm R}}^4}{128\pi^2}
\left[1-2\ln \left(\frac{m_{_{\rm R}}^2}{M_{_{\rm C}}^2}\right)
\right], 
\end{equation}
where we have used Eq.~(\ref{eq:defmc}). Even more remarkable is that
the above equation is very similar to Eq.~(\ref{eq:rhovacrenorm}). In
particular, the vacuum energy is proportional to fourth power, not of
the cut-off, but of the mass of the corresponding particle and the
scale $M_{_{\rm C}}$ is present as a logarithmic
correction. Therefore, we conclude that our analysis based on the
Gaussian effective potential has totally confirmed the results
presented in the previous sections. We end this section by noticing
that the vacuum energy predicted by the Coleman Weinberg potential is,
see Eq.~(\ref{eq:potcw}), $\rho_{_{\rm vac}}=I_1(m_{_{\rm B}})$ and is
therefore different from the one obtained before, see
Eq.~(\ref{eq:rhovacgegrenorm}). Moreover, it seems that this
expression remains divergent.

\section{Contribution from Other Fields}
\label{sec:otherfield}

So far, for simplicity, we have only treated the case of a scalar
field. However, in order to be more realistic, it is clear that one
must also evaluate the vacuum energy for other type of fields. This
question is the purpose of the present section. As we will see, the
corresponding calculations are in fact very similar. In fact, the
important question is the dependence of the overall coefficient which
multiplies $\rho_{_{\rm vac}}$ with the spin of the particle. As is
well-known, this is at the origin of a remarkable cancellation which
leads to the concept of super-symmetry. Therefore, in the following, we
will pay special attention to this issue.

\subsection{Fermion Fields}
\label{subsec:fermion}

We start with the case of a Dirac spinor field describing a spin $1/2$
particle~\cite{Peskin:1995ev,Greiner:1996zu,Itzykson:1980rh,
  Bailin:1994qt,LeBellac:1991cq,Ryder:1985wq,Mandl:1985bg}. For
convenience, we quickly remind basics fact about the quantum field
theory of a spin $1/2$ field. Then, we turn to the calculation of
$\rho_{_{\rm vac}}$.

\par

The Lagrangian of the free model is [we treated the mass-less case in
Eq.~(\ref{eq:diracmasslessL})]
\begin{equation}
\label{eq:diracl}
{\cal L}_{\rm Dirac}=-\overline{\Psi}\left(i\gamma ^{\mu}\partial _{\mu}-m
\right)\Psi,
\end{equation}
where $\gamma ^{\mu}$ are matrices satisfying
\begin{equation}
\label{eq:diracanticom}
\left\{\gamma ^{\mu},\gamma ^{\nu}\right\}=-2g^{\mu \nu}.
\end{equation}
The above equation is in fact valid for any metric tensor. Here, we
just consider the flat space-time case and, therefore, we will always
consider that $g_{\mu \nu}=\eta_{\mu \nu}$. The quantity $\Psi(t,{\bm
  x})$ represents a four components Dirac spinor and
$\overline{\Psi}\equiv \Psi^{\dagger}\gamma^0$ [this definition was
already mentioned just after Eq.~(\ref{eq:spinorrl})]. Here, we use
the Dirac representation which means that the $\gamma ^{\mu}$ matrices
can be taken to be equal to [recall that, in
Eq.~(\ref{eq:chiralrepresentation}), we used the chiral
representation]
\begin{eqnarray}
\label{eq:diracmatrices}
\gamma ^0 &=& \begin{pmatrix} \mathbb{I}_2 & 0 \cr 0& -\mathbb{I}_2 
\end{pmatrix},
\quad 
\gamma ^i=\begin{pmatrix} 0 & \sigma ^i\cr -\sigma ^i & 0\end{pmatrix},
\nonumber \\  
\gamma ^5&=& \begin{pmatrix} 0 & \mathbb{I}_2 \cr \mathbb{I}_2 & 0 
\end{pmatrix}, 
\end{eqnarray}
where $\sigma^i$ are the standard two-dimensional Pauli matrices. The
equation of motion for the spinor $\Psi$ is easily obtained from
Eq.~(\ref{eq:diracl}). Varying the Dirac Lagrangian, it is
straightforward to show that
\begin{equation}
\label{eq:diraceq}
\left(i\gamma ^{\mu}\partial _{\mu}-m\right)\Psi=0,
\end{equation}
which is of course nothing but the Dirac equation. Then, as we have
done for the scalar field case, we now expand the $\Psi$ operator in
Fourier modes according to
\begin{eqnarray}
\label{eq:expansionpsi}
\Psi\left(t,{\bm x}\right) &=& \frac{1}{(2\pi)^{3/2}}
\int \frac{{\rm d}{\bm k}}{\sqrt{2\omega(k)}}
\sum _{r=1}^{2}\biggl[b_{\bm k}^r u({\bm k},r)
{\rm e}^{-i\omega t+i{\bm k}\cdot
{\bm x}}\nonumber \\ & &+
\left(d_{\bm k}^r\right)^{\dagger} v({\bm k},r)
{\rm e}^{i\omega t-i{\bm k}\cdot
{\bm x}}\biggr].
\end{eqnarray}
As appropriate for a spinor field, the operators $b_{\bm k}^r$ and
$d_{\bm k}^r$ satisfy anti-commutation rules. The index $r$ represents
the two polarization states. Inserting the
expansion~(\ref{eq:expansionpsi}) into the Dirac
equation~(\ref{eq:diraceq}) leads to two equations, namely
\begin{eqnarray}
\left(\gamma ^0\omega -\gamma ^ik_i-m\right)u({\bm k},r) &=& 0,\\
\left(\gamma ^0\omega -\gamma ^ik_i+m\right)v({\bm k},r) &=& 0.
\end{eqnarray}
Upon using the explicit form~(\ref{eq:diracmatrices}), the first
equation can be re-written as
\begin{equation}
\label{eq:eomu}
\begin{pmatrix} \left(m-\omega\right)\mathbb{I}_2 & \sigma ^ik_i \cr
-\sigma ^i k_i & \left (m+\omega\right)\mathbb{I}_2 
\end{pmatrix}u({\bm k},r)=0.
\end{equation}
In order to have a non-trivial solution, the determinant of this
matrix should vanish and this leads to
\begin{equation}
w(k)=\sqrt{k^2+m^2},
\end{equation}
where we have used the following properties of the Pauli matrices,
$\sigma^ik_i\sigma^jk_j=k^2$. On the other hand, Eq.~(\ref{eq:eomu})
also implies that
\begin{equation}
u({\bm k},r)=\begin{pmatrix}
\psi \cr \displaystyle \frac{\sigma ^ik_i}{m+\omega }\psi 
\end{pmatrix},
\end{equation}
where $\psi$ is an arbitrary two components spinor. Choosing the
solution to be an eigenstate of the spin operator along the
$z$-direction (in the rest frame of the particle), the two components
of $u({\bm k},r)$ can be expressed as
\begin{equation}
u({\bm k},1)=\sqrt{m+\omega}\begin{pmatrix}
\begin{pmatrix} 1 \cr 0\end{pmatrix} \cr 
\displaystyle \frac{\sigma ^ik_i}{m+\omega }
\begin{pmatrix} 1 \cr 0\end{pmatrix}
\end{pmatrix},
\end{equation}
and 
\begin{equation}
u({\bm k},2)=\sqrt{m+\omega}\begin{pmatrix}
\begin{pmatrix} 0 \cr 1\end{pmatrix} \cr 
\displaystyle \frac{\sigma ^ik_i}{m+\omega }
\begin{pmatrix} 0 \cr 1\end{pmatrix}
\end{pmatrix}.
\end{equation}
It is easy to check that this spinor is normalized according to
\begin{equation}
\label{eq:normu}
\overline{u}({\bm k},r)u({\bm k},s)=2m\delta _{sr}.
\end{equation}
In the same manner, the solution $v({\bm k},r)$, corresponding to the
other branch of the Fourier expansion, can be written as
\begin{equation}
\label{eq:v1}
v({\bm k},1)=\sqrt{m+\omega}\begin{pmatrix}
\displaystyle \frac{\sigma ^ik_i}{m+\omega }
\begin{pmatrix} 1 \cr 0\end{pmatrix} \cr
\begin{pmatrix} 1 \cr 0\end{pmatrix} 
\end{pmatrix},
\end{equation}
and 
\begin{equation}
\label{eq:v2}
v({\bm k},2)=\sqrt{m+\omega}\begin{pmatrix}
\displaystyle \frac{\sigma ^ik_i}{m+\omega }
\begin{pmatrix} 0 \cr 1\end{pmatrix} \cr
\begin{pmatrix} 0 \cr 1\end{pmatrix} 
\end{pmatrix}.
\end{equation}
The normalization of this spinor satisfies a relation almost identical
to that of $u({\bm k},r)$ except for the presence of a minus sign
\begin{equation}
\label{eq:normv}
\overline{v}({\bm k},r)v({\bm k},s)=-2m\delta _{sr}.
\end{equation}
Finally, we also have
\begin{equation}
\label{eq:normuv}
\overline{u}({\bm k},r)v({\bm k},s)=0,
\end{equation}
which can easily be obtained from the above expressions.

\par

All the previous considerations are standard textbooks
calculations. We now turn to our main goal, namely the calculation of
the vacuum energy. Therefore, the next step consists in calculating
the energy momentum tensor. Either using the Noether procedure or a
variation of the action with respect to vierbeins lead to the equation
\begin{equation}
\label{eq:stressfermion}
T_{\mu \nu}=-\frac{i}{4}\left(\overline{\Psi} \gamma _{\mu}\partial _{\nu}
\Psi+\overline{\Psi} \gamma _{\nu}\partial _{\mu}\Psi
-\partial _{\mu}\overline{\Psi} \gamma _{\nu}\Psi
-\partial _{\nu}\overline{\Psi }\gamma _{\mu}\Psi\right).
\end{equation}
Of course, this expression is symmetric in $\mu $ and $\nu$ as it
should. Endowed with this formula, one can now evaluate the various
components of the stress energy tensor in the vacuum state. Let us
first calculate the energy density. One has
\begin{equation}
\langle \rho \rangle =-\frac{i}{4}\langle 0\left \vert 
\left(-2\Psi^{\dagger}\partial _0\Psi
+2\partial _0\Psi^{\dagger} \Psi\right)\right \vert 0\rangle, 
\end{equation} 
since $\gamma ^0=-\gamma _0$ and $\left(\gamma
  ^0\right)^2=\mathbb{I}_4$. Inserting the Fourier
expansion~(\ref{eq:expansionpsi}) into the above expression, one
arrives at
\begin{equation}
  \langle \rho \rangle =-\frac{1}{(2\pi)^3}\frac{1}{2}
  \int {\rm d}{\bm k}\sum _{s=1}^2v^{\dagger}({\bm k},s)v({\bm k},s).
\end{equation}
This equation can be further simplified. Upon using the explicit form
of the spinors given above, one can show that
\begin{equation}
\sum _{s=1}^2v^{\dagger}({\bm k},s)v({\bm k},s)=4\omega(k),
\end{equation}
and, therefore, we obtain the following expression for the vacuum
energy
\begin{equation}
\label{eq:rhovacfermion}
  \langle \rho \rangle =-\frac{1}{(2\pi)^3}\frac{4}{2}
  \int {\rm d}{\bm k}\, \omega (k).
\end{equation}
Several comments are in order at this point. From the above
expression, we see that we find minus four times the corresponding
result for a scalar field, see Eq.~(\ref{eq:rhovac}). We have a factor
four because we have two particles (the particle and its
anti-particle) with two polarization state each. Of course, the most
striking aspect of the above equation is that $\langle \rho \rangle $
is negative. It is therefore interesting to understand the origin of
the minus sign in more details. For this purpose, let us calculate the
Hamiltonian operator associated to the Dirac field
\begin{equation}
H=\int {\rm d}^3{\bm x} \, T_{00}.
\end{equation}
Inserting the Fourier expansion~(\ref{eq:expansionpsi}) of the spinor
$\Psi$ into the above equation, it is straightforward to arrive
at~\footnote{For convenience, let us notice that the following
  relationships are useful for this calculation
\begin{eqnarray}
u^{\dagger}({\bm k},s)u({\bm k},r)&=& 2\omega \delta_{rs}, 
\\
v^{\dagger}({\bm k},s)v({\bm k},r)&=&2\omega \delta_{rs},\\ 
v^{\dagger}({\bm k},s)v(-{\bm k},r)&=& 0.
\end{eqnarray}
These three equations are responsible for the fact that only the
quantity $\omega(k)$ appears into the expression of the
Hamiltonian~(\ref{eq:Hspinor}).}
\begin{equation}
\label{eq:Hspinor}
H=\int {\rm d}{\bm k}\sum _{s=1}^2\omega (k)\left[
\left(b_{\bm k}^s\right)^{\dagger}b_{\bm k}^s-
d_{\bm k}^s\left(d_{\bm k}^s\right)^{\dagger}\right].
\end{equation}
However, as we have already recalled, a spinor field is a quantity
which anti-commutes and, therefore, one has $\left\{d_{\bm
    k}^s,\left(d_{\bm p}^r\right)^{\dagger}\right\}= \delta \left({\bm
    k}-{\bm p}\right)\delta _{rs}.  $ As a consequence, the
Hamiltonian~(\ref{eq:Hspinor}) can be re-written as
\begin{equation}
H=\int {\rm d}{\bm k}\sum _{s=1}^2\omega (k)\left[
N_b+N_d-\delta(0)\right],
\end{equation}
where $N_b\equiv \left(b_{\bm k}^s\right)^{\dagger}b_{\bm k}^s$ and
$N_d\equiv \left(d_{\bm k}^s\right)^{\dagger}d_{\bm k}^s$ are the
particle and anti-particle number operators respectively. Of course,
their mean value vanishes in the vacuum state which contains no
particle. The zero point energy is thus given by the formally infinite
term $\delta(0)$. Since this term appears with a minus sign in the
Hamiltonian, the corresponding energy density is indeed negative. The
origin of this minus sign is the anti-commuting properties of the
creation and annihilation operators. We conclude that the vacuum
energy density is negative because we deal with fermions which are
anti-commuting objects.

\par

Let us now turn to the calculation of the pressure. Upon using
Eq.~(\ref{eq:stressfermion}), one has
\begin{eqnarray}
\langle p\rangle &=&\frac13\perp^{\mu \nu}T_{\mu \nu}\\
&=& -\frac{i}{6}g^{00}\bigl(\overline{\Psi}\gamma _0\partial _0\Psi
-\partial _0\overline{\Psi}\gamma _0 \Psi\bigr)
-\frac{i}{6}g^{ij}\bigl(\overline{\Psi}\gamma _i\partial _j\Psi
\nonumber \\ &-&
2\partial _i\overline{\Psi}\gamma _j \Psi\bigr)
-\frac{i}{6}u^{0}u^0\bigl(\overline{\Psi}\gamma _0\partial _0\Psi
-\partial _0\overline{\Psi}\gamma _0 \Psi\bigr)\\
&=&-\frac{i}{6}\delta^{ij}\left(\overline{\Psi}\gamma _i\partial _j\Psi
-2\partial _i\overline{\Psi}\gamma _j \Psi\right).
\end{eqnarray}
From the above expression, we see that we have two terms to
calculate. Using the definition of $\overline{\Psi}$, the first one
reads
\begin{eqnarray}
\langle 0\left \vert\Psi^{\dagger}\gamma ^0\gamma ^i\partial _i\Psi 
\right\vert 0\rangle &=&-\frac{i}{(2\pi)^3}
\int \frac{{\rm d}{\bm k}}{\sqrt{2\omega(k)}}k_i
\nonumber \\ &\times& 
\sum _{s=1}^2
v^{\dagger}({\bm k},s)
\gamma^0\gamma ^iv({\bm k},s),
\end{eqnarray}
while the second one takes the form
\begin{equation} 
\langle 0\left \vert\Psi^{\dagger}\gamma ^0\gamma ^i\partial _i\Psi 
\right\vert 0\rangle=-
\langle 0\left \vert\partial _i\Psi^{\dagger}\gamma ^0\gamma ^i\Psi 
\right\vert 0\rangle.
\end{equation}
As a consequence, one finds that the vacuum pressure is given by the
following expression
\begin{eqnarray}
\langle p\rangle &=&-\frac13\frac{1}{(2\pi)^3}
\int \frac{{\rm d}{\bm k}}{\sqrt{2\omega(k)}}
\nonumber \\ &\times& 
\sum _{s=1}^2
v^{\dagger}({\bm k},s)
\gamma^0\gamma ^ik_iv({\bm k},s).
\end{eqnarray}
One must now evaluate the sum present in the above integral. We have
seen that the spinors $v$ obeys the Dirac equation $\left(\gamma
  ^0\omega -\gamma ^ik_i+m\right)v({\bm k},s)=0$. Using this last
equation, one deduces that
\begin{eqnarray}
v^{\dagger}({\bm k},s)\gamma ^0\gamma ^ik_iv({\bm k},s)&=&
\omega v^{\dagger}({\bm k},s)v({\bm k},s)\nonumber \\
&+& v^{\dagger}({\bm k},s)\gamma ^0mv({\bm k},s).
\end{eqnarray}
Then, one can sum over the polarization and use the normalization of
the spinors established before, see for instance
Eqs.~(\ref{eq:normu}), (\ref{eq:normv}) and~(\ref{eq:normuv}), to
obtain that
\begin{eqnarray}
& & \sum _{s=1}^2v^{\dagger}({\bm k},s)\gamma ^0\gamma ^ik_iv({\bm k},s)=
4\omega^2 \nonumber \\ & & 
+\sum _{s=1}^2v^{\dagger}({\bm k},s)\gamma ^0mv({\bm k},s).
\end{eqnarray}
On the other hand, with the explicit expressions~(\ref{eq:v1})
and~(\ref{eq:v2}), it is easy to directly check that $\sum
_{s=1}^2v^{\dagger}({\bf k},s)\gamma ^0mv({\bf k},s) =-4m^2 $ from
which one can establish that
\begin{equation}
\sum _{s=1}^2v^{\dagger}({\bm k},s)\gamma ^0\gamma ^ik_iv({\bm  k},s)=
4\omega ^2-4m^2=4k^2,
\end{equation}
which is exactly the sum needed to evaluate the pressure. Therefore,
our final expression for $\langle p\rangle$ can be written as
\begin{eqnarray}
\langle p\rangle &=&-\frac{1}{(2\pi)^3}\frac{4}{6}
\int {\rm d}{\bm k}\frac{k^2}{\omega(k)}.
\end{eqnarray}
We see that, again, we recover exactly the same result as for the
scalar field but with the additional multiplicative factor $-4$, see
Eq.~(\ref{eq:pressurevac}). This means that all the previous
considerations discussed in the scalar field case also apply to the
fermion case. In particular, the regularization of the above divergent
integrals would proceed in the same way. Introducing a cut-off would
break Lorentz invariance and would lead to an incorrect equation of
state. On the other hand, using, say, dimensional regularization
produces the correct vacuum equation of state and leads to a
regularized vacuum energy which is minus four times the regularized
scalar field vacuum energy. We do not need to repeat all these
calculations here and we now move to the case of a vector field.

\subsection{Proca Fields}
\label{subsec:proca}

Having treated the case of a spin $1/2$ field in the previous
sub-section, we now turn to the case of a Proca
field~\cite{Peskin:1995ev,Greiner:1996zu,Itzykson:1980rh,
  Bailin:1994qt,LeBellac:1991cq,Ryder:1985wq,Mandl:1985bg}. A Proca
field $A_{\mu}(t,{\bm x})$ is a massive vector, spin $1$, field. The
corresponding action can be written as
\begin{equation}
\label{eq:actionproca}
S[A_{\mu}]\equiv -\int {\rm d}^4x\sqrt{-g}\left(\frac14F_{\mu \nu}F^{\mu \nu}
+\frac12m^2A_{\mu}A^{\mu}\right),
\end{equation}
where $F_{\mu \nu}\equiv \partial _{\mu}A_{\nu}-\partial_{\nu
}A_{\mu}$ and $m$ is the mass of the particle. Again, we will restrict
our considerations to flat space-time and, therefore, we will take for
the metric determinant $g=\eta =1$.

\par

Before turning to the calculation of the vacuum stress energy
momentum, we quickly remind some basics fact about the quantization of
this type of fields. Varying with respect to $A_{\mu}$, one obtains
the equation of motion
\begin{equation}
\partial _{\mu}F^{\mu \nu}-m^2A^{\nu}=0.
\end{equation}
Taking the derivative of this expressions and noting that $\partial _{\nu}
\partial _{\mu}F^{\mu \nu}=0$ (since this is the contraction of a
symmetric expression with an anti-symmetric one), one arrives at
\begin{equation}
\partial _{\nu}A^{\nu}=0,
\end{equation}
provided that $m\neq 0$, which we assume in this section. As a
consequence, the equation of motion of the vector field is given by
\begin{equation}
\partial _{\nu}\partial ^{\nu}A_{\mu}-m^2A_{\mu}=0,
\end{equation}
which is nothing but a Klein-Gordon equation for each component of
$A_{\mu}$. As usual, we Fourier expand the field in terms of creation
and annihilation operators. This leads to the following expression
\begin{eqnarray}
A_{\mu}\left(t,{\bm x}\right)&=&\frac{1}{\left(2\pi\right)^{3/2}}
\int \frac{{\rm d}{\bm k}}{\sqrt{2\omega (k)}}\sum _{\alpha=0}^4
\epsilon_{\mu}^{\alpha}({\bm k})
\biggl[a_{\bm k}^{\alpha}{\rm e}^{-i\omega t+i{\bm k}\cdot {\bm x}}\nonumber \\ 
& &
+\left(a_{\bm k}^{\alpha}\right)^{\dagger}
{\rm e}^{i\omega t-i{\bm k}\cdot {\bm x}}\biggr],
\end{eqnarray}
where, as usual, one has $\omega (k)=\sqrt{k^2+m^2}$. Of course, the
operators $a_{\bm k}^{\alpha}$ satisfy commutation rules since a
vector field is a bosonic field. The sum over $\alpha$ is a sum over 
four vectors that span space-time and form a basis. At
this point, $A_{\mu}$ satisfies the equation of motion but not yet the
constraint $\partial _{\nu}A^{\nu}=0$. Working out this constraint, it is
easy to show that this amounts to $\epsilon_{\mu}^{\alpha}k^{\mu}a_{\bm
  k}^{\alpha }=0$. If we now choose a frame where the wave-vector is
aligned along the $z$-direction, then the basis can be taken to be
\begin{eqnarray}
\label{eq:vecbasis1}
\epsilon_{\mu}^0&=&\begin{pmatrix} -\omega/m \cr 0 \cr 0 \cr k/m
\end{pmatrix},\quad
\epsilon_{\mu}^1=\begin{pmatrix} 0 \cr \hat{\epsilon}_x^1 \cr 
\hat{\epsilon}_y^1 \cr 0
\end{pmatrix},\\
\label{eq:vecbasis2}
\epsilon_{\mu}^2&=&\begin{pmatrix} 0 \cr \hat{\epsilon}_x^2 \cr 
\hat{\epsilon}_y^2 \cr 0
\end{pmatrix}, \quad
\epsilon_{\mu}^3=\begin{pmatrix} -k/m \cr 0 \cr 0 \cr \omega/m
\end{pmatrix},
\end{eqnarray}
where $\hat{\epsilon}^{1,2}$ are vectors perpendicular to the
wave-vector (the time components of $\epsilon^0$ and $\epsilon^3$
carry a minus sign because we show here the covariant vectors). But we
have $\epsilon_{\mu}^0k^{\mu}\neq 0$ despite the fact that this is the
only choice such that $\epsilon^0_{\mu}\epsilon^{\mu 3}=0$. Therefore,
the only possibility left in order to satisfy the constraint is to
take $a_{\bm k}^0=0$. As a consequence, the Fourier expansion of the
vector potential is now given by
\begin{eqnarray}
\label{eq:expansionproca}
A_{\mu}\left(t,{\bm x}\right)&=&\frac{1}{\left(2\pi\right)^{3/2}}
\int \frac{{\rm d}{\bm k}}{\sqrt{2\omega (k)}}\sum _{\alpha=1}^3
\epsilon_{\mu}^{\alpha}({\bm k})
\biggl[a_{\bm k}^{\alpha}{\rm e}^{-i\omega t+i{\bm k}\cdot {\bm x}}
\nonumber \\ 
& &
+\left(a_{\bm k}^{\alpha}\right)^{\dagger}
{\rm e}^{i\omega t-i{\bm k}\cdot {\bm x}}\biggr],
\end{eqnarray}
where the three vectors appearing in this decomposition are still 
given in Eqs.~(\ref{eq:vecbasis1}) and~(\ref{eq:vecbasis2}).

\par

Let us now come to our main subject and calculate the stress-energy tensor. 
There are several ways to carry out this calculation. Here, we choose 
to return to Eq.~(\ref{eq:actionproca}) and to vary this action with respect 
to the space-time metric tensor. Then, in order to work with the stress 
energy tensor in flat space-time, we will take $g_{\mu \nu}=\eta_{\mu \nu}$ 
but, of course, only after having performed the variation. Straightforward
manipulations lead to the following expression
\begin{eqnarray}
T_{\mu \nu} &\equiv& -\frac{2}{\sqrt{-g}}
\frac{\delta S[A_{\mu}]}{\delta g^{\mu \nu}}
\\
&=& -g_{\mu \nu}\left(\frac14 g^{\lambda \alpha}g^{\kappa \beta}
F_{\lambda \kappa}F_{\alpha \beta}+
\frac12m^2g^{\alpha \beta }A_{\alpha }A_{\beta}
\right)
\nonumber \\ & & 
+g^{\alpha \beta}F_{\mu \alpha}F_{\nu \beta}+m^2A_{\mu}A_{\nu}.
\end{eqnarray}
We are now in a position where we can compute the various components of 
$T_{\mu \nu}$. Let us start with the energy density. From the above 
expression, one deduces that
\begin{equation}
\label{eq:timetimeproca}
T_{00}=\frac12\delta ^{ij}F_{0i}F_{0j}+\frac14 \delta ^{ij}\delta^{kl}
F_{ik}F_{jl}+\frac12m^2A_0^2+\frac12m^2\delta^{ij}A_iA_j.
\end{equation}
Then, in order to evaluate the above equation, we have to separately calculate 
the four terms that appear in this expression. Let us first give the relevant 
results needed to calculate the first term. Upon using 
Eq.~(\ref{eq:expansionproca}), it is easy to establish that
\begin{eqnarray}
\delta ^{ij}\langle 0\vert \partial _0A_i\partial _0A_j\vert 0\rangle &=& 
\frac{1}{(2\pi)^3}\int \frac{{\rm d}{\bm k}}{2\omega (k)}
\delta ^{ij}\nonumber \\ &\times& 
\sum _{\alpha=1}^3\epsilon_i^{\alpha }\epsilon_j^{\alpha}\omega ^2,
\\
\delta ^{ij}\langle 0\vert \partial _0A_i\partial _jA_0\vert 0\rangle &=& 
-\frac{1}{(2\pi)^3}\int \frac{{\rm d}{\bm k}}{2\omega (k)}
\delta ^{ij}\nonumber \\ &\times& 
\sum _{\alpha=1}^3\epsilon_i^{\alpha }\epsilon_0^{\alpha}\omega k_j,
\end{eqnarray}
and, 
\begin{eqnarray}
\delta ^{ij}\langle 0\vert \partial _iA_0\partial _jA_0\vert 0\rangle &=& 
\frac{1}{(2\pi)^3}\int \frac{{\rm d}{\bm k}}{2\omega (k)}
\delta ^{ij}\nonumber \\ &\times & 
\sum _{\alpha=1}^3\epsilon_0^{\alpha }\epsilon_0^{\alpha}k_i k_j.
\end{eqnarray}
In the same manner, the second term, containing the space space component 
of the field strength, can be expressed as
\begin{align}
\delta ^{ij}\delta^{kl}\langle 0\vert F_{ik}F_{jl}&\vert 0\rangle =
\frac{2}{(2\pi)^3}\int \frac{{\rm d}{\bm k}}{2\omega (k)}
\delta ^{ij}\delta^{kl}\nonumber \\ &\times
\sum _{\alpha=1}^3
\left(\epsilon_k^{\alpha }\epsilon_l^{\alpha}k_i k_j
-\epsilon_k^{\alpha }\epsilon_j^{\alpha}k_i k_l\right).
\end{align}
The third term can be written as
\begin{eqnarray}
\langle 0\vert A_0^2\vert 0\rangle &=& 
\frac{1}{(2\pi)^3}\int \frac{{\rm d}{\bm k}}{2\omega (k)}
\sum _{\alpha=1}^3\epsilon_0^{\alpha }\epsilon_0^{\alpha},
\end{eqnarray}
and, finally, the fourth and last term takes the form
\begin{eqnarray}
\delta ^{ij}\langle 0\vert A_iA_j\vert 0\rangle &=& 
\frac{1}{(2\pi)^3}\int \frac{{\rm d}{\bm k}}{2\omega (k)}
\sum _{\alpha=1}^3\delta ^{ij}\epsilon_i^{\alpha }\epsilon_j^{\alpha}.
\end{eqnarray}
Having calculated the four terms appearing in the expression of
$T_{00}$, one must now combine them. Using the relationships satisfied
by the vectors basis\footnote{For convenience, let us give the two
  equations needed to complete the calculation. It involves simple
  combination of the polarization vectors. Using the explicit
  representation given by Eqs.~(\ref{eq:vecbasis1})
  and~(\ref{eq:vecbasis2}), one arrives at
\begin{eqnarray}
\delta^{ij}\sum_{\alpha =1}^3\epsilon_i^{\alpha}\epsilon_j^{\alpha}
&=&2+\frac{\omega^2}{m^2},\\
\sum_{\alpha =1}^3\epsilon_0^{\alpha}\epsilon_0^{\alpha}
&=&\frac{k^2}{m^2}.
\end{eqnarray}
Then, these two expressions can be combined to lead to the following 
relation which is what is needed in order to simplify the expression 
of the vacuum energy density
\begin{eqnarray}
\sum _{\alpha=1}^3
\left(\epsilon_k^{\alpha }\epsilon_l^{\alpha}k_i k_j
-\epsilon_k^{\alpha }\epsilon_j^{\alpha}k_i k_l\right)&=&
k^2\left(2+\frac{\omega^2}{m^2}\right)-\frac{k^2\omega^2}{m^2}
\\
\nonumber 
&=& 2 .
\end{eqnarray}}, one obtains the following expression for the vacuum 
energy density
\begin{equation}
\langle \rho \rangle =\frac{1}{(2\pi)^3}\frac{3}{2}
\int {\rm d}{\bm k}\, \omega (k).
\end{equation}
Based on our experience with the calculation of the vacuum energy of a
spin $1/2$ field, this expression is totally expected. As before, it
is equal to the scalar field energy density times a multiplicative
factor which takes into account the spin of the particle. Here, for a
massive vector field, there are three polarization states and, as a
consequence, the multiplicative factor is three.

\par

In order to be exhaustive, let us now evaluate the pressure. The calculation 
proceeds exactly as before and it is easy to show that
\begin{equation}
\label{eq:linkrhopmassivevector}
\perp^{\mu \nu}T_{\mu \nu}=T_{00}+m^2A_0^2-m^2\delta^{ij}A_iA_j,
\end{equation}
from which, using the relations already established before, we immediately 
deduce that
\begin{equation}
\langle p \rangle =\frac{1}{(2\pi)^3}\frac{3}{6}
\int {\rm d}{\bm k}\, \frac{k^2}{\omega (k)}.
\end{equation}
Of course, this is again the expected formula, namely the scalar field
pressure times a multiplicative factor equals to three. Clearly, there
is no need to repeat the discussion about the regularization of the
above divergent integrals. It proceeds exactly as explained before.

\subsection{Mass-less Vector Field}
\label{subsec:masslessvector}

Finally, we consider the case where the vector field is mass-less,
$m=0$~\cite{Peskin:1995ev,Greiner:1996zu,Itzykson:1980rh,
  Bailin:1994qt,LeBellac:1991cq,Ryder:1985wq,Mandl:1985bg}. This is of
course necessary if we want to realistically evaluate the vacuum
energy density since we know that this type of field exist in
Nature. If the mass vanishes, this means that the dispersion relation
is now given by
\begin{equation}
\omega(k)=k,
\end{equation}
as appropriate for photons. In that case, contrary to the massive case, we 
can no longer establish that $\partial _{\nu}A^{\nu}=0$. Therefore, the 
equation of motion takes the form
\begin{equation}
\partial _{\mu}\partial ^{\mu}A^{\nu}-\partial ^{\nu}\left(\partial _{\mu}
A^{\mu}\right)=0.
\end{equation}
At this stage, we need to fix a gauge. As is well-known, it is convenient 
to work in the Coulomb gauge defined by the following conditions
\begin{equation}
A_0=0, \quad \partial _iA^i=0.
\end{equation}
In particular, the last condition implies that $A_i(t,{\bm x})$ is a transverse
field. As a consequence, its Fourier expansion can be written as
\begin{eqnarray}
\label{eq:expansionphoton}
A_{i}\left(t,{\bm x}\right)&=&\frac{1}{\left(2\pi\right)^{3/2}}
\int \frac{{\rm d}{\bm k}}{\sqrt{2\omega (k)}}\sum _{\alpha=1}^2
\epsilon_{i}^{\alpha}({\bm k})
\biggl[a_{\bm k}^{\alpha}{\rm e}^{-i\omega t+i{\bm k}\cdot {\bm x}}\nonumber \\ 
& &
+\left(a_{\bm k}^{\alpha}\right)^{\dagger}
{\rm e}^{i\omega t-i{\bm k}\cdot {\bm x}}\biggr].
\end{eqnarray}
The two three-vectors $\epsilon^1$ and $\epsilon^2$ are perpendicular
to the wave-vectors ${\bm k}$. In the mass-less case, only two degrees
of freedom remain.

\par

The calculation of the stress-energy tensor is now straightforward. Using 
the calculation of the previous section, in particular the expression 
of the time-time component of the stress-energy 
tensor~(\ref{eq:timetimeproca}), one finds that
\begin{equation}
T_{00}=\frac12\delta^{ij}\partial _0A_i\partial _0A_j+\frac14 \delta ^{ij}
\delta ^{kl}F_{ik}F_{jl},
\end{equation}
and, since we now have, 
\begin{eqnarray}
\delta ^{ij}\sum _{\alpha=1}^2\epsilon_i^{\alpha}\epsilon_j^{\alpha}&=&2, \\
\sum _{\alpha=1}^2
\left(\epsilon_k^{\alpha }\epsilon_l^{\alpha}k_i k_j
-\epsilon_k^{\alpha }\epsilon_j^{\alpha}k_i k_l\right)&=& 2k^2,
\end{eqnarray}
one finally arrives at the following expression for the vacuum energy density
\begin{equation}
\langle \rho \rangle =\frac{1}{(2\pi)^3}\frac{2}{2}
\int {\rm d}{\bm k}\, \omega (k),
\end{equation}
that is to say the scalar field energy density times two, as expected
since there are two polarization states. For the pressure, using the
same approach, one immediately obtains
\begin{equation}
\langle p \rangle =\frac{1}{(2\pi)^3}\frac{2}{6}
\int {\rm d}{\bm k}\, \frac{k^2}{\omega (k)},
\end{equation}
where, as usual, the factor two in front of the whole expression originates
from the the two polarization states of a mass-less vector particle.

\par

We conclude by stating the main result established in this
section. For any type of fields, the vacuum energy density can be
written as
\begin{equation}
\langle \rho \rangle =\frac{1}{(2\pi)^3}\frac{s}{2}
\int {\rm d}{\bm k}\, \omega (k),
\end{equation}
where $s$ represents the number of polarization states ($s=1$ for a
scalar field, $s=4$ for a spinor field, $s=3$ for a massive vector
field and $s=2$ for a mass-less vector field). The pressure is given by
a similar expression. When these divergent integrals are regularized,
one can show that, again for any type of field, this leads to the
Lorentz invariant form of the vacuum equation of state, namely
$\langle p\rangle =-\langle \rho \rangle$. In principle, the results
of this section makes the accurate evaluation of the vacuum energy
density in the standard model of particle physics possible. However,
before discussing this point, we need to say something about the fact
that $\rho_{_{\rm vac}}$ is positive for bosons and negative for
fermions.

\subsection{Application: Super-Symmetry}
\label{subsec:susy}

\subsubsection{Motivation}
\label{subsubsec:susymotiv}

The fact that the sign of the zero-point energy density is different
for fermions and bosons, see Eqs.~(\ref{eq:rhovac})
and~(\ref{eq:rhovacfermion}), suggests a very simple way of solving
the cosmological constant problem. It is clear that one can design a
theory where the two contributions are equal in absolute value such
that the final result is exactly zero. Let us see how it works in
practice. For this purpose, let us consider again the Lagrangian of a
massive real scalar field, see also Eq.~(\ref{eq:actionsf})
\begin{equation} {\cal L}{_\phi}=- \frac{\alpha}{2}\eta^{\mu
    \nu}\partial _{\mu}\Phi\partial _{\nu}\Phi -\frac{1}{2}m^2\Phi^2,
\end{equation}
where we have introduced a constant $\alpha$ in front of the kinetic
term. Before we took $\alpha =1$ and it is interesting to discuss why
we made this choice (after all, how do we know that $\alpha =1$?). As
usual, see Eq.~(\ref{eq:fourierfield}), one Fourier expands the scalar
field as
\begin{eqnarray}
\label{eq:expandfieldalpha}
\Phi\left(t,\bm{x}\right)&=&\frac{1}{\left(2\pi\right)^{3/2}}
\int \frac{{\rm d}{\bm k}}{\sqrt{2\omega(k,\alpha)}}\biggl(c_{\bm k}
{\rm e}^{-i\omega t +i{\bm k}\cdot {\bm x}}
\nonumber \\ 
& & +
c_{\bm k}^{\dagger}{\rm e}^{i\omega t -i{\bm k}\cdot {\bm x}}
\biggr),
\end{eqnarray}
with a new expression for the quantity $\omega$, compare to
Eq.~(\ref{eq:defomegasf}), namely
\begin{equation}
\omega (k,\alpha)=\sqrt{k^2+\frac{m^2}{\alpha^2}},
\end{equation}
since this is the condition for Eq.~(\ref{eq:expandfieldalpha}) to be
a solution of the ``new'' equation of motion [see also
Eq.~(\ref{eq:kgequation})]
\begin{equation}
-\alpha \ddot{\phi}+\alpha\delta ^{ij}\partial _i\phi\partial_j\phi
-m^2\phi=0.
\end{equation}
The new dispersion relation can easily be understood. We see that the
quantity $\alpha $ simply renormalizes the value of the mass $m$. This
makes sense since, if $m=0$, we clearly see that the relativistic
relation $\omega =k$ is not affected by the presence of the constant
$\alpha$.

\par

The next step is to derive the commutation relations for the creation
and annihilation operators. Let us assume for the moment that we have
the standard relation~(\ref{eq:comsf})
\begin{equation}
\label{eq:comutcc}
\left[c_{\bm k},c_{\bm p}^{\dagger}\right]=\delta\left({\bm k}-{\bm p}\right).
\end{equation}
Then, from the expression of the Lagrangian, one can calculate the
conjugate momentum. It reads $\Pi(t,{\bm x})=-\alpha \dot{\Phi}(t,{\bm
  x})$. This implies that the commutation relation between the field
and its conjugate momentum now reads
\begin{equation}
\label{eq:comutphipi}
\left[\phi(t,{\bm x}),\pi(t,{\bm y})\right]=
i\alpha \delta ({\bm x}-{\bm y}),
\end{equation}
an expression which reduces to the standard commutation relation only
if $\alpha =1$. Therefore, if one wants to have
$[\Phi,\Pi]=i\delta({\bm x}-{\bm y})$ and, at the same time, $[c_{\bm
  k},c_{\bm p}^{\dagger}]=\delta\left({\bm k}-{\bm p}\right)$ there is
no other choice that taking $\alpha =1$. It is worth noticing that
working with the two previous commutation
relations~(\ref{eq:comutcc}), (\ref{eq:comutphipi}) and $\alpha \neq
1$ would lead to a different prediction for the energy density of the
vacuum. Indeed, the stress-energy tensor is now given by
\begin{equation}
T_{\mu \nu}=\alpha \partial _{\mu}\Phi\partial _{\nu}\Phi
-g_{\mu \nu}\left(\frac{\alpha}{2}g^{\alpha \beta}\partial_{\alpha}\Phi
\partial _{\beta }\Phi+\frac12m^2\Phi^2\right),
\end{equation}
which implies that
\begin{equation}
T_{00}=\frac{\alpha}{2}\dot{\Phi}^2+\frac{\alpha}{2}\delta ^{ij}\partial _i
\Phi \partial _j\Phi+\frac{1}{2}m^2\Phi^2.
\end{equation}
From this expression, it is easy to show that
\begin{equation}
\langle \rho\rangle
= \frac{\alpha}{\left(2\pi\right)^3}
\frac12 \int {\rm d}{\bm k}\, \omega (k,\alpha),
\end{equation}
an expression that should be compared to Eq.~(\ref{eq:rhovac}).

\par

Of course, one can also demonstrate that the consistency can be
re-establish if the commutation relations for the creation and
annihilation operators are taken to be $[c_{\bm k},c_{\bm
  p}^{\dagger}]=\delta\left({\bm k}-{\bm p}\right)/\alpha$. In this
case the commutation relation of the field and its conjugate momentum
have the correct form. Moreover, one has
\begin{eqnarray}
\langle 0\vert \dot{\Phi}^2\vert 0\rangle
&=&\frac{1}{(2\pi)^3}\int \frac{{\rm d}{\bm k}}{2\omega(k,\alpha)}\, 
\frac{\omega ^2(k,\alpha)}{\alpha},
\\
\langle 0\vert \delta ^{ij}\partial _i\Phi\partial _j\Phi \vert 0\rangle
&=&\frac{1}{(2\pi)^3}\int \frac{{\rm d}{\bm k}}{2\omega(k,\alpha)}\, 
\frac{k^2}{\alpha},
\\
\langle 0\vert \Phi^2\vert 0\rangle
&=&\frac{1}{(2\pi)^3}\int \frac{{\rm d}{\bm k}}{2\omega(k,\alpha)}
\frac{1}{\alpha},
\end{eqnarray}
[these formulas should be compared to Eqs.~(\ref{eq:phidot2}), 
(\ref{eq:partialphi2}) and~(\ref{eq:phi2}), which implies that
\begin{equation}
\langle \rho\rangle
= \frac{1}{\left(2\pi\right)^3}
\frac12 \int {\rm d}{\bm k}\, \omega (k,\alpha),
\end{equation}
that is to say the ``standard'' result.

\par

Therefore, in conclusion, the ``$1/2$'' coefficient in the expression
of the vacuum energy density is only obtained if the commutation
relations and the normalization of the field are properly and
consistently chosen. In particular, if one chooses to work with a
coefficient ``$1/2$'' in front of the kinetic term (canonical
normalization), then there is no other choice than working with
$[c_{\bm k},c_{\bm p}^{\dagger}]=\delta\left({\bm k}-{\bm
    p}\right)$. Or, in other words, the previous commutation relation
implies that a ``$1/2$'' in front of the kinetic term must be present.

\par

We now apply the previous discussion to the cases of a complex scalar
field and to a Majorana spinor. The Lagrangian for a complex scalar
field should be written as
\begin{equation}
{\cal L}=-g^{\mu \nu}\partial _{\mu}\Phi^*\partial _{\nu}\Phi
-m^2\Phi^*\Phi,
\end{equation}
with, this time, no ``$1/2$'' in front of the whole expression. Indeed, 
if one writes that 
\begin{eqnarray}
\phi&=&\frac{1}{\sqrt{2}}\left(A_1+iA_2\right),
\end{eqnarray}
then the Lagrangian takes the form of the sum of two Lagrangian for
the two real canonically normalized fields $A_1$ and $A_2$.

\par 

The previous discussion has also important implications for the Lagrangian 
of a Majorana spinor. A Majorana spinor is a spinor satisfying
\begin{equation}
\Psi ^{\rm c}_{_{\rm M}}=\Psi_{_{\rm M}},
\end{equation}
where the charge conjugate spinor $\Psi ^{\rm c}$ is defined by 
\begin{equation}
\Psi ^{\rm c}\equiv C\overline{\Psi}^{_{\rm T}}.
\end{equation}
In this expression the symbol ``${\rm T}$'' denotes the transpose
matrix. The matrix $C$ is the charge conjugation operator that can be
represented by $i\gamma ^2\gamma ^0$ and satisfies $C\gamma
_{\mu}^{_{\rm T}}C^{-1}=-\gamma_{\mu}$. We have of course $\left(\Psi
  ^{\rm c}\right)^{\rm c}=\Psi$. From a Dirac spinor $\Psi$ one can
always write two Majorana spinors
\begin{eqnarray}
\Psi_{_{\rm M1}}&=& \frac{1}{\sqrt{2}}\left(\Psi+\Psi^{\rm c}\right),
\\
\Psi_{_{\rm M2}}&=& \frac{1}{i\sqrt{2}}\left(\Psi-\Psi^{\rm c}\right).
\end{eqnarray}
Of course, the above expressions can always be inverted and re-written
as
\begin{eqnarray}
\Psi &=& \frac{1}{\sqrt{2}}\left(\Psi_{_{\rm M1}}+i\Psi_{_{\rm M2}}\right),
\\
\Psi ^{\rm c}&=& \frac{1}{\sqrt{2}}\left(\Psi_{_{\rm M1}}-i\Psi_{_{\rm M2}}
\right).
\end{eqnarray}
The two previous expression can be inserted in the Dirac
Lagrangian~(\ref{eq:diracl}) (taking $m=0$). This leads to the
following expression
\begin{eqnarray}
{\cal L}_{\rm Dirac} &=& -\frac{1}{\sqrt{2}}\left(\overline{\Psi}_{_{\rm M1}}
-i\overline{\Psi}_{_{\rm M2}}\right)i\gamma ^{\mu}\partial _{\mu}
\frac{1}{\sqrt{2}}\left(\overline{\Psi}_{_{\rm M1}}
+i\overline{\Psi}_{_{\rm M2}}\right) \nonumber \\
&=& -\frac{i}{2}\overline{\Psi}_{_{\rm M1}}\gamma ^{\mu}\partial _{\mu}
\Psi_{_{\rm M1}}-\frac{i}{2}\overline{\Psi}_{_{\rm M2}}\gamma ^{\mu}\partial _{\mu}
\Psi_{_{\rm M2}}\nonumber \\ & &
-\frac{i}{2}\left(-i\overline{\Psi}_{_{\rm M2}}\gamma ^{\mu}\partial _{\mu}
\Psi_{_{\rm M1}}+i\overline{\Psi}_{_{\rm M1}}\gamma ^{\mu}\partial _{\mu}
\Psi_{_{\rm M2}}\right).
\end{eqnarray}
But, using partial integrations, one easily shows that
\begin{equation}
\overline{\Psi}_{_{\rm M2}}\gamma ^{\mu}\partial _{\mu}
\Psi_{_{\rm M1}}=-\left(\partial _{\mu}\overline{\Psi}_{_{\rm M2}}\right)\gamma ^{\mu}
\Psi_{_{\rm M1}}=+\overline{\Psi}_{_{\rm M1}}\gamma^{\mu}\partial_{\mu}
\Psi_{_{\rm M2}},
\end{equation}
so that the last term vanishes. Therefore, one conclude that the
Lagrangian for a Majorana spinor is given by
\begin{equation}
  \label{eq:LagragianMajorana}
  {\cal L}_{\rm Majorana}=-\frac{1}{2}
  \overline{\Psi}_{_{\rm M}}\left(i\gamma ^{\mu}\partial _{\mu}-m\right)
  \Psi_{_{\rm M}},
\end{equation}
and that the expansion of a Majorana spinor operator in Fourier can be
expressed as, see also Eq.~(\ref{eq:expansionpsi})
\begin{eqnarray}
\Psi_{_{\rm M}}\left(t,{\bm x}\right) &=& \frac{1}{(2\pi)^{3/2}}
\int \frac{{\rm d}{\bm k}}{\sqrt{2\omega(k)}}
\sum _{r=1}^{2}\biggl[b_{\bm k}^r u({\bm k},r){\rm e}^{-i\omega t+i{\bm k}\cdot
{\bm x}}\nonumber \\ & &+
\left(b_{\bm k}^r\right)^{\dagger} v({\bm k},r){\rm e}^{i\omega t-i{\bm k}\cdot
{\bm x}}\biggr],
\end{eqnarray}
where we notice that only the operators $b_{\bm k}^r$ are present
which reflects the fact that, for a Majorana spinor, the particle and
the anti-particle are the same. But the most important aspect is the
coefficient ``$1/2$'' in Eq.~(\ref{eq:LagragianMajorana}). Clearly, it
means that if we perform the calculation of $\langle \rho\rangle$
again, we are going to find
\begin{equation}
\langle \rho \rangle =-\frac{1}{(2\pi)^3}\frac{2}{2}
\int {\rm d}{\bm k}\, \omega (k),
\end{equation}
that is to say an expression with an overall factor $2$ instead of $4$
for a Dirac fermion, see Eq.~(\ref{eq:rhovacfermion}).

\par

We are now in a position to address the main point of this section. Let
us consider the following model described by a complex scalar field
and a Marojana spinor
\begin{equation}
\label{eq:susylagrangemajorana}
{\cal L}=-g^{\mu \nu}\partial _{\mu}\Phi^*\partial _{\nu}\Phi
-m^2_{_{\rm B}}\phi^*\phi-\frac{1}{2}
\overline{\Psi}_{_{\rm M}}\left(i\gamma ^{\mu}\partial _{\mu}-m_{_{\rm F}}\right)
\Psi_{_{\rm M}},
\end{equation}
where $m_{_{\rm B}}$ is the mass of the bosonic field and $m_{_{\rm
    F}}$ is the mass of the fermionic field. From the above
considerations, it is obvious to calculate the vacuum energy
density. It reads
\begin{eqnarray}
\label{eq:rhosusy}
\langle \rho \rangle &=&\frac{1}{(2\pi)^3}\frac{2}{2}
\int {\rm d}{\bm k}\, \sqrt{k^2+m_{_{\rm B}}^2}\nonumber \\
& &
-\frac{1}{(2\pi)^3}\frac{2}{2}
\int {\rm d}{\bm k}\, \sqrt{k^2+m_{_{\rm F}}^2}.
\end{eqnarray}
We have a factor two for the complex scalar field since we have seen
that a complex scalar field is equivalent to two real scalar fields
and a factor $-2$ for the Marojana spinor as shown before. In general,
the quantity $\langle \rho \rangle$ is not zero because $m_{_{\rm
    B}}\neq m_{_{\rm F}}$. But if we now assume that
\begin{equation}
m_{_{\rm B}}=m_{_{\rm F}},
\end{equation}
then the zero-point energy density exactly cancels out and we have
solved the cosmological constant problem. The above model is in fact
nothing but the simplest super-symmetric model, the so-called
Wess-Zumino
model~\cite{Wess:1974tw,Coleman:1969sm,Callan:1969sn,Wess:1973kz,Bailin:1994qt,Aitchison:2005cf}.

\subsubsection{The Vacuum Energy in Super-Symmetry}

Having motivated the idea of super-symmetry, we now study in more
detail its implications. In particular, we would like to check that
the type of cancellation studied before is generic in
super-symmetry. In order to simplify the discussion, we start with
considering the Lagrangian~(\ref{eq:susylagrangemajorana}) with
$m_{_{\rm B}}=m_{_{\rm F}}=0$. We will now use the standard
super-symmetric notation with the dotted two-components Weyl
spinors~\cite{Bailin:1994qt,Aitchison:2005cf}. In this formalism, the
Majorana spinor can be written as
\begin{equation}
\Psi_{_{\rm M}}=\begin{pmatrix}
\Psi _{\alpha} \cr \overline{\Psi}^{\dot{\alpha}}
\end{pmatrix}.
\end{equation}
Then, using that the Dirac matrices can be expressed as
\begin{equation}
\gamma ^{\mu}=\begin{pmatrix}
0 & \left(\sigma ^{\mu}\right)_{\alpha \dot{\alpha}} \cr
\left(\overline{\sigma }^{\mu}\right)^{\dot{\alpha}\alpha } & 0
\end{pmatrix},
\end{equation}
we find that the Lagrangian~(\ref{eq:LagragianMajorana}) can be
re-written as
\begin{equation}
\label{eq:newmajoranal}
{\cal L}_{\rm Majorana}=-\frac{i}{2}\overline{\Psi}_{\dot{\alpha}}
\left(\overline{\sigma}^{\mu}
\right)^{\dot{\alpha}\alpha}\partial _{\mu}\Psi_{\alpha}
-\frac{i}{2}\Psi^{\alpha}\left(\sigma^{\mu}
\right)_{\alpha \dot{\alpha}}\partial _{\mu}\overline{\Psi}^{\dot{\alpha}},
\end{equation}
where we used that (by definition)
$\overline{\Psi}_{\dot{\alpha}}=(\Psi_{\alpha})^*$ and
$\Psi^{\alpha}=(\overline{\Psi}^{\dot{\alpha}})^*$. The next step
consists in using the relation $\Psi^{\alpha}\left(\sigma^{\mu}
\right)_{\alpha \dot{\alpha}}\partial
_{\mu}\overline{\Psi}^{\dot{\alpha}}=-\partial
_{\mu}\overline{\Psi}_{\dot{\alpha}}
\left(\overline{\sigma}^{\mu}\right)^{\dot{\alpha}\alpha}
\Psi_{\alpha}$ (the spinor components are Grassmann variables and,
therefore, anti-commute, see below) and, then, to transform the
resulting expression by integration by part. In this way, one shows
that the Lagrangian~(\ref{eq:newmajoranal}) is in fact twice the first
term. As a consequence, the Wess-Zumino model is described by
\begin{equation}
\label{eq:susylagrange}
{\cal L}_{_{\rm W-Z}}=-\eta^{\mu \nu}\partial _{\mu}\Phi^{\dagger}
\partial _{\nu}\Phi
-i\overline{\Psi}_{\dot{\alpha}}
\left(\overline{\sigma}^{\mu}
\right)^{\dot{\alpha}\alpha}\partial _{\mu}\Psi_{\alpha} .
\end{equation}
Let us now consider a super-symmetric transformation. In order to
understand what it really means, let us use the analogy with a gauge
transformation~\cite{Bailin:1994qt,Aitchison:2005cf}. For
definiteness, we consider a ${\rm SU}(2)$ transformation. Let $v$ be a
${\rm SU}(2)$ doublet of spinor fields. After an infinitesimal
transformation, this doublet $v$ becomes a new doublet $v'$ such that
\begin{equation}
v'={\rm e}^{i\epsilon^j\sigma _j/2}v \simeq \left(\mathbb{I}_2
+i\epsilon^j\frac{\sigma_j}{2}\right)v,
\end{equation}
or
\begin{equation}
\delta v=i\epsilon^j\frac{\sigma_j}{2}v,
\end{equation}
where $\sigma_j$ are the Pauli matrices, the generators of the
group. The quantity $\epsilon^j$ is a vector (or a collection of three
parameters) specifying the transformation. In super-symmetry, the
transformation is characterized by a Weyl spinor $\xi ^{\alpha}$
which, therefore, plays the role of the quantity $\epsilon^j$. A
super-symmetric transformation transforms a boson into a fermion and
vice-versa. As a consequence, one can write
\begin{eqnarray}
\delta \Phi &=& \sqrt{2}\xi^{\alpha}\Psi_{\alpha},\\
\delta \Psi_{\alpha} &=& -i\sqrt{2}\left(\sigma^{\mu}\right)_{\alpha\dot{\beta}}
\overline{\xi}^{\dot{\beta}}\partial _{\mu}\Phi,
\end{eqnarray}
where the factors $i$ or $\sqrt{2}$ are introduced for future
convenience. Before showing that the Wess-Zumino Lagrangian is indeed
invariant under a super-symmetric transformation, one needs to
introduce a last result. We would like the Lagrangian to be invariant
without using the equation of motion (in other words, we would like
the Lagrangian to be invariant ``off-shell''). In this case, one needs
to add a term such that the new Lagrangian reads
\begin{equation}
\label{eq:susylagrangeoff}
{\cal L}_{_{\rm W-Z}}=-\eta^{\mu \nu}\partial _{\mu}
\Phi^{\dagger}\partial _{\nu}\Phi
-i\overline{\Psi}_{\dot{\alpha}}
\left(\overline{\sigma}^{\mu}
\right)^{\dot{\alpha}\alpha}\partial _{\mu}\Psi_{\alpha}-F^{\dagger}F,
\end{equation}
where $F$ is a complex scalar field. It is easy to see that this term
is harmless since the corresponding equation of motion is just
$F=0$. But this term is necessary if one wants to match the bosonic
and fermionic degrees of freedom. A Weyl spinor has two-complex
components and we now have two complex scalar fields. Obviously, we
also need to modify our super-symmetric transformation and they now
take the form
\begin{eqnarray}
\label{eq:susytransform1}
\delta \Phi &=& \sqrt{2}\xi^{\alpha}\Psi_{\alpha},\\
\label{eq:susytransform2}
\delta \Psi_{\alpha} &=& \sqrt{2}\xi_{\alpha}F
-i\sqrt{2}\left(\sigma^{\mu}\right)_{\alpha\dot{\beta}}
\overline{\xi}^{\dot{\beta}}\partial _{\mu}\Phi, \\
\label{eq:susytransform3}
\delta F&=&i\sqrt{2}\left(\partial _{\mu}\Psi^{\alpha}\right)
\left(\sigma^{\mu}\right)_{\alpha \dot{\beta}}\overline{\xi}^{\dot{\beta}}.
\end{eqnarray}
Our goal is now to show that the Lagrangian~(\ref{eq:susylagrangeoff})
is indeed invariant under the three
transformations~(\ref{eq:susytransform1}), (\ref{eq:susytransform2})
and~(\ref{eq:susytransform3}).

\par

In fact, the key to show that the Lagrangian is invariant is to be
able to evaluate the above transformations for the complex conjugate
quantities. We have
\begin{eqnarray}
\label{eq:deltaphidag}
\delta \Phi^{\dagger}&=&\sqrt{2}\left(\xi^{\alpha}\Psi_{\alpha}\right)^{\dagger}
=\sqrt{2}\left(\Psi_{\alpha}\right)^{\dagger}\left(\xi^{\alpha}\right)^{\dagger}
\nonumber \\
&=&\sqrt{2}\left(\Psi_{\alpha}\right)^*\left(\xi^{\alpha}\right)^*
=\sqrt{2}\, \overline{\Psi}_{\dot{\alpha}} \overline{\xi}^{\dot{\alpha}}
\nonumber \\
&=& \sqrt{2}\, \overline{\xi}_{\dot{\alpha}}\overline{\Psi}^{\dot{\alpha}}
=-\sqrt{2}\, \overline{\xi}^{\dot{\alpha}}\overline{\Psi}_{\dot{\alpha}}.
\end{eqnarray}
We cannot review here the complete dotted spinors formalism in detail,
see Ref.~\cite{Aitchison:2005cf}, but in order to make the
calculations reasonably self-consistent, it is nevertheless
interesting to recall a few things. The first point to discuss is the
definition of the dagger symbol for Grassmann variables. In fact, a
very clear discussion can be found in p.~35 of
Ref.~\cite{Aitchison:2005cf} and there is no need here to entirely
repeat it. In brief, one needs to remember that we deal with quantum
fields. Usually, the dagger symbols means complex conjugation and
transposition. For a scalar field, it simply amounts to take the
complex conjugate of the mode function and put (or remove) a dagger to
the creation and annihilation operators. For the spinor component
$\Psi _{\alpha}$, this is the same (here it is important to remember
that one discusses one component only and not the spinor viewed as a
column and, therefore, we ``cannot'' take the transpose of the matrix
-there is no matrix- !)  and one defines $\Psi _{\alpha}^{\dagger}$ as
the complex conjugate of the corresponding spinor component times the
creation and annihilation operators with or without the dagger
symbol. However, the question arises as how to evaluate the Hermitian
conjugate of a product of Grassmann variables. Here,
Ref.~\cite{Aitchison:2005cf} shows that the most convenient way is to
define $\left(\xi ^{\alpha
  }\Psi_{\alpha}\right)^{\dagger}=\Psi_{\alpha
}^{\dagger}\xi^{\alpha}{}^{\dagger}$, that is to say something similar
to what we are used to in the case of two matrices even if, again, we
do not deal with matrices here but with components of Weyl
spinors. This definition justifies the first step in the previous
equation.

\par

Then, it is also interesting to discuss how the two last equations are
obtained. Let us write $\overline{\Psi}_{\dot{1}}=a$,
$\overline{\Psi}_{\dot{2}}=b$ and $\overline{\xi}^{\dot{1}}=c$,
$\overline{\xi}^{\dot{2}}=d$, where $a$, $b$, $c$ and $d$ are
(complex) Grassman variables. As a consequence, one has
$\overline{\Psi}_{\dot{\alpha}}
\overline{\xi}^{\dot{\alpha}}=ac+bd$. On the other hand,
$\overline{\xi}_{\dot{\alpha}}\equiv
\epsilon_{\dot{\alpha}\dot{\beta}} \overline{\xi}^{\dot{\beta}}$,
where
\begin{equation}
\epsilon_{\dot{\alpha}\dot{\beta}}=\begin{pmatrix}
 0 & 1 \cr -1 & 0 
\end{pmatrix},
\end{equation}
from which we deduce that $\overline{\xi }_{\dot{1}}=d$ and
$\overline{\xi }_{\dot{2}}=-c$. In the same manner, one has
$\overline{\Psi}^{\dot{\alpha}}=\epsilon^{\dot{\alpha}\dot{\beta}}
\overline{\Psi}_{\dot{\beta}}$ with
\begin{equation}
\epsilon^{\dot{\alpha}\dot{\beta}}=\begin{pmatrix}
 0 & -1 \cr 1 & 0 
\end{pmatrix},
\end{equation}
from which one obtains $\overline{\psi}^{\dot{1}}=-b$ and
$\overline{\psi}^{\dot{2}}=a$. Using these results, one has
$\overline{\xi}_{\dot{\alpha}}
\overline{\Psi}^{\dot{\alpha}}=-db-ca=ac+bd=\overline{\Psi}_{\dot{\alpha}}
\overline{\xi}^{\dot{\alpha}}$, which is the wanted result. But, using
again the expressions of the components obtained above, one also
notices that $\overline{\xi}_{\dot{\alpha}}
\overline{\Psi}^{\dot{\alpha}}=-\overline{\xi}^{\dot{\alpha}}
\overline{\Psi}_{\dot{\alpha}}$ which is the last result used to
establish the expression~(\ref{eq:deltaphidag}) of $\delta
\Phi^{\dagger}$. In a more compact notation, one has
\begin{eqnarray}
\overline{\xi}_{\dot \alpha}\overline{\Psi}^{\dot \alpha}
&=&\epsilon_{\dot \alpha \dot \beta}\overline{\xi}^{\dot \beta}
\overline{\Psi}^{\dot \alpha}=\overline{\xi}^{\dot \beta}\epsilon_{\dot \alpha 
\dot \beta}\overline{\Psi}^{\dot \alpha}
=-\overline{\xi}^{\dot \beta}
\epsilon_{\dot \beta \dot \alpha}\overline{\Psi}^{\dot \alpha}
\nonumber \\
&=&-\overline{\xi}^{\dot \beta}\overline{\Psi}_{\dot \beta },
\end{eqnarray}
the key point being the fact that the matrix $\epsilon_{\dot \alpha
  \dot \beta }$ is antisymmetric.

\par

We now need the transformation of $\left(\delta
  \Psi_{\alpha}\right)^*=\delta \overline{\Psi}_{\dot{\alpha}}$. one has 
\begin{eqnarray}
\delta \overline{\Psi}_{\dot{\alpha}}&=&\sqrt{2}\, 
\overline{\xi}_{\dot{\alpha}}F^{\dagger}
+i\sqrt{2}\left[\left(\sigma^{\mu}\right)_{\alpha\dot{\beta}}\right]^*
\xi^{\beta}\partial _{\mu}\phi^{\dagger}, 
\\
&=&\sqrt{2}\, \overline{\xi}_{\dot{\alpha}}F^{\dagger}
+i\sqrt{2}\left(\sigma^{\mu}\right)_{\beta\dot{\alpha}}
\xi^{\beta}\partial _{\mu}\Phi^{\dagger}.
\end{eqnarray}
Finally, following the same lines, the transformation of $\delta
F^{\dagger}$ can be expressed as
\begin{eqnarray}
\delta F^{\dagger}&=&-i\sqrt{2}(\overline{\xi}^{\dot{\beta}})^{\dagger}
\left(\partial _{\mu}\Psi^{\alpha}\right)^{\dagger}\left[\left(\sigma ^{\mu}\right)
_{\alpha \dot{\beta}}\right]^*\\
&=& 
-i\sqrt{2}\xi^{\beta}\left(\partial _{\mu}
\overline{\Psi}^{\dot{\alpha}}\right)
\left(\sigma^{\mu}\right)_{\beta\dot{\alpha}}.
\end{eqnarray}
We are now in a position where one can compute the variation of the
super-symmetric Lagrangian. Collecting the above results, one obtains
\begin{widetext}
\begin{eqnarray}
\label{eq:varlwz}
\delta {\cal L}_{_{\rm W-Z}} &=& -\eta ^{\mu \nu}\partial _{\mu}\delta \Phi^{\dagger}
\partial _{\nu }\Phi
-\eta ^{\mu \nu}\partial _{\mu}\Phi^{\dagger}\partial _{\nu}\delta \Phi
-i\delta\overline{\Psi}_{\dot{\alpha}}
\left(\overline{\sigma}^{\mu}
\right)^{\dot{\alpha}\alpha}\partial _{\mu}\Psi_{\alpha}
-i\overline{\Psi}_{\dot{\alpha}}
\left(\overline{\sigma}^{\mu}
\right)^{\dot{\alpha}\alpha}\partial _{\mu}\delta\Psi_{\alpha}
-\delta F^{\dagger}F-F^{\dagger}\delta F \\
\label{eq:varlwz2}
&=& -i\sqrt{2}F^{\dagger}\overline{\xi}_{\dot{\alpha}}
\left(\overline{\sigma}^{\mu}\right)^{\dot{\alpha}\alpha}\partial _{\mu}\Psi_{\alpha}
-i\sqrt{2}F^{\dagger}\left(\partial _{\mu}\Psi^{\alpha}\right)
\left(\sigma^{\mu}\right)_{\dot{\alpha}\beta}\overline{\xi}^{\dot{\beta}}
-i\sqrt{2}\, \overline{\Psi}_{\dot{\alpha}}
\left(\overline{\sigma}^{\mu}\right)^{\dot{\alpha}\alpha}\xi_{\alpha}\partial _{\mu}F
\nonumber \\
& & +i\sqrt{2}\xi^{\beta}\left(\sigma^{\mu}\right)_{\beta\dot{\alpha}}
\left(\partial_{\mu}\overline{\Psi}^{\dot{\alpha}}\right)F
+\sqrt{2}\eta^{\mu \nu}\overline{\xi}^{\dot{\alpha}}
\partial_{\mu}\overline{\Psi}_{\dot{\alpha}}
\partial_{\nu}\Phi-\sqrt{2}\left(\overline{\sigma}^{\mu}\right)^{\dot{\alpha}\alpha}
\left(\sigma^{\nu}\right)_{\alpha\dot{\beta}}\overline{\Psi}_{\dot{\alpha}}
\overline{\xi}^{\dot{\beta}}\partial_{\mu }\partial_{\nu}\Phi
-\sqrt{2}\eta^{\mu \nu}\xi^{\alpha}\partial_{\nu}\Psi_{\alpha}
\partial_{\mu}\Phi^{\dagger}\nonumber \\ & &
+\sqrt{2}\left(\sigma^{\nu}\right)_{\beta\dot{\alpha}}
\left(\overline{\sigma}^{\mu}\right)^{\dot{\alpha}\alpha}
\xi^{\beta}\partial_{\mu }\Psi_{\alpha}\partial_{\nu}\Phi^{\dagger}.
\end{eqnarray}
\end{widetext}
Each term in this expression needs to be studied in detail. The first
term is also equal to $i\sqrt{2}F^{\dagger}\partial
_{\mu}\Psi^{\alpha}\left(\sigma^{\mu}\right)_{\alpha
  \dot{\alpha}}\overline{\xi}^{\dot{\alpha}}$ and, therefore, exactly
cancels the second one. The third one is, after integration by parts,
$i\sqrt{2}\left(\partial _{\mu}\overline
  \Psi_{\dot{\alpha}}\right)\left(\overline{\sigma}
\right)^{\dot{\alpha}\alpha}\xi_{\alpha}F$, which is also equal to
$-i\sqrt{2}\xi^{\alpha}\left(\sigma ^{\mu}\right)_{\alpha
  \dot{\alpha}}\left(\partial_{\mu}\overline{\Psi}^{\dot{\alpha}}\right)F$. This
cancels exactly the fourth term. As a consequence, all the term
involving the field $F$ give a total derivative contribution only. It
remains the four terms involving the scalar field $\Phi$ and
$\Phi^{\dagger}$. We first focus on the terms involving $\Phi $
only. For this purpose, we use the following
relation~\cite{Bailin:1994qt,Aitchison:2005cf}
\begin{equation}
\left(\overline{\sigma }^{\mu}\right)^{\dot{\alpha}\alpha}
\left(\sigma ^{\nu}\right)_{\alpha \dot{\beta}}=\eta^{\mu \nu}\delta ^{\dot{\alpha}}
{}_{\dot{\beta}}+2
\left(\overline{\sigma }^{\mu \nu}\right)^{\dot{\alpha}}{}_{\dot{\beta}},
\end{equation}
where 
\begin{equation}
\overline \sigma ^{\mu \nu}\equiv \frac14\left(\overline \sigma ^{\mu}
\sigma ^{\nu}-\overline \sigma ^{\nu}\sigma ^{\mu}\right)
=-\overline \sigma ^{\nu \mu}.
\end{equation}
Therefore, the second term involving $\Phi$ [\ie the sixth term in the
above formula~(\ref{eq:varlwz2})] can be re-written as
\begin{eqnarray}
& &-\sqrt{2}\left(\overline \sigma^{\mu}\right)^{\dot{\alpha}\alpha}
\left(\sigma^{\nu}\right)_{\alpha\dot{\beta}}\overline\Psi_{\dot{\alpha}}
\overline \xi^{\dot{\beta}}\partial_{\mu }\partial_{\nu}\Phi
\nonumber \\
&=& -\sqrt{2}\left[\eta^{\mu \nu}\delta ^{\dot{\alpha}}
{}_{\dot{\beta}}+2\left(\overline\sigma ^{\mu \nu}\right)^{\dot{\alpha}}{}_{\dot{\beta}}
\right]\overline\Psi_{\dot{\alpha}}
\overline \xi^{\dot{\beta}}\partial_{\mu }\partial_{\nu}\Phi
\nonumber \\ 
&=& -\sqrt{2}\eta ^{\mu \nu}\overline \Psi_{\dot{\alpha}}
\overline  \xi^{\dot{\alpha}}\partial_{\mu }\partial_{\nu}\Phi,
\end{eqnarray}
the second term in this equation leading to a vanishing contribution
since $\overline \sigma ^{\mu \nu}$ is anti-symmetric. The next
manipulation is to integrate by part. This gives
\begin{eqnarray}
& &-\sqrt{2}\left(\overline \sigma^{\mu}\right)^{\dot{\alpha}\alpha}
\left(\sigma^{\nu}\right)_{\alpha\dot{\beta}}\overline\Psi_{\dot{\alpha}}
\overline \xi^{\dot{\beta}}\partial_{\mu }\partial_{\nu}\Phi
\nonumber \\
&=& -\sqrt{2}\eta ^{\mu \nu}\overline\Psi_{\dot{\alpha}}
\overline \xi^{\dot{\alpha}}\partial_{\mu }\partial_{\nu}\Phi 
=\sqrt{2}\eta ^{\mu \nu}\partial _{\mu} \overline\Psi_{\dot{\alpha}}
\overline \xi^{\dot{\alpha}}\partial_{\nu}\Phi 
\nonumber \\  
&=&\sqrt{2}\eta ^{\mu \nu}\overline \xi_{\dot{\alpha}} 
\partial _{\mu}\overline\Psi^{\dot{\alpha}}
\partial_{\nu}\Phi 
=-\sqrt{2}\eta ^{\mu \nu}\overline \xi^{\dot{\alpha}} \partial _{\mu}
\overline\Psi_{\dot{\alpha}}
\partial_{\nu}\Phi. \nonumber \\
\end{eqnarray}
Therefore, this term exactly cancels the first term involving $\Phi$
[or the fifth term in the formula~(\ref{eq:varlwz2})]. Clearly, this
also works for the two terms involving $\Phi^{\dagger}$. We have thus
shown, without using the equation of motion, that the variation of the
Wess-Zumino Lagrangian~(\ref{eq:susylagrangeoff}) is a total
derivative. As a consequence, the Wess-Zumino model is indeed
super-symmetric, that is to say invariant under the super-symmetric
transformations~(\ref{eq:susytransform1}), (\ref{eq:susytransform2})
and (\ref{eq:susytransform3}).

\par

We have seen before that the constant spinor $\xi^{\alpha}$ was the
equivalent of the vector $\epsilon^j$ in the case of the ${\rm SU}(2)$
transformation considered at the beginning of this section. But what
is the equivalent of the generator $\sigma_j/2$ for a super-symmetric
transformation? Moreover, we know that the generators obey the formula
$\left[\sigma _i/2,\sigma _j/2\right]=i\epsilon_{ijk}\sigma_k/2$. What
is the corresponding equation satisfied by the super-symmetric
generators? We now turn to these questions.

\par

Using again the analogy with the ${\rm SU}(2)$ transformation, we
expect that an infinitesimal super-symmetric transformation to have the
form
\begin{equation}
\label{eq:infsusytrans}
\delta _{\xi}\Phi \simeq i\left(\xi ^{\alpha}Q_{\alpha}+
\overline \xi _{\dot{\alpha}}\overline Q^{\dot{\alpha}} \right)\Phi.
\end{equation}
Here $Q_{\alpha}$ is the super-symmetric generator, the equivalent of
$\sigma _j/2$. Clearly, since $\xi ^{\alpha}Q_{\alpha}$ must be a
scalar, $Q_{\alpha}$ is a Weyl spinor as the notation indicates.

\par

In order to find the super-symmetric algebra, we first compute the
quantity $\left(\delta _{\eta}\delta
  _{\xi}-\delta_{\xi}\delta_{\eta}\right)\Phi$ using
Eqs.~(\ref{eq:susytransform1}), (\ref{eq:susytransform2}) and
(\ref{eq:susytransform3}). It is straightforward to show that
\begin{eqnarray}
& & \left(\delta _{\eta}\delta _{\xi}-\delta_{\xi}\delta_{\eta}\right)\Phi
=2\xi^{\alpha}\eta_{\alpha}F-2i\xi^{\alpha}
\left(\sigma ^{\mu}\right)_{\alpha \dot{\beta}}
\overline \eta ^{\dot{\beta }}\partial _{\mu}\Phi
\nonumber \\ & & 
-2\eta^{\alpha}\xi_{\alpha}F+2i\eta ^{\alpha}
\left(\sigma ^{\mu}\right)_{\alpha \dot{\beta}}
\overline \xi ^{\dot{\beta }}\partial _{\mu}\Phi
\\
\label{eq:com1} 
& & =-2i\xi^{\alpha}\left(\sigma ^{\mu}\right)_{\alpha \dot{\beta}}
\overline \eta ^{\dot{\beta }}\partial _{\mu}\Phi
+2i\eta ^{\alpha}\left(\sigma ^{\mu}\right)_{\alpha \dot{\beta}}
\overline \xi ^{\dot{\beta }}\partial _{\mu}\Phi,
\end{eqnarray}
since $\xi^{\alpha}\eta _{\alpha}=\eta ^{\alpha}\xi_{\alpha}$. But, of
course, we can evaluate the same commutator using the
expression~(\ref{eq:infsusytrans}) for the infinitesimal
super-symmetric transformation. This leads to
\begin{eqnarray}
  & & \left(\delta _{\eta}\delta _{\xi}-\delta_{\xi}\delta_{\eta}\right)\Phi
  =-\left(\eta ^{\alpha}Q_{\alpha}+
    \overline \eta _{\dot{\alpha}}\overline Q^{\dot{\alpha}}\right)
  \nonumber \\ & & \times 
  \left(\xi ^{\beta}Q_{\beta}+
    \overline \xi _{\dot{\beta}}\overline Q^{\dot{\beta}}\right)\Phi
  +\left(\xi ^{\alpha}Q_{\alpha}+
    \overline \xi _{\dot{\alpha}}\overline Q^{\dot{\alpha}}\right)
  \nonumber \\ & & \times
  \left(\eta ^{\beta}Q_{\beta}+
    \overline \eta _{\dot{\beta}}\overline Q^{\dot{\beta}}\right)\Phi.
\end{eqnarray}
Then, expanding this expression (carefully taking into account the
fact that the spinor components are Grassmann variables), we arrive at
\begin{eqnarray}
\label{eq:com2}
\left(\delta _{\eta}\delta _{\xi}-\delta_{\xi}\delta_{\eta}\right)\Phi
&=&\eta ^{\alpha}\left(Q_{\alpha}Q_{\beta}+Q_{\beta}Q_{\alpha}\right)\xi^{\beta}\Phi
\nonumber \\ 
&+&\overline \eta _{\dot \alpha}\left(\overline Q^{\dot \alpha}
\overline Q^{\dot \beta}+\overline Q^{\dot \beta}\overline Q^{\dot \alpha}
\right)\overline \xi_{\dot \beta}\Phi 
\nonumber \\  
&+&\eta ^{\alpha}\left(Q_{\alpha}
\overline Q^{\dot \beta}+\overline Q^{\dot \beta}Q_{\alpha}
\right)\overline \xi_{\dot \beta}\Phi
\nonumber \\ 
&-&\xi ^{\alpha}\left(Q_{\alpha}
\overline Q^{\dot \beta}+\overline Q^{\dot \beta}Q_{\alpha}
\right)\overline \eta_{\dot \beta}\Phi.
\end{eqnarray}
One must now compare Eq.~(\ref{eq:com1}) with
Eq.~(\ref{eq:com2}). First of all, we notice that, in
Eq.~(\ref{eq:com1}), there is no term with two undotted or two dotted
spinors $\xi$ and $\eta$. This implies that
\begin{equation}
\left\{Q_{\alpha},Q_{\beta}\right\}
=\left\{Q^{\dot{\alpha}},Q^{\dot{\beta}}\right\}=0.
\end{equation}
Then, a comparison of the first ``cross-terms'' leads to 
\begin{equation}
\left\{Q_{\alpha},\overline Q_{\dot \beta}\right\}\Phi
=-2i\left(\sigma ^{\mu}
\right)_{\alpha \dot \beta}\partial _{\mu}\Phi,
\end{equation}
and a similar relation for the second cross-term. Remembering that
$P_{\mu}=-i\partial _{\mu}$ (the minus sign is in fact just a
convention; the problem with super-symmetry is that the literature
contains many different conventions (A nice treatment of this question
can be found in the first problem of Ref.~\cite{Binetruy:2006ad},
Appendix C), one arrives at
\begin{equation}
\left\{Q_{\alpha},\overline Q_{\dot{\beta}}\right\}
=2\left(\sigma ^{\mu}\right)_{\alpha \dot{\beta}}P_{\mu}.
\end{equation}
This equation shows that two successive super-symmetric transformations
are in fact equivalent to a space-time translation. As we will see,
this has far-reaching consequences. However, for the subject discussed
here, the above formula has also very important implications. Indeed,
working it out explicitly, one has
\begin{eqnarray}
Q_1\overline Q_{\dot{1}}+\overline Q_{\dot{1}}Q_1 &=& 2P_3+2P_0, \\
Q_2\overline Q_{\dot{2}}+\overline Q_{\dot{2}}Q_2 &=& -2P_3+2P_0,
\end{eqnarray}
from which we deduce
\begin{equation}
4P_0=Q_1\overline Q_{\dot{1}}+\overline Q_{\dot{1}}Q_1+Q_2\overline Q_{\dot{2}}
+\overline Q_{\dot{2}}Q_2.
\end{equation}
Therefore, since the Hamiltonian is nothing but the time translation
generator $P_0$, this implies that
\begin{equation}
\left\langle 0\left \vert H\right \vert 0\right \rangle=
\frac14
\left\langle 0\left \vert 
Q_1\overline Q_{\dot{1}}+\overline Q_{\dot{1}}Q_1+Q_2\overline Q_{\dot{2}}
+\overline Q_{\dot{2}}Q_2\right \vert 0\right \rangle \ge 0.
\end{equation}
We have learned two things: firstly, if the vacuum state is not
super-symmetric, \ie if super-symmetry is spontaneously broken (the
underlying theory is super-symmetric but the solution or the state in
which the system is placed is not), then the vacuum energy is
necessarily positive. Secondly, another remarkable consequence of the
above results is that, when the vacuum state is super-symmetric, that
is to say when
 \begin{equation}
Q_{\alpha }\vert 0 \rangle =0,
\end{equation}
one has $\left\langle 0\left \vert H\right \vert 0\right \rangle=0$,
\ie the vacuum energy is automatically zero. We have thus proven that
the cancellation discussed at the beginning of this section is in fact
not a coincidence. Moreover, we have identified under which conditions
it is valid: it is always true for a super-symmetric system. This is
the reason why super-symmetry has the potential to solve the
cosmological constant problem. It gives a concrete example of a
symmetry which forces the vacuum energy to vanish.

\par

Unfortunately, we know that super-symmetry must be broken in the real
world. This comes from the fact that each boson (fermion) of the
standard model is not observed to have a super-symmetric partner of
the same mass. This means that vacuum energy is now given by
Eq.~(\ref{eq:rhosusy}). Let us evaluate it with our ``incorrect''
method which consists in regularizing the divergent integrals 
with a cut-off $M$, see Eq.~(\ref{eq:rhocut}). We find 
\begin{eqnarray}
\langle \rho \rangle_{_{\rm SUSY}}&=&
\frac{M^4}{8 \pi^2}\left(1+\frac{m^2_{_{\rm B}}}{M^2}
+\cdots \right)\nonumber \\ & & 
-\frac{M^4}{8 \pi^2}\left(1+\frac{m^2_{_{\rm F}}}{M^2}
+\cdots \right)\\
&=& \frac{M^2}{8\pi^2}\left(m^2_{_{\rm B}}-m^2_{_{\rm F}}\right).
\end{eqnarray}
We see that, even if super-symmetry is broken, the most divergent part
$\propto M^4$ of the energy density cancel out. If we now properly
estimate the energy density by means of dimensional
regularization~(\ref{eq:rhovacrenorm}), one obtains that
\begin{eqnarray}
\langle \rho \rangle_{_{\rm SUSY}}&=& \frac{m^4_{_{\rm B}}}{32\pi^2}
\ln \left(\frac{m^4_{_{\rm B}}}{\mu^2}\right)
-\frac{m^4_{_{\rm F}}}{32\pi^2}
\ln \left(\frac{m^4_{_{\rm F}}}{\mu^2}\right).
\end{eqnarray}
Therefore, roughly speaking, one can estimate that vacuum energy is
now given by $\simeq M_{_{\rm SUSY}}^4$ where $M_{_{\rm SUSY}}$ is the
typical difference between the mass of the super-partners, which is
also the super-symmetric breaking mass. 

\par

A last remark is in order. In this section we have considered
super-symmetry in flat space-time while the cosmological constant
problem is formulated in curved space-time. In order to be consistent,
we should therefore treat super-symmetry in this last context. As is
well-known, this leads to
super-gravity~\cite{Bailin:1994qt,Binetruy:2006ad}. Regarding the
vacuum energy question, super-gravity presents important differences
compared to super-symmetry. In particular, one loses the result that
vacuum energy is always positive. But, this also shows that discussing
the zero-point energy density issue in curved space-time can bring new
aspects to the problem. This is the reason why, in the next section,
we address this question.

\section{The Vacuum Energy Density in Curved Space-time}
\label{sec:vaccurve}

\begin{figure*}
\begin{center}
\includegraphics[width=14cm,height=5cm]{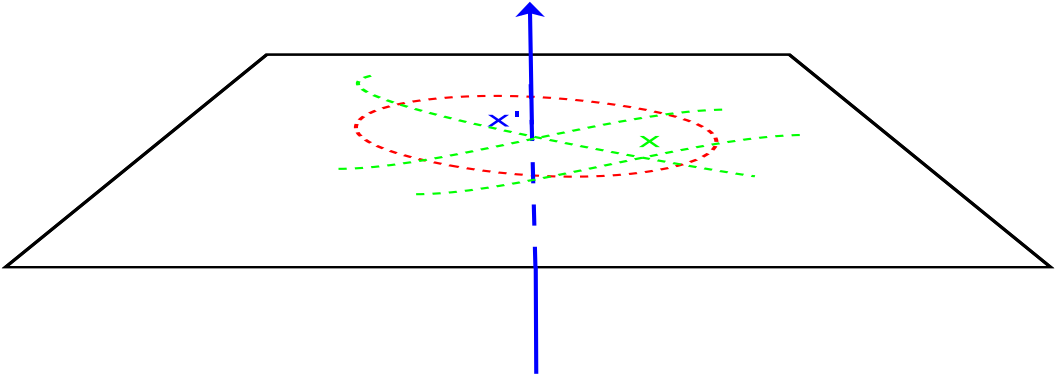}
\caption{Sketch of the Riemann coordinates used in order to derive an
  approximate expression for the Green function, see
  Eq.~(\ref{eq:propergreen}). The plot aims at illustrating the fact
  that the Riemann coordinates are a local concept. Since it is
  sufficient to establish the Green function locally (and not in the
  entire curved manifold), one can always endow the neighborhood of a
  point $x'$ (denoted here by the region surrounded by the closed
  dotted line) with special coordinates $x^{\mu}$ such that the
  calculation is simplified. It is important to notice that this does
  not restrict the generality of the obtained results.}
\label{fig:riemanncoordinates}
\end{center}
\end{figure*}

In the previous sections, we have computed the vacuum energy density
in flat space-time. This may seem inappropriate since we have argued
that the problem occurs only in curved space-time. Here we show how to
justify the previous approach. Intuitively, this is justified because
we deal with ultra-violet divergences for which the large scale
structure of the curved manifold should not play a too important
role. In other words, in the ultra-violet regime, one only probes the
local properties of space-time and, locally, one cannot distinguish a
curved manifold from the Minkowski space-time. However, we need to put
this intuitive reasoning on solid grounds. For this purpose, let us
consider our toy model again where matter is simply represented by a
scalar field. The only difference with the previous sections is that,
now, this quantum field lives in a curved space-time. Therefore, in
order to have meaningful Einstein equations, one must assume that the
quantum average of the stress energy tensor sources the Einstein
equations. In other words, we start from~\cite{Birrell:1982ix}
\begin{equation}
R_{\mu \nu}-\frac{1}{2}Rg_{\mu \nu}+\Lambda_{_{\rm B}} g_{\mu \nu}
=\kappa \langle T_{\mu \nu}\rangle\, ,
\end{equation}
Then, by analogy with Eq.~(\ref{eq:defstresstensor}), we define the
effective action $W$ to be
\begin{equation}
\langle T_{\mu \nu}\rangle 
=-\frac{2}{\sqrt{-g}}\frac{\delta W}{\delta g^{\mu \nu}}.
\end{equation}
The effective action can be written in term of the sourceless
generating functional $Z[0]$, namely
\begin{equation}
W=-i \ln Z[0],
\end{equation}
where we recall that the explicit expression of $Z[0]$ is given by,
see also Eq.~(\ref{eq:defZ}),
\begin{equation}
Z[0]=\int {\cal D}\phi \, {\rm e}^{iS_{\rm matter}[\phi]},
\end{equation}
the quantity $S_{\rm matter}$ being the action of the scalar
field. The above result can be proven in the following way. Varying
$Z[0]$, one obtains
\begin{eqnarray}
\delta Z[0]&=&\int {\cal D}\phi \, i\delta S_{\rm matter}
\, {\rm e}^{iS_{\rm matter}[\phi]} \\
&=& \left\langle 0\left\vert i\delta S_{\rm matter}
\right\vert 0\right\rangle .
\end{eqnarray}
As a consequence,
\begin{eqnarray}
-\frac{2}{\sqrt{-g}}
\frac{\delta W}{\delta g^{\mu \nu}}&=&
-\frac{2}{\sqrt{-g}}
\frac{-i}{Z[0]}\frac{\delta Z[0]}{\delta g^{\mu \nu}}
\\
&=&\frac{1}{Z[0]}\left\langle 0\left \vert 
-\frac{2}{\sqrt{-g}}
\frac{\delta S_{\rm matter}}{\delta g^{\mu \nu}}
\right \vert 0\right \rangle 
\\
&=&\frac{1}{Z[0]}
\left\langle 0\left \vert 
T_{\mu \nu}
\right \vert 0\right \rangle 
=\left \langle T_{\mu \nu}\right \rangle,
\end{eqnarray} 
since the mean value is defined by $\langle \cdots \rangle \equiv
\langle 0\vert \cdots \vert 0\rangle/\langle 0\vert 0\rangle$. 

\par

The next step is to calculate $W$ explicitly in our case, \ie for a
scalar field. We now show that the effective action can be expressed
in terms of the Green function. The Green function is defined by the
following expression~\cite{Birrell:1982ix}
\begin{equation}
\label{eq:defgreen}
\left(-g^{\mu \nu}\nabla_{\mu}\nabla_{\nu}+m^2\right)G_{_{\rm F}}(x,y)
=\frac{1}{\sqrt{-g}}\delta (x-y).
\end{equation}
It is easy to show that $G_{_{\rm F}}(x,y)=iD_{_{\rm F}}(x-y)$, where
the propagator has been defined in
Eq.~(\ref{eq:propagator}). Integrating by part the action of the
scalar field $\Phi(x)$, the generating function can be written as
\begin{widetext}
\begin{eqnarray}
\label{eq:curvedZ}
Z[J]&=&\int {\cal D}\Phi 
\exp\left[-\frac{i}{2}\int {\rm d}^dx\sqrt{-g}
\Phi(x)\left(-g^{\mu \nu}\nabla _{\mu}\nabla_{\nu} 
+m^2\right)\Phi(x)
+i\int {\rm d}^dx\sqrt{-g}J(x)\Phi(x)\right],
\end{eqnarray}
\end{widetext}
where we have considered the $d$-dimensional case for future
convenience. This path integral is a Gaussian integral and, therefore,
can be evaluated using the standard techniques. For $J=0$ one has
\begin{eqnarray}
Z[0]&=&(2\pi)^{\infty/2}\left (\det K\right)^{-1/2}
\\
&=&(2\pi)^{\infty/2}{\rm e}^{\frac12 {\rm tr} \left(\ln K^{-1}\right)},
\end{eqnarray}
where $K$ is the following operator, appearing in
Eq.~(\ref{eq:curvedZ}),
\begin{equation}
\label{eq:defK}
K(x,y)\equiv i\sqrt{-g}\left(-g^{\mu \nu}\nabla _{\mu}\nabla_{\nu} 
+m^2\right)\delta ^d(x-y).
\end{equation}
But the inverse of the operator $K$ is in fact the propagator
$D_{_{\rm F}}(x,y)$. A pedagogical, but simplistic, demonstration of
this fact is the following one. The generating functional can be
written as
\begin{equation}
Z[0]=\int \Pi _{x} {\rm d}\Phi_i 
{\rm e}^{-\frac12 \sum_{x,y}\Phi_xK_{xy}\Phi_y},
\end{equation}
where, in order to reproduce Eq.~(\ref{eq:defK}) (or, rather, to
mimic this formula), the matrix $K_{xy}$ is taken to be
$K_{xy}=K\delta_{xy}$ and the ``number $K$ is defined'' by $K\equiv
i\sqrt{-g}\left(-g^{\mu \nu}\nabla _{\mu}\nabla_{\nu}
  +m^2\right)$. Then, the inverse of the matrix $K$ must obey
\begin{equation}
\sum_z K_{xz}K^{-1}_{zy}=\delta_{xy},
\end{equation}
that is to say, using the definition of $K_{xy}$,
\begin{equation}
K K^{-1}_{xy}=\delta_{xy}.
\end{equation}
But this equation is nothing but Eq.~(\ref{eq:defgreen}). This shows
that, as announced, $K^{-1}_{xy}=\left(D_{_{\rm F}}\right)_{xy}$. As a
consequence, one can write
\begin{eqnarray}
Z[0]
&=&(2\pi)^{\infty/2}{\rm e}^{\frac12 {\rm tr} \left(\ln D_{_{\rm F}}\right)},
\end{eqnarray}
or,
\begin{equation}
\label{eq:effectiveactioninter}
W=-\frac{i}{2}{\rm tr}\left(\ln D_{_{\rm F}}\right)-i\ln 
\left[\left(2\pi\right)^{\infty/2}\right].
\end{equation}
We have reduced the evaluation of the effective action to the
evaluation of the Green function (or the propagator) in curved
space-time. We now turn to this question.

\par

For this purpose, we introduce the Riemann normal coordinates
$x^{\mu}$ with origin at the point $x'$, see
Fig.~\ref{fig:riemanncoordinates} and
Refs.~\cite{Parker:1978gh,Bunch:1979uk,Poisson:2011nh}. As it will
become gradually clear, the motivation for such a choice is as
follows. Since we want to see how the Einstein equations are modified,
one only needs to evaluate the Green function locally. In this case,
one can introduce coordinates that drastically simplify the problem,
\ie the calculation. This is exactly the role played by the Riemann
normal coordinates. In these coordinates, the metric can be written
as~\cite{Parker:1978gh,Bunch:1979uk}
\begin{widetext}
\begin{eqnarray}
g_{\mu \nu}&=&\eta_{\mu \nu}-\frac13 R_{\mu\alpha\nu\beta}x^{\alpha}x^{\beta}
-\frac16R_{\mu \alpha\nu \beta;\gamma}x^{\alpha}x^{\beta}x^{\gamma}
+
\left(-\frac{1}{20}R_{\mu \alpha\nu \beta;\gamma\delta}
+\frac{2}{45}R_{\alpha\mu \beta\lambda}R^{\lambda}{}_{\gamma\nu\delta}\right)
x^{\alpha}x^{\beta}x^{\gamma}x^{\delta}+\cdots 
\end{eqnarray}
Then, we insert this metric into the Green function
equation~(\ref{eq:defgreen}) and retain terms with coefficients
involving four derivatives of the metric or fewer. A lengthy
calculation leads to
\begin{eqnarray}
\label{eq:greencurved}
& & \eta^{\mu \nu}\partial _{\mu }\partial_{\nu}\overline{G}
-\left(m^2-\frac16\right)
\overline{G}
-\frac13 R_{\alpha}{}^{\nu}\partial _{\nu}\overline{G}
+\frac13 R^{\mu}{}_{\alpha}{}^{\nu}{}_{\beta}x^{\alpha}x^{\beta}
\partial_{\mu}\partial_{\nu}\overline{G}
+\frac16R_{;\alpha}x^{\alpha}\overline{G}
+\left(-\frac13R_{\alpha}{}^{\nu}{}_{;\beta}+\frac16R_{\alpha\beta}{}^{;\nu}\right)
x^{\alpha }x^{\beta}\partial _{\nu}\overline{G}
\nonumber \\
&+&\frac16R^{\mu}{}_{\alpha}{}^{\nu}{}_{\beta;\gamma}x^{\alpha}x^{\beta}x^{\gamma}
\partial_{\mu}\partial_{\nu}\overline{G}
+\frac{1}{12}R_{;\alpha \beta}x^{\alpha}x^{\beta}\overline{G}
+\Biggl(-\frac{1}{30}R_{\alpha}{}^{\lambda}R_{\lambda \beta}
+\frac{1}{60}R^{\kappa}{}_{\alpha}{}^{\lambda}{}_{\beta}R_{\kappa \lambda}
+\frac{1}{60}R^{\lambda \mu \kappa}{}_{\alpha}
R_{\lambda \mu \kappa \beta}
-\frac{1}{120}R_{;\alpha \beta}
\nonumber \\
&+&\frac{1}{40}\eta^{\mu \nu}\partial_{\mu}\partial_{\nu}R_{\alpha \beta}\Biggr)
x^{\alpha}x^{\beta}\overline{G}
+\left(-\frac{3}{20}R^{\nu}{}_{\alpha;\beta \gamma}
+\frac{1}{10}R_{\alpha \beta}{}^{;\nu}{}_{\gamma}
-\frac{1}{60}
R^{\kappa}{}_{\alpha}{}^{\nu}{}_{\beta}R_{\kappa \lambda}
+\frac{1}{15}R^{\kappa}{}_{\alpha \lambda \beta}R_{\kappa}{}^{\nu}{}_{\gamma}{}^{\lambda}
\right)x^{\alpha}x^{\beta}x^{\gamma}\partial _{\nu}\overline{G}
\nonumber \\ 
&+&\left(\frac{1}{20}R^{\mu}{}_{\alpha}{}^{\nu}{}_{\beta;\gamma \delta}
+\frac{1}{15}
R^{\mu}{}_ {\alpha \lambda\beta }
R^{\lambda}{}_{\gamma}{}^{\nu}{}_{\delta}\right)x^{\alpha}x^{\beta}x^{\gamma}
x^{\delta}\partial_{\mu}\partial_{\nu}\overline{G}=-\delta^d\left(x\right),
\end{eqnarray}
where all the indices are raised or lowered by the Minkowski metric
and where we have defined a rescaled Green function according to
$G_{_{\rm F}}(x,y)\equiv g^{-1/4}\overline{G}(x,y)$. The next step is
to Fourier transform the above equation. In this way the local
expansion which, in real space, looks like an expansion in the
coordinates $x^{\mu}$ is transformed, in momentum space, as an
expansion in terms of the inverse of $k^{\mu}$, as appropriate for a
local expansion. More precisely, defining 
\begin{equation}
\label{eq:greenfouriercurved}
\overline{G}(x,x')=\frac{1}{(2\pi)^{d}}\int {\rm d}^dk\, 
{\rm e}^{i\eta^{\alpha \beta}k_{\alpha }x_{\beta}}\overline{G}(k),
\end{equation}
we can expand the Green function as
$\overline{G}(k)=\overline{G}_0(k)+\overline{G}_1(k)+\cdots $ where
$\overline{G}_i(k)$ is the quantity appearing in front of a
coefficient involving the $i$-th derivatives of the metric
tensor~\cite{Parker:1978gh,Bunch:1979uk}. Let us notice that the Green
function is a function of $x'$ in the sense that $x^{\mu}$ represents
the coordinates of a point $x$ defined such $x'$ is at the
origin. Another remark is that we define the Fourier transform of the
Green function with an overall factor $(2\pi)^{-d}$ such that the
coefficient of the Dirac function in the right hand side of
Eq.~(\ref{eq:greencurved}) is minus one and does not contain factors
of $\pi$, thanks to the formula $\delta^d(x)=1/(2\pi)^d\int {\rm d}^d
k \, {\rm e}^{ikx}$.

\par

Introducing this expansion into Eq.~(\ref{eq:greencurved}) and
consistently identifying the terms of a given order leads to the
desired result. This iterative procedure gives
\begin{eqnarray}
\overline{G}(k)&=&\frac{1}{k^2+m^2}+\frac{R}{6}\frac{1}{\left(k^2+m^2\right)^2}
+\frac{i}{6}R_{;\alpha}\frac{1}{k^2+m^2}
\frac{\partial }{\partial k_{\alpha}}
\left[\left(k^2+m^2\right)^{-1}\right]
+\frac{R^2}{36}\frac{1}{\left(k^2+m^2\right)^3}
\nonumber \\ & & +
a_{\alpha \beta}\frac{1}{k^2+m^2}
\frac{\partial ^2}{\partial k_{\alpha}\partial k_{\beta}}
\left[\left(k^2+m^2\right)^{-1}\right]+\cdots ,
\\ 
&=& \frac{1}{k^2+m^2}+\frac{R}{6}\frac{1}{\left(k^2+m^2\right)^2}
+\frac{i}{12}R_{;\alpha}\frac{\partial }{\partial k_{\alpha}}
\left[\left(k^2+m^2\right)^{-2}\right]
+\frac{1}{3}a_{\alpha \beta}
\partial ^{\alpha}\partial ^{\beta}\left[\left(k^2+m^2\right)^{-2}\right]
\nonumber \\ & &
+\left(\frac{R^2}{36}-\frac23 a^{\lambda}{}_{\lambda}\right)
\frac{1}{\left(k^2+m^2\right)^3}+\cdots ,
\end{eqnarray}
where the coefficient $a_{\alpha \beta }$ can be expressed as
\begin{eqnarray}
a_{\alpha \beta}&=&-\frac{3}{40}R_{;\alpha \beta}-\frac{1}{40}\eta ^{\mu \nu}
\partial _{\mu}\partial _{\nu}R_{\alpha \beta}
+\frac{1}{30}R_{\alpha}{}^{\lambda}R_{\lambda \beta}
-\frac{1}{60}R^{\kappa}{}_{\alpha}{}^{\lambda}{}_{\beta}
R_{\kappa \lambda}
+\frac{1}{60}R^{\lambda \mu \kappa}{}_{\alpha}
R_{\lambda \mu \kappa \beta}.
\end{eqnarray}
Inserting this last expression into Eq.~(\ref{eq:greenfouriercurved})
leads to
\begin{equation}
\overline{G}(x,x')=\frac{1}{(2\pi)^{d}}
\int {\rm d}^nk\, 
{\rm e}^{i\eta^{\alpha \beta}k_{\alpha }x_{\beta}}
\left[1+f_1(x,x')\left(-\frac{\partial }{\partial m^2}\right)
+f_2(x,x')\left(-\frac{\partial }{\partial m^2}\right)^2+\cdots 
\right]\frac{1}{k^2+m^2},
\end{equation}
\end{widetext}
with
\begin{eqnarray}
f_1(x,x') &=& \frac{R}{6}+\frac{1}{12}R_{;\alpha}x^{\alpha}
-\frac13 a_{\alpha \beta}x^{\alpha \beta}, \\
f_2(x,x') &=& \frac{R^2}{72}-\frac13 a^{\lambda}{}_{\lambda}
\end{eqnarray}
The final step of the calculation consists in expressing the Green
function in the so-called proper time
formalism~\cite{Parker:1978gh,Bunch:1979uk}. Firstly, we replace the
quantity $(k^2+m^2)^{-1}$ with
\begin{equation}
\label{eq:inverse}
\frac{1}{k^2+m^2}=\int_0^{\infty}i{\rm d}s{\rm e}^{-is\left(k^2+m^2\right)},
\end{equation}
and, secondly, we make use of the result
\begin{eqnarray}
& &\frac{1}{(2\pi)^{d}}\int {\rm d}^nk\, 
{\rm e}^{i\eta^{\alpha \beta}k_{\alpha }x_{\beta}-is\left(k^2+m^2\right)}
\nonumber \\ & &
=\frac{i}{(4\pi)^{d/2}}\left(is\right)^{-d/2}{\rm e}^{-ism^2+i\sigma/(2s)},
\end{eqnarray}
where $\sigma =\eta^{\alpha \beta}x_{\alpha}x_{\beta}/2$. This leads
to our final expression for the Green function, namely
\begin{eqnarray}
\label{eq:propergreen}
\overline{G}(x,x')&=&\frac{i}{(4\pi)^{d/2}}
\int _0^{\infty}\frac{i{\rm d}s}{\left(is\right)^{d/2}}
{\rm e}^{-im^2s-\sigma/(2is)}
\nonumber \\ & & \times
\left[1+is f_1+\left(is\right)^2f_2+\cdots \right].
\end{eqnarray}
Of course, one could always push the expansion to higher orders in
inverse of the momenta in order to obtain higher orders in $s$. But,
in the present context, this will not be necessary.

\par

We now have all the tools to evaluate the effective
action~(\ref{eq:effectiveactioninter}). We now seek an explicit
expression for the quantity $\ln \left(D_{_{\rm F}}\right)$. As
already mentioned, $D_{_{\rm F}}$ can be viewed as the inverse of the
``continuous matrix'' $K$ which means that, formally, one can write
\begin{equation}
D_{_{\rm F}}=\int _0^{\infty}{\rm e}^{-iKs}i{\rm d}s,
\end{equation}
where we have used again Eq.~(\ref{eq:inverse}). This expression
allows us to express the matrix element of ${\rm e}^{-iKs}$. Indeed,
given that $D_{_{\rm F}}=-ig^{-1/4}\overline{G}$, a direct comparison
of Eq.~(\ref{eq:propergreen}) with the previous relation leads to
\begin{eqnarray}
\label{eq:matrixeiK}
\left({\rm e}^{-iKs}\right)_{xx'}&=&\frac{g^{-1/4}}{(4\pi)^{d/2}}
\frac{{\rm e}^{-im^2s-\sigma/(2is)}}{\left(is\right)^{d/2}}
\nonumber \\ &\times &
\left[1+is f_1+\left(is\right)^2f_2+\cdots \right].
\end{eqnarray}
Now, let us consider the following integral
\begin{eqnarray}
  & &\int_{\lambda}^{\infty}{\rm e}^{-iKs}\frac{i{\rm d}s}{is}=
  -\int _{i\lambda K}^{\infty}\frac{{\rm e}^{-t}}{t}{\rm d}t
  \nonumber \\
  &=& -\gamma -\ln \left(i\lambda K\right)
  -\sum_{k=1}^{\infty}\frac{(-1)^{k+1}(i\lambda K)^k}{k k!}.
\end{eqnarray}
In the limit $\lambda \rightarrow 0$, up to an infinite constant that
can be ignored, only the logarithm survives. Therefore, we can define
the logarithm of the operator $K$ as
\begin{equation}
-\ln K=\ln D_{_{\rm F}}=\int_{0}^{\infty}{\rm e}^{-iKs}\frac{i{\rm d}s}{is}.
\end{equation}
Then, using Eq.~(\ref{eq:matrixeiK}), one obtains
\begin{eqnarray}
\left(\ln D_{_{\rm F}}\right)_{xx'}&=&\frac{g^{-1/4}(x)}{(4\pi)^{d/2}}
\int _0^{\infty}
\frac{{\rm e}^{-im^2s-\sigma(x,x')/(2is)}}{\left(is\right)^{d/2}}
\nonumber \\ &\times &
\biggl[1+is f_1(x,x')
\nonumber \\ & & 
+\left(is\right)^2f_2(x,x')+\cdots \biggr]
\frac{i{\rm d}s}{is}.
\end{eqnarray}
As a consequence, we have now an explicit expression for the effective
action, namely
\begin{equation}
W=-\frac{i}{2}\int {\rm d}^dx \sqrt{-g}
\lim _{x'\rightarrow x}\left(\ln D_{_{\rm F}}\right)_{xx'},
\end{equation}
or
\begin{eqnarray}
W&=&\int {\rm d}^dx \sqrt{-g}
\Biggl\{-\frac{i}{2}\frac{1}{(4\pi)^{d/2}}\int_0^{\infty}
\frac{{\rm e}^{-im^2s}}{(is)^{d/2}}
\nonumber \\ & & \times
\biggl[1+isf_1
+(is)^2f_2+\cdots \biggr]\frac{i{\rm d}s}{is}\Biggr\}.
\end{eqnarray}
In the coincident limit $x'\rightarrow x$, the determinant of the
metric tensor has been taken to one since it is calculated in $x'$,
that is to say for $x=0$. For the same reason, we also have considered
that $\sigma=0$. The above expression can be evaluated explicitly in
terms of Euler function and one obtains
\begin{eqnarray}
W&=&\int {\rm d}^dx \sqrt{-g}
\Biggl\{-\frac{i}{2}\frac{1}{(4\pi)^{d/2}}
\left(\frac{m}{\mu}\right)^{d-4}
\nonumber \\ & & \times
\Biggl[m^4\Gamma\left(-\frac{d}{2}\right)
+m^2f_1\Gamma\left(1-\frac{d}{2}\right)
\nonumber \\ & &
+f_2\Gamma\left(2-\frac{d}{2}\right)
+\cdots \biggr]\Biggr\},
\end{eqnarray}
where we have introduced the scale $\mu$ in order to maintain the
correct dimension of the effective action. The above expression can be
analyzed in a dimensional regularization scheme. In this case, one is
led to the following expression
\begin{eqnarray}
iW&=&\int {\rm d}^dx \sqrt{-g}
\Biggl\{\Biggl[\frac{2}{\epsilon}-\gamma-\ln 
\left(\frac{m^2}{4\pi \mu^2}\right)\Biggr]
\nonumber \\ & & \times
\left(\frac{m^4}{64\pi^2}-\frac{m^2f_1}{32\pi^2}
+\frac{f_2}{32\pi^2}\right)+\cdots \Biggr\},
\end{eqnarray}
which is our final result. The $i$ factor can be absorbed into a
constant shift so we do not need to worry about it. If we compare the
effective action with the original Lagrangian~(\ref{eq:totalaction}),
we see that the first divergent term in the above expression can be
absorbed into a redefinition of the cosmological constant (the second
term can be viewed as a redefinition of the Newton constant and the
third term leads to the appearance of new terms in the action, see
Ref.~\cite{Birrell:1982ix}). Explicitly, one has
\begin{equation}
\Lambda_{_{\rm eff}}=\Lambda _{_{\rm B}}
+\kappa \frac{m^4}{64\pi^2}
\left[\frac{2}{\epsilon}-\gamma-\ln 
\left(\frac{m^2}{4\pi \mu^2}\right)\right],
\end{equation}
that is to say exactly the same expression as before (except an
unimportant factor $3/2$), see for instance
Eq.~(\ref{eq:rhovaceps}). As a consequence in a $\overline{{\rm MS}}$
scheme~\cite{Peskin:1995ev}, we can remove the divergence and obtain
exactly the same expression derived before, see
Eq.~(\ref{eq:rhovacrenorm}), namely
\begin{equation}
\label{eq:lambdacurve}
\Lambda_{_{\rm eff}}=\Lambda _{_{\rm B}}
+\kappa \frac{m^4}{64\pi^2}
\ln \left(\frac{m^2}{\mu^2}\right).
\end{equation}
The calculations of this section justify our previous approach in flat
space-time. Indeed, we have just proven that all the results derived
before in flat space-time are in fact valid in the more rigorous
approach where the curvature of space-time is properly taken into
account. This confirms that the vacuum energy does not scale as the
cut-off to the power four, see Eqs.~(\ref{eq:rhocut})
and~(\ref{eq:pcut}) but is in fact proportional to the mass of the
particle to the power four times a logarithmic factor depending on the
renormalization scale. In some sense, the calculation in flat
space-time can be considered as a computational trick: it is
sufficient to calculate the vacuum energy in this simple framework
since, from the previous considerations, we know that this is also the
result that consistently emerges from an approach where the curvature
of space-time is properly included. Of course, physically, the reason
for this success is not so surprising and was already mentioned
before. Since we deal with the ultraviolet behavior of the theory, a
local analysis should necessarily lead to a correct and consistent
result.

\par

It is also worth mentioning that Eqs.~(\ref{eq:rhovacrenorm})
and~(\ref{eq:lambdacurve}) recently attracted lot of attention, see
Refs.~\cite{Akhmedov:2002ts,Koksma:2011cq}. In particular, as we
discussed at length in Sec.~\ref{subsec:zeroenergy} and as was
emphasized in these articles, these equations are the consequence of
the fact that the regularization scheme used to tame the divergence of
the vacuum energy must respect Lorentz invariance. In fact, these
equations have been known for a long time, see in particular
Eq.~(6.50) of the standard textbook~\cite{Birrell:1982ix}. Therefore,
the fact that the cosmological constant is proportional to the mass of
the particle to the power four, and not to the cut-off to the power
four as usually claimed, see again Eqs.~(\ref{eq:rhocut})
and~(\ref{eq:pcut}), is in fact not a new result. Of course, it
modifies a lot our estimate of vacuum energy. For instance, the
correct result gives zero for the photons while, for the same
situation, the result based on the wrong regularization scheme gives
infinity. Clearly, this means that the correct regularization scheme
leads to a cosmological constant much smaller than the one obtained
from the wrong approach. As a consequence, one could hope that the
cosmological constant problem is in fact just an artifact due to the
use of an incorrect regularization method. We turn to this question in
the next section but, unfortunately, we will see that this is not the
case.

\section{The Value of the Cosmological Constant}
\label{sec:valuelambda}

\begin{figure*}
\begin{center}
\includegraphics[width=14cm]{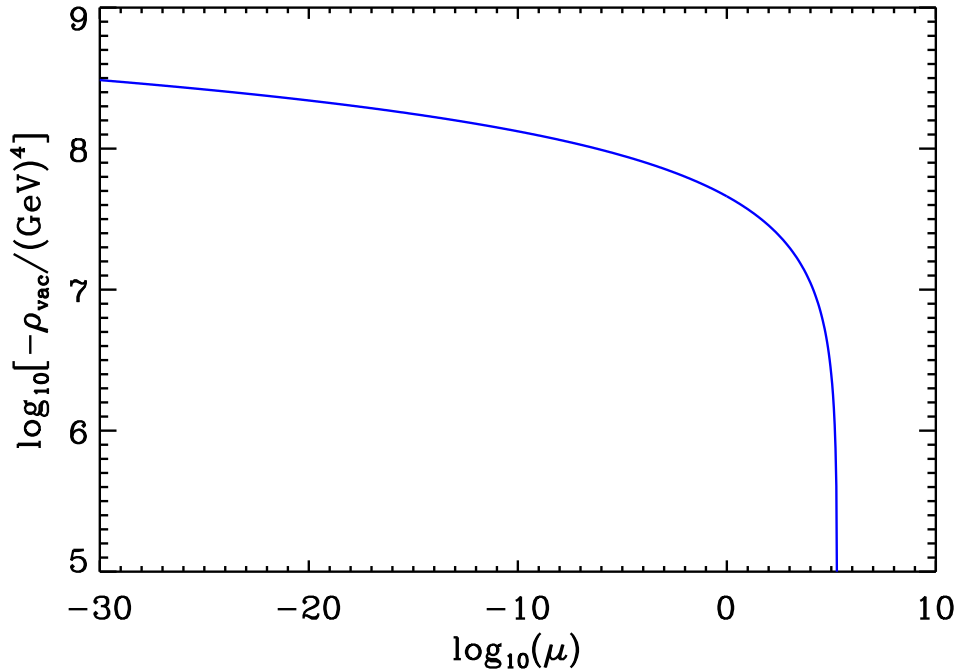}
\caption{Evolution of the vacuum energy density [more precisely the
  first term in Eq.~(\ref{eq:rhovacstandard}) versus the
  renormalization scale $\mu$. In the range considered here, the
  vacuum energy density is negative. The ``divergence'' observed
  around $\log_{10}\mu \simeq 5$ does not correspond to a new physical
  effect but just signal that $\rho_{_{\rm vac}}$ becomes positive.}
\label{fig:lambda}
\end{center}
\end{figure*}

We are now in a position to conclude the first part of this
review. The considerations of the previous sections lead to the
following expression for the vacuum energy~\cite{Koksma:2011cq}
\begin{equation}
\label{eq:rhovacstandard}
\rho_{_{\rm vac}}=\sum _in_i\frac{m_i^4}{64\pi^2}\ln\left(
\frac{m_i^2}{\mu^2}\right)+\rho_{_{\rm B}}+\rho_{_{\rm vac}}^{_{\rm EW}}+
\rho_{_{\rm vac}}^{_{\rm QCD}}+\cdots ,
\end{equation}
where $\rho_{_{\rm B}}$ indicates the contribution coming from the
bared cosmological constant and where the dots means that other phase
transition could contribute. Let us now evaluate the first term more
precisely for the standard model of particles
physics~\cite{Koksma:2011cq}. In this case, one has one scalar field,
the Higgs boson with $n_{_{\rm H}}=1$, $m_{_{\rm H}}\simeq 125$ GeV,
the six quarks (fermions) for which $n_{_{\rm quarks}}=-4$ and $m_{\rm
  t}\simeq 171.2 \mbox{GeV}$, $m_{\rm b}\simeq 4.2 \mbox{GeV}$,
$m_{\rm c}\simeq 1.27 \mbox{GeV}$, $m_{\rm s}\simeq 0.104 \mbox{GeV}$,
$m_{\rm u}\simeq 0.24 \mbox{GeV}$ and $m_{\rm d}\simeq 0.48
\mbox{GeV}$, the leptons (fermions) with $n_{_{\rm leptons}}=4$ and
$m_{\rm e}\simeq 0.511 \mbox{MeV}$, $m_{\rm \mu}\simeq 105
\mbox{MeV}$, $m_{\rm \tau}\simeq 1.77 \mbox{MeV}$, the neutrinos
(fermions) the mass of which is so small that we can ignore them and,
finally, the gauge bosons (massive vector fields), $n_{_{\rm Z}}=3$,
$m_{_{\rm Z}}\simeq 91 \mbox{GeV}$ and $n_{_{\rm W^\pm}}=3$, $m_{_{\rm
    W^\pm}}\simeq 80 \mbox{GeV}$. Of course, we also have the photon
but since $m_{\gamma}=0$, as already mentioned before, it does not
contribute to the vacuum energy. The only piece missing in order to
calculate $\rho_{_{\rm vac}}$ in Eq.~(\ref{eq:rhovacstandard}) is the
renormalization scale $\mu$. It was argued in
Ref.~\cite{Koksma:2011cq} that, because we use the photons coming from
the supernovae to determine the cosmological constant and because
these photons couple to the metric the expansion rate of which is
characterized by the Hubble constant, one should take $\mu \sim
\sqrt{E_{\gamma }E_{_{\rm grav}}}$ with $E_{_{\rm grav}}\simeq
H_0\simeq 3.7 \times 10^{-41}\, \mbox{GeV}$ and with the energy of the
photons corresponding to the wavelength $\lambda \simeq 500\, 
\mbox{nm}$. This leads to $\mu \simeq 3\times 10^{-25}\, \mbox{GeV}$
and implies that Eq.~(\ref{eq:rhovacstandard}) reads
\begin{equation}
\label{eq:rhovacnum}
\rho_{_{\rm vac}}\simeq -2\times 10^8 \, \mbox{GeV}^4 
+\rho_{_{\rm B}}+\rho_{_{\rm vac}}^{_{\rm EW}}+
\rho_{_{\rm vac}}^{_{\rm QCD}}+\cdots .
\end{equation}
If one does not want to estimate $\mu$, one can simply plot
$\rho_{_{\rm vac}}$ as a function of $\mu$ as done in
Fig.~\ref{fig:lambda}. Clearly, regardless of the precise value of
$\mu$, we are very far from $\rho_{_{\rm vac}}\simeq 10^{72}\,
\mbox{GeV}^4$ always mentioned in the literature.

\par

At this stage, Eq.~(\ref{eq:rhovacnum}) can be seen as the
``prediction'' of the standard model for the vacuum energy. At first
sight, the calculation is straightforward and, in order to obtain the
number in Eq.~(\ref{eq:rhovacnum}), we have used techniques of
regularization that are known to work in similar contexts (\ie when
one computes a cross-section in particle physics) and to lead to a
very good agreement with experiments. Therefore, we now turn to the
question of what is known experimentally about the vacuum energy.

\section{Measuring the Cosmological Constant in Cosmology}
\label{sec:measuringlambda}

Having discussed how the vacuum energy can be calculated, we now
address the question of how it is determined and/or constrained
experimentally/observationally. Here, we should distinguish the case
of cosmology, treated in this section, and the other experiments,
treated in the following sections. The difference is that, in
cosmology, one claims a detection of $\Lambda$ while, as we will see,
the other experiments can only put an upper limit on its
value. Moreover, in cosmology, there are in fact many different and
independent observables that can be used to measure the cosmological
constant~\cite{Perlmutter:1998np,Riess:1998cb,Scranton:2003in,Solevi:2004tk,Spergel:2006hy,Tegmark:2003ud}. All
these techniques are reviewed in other
articles~\cite{Astier:2012ba,Kunz:2012aw} and, therefore, it is not our
purpose here to discuss this question in details. However, for
completeness and also because this has some impact on the theoretical
status of the cosmological constant problem, we now discuss how
$\Lambda$ can be determined. We will do so from a theoretical
perspective leaving, as mentioned before, the technical aspects to
other articles~\cite{Astier:2012ba,Kunz:2012aw}.

\subsection{The Accelerating Universe}
\label{subsec:acceleratingu}

As is well-known, the claim that vacuum energy has been measured (and
not only constrained) was first established from the discovery that
the expansion of the universe is accelerated (although one can find
many claims about the value of the cosmological constant in the
history of cosmology)~\cite{Perlmutter:1998np,Riess:1998cb}. Since the
above mentioned claim has clearly very important consequences for
physics, one needs to be very precise at this stage and this is why,
in what follows, we will review the exact origin of this result.

\par

In order to construct a cosmological model we apply general relativity
to the Universe as a whole~\cite{Kolb:1990aa}. The cosmological
principle implies that the Universe is, on large scales, homogeneous
and isotropic. As a consequence, the metric is given by the following
expression
\begin{equation}
\label{eq:metric0}
{\rm d}s^2=-{\rm d}t^2+a^2(t)\gamma _{ij}^{(3)}{\rm d}x^i{\rm d}x^j\, ,
\end{equation}
where $\gamma _{ij}^{(3)}$ is the metric of the three-dimensional
space-like sections and where $a(t)$ is the scale factor. The expansion
is characterized by the Hubble parameter, $H=\dot{a}/a$ where a dot
means a derivative with respect to cosmic time. In the standard model
of cosmology, matter is assumed to be a collection of $N$ perfect
fluids (at least in the simplest version) and, as a consequence, its
stress-energy tensor is given by the following expression
\begin{equation}
T_{\mu \nu }=\sum _{i=1}^NT_{\mu \nu}^{(i)}=(\rho _{_{\rm T}}
+p_{_{\rm T}})u_{\mu }u_{\nu}+p_{_{\rm T}}g_{\mu \nu}\, ,
\end{equation}
where $\rho _{_{\rm T}}$ is the (total) energy density and $p_{_{\rm
    T}}$ the (total) pressure. These two quantities are related by the
equation of state
\begin{equation}
p_{_{\rm T}}=w(\rho _{_{\rm T}}), 
\end{equation}
[in general, there is an equation of state per fluid considered \ie 
$p_i=w_i(\rho _i)$]. The vector $u_{\mu }$ is the four
velocity and satisfies the relation $u_{\mu }u^{\mu }=-1$. In terms of
cosmic time this means that $u^{\mu }=(1,\bm{0})$. The
fact that the stress-energy tensor is conserved, $\nabla ^{\alpha
}T_{\alpha \mu}=0$, amounts to 
\begin{equation}
\label{eq:conservation}
\dot{\rho}_{_{\rm T}}+3\frac{\dot{a}}{a}
\left(\rho_{_{\rm T}} +p_{_{\rm T}}\right)=0, .
\end{equation}
This expression is obtained when one takes $\mu =0$. The case $\mu =i$
does not lead to an interesting equation for the background. Moreover,
if we assume that the fluids are separately conserved, then the above
relation is in fact true for each fluid, \ie
$\dot{\rho}_i+3H(\rho_i+p_i)=0$.

\par

We are now in a position to write down the Einstein
equations~(\ref{eq:einsteineqs}) for the Universe. We arrive at the
following non-linear, second order, differential equations
\begin{eqnarray}
\frac{\dot{a}^2}{a^2}+\frac{k}{a^2} &=&
\frac{\kappa }{3}\sum _{i=1}^{N}\rho _i+\frac{\Lambda_{_{\rm eff}}}{3}\, , 
\\
-\biggl(2\frac{\ddot{a}}{a}+\frac{\dot{a}^2}{a^2}+\frac{k}{a^2}\biggr) 
&=&\kappa \sum _{i=1}^Np_i-\Lambda_{_{\rm eff}} \, ,
\end{eqnarray}
From the previous formulas, as already signaled in
Eq.~(\ref{eq:deflambdaeff}), we see that we can also define $\rho
_{_{\Lambda_{_{\rm B}}}}$ by $\rho _{_{\Lambda _{_{\rm
        B}}}}=\Lambda_{_{\rm B}}/\kappa $ and $p_{_{\Lambda _{_{\rm
        B}}}}$ by $p_{_{\Lambda _{_{\rm B}}}}=-\rho _{_{\Lambda
    _{_{\rm B}}}}=-\Lambda _{_{\rm B}}/\kappa $. In this case, the
cosmological constant is described as fluid with a constant energy
density and pressure and with an equation of state $p_{_{\Lambda
    _{_{\rm B}}}}+\rho _{_{\Lambda _{_{\rm B}}}}=0$, i.e.
\begin{equation}
w_{_{\rm vac}}\equiv 
\frac{p_{_{\Lambda _{_{\rm B}}}}}{\rho _{_{\rm \Lambda _{_{\rm B}}}}}=-1\, .
\end{equation}
This is of course consistent with Eq.~(\ref{eq:qmvac}). It is worth
noticing that the vacuum energy is described by a fluid with a
negative effective pressure.

\par

Combining the two Einstein equations, one gets an equation which
permits to calculate the acceleration of the scale factor
\begin{equation}
\label{eq:accela}
\frac{\ddot{a}}{a}=-\frac{\kappa }{6}\sum _{i=1}^N(\rho _i
+3p_i)+\frac{2}{3}\Lambda_{_{\rm eff}} .
\end{equation}
Let us notice that, in order to obtain the above formula, we have not
assumed anything about the curvature $k$. This equation is especially
interesting because it gives the condition leading to an accelerated
expansion, namely
\begin{equation}
\rho _{_{\rm T}}+3p_{_{\rm T}}<0 \, .
\end{equation}
Since the energy density of matter must be positive, we see that the
above condition requires a negative pressure, \ie some exotic form of
matter. But, this is of course exactly what happens when the vacuum
energy is the dominant fluid in the universe. Then,
Eq.~(\ref{eq:accela}) can be written as $\ddot{a}/a\simeq
2\Lambda_{_{\rm eff}}/3$ which is positive for a positive effective
cosmological constant. Therefore, in the framework of the standard
model described before, the discovery that $\ddot{a}>0$ means that a
fluid with a negative pressure is driving the expansion.

\subsection{The Hubble Diagram}
\label{subsec:hubble}

In this section, we explain how the measurement of the Hubble diagram
can be used to infer the value of the cosmological constant. 

\par

Let us first consider a source in flat space-time whose absolute
luminosity is given by $L$ (energy emitted per unit of time). This
quantity can be written as
\begin{equation}
L=\frac{n_{_{\rm E}}\hbar \nu }{\Delta t}\, ,
\end{equation}
where $n_{_{\rm E}}$ is the number of photons emitted during the time
interval $\Delta t$. Suppose that we observe the source from a
distance $d_{_{\rm L}}$ using a telescope, the surface of the mirror
being $\Delta S$. The number of photons received is given by 
\begin{equation}
n_{_{\rm R}}=n_{_{\rm E}}\biggl(
\frac{\Delta S}{4\pi d_{_{\rm L}}^2}\biggr)\, .
\end{equation}
On the other hand, the flux of photons (or apparent luminosity \ie
energy per unit of time and surface) through the telescope is, by
definition,
\begin{equation}
\Phi _{_{\rm R}}=\frac{n_{_{\rm R}}\hbar \nu }{\Delta S \Delta t}\, ,
\end{equation}
from which we deduce that
\begin{equation}
\label{eq:defd}
\Phi _{_{\rm R}}=\frac{L}{4\pi d_{_{\rm L}}^2}\, .
\end{equation}
We have just recovered the well-known fact that the apparent
luminosity of a source decreases as the inverse squared of the
distance. On the other hand, we can use this relation as an
operational definition of the distance, namely $d_{_{\rm
    L}}=\sqrt{L/(4\pi \Phi _{_{\rm R}})}$. 

\par

The previous considerations can be used as a definition of distance in
cosmology. For this purpose, let us now consider the same situation
but in a curved space-time with the following metric [a form more
explicit than the one given in Eq.~(\ref{eq:metric0})]
\begin{equation}
{\rm d}s^2=-{\rm d}t^2+a^2(t)({\rm d}r^2+r^2{\rm d}\Omega _2^2)\, .
\end{equation}
We assume that the source is located at the co-moving coordinate $r_1$
while the observer is located at the origin of the coordinates. In
order to reproduce the above flat space-time calculation in this new
situation, we first need to calculate the surface of the sphere
$t=t_{_{\rm R }}=$const., $r=r_1=$const. surrounding the source. The
surface is just given by the integral of the covariant volume element,
namely
\begin{equation}
S=\int \sqrt{g}\, {\rm d}^2x=\int _0^{\pi }\int _0^{2\pi }
\sqrt{a^4_{_{\rm R}}r_1^4}\sin \theta {\rm d}\theta 
{\rm d}\varphi =4\pi a_{_{\rm R}}^2r_1^2\, .
\end{equation}
This means that the number of photons received in the telescope 
can now be expressed as
\begin{equation}
n_{_{\rm R}}=n_{_{\rm E}}\biggl(
\frac{\Delta S}{4\pi a_{_{\rm R}}^2r_1^2}\biggr)\, .
\end{equation}
But we have also to take into account the two following
phenomena. Firstly, the energy of the photons has changed during the
propagation from the source to the observer. If $\lambda $ is the
wavelength, we have $\lambda =(2\pi /k)a(t)\propto \nu^{-1} $, from
which we deduce that $E_{_{\rm R}}/E_{_{\rm E}}=\nu _{_{\rm R}}/\nu
_{_{\rm E}}=a_{_{\rm E}}/a_{_{\rm R}}$, that is to say we have the
redshift due to the expansion. Secondly, if the photons are emitted as
a burst of duration $\Delta t_{_{\rm E}}$, they will be received as
burst of duration $\Delta t_{_{\rm R}}=\Delta t_{_{\rm E}}a_{_{\rm
    R}}/a_{_{\rm E}}$. We are now in a position where we can calculate
the apparent luminosity. We obtain
\begin{equation}
\Phi _{_{\rm R}}=\frac{L}{4\pi a_{_{\rm R}}^2r_1^2}\biggl(
\frac{a_{_{\rm E}}}{a_{_{\rm R}}}\biggr)^2=
\frac{L}{4\pi a_{_{\rm R}}^2r_1^2}\frac{1}{(1+z)^2}\, , 
\end{equation}
where $z$ is the redshift of the source. Now if we use the operational
definition of the luminosity distance~(\ref{eq:defd}), we find
\begin{equation}
\label{eq:defdl}
d_{_{\rm L}}(z)=a_{_{\rm R}}r_1(1+z)=a_{_{\rm R}}(1+z)\int _{t_{_{\rm E}}}
^{t_{_{\rm R}}}\frac{c{\rm d}\tau }{a(\tau )}\, .
\end{equation}
We see from the last expression that the distance of a source depends
on the behavior of the scale factor between the time of emission and
the time of reception. On the contrary, if we measure the luminosity
distance versus the redshift, we can learn about this
behavior. Measuring the apparent luminosity is quite easy and,
clearly, the difficulty lies in estimating the absolute luminosity of
the source. It is also convenient to work in terms of the distance
modulus, $m-M$, since the data are often presented in this way. The
quantity $m$ (or sometimes $m_{\rm bol}$) is the apparent bolometric
magnitude. It is related to the apparent bolometric luminosity $\ell $
by
\begin{equation}
\ell =10^{-2m/5}\times 2.52\times 10^{-5}\,
\mbox{erg}/\mbox{cm}^2\times \mbox{sec}\, .
\end{equation}
For instance the sun has an apparent bolometic magnitude of
$-26.85$. On the other hand, the absolute bolometric magnitude $M$ is
the apparent bolometric magnitude that the source would have at a
distance of $10$ pc and is given by
\begin{equation}
L=10^{-2M/5}\times 3.02\times 10^{35}\,
\mbox{erg}/\mbox{sec}\, .
\end{equation}
For instance, the absolute bolometric magnitude of the sun is $4.72$.
As a consequence, using these two definitions, the distance modulus is
defined by
\begin{equation}
d_{_{\rm L}}(z)\equiv \frac{c}{H_0}\bar{d}_{_{\rm L}}(z)\equiv
10^{1+(m-M)/5}\times 10^{-6} \mbox{Mpc}\, ,
\end{equation}
such that $\bar{d}_{_{\rm L}}$ is dimensionless and $c/H_0=3000\,
h^{-1}\mbox{Mpc}$ is of course nothing but the Hubble length (we have
re-established the speed of light for convenience). From the above
expression, one immediately deduces that
\begin{equation}
  \mu(z)\equiv m-M=5 \log _{10}\bar{d}_{_{\rm L}}(z)+25+5\log
  _{10}\left(\frac{c}{H_0}\right)\, ,
\end{equation}
where $c$ is expressed in $\mbox{km} \times \mbox{s}^{-1}$, $H_0$ in
$\mbox{km}\times $s$^{-1}\times \mbox{Mpc}^{-1}$ and $d_{_{\rm L}}$ in
$\mbox{Mpc}$. 

\par

\begin{figure*}[t]
  \includegraphics[width=.45\textwidth,height=.35\textwidth]{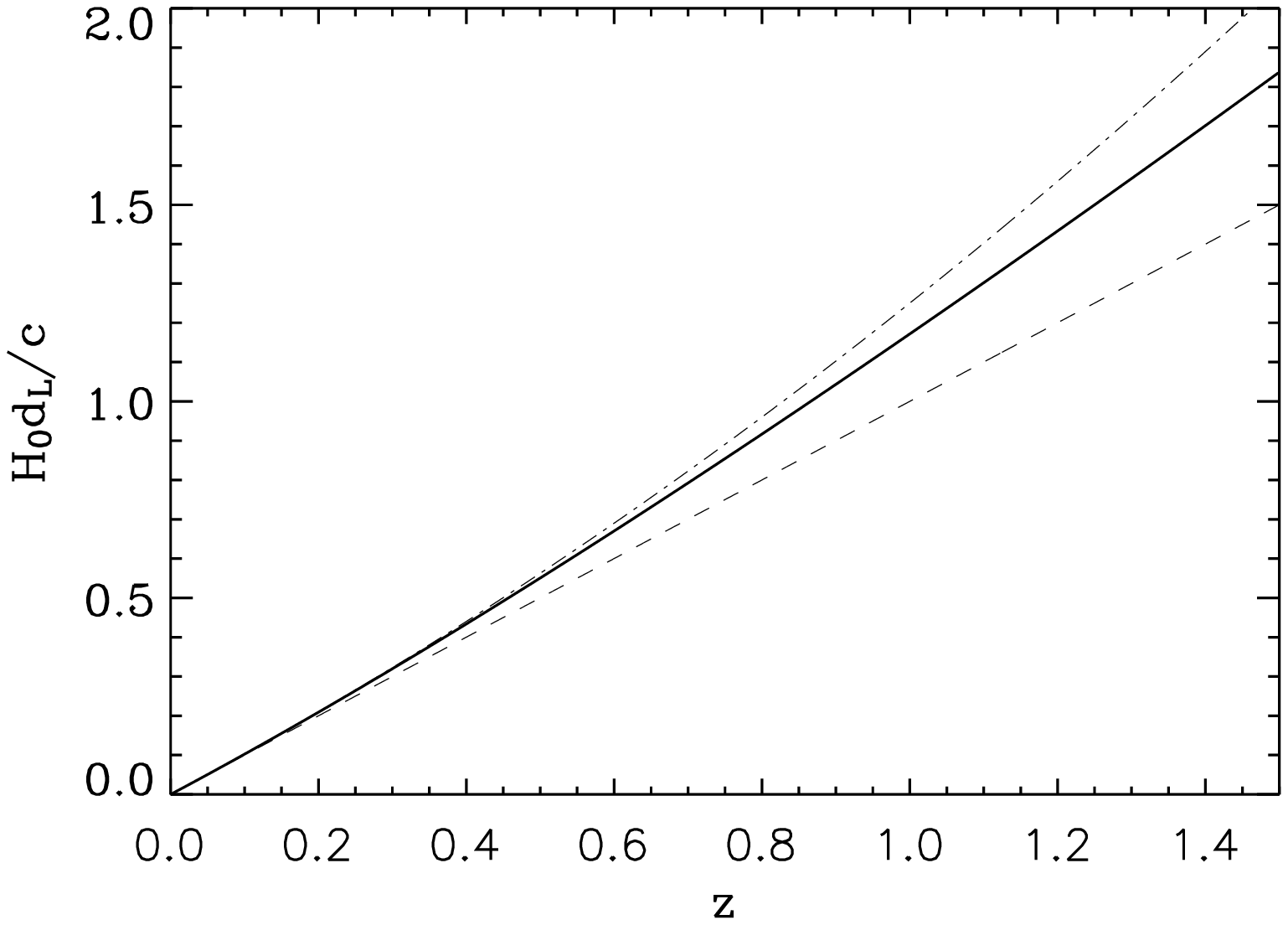}
 \includegraphics[width=.45\textwidth,height=.35\textwidth]{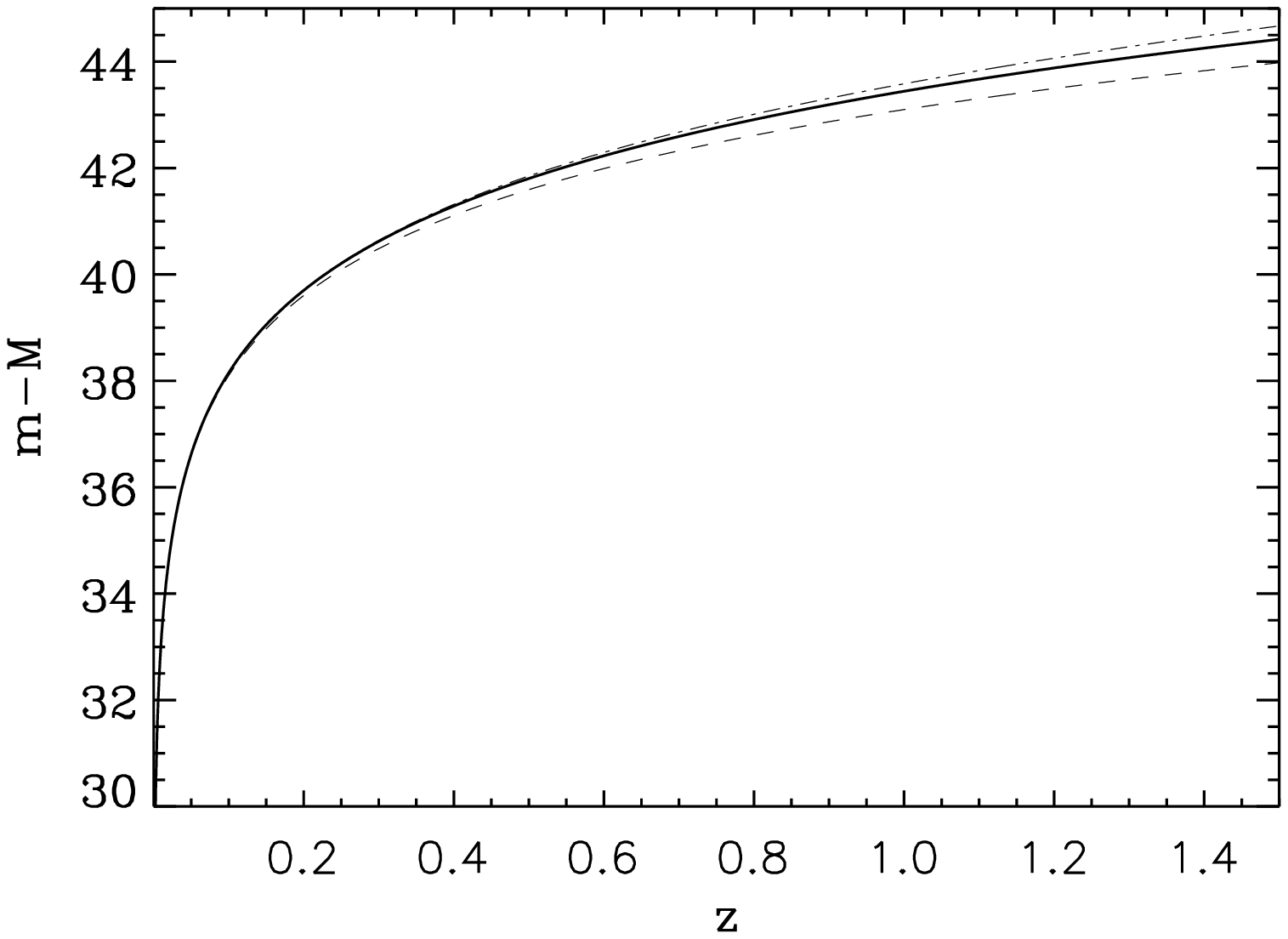}
 \caption{Luminosity distance $d_{_{\rm L}}(z)$ (left
   panel) and distance modulus $\mu(z)$ (right panel) in a matter
   dominated Universe. The solid line corresponds to the exact
   result. The dashed line is the linear law $d_{_{\rm L}}\simeq
   cz/H_0$ and the dotted dashed line represents the approximated
   result at second order, see Eq.~(\ref{dLtaylor}). We have used
   $h=0.72$.}
\label{fig:dl}
\end{figure*}

In practice we only have access to $d_{_{\rm L}}$ over a limited range
of redshifts. Therefore, one can derive a general expression by Taylor
expanding the scale factor. For small redshifts, we have
\begin{widetext}
\begin{equation}
a(t)=a(t_{_{\rm R}})+\dot{a}(t_{_{\rm R}})(t-t_{_{\rm R}})
+\frac{1}{2}\ddot{a}(t_{_{\rm R}})(t-t_{_{\rm R}})^2+\cdots
=a(t_{_{\rm R}})\biggl[1+H_{_{\rm R}}(t-t_{_{\rm R}})
-\frac{1}{2}H_{_{\rm R}}^2q_{_{\rm R}}(t-t_{_{\rm R}})^2
+\cdots\biggr]\, ,
\end{equation}
where $q\equiv -a\ddot{a}/(\dot{a})^2$ is the acceleration
parameter. A positive acceleration parameter corresponds to a
decelerating universe whether a negative one corresponds to an
accelerating universe. Using the previous expression, the luminosity
distance becomes
\begin{eqnarray}
d_{_{\rm L}}(z) &=& c(1+z)\int _{t_{_{\rm E}}}
^{t_{_{\rm R}}}{\rm d}\tau 
\biggl[1-H_{_{\rm R}}(\tau-t_{_{\rm R}})
+\frac{1}{2}H_{_{\rm R}}^2q_{_{\rm R}}(\tau -t_{_{\rm R}})^2
+H_{_{\rm R}}^2q_{_{\rm R}}(\tau -t_{_{\rm R}})^2
+\cdots\biggr]
\\
\label{dl}
&=& c(1+z)\biggl[-(t_{_{\rm E}}-t_{_{\rm R}})+\frac{1}{2}H_{_{\rm R}}
(t_{_{\rm E}}-t_{_{\rm R}})^2-\frac{1}{6}H_{_{\rm R}}^2q_{_{\rm R}}
(t_{_{\rm E}}-t_{_{\rm R}})^3
-\frac{1}{3}H_{_{\rm R}}^2q_{_{\rm R}}
(t_{_{\rm E}}-t_{_{\rm R}})^3
+\cdots \biggr]
\, .
\end{eqnarray}
\end{widetext}
As expected the result only depends on the time of flight $t_{_{\rm
E}}-t_{_{\rm R}}$. In order to obtain our final result, we must
express this time of flight in terms of the redshift. The redshift of
the source is defined by $z=a(t_{_{\rm R}})/a(t_{_{\rm E}})-1$. Using 
the Taylor expansion of the scale factor, we end up with
\begin{equation}
z=-H_{_{\rm R}}(t_{_{\rm E}}-t_{_{\rm R}})+H_{_{\rm R}}^2\biggl(1+\frac{1}{2}
q_{_{\rm R}}\biggr)(t_{_{\rm E}}-t_{_{\rm R}})^2 +\cdots \, ,
\end{equation}
which can be easily inverted to give the expression of the time of
flight, namely
\begin{equation}
t_{_{\rm E}}-t_{_{\rm R}}=-\frac{z}{H_{_{\rm R}}}+\frac{1}{H_{_{\rm R}}}
\biggl(1+\frac{1}{2}q_{_{\rm R}}\biggr)z^2+\cdots \, .
\end{equation}
The last step is to replace the above expression into the
Eq.~(\ref{dl}). This gives
\begin{equation}
\label{dLtaylor}
d_{_{\rm L}}(z)=\frac{c}{H_0}\biggl[z+
\frac{1}{2}(1-q_0)z^2+\cdots \biggr]\, ,
\end{equation}
where we have taken into account that the time of reception is the
present time (denoted with the subscript ``0''). We see that the
measurement of the luminosity distance versus the redshift allows us
to determine the Hubble parameter today. If one goes sufficiently far
in redshifts, one can determine whether the universe is accelerating
or not via the accelerating parameter $q_0$. It is worth noticing that
this last conclusion is independent of the underlying gravity theory
(provided this is a metric theory). The typical behavior of the
luminosity distance and of the modulus is illustrated in
Fig.~\ref{fig:dl} for a matter dominated universe. Let us also notice
that the method that we have just presented for pedagogical reasons is
of course not used in practice because we have now measured up to
redshifts of $z\simeq 1.4$ for which the Taylor expansion is not valid
(and, of course, in the real world the analysis is always more
complicated that the simplified version discussed above). In this
case, one must use an exact expression, that is to say one must
compute the integral in Eq.~(\ref{eq:defdl}) exactly.

\par

If we now use general relativity (\ie a specific theory for gravity),
one can go further and relates the value of the acceleration parameter
to the matter content of the Universe. The critical energy density
$\rho _{\rm cri}$, defined by $\rho _{\rm cri}\equiv 3H^2/\kappa $, is
used to evaluate the contribution of each species through the
parameter $\Omega _i\equiv \rho _i/\rho _{\rm cri }$ where $\rho _i$
is the energy density of the species $i$ at present time. Then, the
Friedman equation can be re-written as
\begin{equation}
\sum _{i=1}^{N}\Omega _i=1+\frac{k}{a^2H^2}\, ,
\end{equation}
and the acceleration parameter can be re-expressed according to
\begin{equation}
\label{paraq}
q_0=\frac12\sum _{i=1}^{N}\Omega _i\left(1+3w_i\right)\, .
\end{equation}
The universe contains pressure-less gas (cold dark matter and ordinary
baryonic matter) for which $w=0$ but also radiation ($w=1/3$) and,
finally, vacuum energy ($w=-1$). In this case, one obtains
\begin{equation}
q_0=\frac{1}{2}\Omega _{\rm m}+\Omega _{\gamma}-\Omega _{_{\rm vac}}\,
.
\end{equation}
Today, the contribution of radiation is negligible since we know from
the measurement of the CMB that $\Omega _{\gamma}h^2\simeq 2.48\times
10^{-5}$~\cite{Spergel:2006hy}. Therefore, the acceleration parameter
is in fact given by $q_0\simeq \Omega_{\rm m}/2-\Omega_{_{\rm
    vac}}$. On the other, we also know that the universe is spatially
flat which means that $\Omega_{\rm m}+\Omega_{_{\rm vac}}\simeq
1$~\cite{Spergel:2006hy}. As a consequence, from the measurement of
$q_0\simeq -0.67\pm 0.25$~\cite{Perlmutter:1998np,Riess:1998cb}, one
can deduce the two quantities $\Omega_{\rm m}$ and $\Omega_{_{\rm
    vac}}$. On finds that $\Omega_{\rm m}\simeq 0.3$ and
$\Omega_{_{\rm vac}}\simeq 0.7$. This is how one reaches the
conclusion that our universe is accelerating and that vacuum energy
dominates the present energy content. Remarkably, as announced at the
beginning of this section, it is a detection and a measurement of
$\rho_{_{\rm vac}}$. One finds
\begin{equation}
\label{eq:rhovaccosmology}
\rho_{_{\rm vac}}\simeq \Omega_{_{\rm vac}}\rho_{\rm cri}\simeq 10^{-47}\, 
\mbox{GeV}^4 ,
\end{equation}
a number that should be compared with Eq.~(\ref{eq:rhovacnum}).

\par

At this stage, several remarks are in order. Firstly, we notice the
large mismatch between the theoretical
expectation~(\ref{eq:rhovacnum}) and the above, observationally
determined, number~(\ref{eq:rhovaccosmology}), something like $54$
orders of magnitude (but much less than the $122$ orders of magnitude
often quoted in the literature). This is of course nothing but the
cosmological constant problem although we will be more accurate with
regards to its definition in the subsequent sections. Something is
clearly wrong but, as will be discussed, it is difficult to identify
where the mistake is. Secondly, it is clear that the above conclusion
rests on many assumptions. For instance, we have assumed that the
universe is homogeneous and isotropic~\cite{Clarkson:2012bg} and that
gravity is well described by general relativity. This means that, a
priori, the cosmological constant problem is present in this context
only. Thirdly, we have also implicitly assumed that the reason for the
accelerated expansion is vacuum energy. This is clearly a natural idea
since there is a term in our theory that must be present since it
satisfies all the required properties (covariance, energy
conservation) and that precisely leads to the observed phenomenon. A
priori, nothing more can be asked to a new theoretical framework. But,
given the mismatch discussed before, one can also doubt the
identification of the source of the accelerated expansion with vacuum
energy. As a matter of fact, it is possible to construct models where
the universe accelerates because of some new source of matter (often
named ``dark energy''). This is for instance the case of quintessence
and/or galileons~\cite{deRham:2012az} models where a scalar field is
responsible for the acceleration. In this case, one can no longer
claim that we have detected the vacuum energy and that its value is
given by Eq.~(\ref{eq:rhovaccosmology}). However, this does not solve
the cosmological constant problem because, even if the source of the
acceleration is another fluid, we still have the constraint
$\rho_{_{\rm vac}}<\rho_{\rm cri}$. Clearly, the
number~(\ref{eq:rhovacnum}) does not satisfy this inequality which
means that the problem is still present. In this sense, the constraint
deduced from cosmology is very important: even if dark energy is not
the cosmological constant, the observation that the energy density
today is the critical energy density severely limits the value of
$\Lambda_{_{\rm eff}}$. Fourthly, in principle, it is possible to
check experimentally whether the reason for the accelerated expansion
is vacuum energy or something else like
quintessence~\cite{Ratra:1987rm,Brax:1999gp,Brax:1999yv,Brax:2000yb,Brax:2001ah,Martin:2004ba,Martin:2005bp,Brax:2006np,Brax:2006kg,Brax:2006dc,Brax:2009kd,Kolda:1998wq,Carroll:1998zi,Doran:2002qd,Doran:2004dc,Garny:2006wc}
(more recently, in Ref.~\cite{Ringeval:2010hf}, the current
acceleration was also explained by a scalar field experiencing quantum
fluctuations during inflation). Measuring the equation of state
parameter $w$ is the clue since, if this is a constant equals to $-1$,
then one can be sure that the source of the acceleration is the
cosmological constant. Unfortunately, this measurement is
difficult~\cite{Maor:2000jy}.

\par

One more general grounds, it is clear that it would be better to have
constraints, or even a measurement, of vacuum energy in an
experimental context which is is not cosmology where, as reminded
before, measurements are difficult and always subject to many biases
and systematic errors. In the next sections, we explore whether this
is possible.

\section{Measuring the Cosmological Constant Elsewhere Than In
  Cosmology?}
\label{sec:lambdaelsewhere}

In the previous section, we have written the Einstein equation in an
homogeneous and isotropic universe. Here, we investigate the influence
of the cosmological term in a spherically symmetric situation. This
will allow us to study how the presence of $\Lambda_{_{\rm eff}}$
affects the motion of the planets in our solar system and how it
modifies the energy levels of the Hydrogen atom.

\subsection{The Static and Spherically Symmetric Gravitational Field
  in Presence of a Cosmological Constant}
\label{subsec:schwarlambda}

Let us analyze the influence of the cosmological constant in a static
and spherically symmetric situation~\cite{weinberg:1972}. This means
that, in spherical coordinates, one can write the metric tensor
as~\cite{weinberg:1972}
\begin{equation}
\label{eq:metricsphere}
{\rm d}s^2=-B(r){\rm d}t^2+A(r){\rm d}r^2+r^2\left({\rm d}\theta^2
+\sin ^2\theta {\rm d}\varphi^2\right),
\end{equation}
where $A(r)$ and $B(r)$ are two free functions. We seek solutions of
the Einstein equations~(\ref{eq:einsteineqs}) with a vanishing stress
energy tensor. It is easy to show that these equations can be
expressed as
\begin{equation}
\label{eq:Einsteinvac}
R_{\mu \nu}-\Lambda _{_{\rm eff}}g_{\mu \nu}=0.
\end{equation}
The next step is to calculate the components of the Ricci tensor for
the metric tensor given by
Eq.~(\ref{eq:metricsphere}). Straightforward calculations lead to the
following expressions
\begin{eqnarray}
R_{tt} &=& \frac{B}{A}\left(\frac12 \frac{B''}{B}-\frac14 \frac{B'^2}{B^2}
-\frac14 \frac{A'B'}{AB}+\frac{1}{r}\frac{B'}{B}\right),
\\
R_{rr} &=& -\frac12 \frac{B''}{B}+\frac14 \frac{B'^2}{B^2}
+\frac14 \frac{A'B'}{AB}+\frac{1}{r}\frac{B'}{B},
\\
R_{\theta \theta}&=& \frac{1}{A}\left(\frac{r}{2}\frac{A'}{A}
-\frac{r}{2}\frac{B'}{B}-1\right)+1,
\\
R_{\varphi \varphi}&=& R_{\theta \theta}\sin ^2\theta ,
\end{eqnarray}
the other components being zero. In the above equations, a prime
denotes a derivative with respect to the radial coordinate $r$. 

\par

We are now in a position where one can determine the solution of the
field equations. If we take the combination
$R_{tt}/B+R_{rr}/A=-\Lambda_{_{\rm eff}}B/B+\Lambda_{_{\rm eff}}A/A=0$
[from the Einstein equations~(\ref{eq:Einsteinvac})], one obtains
\begin{equation}
\frac{1}{rA}\left(\frac{B'}{B}+\frac{A'}{A}\right)=0,
\end{equation}
which implies that $B(r)=C/A(r)$, where $C$ is a constant. Using this
expression, one can re-express $R_{\theta \theta}$ as
\begin{equation}
R_{\theta \theta}=\frac{B(r)}{C}\left(-r\frac{B'}{B}-1\right)+1,
\end{equation}
and, as a consequence, the corresponding Einstein equation now reads
\begin{equation}
rB'(r)+B(r)=C-C\Lambda_{_{\rm eff}}r^2.
\end{equation}
This equation can easily be integrated and one obtains
\begin{equation}
B(r)=D+\frac{C}{r}-D\Lambda_{_{\rm eff}}\frac{r^2}{3},
\end{equation} 
where $D$ is another integration constant. The constants $C$ and $D$
are fixed in such a way that, in absence of a cosmological constant,
one recovers the standard Schwarschild solution. This amounts to take
$D=1$ and $C=2GM$, where $M$ is the mass of the central body and $G$
the Newton constant. This completes our calculation since the two
functions $A(r)$ and $B(r)$ are now completely specified, namely
\begin{equation}
\label{eq:solspherelambda}
A(r)=\frac{1}{B(r)}, \quad 
B(r)=1-\frac{2GM}{r}-\frac{\Lambda _{_{\rm eff}}r^2}{3}.
\end{equation}
We can now compare this solution to the standard Schwarschild
solution. The fact that the time-time component of the metric tensor
is inversely proportional to the $r$-$r$ component is still true but
the radial dependence of the function $B(r)$ is modified by the
presence of a cosmological constant. In the Newtonian limit, we have
$g_{tt}\simeq 1+2V(r)$, where $V(r)$ is the Newtonian potential. In
the present context, this implies that
\begin{equation}
\label{eq:potlambda}
V(r)\simeq -\frac{GM}{r}-\frac{\Lambda_{_{\rm eff}}r^2}{6}.
\end{equation}
Therefore, we obtain the usual gravitational potential corrected by a
term proportional to the cosmological constant. This term looks like
an inverted harmonic oscillator. This corresponds to a force felt by a
test body of mass $m$ given by
\begin{equation}
\label{eq:forcelambda}
\bm{F}=\frac{GMm}{r^2}\bm{u}-\frac{m\Lambda_{_{\rm eff}}r}{3}\bm{u},
\end{equation}
where $\bm{u}$ is a vector directed towards the central body of mass
$M$. The cosmological constant force is directed in the other
direction and can therefore be viewed as a kind of ``anti-gravity
force''. We also notice that it is proportional to $r$ which means
that its effect will be more important on large scales. This last
property is the clue to understand why measuring the cosmological
constant is more efficient on large scales (the best example being of
course cosmology). This means that the modification of the planet
orbits and/or the energy levels of the atoms will certainly be a small
effect. On the other hand, we know that measurements in, say, atomic
physics are extremely accurate and, a priori, there is the hope that
this accuracy could compensate the smallness of the
effect. Unfortunately, as we are now going to study, this will not be
the case.

\subsection{Planets Orbits}
\label{subsec:planets}

\begin{figure*}
\begin{center}
\includegraphics[width=10cm]{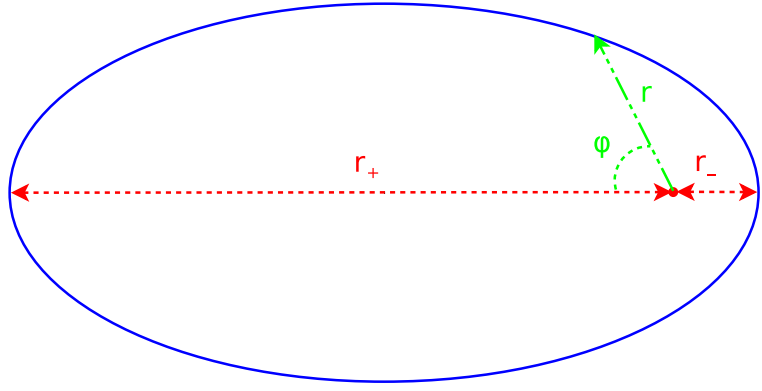}
\caption{Sketch of the elliptic planet orbit with the aphelion and
  perihelion. The orbit is planar and the position of the planet can
  be characterized by the coordinates $(r,\varphi)$.}
\label{fig:orbit}
\end{center}
\end{figure*}

Our goal in this section is to study how the planet orbits are
modified by the presence of vacuum
energy~\cite{Wright:1998bc,Kerr:2003bp,Jetzer:2007cr}. We will follow
the treatment of Ref.~\cite{Wright:1998bc}. We assume that the
gravitational field created by the sun and the cosmological constant
is given by Eqs.~(\ref{eq:metricsphere})
and~(\ref{eq:solspherelambda}) with $M=M_{\odot}$. The motion of a
planet is represented by the motion of a test body in this background
gravitational field. The corresponding geodesic equation can be
written as
\begin{equation}
\frac{{\rm d}^2x^{\mu}}{{\rm d}\tau^2}+\Gamma ^{\mu}_{\nu \lambda}
\frac{{\rm d}x^{\nu}}{{\rm d}\tau}
\frac{{\rm d}x^{\lambda}}{{\rm d}\tau}=0,
\end{equation}
where $\tau$ is an affine parameter (not necessarily the proper
time). Then, after having evaluated the Christoffel symbols for the
metric~(\ref{eq:metricsphere}), one can express these equations
explicitly. One obtains
\begin{widetext}
\begin{eqnarray}
\label{eq:tgeo}
\frac{{\rm d}^2t}{{\rm d}\tau^2}+\frac{B'}{B}\frac{{\rm d}t}{{\rm d}\tau}
\frac{{\rm d}r}{{\rm d}\tau}&=& 0,\\
\label{eq:rgeo}
\frac{{\rm d}^2r}{{\rm d}\tau^2}+\frac{B'}{2A}
\left(\frac{{\rm d}t}{{\rm d}\tau}\right)^2
+\frac{A'}{2A}
\left(\frac{{\rm d}r}{{\rm d}\tau}\right)^2
-\frac{r}{A}
\left(\frac{{\rm d}\theta}{{\rm d}\tau}\right)^2
-\frac{r\sin^2\theta}{A}
\left(\frac{{\rm d}\varphi}{{\rm d}\tau}\right)^2&=&0, \\
\label{eq:thetageo}
\frac{{\rm d}^2\theta}{{\rm d}\tau^2}+\frac{2}{r}\frac{{\rm d}r}{{\rm d}\tau}
\frac{{\rm d}\theta}{{\rm d}\tau}
-\cos \theta \sin \theta
\left(\frac{{\rm d}\varphi}{{\rm d}\tau}\right)^2&=& 0,\\
\label{eq:varphigeo}
\frac{{\rm d}^2\varphi}{{\rm d}\tau^2}+\frac{2}{r}\frac{{\rm d}r}{{\rm d}\tau}
\frac{{\rm d}\varphi}{{\rm d}\tau}
+2\frac{\cos \theta }{\sin \theta}
\frac{{\rm d}\theta}{{\rm d}\tau}
\frac{{\rm d}\varphi}{{\rm d}\tau}&=& 0,
\end{eqnarray}
\end{widetext}
Let us now solve these equations. Firstly, we can assume that the
orbit stays in the $\theta=\pi/2$ plane. This automatically solves
Eq.~(\ref{eq:thetageo}). Then, if we divide Eq.~(\ref{eq:tgeo}) by
${\rm d}t/{\rm d}\tau$ and Eq.~(\ref{eq:varphigeo}) by ${\rm
  d}\varphi/{\rm d}\tau$, we obtain two new equations, namely
\begin{eqnarray}
\frac{{\rm d}}{{\rm d}\tau}\left[
\ln \left(\frac{{\rm d}t}{{\rm d}\tau}\right)
+\ln B\right] &=& 0,\\
\frac{{\rm d}}{{\rm d}\tau}\left[
\ln \left(\frac{{\rm d}\varphi}{{\rm d}\tau}\right)
+\ln r^2\right] &=& 0.
\end{eqnarray}
The first equation can be used to define the affine parameter. We
choose it such that it satisfies ${\rm d}t/{\rm d}\tau=1/B$. Then, the
second equation tells us that
\begin{equation}
r^2\frac{{\rm d}\varphi}{{\rm d}\tau}=J,
\end{equation}
where $J$ is a constant. Finally, Eq.~(\ref{eq:rgeo}) can be
re-written as
\begin{equation}
\frac{{\rm d}^2r}{{\rm d}\tau^2}+\frac{A'}{2A}\left(\frac{{\rm d}r}
{{\rm d}\tau}\right)^2-\frac{J^2}{Ar^3}+\frac{B'}{2AB^2}=0.
\end{equation}
If we multiply this last equation by $2A{\rm d}r/{\rm d}\tau$, one
arrives at
\begin{equation}
\frac{{\rm d}}{{\rm d}\tau}
\left[A\left(\frac{{\rm d}r}
{{\rm d}\tau}\right)^2+\frac{J^2}{r^2}-\frac{1}{B}\right]=0.
\end{equation}
This means that the quantity inside the square brackets is a
constant. Let us call this constant $-E$. Then, this gives an equation
expressing how the radial coordinate $r$ varies with the angle
$\varphi$,
\begin{equation}
\label{eq:rphi}
\left(\frac{{\rm d}r}
{{\rm d}\varphi}\right)^2
=\frac{r^4}{A(r)}\left[\frac{1}{J^2B(r)}-\frac{1}{r^2}
-\frac{E}{J^2}\right].
\end{equation}
It is easy to express the constants $E$ and $J$ in terms of the
aphelion and perihelion $r_+$ and $r_-$, which are the two points on
the orbit such that ${\rm d}r/{\rm d}\varphi=0$, see
Fig.~\ref{fig:orbit} (where the elements of the orbits are
represented). This gives
\begin{eqnarray}
\label{eq:Jdeux}
\frac{1}{J^2} &=& \frac{r_+^{-2}-r_-^{-2}}{B_+^{-1}-B_-^{-1}}, \\
\label{eq:E}
E&=& \frac{r_+^2/B_+-r_-^2/B_-}{r_+^2-r_-^2},
\end{eqnarray}
with $B_{\pm}$ defined by $B_{\pm}\equiv B(r_\pm)$. 

\par

So far, we have been very general and we have never specified the
function $B(r)$. From now on, we are going to use the explicit form of
this function, see Eq.~(\ref{eq:solspherelambda}). Let us notice in
particular that, in absence of a cosmological constant and in the
limit where $GM/r\ll 1$, the two above expressions reduce to
\begin{equation}
\label{eq:limitnewton}
\frac{E-1}{2}\simeq \frac{GM}{r_++r_-}, \quad J^2\simeq 2GMD,
\end{equation}
where $D\equiv r_+r_-/(r_++r_-)$ (of course, not to be confused with
the integration constant introduced in the previous sub-section). As
is shown in the following footnote, these are exactly the relations
obtained in the Newtonian case. More generally, Eq.~(\ref{eq:rphi})
takes the form
\begin{eqnarray}
\label{eq:drdvarphi}
\left(\frac{{\rm d}r}
{{\rm d}\varphi}\right)^2&=&\frac{1-E}{J^2}r^4+\frac{2GME}{J^2}r^3
-r^2+2GMr \nonumber \\ & & 
+\frac{\Lambda_{_{\rm eff}}r^4}{3}+\frac{E\Lambda_{_{\rm eff}}r^6}{3J^2}.
\end{eqnarray}
Performing the standard change of variable, $u\equiv 1/r$ and
differentiating with respect to $\varphi$, we obtain~\footnote{ It is
  interesting to compare this result with the Newtonian analysis. In
  this case, the Lagrangian is given by
\begin{equation}
L=\frac{m}{2}\left(\dot{r}^2+r^2\dot{\varphi}^2\right)+\frac{GMm}{r},
\end{equation}
where a dot denotes a derivative with respect to time. Again we have
chosen the orbit to be in the plane $\theta=\pi/2$. This leads to the
following equations of motion
\begin{eqnarray}
m\ddot{r}-mr\dot{\varphi}^2+\frac{GMm}{r^2}&=&0,\\
mr^2\ddot{\varphi}+2mr\dot{r}\dot{\varphi}&=&0.
\end{eqnarray}
The second equation gives $\dot{\varphi}=J/r^2$ while the first one can 
be re-written as
\begin{equation}
\frac{{\rm d}}{{\rm d}t}\left(\frac{\dot{r}^2}{2}+\frac{J^2}{2r^2}
-\frac{GM}{r}\right)=0
\end{equation}
The quantity between the parenthesis is a constant and we call it
$-E$. Then, this equation can be re-expressed as
\begin{equation}
\left(\frac{{\rm d}r}{{\rm d}\varphi}\right)^2=-\frac{2E}{J^2}r^4
+\frac{2GM}{J^2}r^3-r^2.
\end{equation}
From this equation one can deduce an expression for $E$ and $J$ in terms 
of the aphelion and the perihelion. One gets
\begin{equation}
E=\frac{GM}{r_++r_-}, \quad J^2=2GMD.
\end{equation}
Introducing the variable $u=1/r$ and differentiating once more with
respect to $\varphi$, one arrives at
\begin{equation}
\label{eq:eomnewton}
  \frac{{\rm d}^2u}{{\rm d}\varphi^2}+u=\frac{GM}{J^2}.
\end{equation}
This equation can easily be integrated. The solution reads
\begin{equation}
u(\varphi)=F\cos\left(\varphi-\varphi_0\right) +\frac{GM}{J^2}
=\frac{GM}{J^2}\left(1-e\cos\varphi\right),
\end{equation}
where $F$ and $\varphi_0$ are two arbitrary constants. Clearly, one
can always take $\varphi_0=0$. The above trajectory represents, as is
well-known, an ellipse with eccentricity $e$.}
\begin{equation}
\label{eq:masteru}
\frac{{\rm d}^2u}{{\rm d}\varphi^2}+u=\frac{E}{J^2}GM
+3GMu^2-\frac{E}{J^2}\frac{\Lambda_{_{\rm eff}}}{3u^3}.
\end{equation}
This is the differential equation that needs to be solved in order to
find the trajectory of the planet. It should be compared with its
Newtonian counterpart, see Eq.~(\ref{eq:eomnewton}). One can check
that, in this regime (without a cosmological constant), the two
equations are indeed identical since according to
Eq.~(\ref{eq:limitnewton}), $E\rightarrow 1$, $\Lambda_{_{\rm eff}}=0$
and the term $3GMu^2$ can be neglected. The relativistic term $3GMu^2$
is responsible for the precession of the perihelion and the last term
represents the correction due to the cosmological constant.

\par

We now turn to the solution of Eq.~(\ref{eq:masteru}). Let us first
review how one calculates the precession of the perihelion without a
cosmological constant~\cite{weinberg:1972}. The idea is to restart
from Eq.~(\ref{eq:drdvarphi}). For $\Lambda _{_{\rm eff}}=0$, the
polynomial on the right hand side can be written as
\begin{equation}
\frac{1-E}{J^2}r(r-r_+)(r-r_-)(r-\epsilon)
\end{equation}
since we know it has to vanish at $r=r_{\pm}$. The only quantity that
remains to be determined in this expression is $\epsilon$. Comparing
the above relation with Eq.~(\ref{eq:drdvarphi}), we deduce that
\begin{equation}
\frac{1-E}{J^2}r_+r_-\epsilon=2GM.
\end{equation}
Using the expressions derived before, it is easy to show that
$\epsilon=2GM/\left(1-2GM/D\right)$. As a consequence, one can write
\begin{eqnarray}
\label{eq:deltaphi}
\varphi_+-\varphi_-&=&\frac{J}{\sqrt{1-E}}
\int _{r_-}^{r_+}
\frac{1}{\sqrt{(r-r_+)(r-r_-)}}
\nonumber \\ & & \times
\frac{1}{\sqrt{1-\epsilon/r}}\frac{{\rm d}r}{r}
.
\end{eqnarray}
There are several methods to evaluate this integral. As a matter of
fact, it can be integrated exactly in terms of Elliptic
functions~\cite{Abramovitz:1970aa,Gradshteyn:1965aa}. But, being given
that $\epsilon$ is a small quantity, it is better to expand the result
in this parameter. This leads to easier calculations that are,
moreover, more explicit. In addition, the method can be generalized to
the case of a non-vanishing cosmological constant, which is of course
our main goal. In fact, if we restart from Eq.~(\ref{eq:rphi}) and use
that $B^{-1}\simeq 1+2GM/r+\cdots $, we see that the term in the
squared bracket in Eq.~(\ref{eq:rphi}) is a quadratic function of
$1/r$ which vanishes at $r=r_{\pm}$. This means that one necessarily
has
\begin{equation}
\label{eq:defC}
\frac{1}{J^2B(r)}-\frac{E}{J^2}-\frac{1}{r^2}\simeq C\left(\frac{1}{r}
-\frac{1}{r_-}\right)\left(\frac{1}{r}
-\frac{1}{r_+}\right).
\end{equation}
Then, writing Eq.~(\ref{eq:rphi}) in terms of $u$ rather than $r$, one
arrives at
\begin{equation}
\label{eq:dudphi}
\left(\frac{{\rm d}u}{{\rm d}\varphi}\right)^2
\simeq \frac{C}{A(r)}\left(u-u_-\right)\left(u-u_+\right).
\end{equation}
Now, when $\Lambda_{_{\rm eff}}\neq 0$, as is clear from
Eq.~(\ref{eq:drdvarphi}), the term in the square bracket of
Eq.~(\ref{eq:rphi}) is no longer a quadratic function of $1/r$ even if
the expansion of the function $B^{-1}(r)$ is used. However, this term
still vanishes at $r=r_{\pm}$ and the integral will be dominated by
the contributions coming from the vicinity of these points, see also
Eq.~(\ref{eq:deltaphi}). The idea is then to continue to work with
Eq.~(\ref{eq:defC}) and to take into account the cosmological constant
in the definition of $C$~\cite{Wright:1998bc}. This should provide a
reasonable approximation of $\varphi_+-\varphi_-$. Using
Eqs.~(\ref{eq:Jdeux}) and~(\ref{eq:E}), one can write
\begin{eqnarray}
& &\frac{1}{J^2B(r)}-\frac{E}{J^2}-\frac{1}{r^2}\nonumber  \\
&=&\frac{r_-^2\left[B^{-1}(r)-B^{-1}_-\right]
-r_+^2\left[B^{-1}(r)-B_+^{-1}\right]}{r_+^2r_-^2
\left(B^{-1}_+-B^{-1}_-\right)}-\frac{1}{r^2}
\nonumber \\ 
&\simeq &C\left(\frac{1}{r}
-\frac{1}{r_-}\right)\left(\frac{1}{r}
-\frac{1}{r_+}\right).
\end{eqnarray}
Then, if we differentiate both sides twice with respect to $u$, this
gives
\begin{eqnarray}
C&=&\frac{\left(u_+-u_-\right)
\left(u_+-u_-\right)A''\left(u\right)}{2\left(u_+-u_-\right)A'(u)}-1,
\\
&\simeq & -1+\frac{uA''}{A'}\Biggl\vert_{u=L^{-1}},
\end{eqnarray}
where $L\equiv 2D$ is the semilatus rectum. In the above expression,
we have used $A_+-A_-\simeq
\left(u_+-u_-\right)A'\left(L^{-1}\right)$. So far, the method is
completely general provided that $A^{-1}=B$. We now use the explicit
form of $A^{-1}$. Expanding everything in $GMu$ and in $\Lambda_{_{\rm
    eff}}$, one obtains~\cite{Wright:1998bc}
\begin{equation}
\label{eq:expansionc}
C\simeq -1+\frac{4GM}{L}+\frac{\Lambda_{_{\rm eff}}L^3}{GM}+
\frac{4\Lambda_{_{\rm eff}}L^3}{3}+\cdots .
\end{equation}
The final step consists in using the above expression in
Eq.~(\ref{eq:dudphi}). One finds~\cite{Wright:1998bc}
\begin{eqnarray}
\varphi_+-\varphi_-&\simeq&
\int_{u_-}^{u_+}\frac{A^{1/2}(u){\rm d}u}
{\left[C\left(u-u_+\right)\left(u-u_-\right)\right]^{1/2}}
\nonumber \\
&=&\pi\left(1+\frac32\frac{2GM}{L}+\frac{\Lambda_{_{\rm eff}}L^3}{2GM}
+\cdots \right),
\end{eqnarray}
where we have expanded $A^{1/2}C^{-1/2}$ and where we have used the
fact that $\int _{u_-}^{u_+}{\rm
  d}u/[(u-u_+)(u-u_-)]^{1/2}=\ln(u_+-u_-)-\ln(u_+-u_-)+2 i\pi/2$, the
complex $i$ being exactly cancel by the $i$ coming from the fact that
$C$, which appears in a square root, is negative [its expansion starts
with $-1$, see Eq.~(\ref{eq:expansionc})]. As a consequence, we obtain
\begin{equation}
  \Delta \varphi\equiv 2\left(\varphi_+-\varphi_-\right)-2\pi
  \simeq 6\pi \frac{GM}{L}+\Delta \varphi_{_{\Lambda}},
\end{equation}
where the first term is the standard relativistic perihelion
precession while~\cite{Wright:1998bc}
\begin{equation}
\Delta \varphi_{_{\Lambda}}\simeq \frac{\pi \Lambda_{_{\rm eff}}L^3}{GM},
\end{equation}
is the contribution due to the cosmological constant. The above
equation is the main result of this sub-section. It gives the
perihelion precession due to the presence of vacuum energy. Let us
notice that the scaling $\Delta \varphi_{_{\Lambda}}\propto L^3$
indicates that the precession is larger for planets that are far from
the Sun. It is of course totally consistent with the fact that the
effect of the cosmological constant is more important on large scales.

\par

We now use the above result to constrain the amount of vacuum energy
present in our universe. For this purpose, let us now calculate this
perihelion shift for, say, Mercury (as mentioned before, the effect is
larger for other planets but this does not modify the conclusions
obtained below; estimates for Venus, Earth and Mars can be found in
Ref.\cite{Jetzer:2007cr}). In this case, we have $r_+\simeq 68.8\times
10^6\mbox{km}$ and $r_-\simeq 46\times 10^6\mbox{km}$. This implies
that $D\simeq 27.7\times 10^6\mbox{km}$.  We also have $GM\simeq 2.95
\mbox{km}$. Finally, it is interesting to calculate in terms of the
vacuum energy density observed in cosmology, see
Eq.~(\ref{eq:rhovaccosmology}). For the cosmological constant, this
leads to the following number $\Lambda_{_{\rm obs}}\simeq 8.81\,
h^2\times 10^{-48} \mbox{km}^{-2}$. This leads to the following
expression
\begin{equation}
\Delta \varphi_{_{\Lambda}} \simeq 1.6 h^2 \times 10^{-24}
\left(\frac{\Lambda}{\Lambda_{_{\rm obs}}}\right) \quad 
\mbox{rad/revolution},
\end{equation}
or, since Mercury completes $415$ revolutions each century and there
are $360\times 60\times 60/(2\pi)$ arc-seconds per radian,
\begin{equation}
\Delta \varphi_{_{\Lambda}} \simeq 2.7 h^2 \times 10^{-16}
\left(\frac{\Lambda}{\Lambda_{_{\rm obs}}}\right) \quad 
\mbox{arc-second/century},
\end{equation}
which is, of course, completely unobservable. The reason for such a
result has already been mentioned. The solar system has a
characteristic scale (say a few astronomical units) which is too small
for the cosmological constant to leave a sizable imprint. However,
even if one cannot measure the vacuum energy density with this method,
it can nevertheless be used to constrain its value. Indeed, the
uncertainty in the precession of the perihelion of Mercury is about
$0.1''$ par century, which means that
\begin{equation}
\Delta \varphi_{_{\Lambda}} \simeq 2.7 h^2 \times 10^{-16}
\left(\frac{\rho_{_{\Lambda}}}{\rho_{_{\rm cri}}}\right)
<0.1,
\end{equation}
implying that
\begin{equation}
\rho_{_{\rm vac}}\lesssim 10^{15}\rho_{_{\rm cri}}\simeq 3\times 10^{-32}
\, \mbox{GeV}^4.
\end{equation}
Of course, this limit is not competitive with what one has obtained in
cosmology. However, when compared to Eq.~(\ref{eq:rhovacnum}), it
leads to an interesting piece of information which allows us to define
what the cosmological constant problem is more accurately. Indeed, we
see that there is a contradiction between the theoretical
expectation~(\ref{eq:rhovacnum}) and the observations of the planet
trajectories in our solar system. In other words, the predictions of
Eq.~(\ref{eq:rhovacnum}), if true, would lead to orbits drastically
different from what we observe. Therefore, even without the
cosmological observations and only from the motion of the planets in
the solar system, one sees that our calculation of vacuum energy must
be incorrect. Put it differently, it would be incorrect to say that
the cosmological constant problem originates only from the recent
astrophysical observations. In some sense, the problem is much
worse. Only by observing the planets in the solar system, a class of
observations which seems to be much more straightforward than the
cosmological observations, we know that our regularization method of
the vacuum energy must be flawed.

\subsection{Modifications of the Atomic Levels}
\label{subsec:atomlambda}

We have just seen that observing the motion of planets in our solar
system can tell us something about the value of the vacuum energy. A
priori, another possibility to observe the influence of the
cosmological constant is through the modification of the atomic energy
levels that its presence would cause. In this section, we investigate
this question in the case of the Hydrogen atom since this is
technically
simple~\cite{Parker:1980kw,Parker:1982nk,Parker:1981wt,Moradi:2009zzc,Fischbach:1981ne,Boulanger:2006hz}. We
will follow the treatment of Ref.~\cite{Parker:1980kw}. In particular,
since the problem is still spherically symmetric, one can use the
metric tensor~(\ref{eq:metricsphere}) obtained before.

\par

The motion of an electron around the nucleus, living in a static and
spherically symmetric manifold, is described by the following Dirac
equation
\begin{equation}
\left(\Gamma ^{\mu}D_{\mu}+m\right)\Psi=0,
\end{equation}
where $\Psi$ is a Dirac spinor, see Sec.~\ref{subsec:fermion} and $m$
the mass of the electron. In the above equation, $\Gamma _{\mu}$ are
the Dirac matrices satisfying the relation
\begin{equation}
\left\{\Gamma ^{\mu},\Gamma ^{\nu}\right\}=2g^{\mu \nu}.
\end{equation}
Let us notice that we have already presented this formula, see
Eq.~(\ref{eq:diracanticom}). Here, we slightly change the notation and
now denote the Dirac matrices in curved space-time by $\Gamma^{\mu}$
(notice that we have also included a factor $i$ in the definition of
$\Gamma^{\mu}$). The matrices $\Gamma ^{\mu}$ are related to the Dirac
matrices in flat space-time, $\gamma ^{\mu}$, by $\Gamma
_{\mu}=e^{\alpha}{}_{\mu}\gamma _{\alpha}$, where $e^{\alpha}{}_{\mu}$
are the vierbein fields such that $g_{\mu
  \nu}=e^{\alpha}{}_{\mu}e^{\beta}{}_{\nu}\eta_{\alpha \beta}$. The
symbol $D_{\mu}$ denotes the covariant derivative acting on the
spinor, namely
\begin{equation}
D_{\mu}\Psi=\left(\partial_{\mu}-\omega_{\mu}\right)\Psi,
\end{equation}
where the spinorial connection can be expressed as
\begin{equation}
\omega _{\mu}\equiv -\frac14 \gamma _{\alpha}\gamma _{\beta}
e^{\alpha}{}_{\nu}g^{\nu \lambda}\left(\nabla _{\mu}e^{\beta}{}_{\lambda}\right)
+iqA_{\mu},
\end{equation}
with
$\nabla_{\mu}e^{\beta}{}_{\lambda}\equiv \partial_{\mu}e^{\beta}{}_{\lambda}
-\Gamma^{\sigma}_{\mu \lambda}e^{\beta}{}_{\sigma}$ and $A_{\mu}$ is
the vector field describing the electromagnetic field acting on the
electron ($q$ is the charge). The Dirac equation can also be
re-written as
\begin{equation}
i\partial _0\Psi=H\Psi,
\end{equation}
where the effective Hamiltonian is given by
\begin{eqnarray}
\label{eq:effectiveH}
H &=& -i\left(g^{00}\right)^{-1}\Gamma ^0\Gamma ^i\left(\partial_i\Psi
-\omega_i\Psi\right)+i\Gamma _0\Psi
\nonumber \\ & &
-i\left(g^{00}\right)^{-1}\Gamma ^0m\Psi.
\end{eqnarray}
In this case, the Dirac equation looks like an effective Schr\"odinger
equation.

\par

It is clear that the typical dimension of an atom is very small in
comparison to the typical curvature scale of the manifold. Therefore,
as was done for the calculation of the Green function in
Sec.~\ref{sec:vaccurve}, one can perform a local analysis. For this
purpose, we now use the Fermi normal
coordinates~\cite{Poisson:2011nh}. In these coordinates, the metric
tensor takes the form
\begin{eqnarray}
g_{00} &=&-1-R_{0\ell 0m}x^{\ell}x^m+\cdots\, ,\\
g_{0i} &=& -\frac23R_{0\ell im}x^{\ell}x^m+\cdots ,\\
g_{ij} &=& \delta _{ij}-\frac13R_{i\ell jm}x^{\ell}x^m+\cdots \, ,
\end{eqnarray}
where the dots indicate higher order terms. This implies that the
veirbein fields can be expressed as
\begin{eqnarray}
e^{\alpha}{}_0 &=& \delta ^{\alpha}{}_0-\frac12R^{\alpha}{}_{\ell 0 m}x^{\ell}x^m, \\
e^{\alpha}{}_i &=& \delta ^{\alpha}{}_i-\frac12R^{\alpha}{}_{\ell i m}x^{\ell}x^m,
\end{eqnarray}
and the Christoffel symbols can be written as
\begin{eqnarray}
\Gamma ^0_{ij} &=& \frac13\left(R_{0ijm}+R_{0jim}\right)x^m, 
\\ \Gamma ^0_{0i} &=& R_{0iom}x^m, \\
\Gamma ^{i}_{jk}&=& \frac13\left(R_{jikm}+R_{kijm}\right)x^m,
\quad \Gamma ^0_{00}=0,\\
\Gamma _{0j}^i&=&R_{0mji}x^m, \quad \Gamma ^i_{00}=R_{0i0m}x^m.
\end{eqnarray}
Using the above results, this finally leads to the following
expression for the spinorial connection~\cite{Parker:1980kw}
\begin{eqnarray}
\label{eq:spinconnection}
\omega _0 &=& \frac12\gamma _0\gamma _jR^j{}_{00m}x^m+\frac14\gamma _k
\gamma _jR^{kj}{}_{0m}x^m \nonumber \\
& & +iqA_0, \\
\omega_i &=& \frac14\gamma _0\gamma _jR^{0j}{}_{im}x^m+\frac18\gamma _k
\gamma _jR^{kj}{}_{im}x^m
\nonumber \\ & &
+iqA_i.
\end{eqnarray}
The above formulas will allow us to calculate the effective
Hamiltonian~(\ref{eq:effectiveH}) explicitly. But before reaching this
stage, we must also determine vector field $A_{\mu}$. The Lagrangian
of the corresponding gauge field is given by, see also
Eq.~(\ref{eq:actionproca})
\begin{equation}
S=-\int {\rm d}^4x\sqrt{-g}
\left(\frac14 g^{\alpha \beta}g^{\mu \nu}
F_{\mu \alpha}F_{\nu \beta}-j^{\mu}A_{\mu}\right),
\end{equation}
from which we deduce that the equations of motion read
\begin{equation}
g^{\mu \nu}\nabla_{\mu}\nabla_{\nu}A_{\gamma}
-R^{\sigma}{}_{\gamma}A_{\sigma}=-j_{\gamma}.
\end{equation}
At first order in the curvature, these equations can be expressed
as~\cite{Parker:1980kw}
\begin{widetext}
\begin{eqnarray}
\label{eq:eomgaugeatom}
& &\delta^{ij}\partial _i\partial_jA_0
+\frac13R_{i\ell jm}x^{\ell }x^m\partial ^i\partial^jA_0
+\frac53 R_{i00m}x^m\partial ^iA_0+2R^k{}_{i0m}x^m\partial ^iA_k
-\frac23R_{im}x^m\partial ^iA_0 = -j_0 \\
& &\delta^{ij}\partial _i\partial_jA_k
+\frac13R_{i\ell jm}x^{\ell }x^m\partial ^i\partial ^j A_k
+\frac23\delta^{ij}\left(R^{\alpha}{}_{kim}+R^{\alpha}{}_{ikm}\right)x^m
\partial_jA_{\alpha}
-\frac23R^{\sigma}{}_kA_{\sigma}
-\frac13 R^p{}_{00k}A_p
\nonumber \\ & &
-\frac13R_{i00m}x^m\partial ^iA_k 
-\frac23R_{pm}x^m\partial^pA_k
= j_k.
\end{eqnarray}
\end{widetext}
For a one electron atom with a nucleus of charge $Ze$, one can take
\begin{equation}
  j^0=-Ze\, \delta\left(\bm{r}\right),\quad j^k=0.
\end{equation}
In what follows, we determine the vector perturbatively in the
curvature. At zeroth order, in flat space-time, the equation for the
time component reduces to
\begin{equation}
\delta^{ij}\partial _i\partial_jA_0
 = Ze \, \delta\left(\bm{r}\right),
\end{equation}
whose solution is of course
\begin{equation}
\label{eq:solzerogaugeatom}
A_0^{(0)}=-\frac{Ze}{4\pi}\frac{1}{r}.
\end{equation}
Then, we write $A_0=A_0^{(0)}+\delta A_0$, where $\delta A_0$ is first
order in the curvature. By inserting this expression into the full
equation~(\ref{eq:eomgaugeatom}) and using the zeroth order
solution~(\ref{eq:solzerogaugeatom}), one arrives
at~\cite{Parker:1980kw}
\begin{equation}
\delta^{ij}\partial _i\partial_j\delta A_0
+\frac13 \frac{Ze}{4\pi r^3}\left(3R^0{}_{\ell 0m}
-R_{\ell m}\right)x^{\ell}x^m=0.
\end{equation}
The solution of the above equation reads~\cite{Parker:1980kw}
\begin{eqnarray}
\delta A_0 &=& \frac{1}{12}\frac{Ze}{4\pi}\left(R+4R_{00}\right)r
\nonumber \\ & &
+\frac{1}{12}\frac{Ze}{4\pi}\left(3R^0{}_{\ell 0m}-R_{\ell m}\right)
\frac{x^{\ell}x^m}{r}.
\end{eqnarray}
In the same manner, one can determine the space component. Obviously,
at zeroth order, one has $A_k^{(0)}=0$. The first order perturbation
$\delta A_k$ obeys the equation~\cite{Parker:1980kw}
\begin{equation}
\delta^{ij}\partial _i\partial_j\delta A_k
+\frac23 \frac{Ze}{4\pi} R^0{}_{k}\frac{1}{r}
+\frac23 \frac{Ze}{4\pi}
R^0{}_{ikm}\frac{x^{i}x^m}{r^3}=0,
\end{equation}
whose solution can be written as~\cite{Parker:1980kw}
\begin{equation}
\label{eq:solkgaugeatom}
\delta A_k=\frac12 \frac{Ze}{4\pi}R_{0k}r+\frac{1}{6}\frac{Ze}{4\pi}R^0{}_{lkm}
\frac{x^{\ell }x^m}{r}.
\end{equation}
This completes our calculation of the electromagnetic field. 

\par

Having determined the vierbeins fields (\ie the quantities describing
the gravitational force) and the gauge field (\ie the quantity
describing the electromagnetic force), we are now in a position where
we can calculate the Hamiltonian~(\ref{eq:effectiveH}). In order to
make contact with the usual treatment of the Hydrogen atom, it is more
convenient to introduce the standard Dirac matrices $\alpha_i$ and
$\beta$ related to the flat space-time Dirac matrices by
\begin{equation}
\gamma_0=i\beta, \quad \gamma _i=-i\beta\alpha_i.
\end{equation}
As is well-known, they satisfy
\begin{eqnarray}
\alpha_i\alpha_j+\alpha_j\alpha_i&=&2\delta_{ij}, \\
\alpha_i\beta+\beta \alpha_i&=&0, \quad \beta^2=1.
\end{eqnarray}
Then, we use the expressions of the spinorial
connections~(\ref{eq:spinconnection}), of the gauge
field~(\ref{eq:solzerogaugeatom}),~(\ref{eq:solkgaugeatom}) and of the
Dirac matrices in the equation~(\ref{eq:effectiveH}) of the
Hamiltonian. This leads to the following expression
\begin{widetext}
\begin{eqnarray}
H&=& -i\alpha^i\partial_i+m\beta-\frac{\zeta}{r}
-\frac{i}{2}R_{0\ell 0m}x^{\ell }x^m\alpha_i\partial^i
-\frac{i}{6}R_{i\ell jm}x^{\ell }x^m\alpha^j\partial^i
-\frac{i}{6}R_{0\ell jm}x^{\ell}x^m\alpha^j\alpha^i\partial _i
-\frac{i}{2}R_{0\ell im}x^{\ell}x^m\partial ^i
\nonumber \\ & &
+\frac{i}{8}\alpha^i\alpha^j\alpha^kR_{jkim}x^m
+\frac{i}{4}\alpha^i\alpha^jR_{0jmi}x^m
+\frac{i}{4}\alpha^i\alpha^jR_{0mij}x^m
-\frac{i}{2}\alpha^jR_{0j0m}x^m
+\frac{1}{6}\zeta R_{0\ell mi}x^{\ell}x^m\frac{\alpha^i}{r}
+\frac12 \zeta R_{0i}\alpha^ir
\nonumber \\ & & 
+\frac{1}{12}\zeta \left(R+4R_{00}\right)r
-\frac{1}{12}\zeta \left(R_{\ell m}+3R_{0\ell 0m}\right)
\frac{x^{\ell }x^m}{r}
-\frac{m}{6}R_{i\ell 0m}x^{\ell}x^m\beta\alpha^i
+\frac{m}{2}R_{0\ell 0m}x^{\ell }x^m\beta.
\end{eqnarray}
\end{widetext}
In this expression we have defined $\zeta\equiv Ze^2/(4\pi)$ and we
have taken $q=-e$. The three first terms are the usual Dirac
Hamiltonian. The other terms represent the correction due to
space-time curvature. Since the above expression is quite complicated,
it is interesting to estimate the order of magnitude of each term. The
curvature is given by $D^{-2}$ where $D$ is a typical distance. Each
$x^i$ is typically of the order of the Bohr radius, \ie $x^i\simeq
a_0\simeq m^{-1}e^{-2}\simeq m^{-1}\zeta^{-1}$. As a consequence,
$\partial _i$ is of order $ m\zeta$. This also means that the typical
velocity is given by $m\zeta/m\simeq \zeta$. Of course the matrices
$\alpha^i$ and $\beta$ are dimensionless but, in the non relativistic
limit $\alpha^i\partial _i$ must reduce to $\simeq mv^2$ which implies
that $\alpha^i\simeq v\simeq \zeta$.  In the following we consider
that $\zeta=Z\alpha$ where $\alpha $ is the fine structure constant is
a small number. Then, the dominant term in the above expression is the
last one, which means that the Hamiltonian can be approximated as
\begin{eqnarray}
H&=& H_0+H_{_{\rm int}}\simeq \bm{\alpha}\cdot \bm{p}+m\beta-\frac{\zeta}{r}
\nonumber \\ & &
+\frac{1}{2}mR_{0\ell 0m}x^{\ell }x^m\beta,
\end{eqnarray}
where we have used $p_i=-i\partial_i$. If we use the following
representation for the Dirac matrices
\begin{equation}
\alpha^i=
\begin{pmatrix}
0 & \sigma^i \\ \sigma^i & 0
\end{pmatrix}, \quad
\beta =
\begin{pmatrix}
\mathbb{I}_2 & 0 \\ 0 & -\mathbb{I}_2,
\end{pmatrix}
\end{equation}
and write the Dirac spinor as
\begin{equation}
\Psi = {\rm e}^{iE x^0}
\begin{pmatrix}
\phi \\ \chi
\end{pmatrix},
\end{equation}
then the Dirac equation reduces to two equations that can be expressed
as
\begin{eqnarray}
\bm{\alpha}\cdot \bm{p} \, \chi + \left(m-\frac{\zeta}{r}
+\frac{1}{2}mR_{0\ell 0m}x^{\ell }x^m\right) \phi&=& E\phi, \\
\bm{\alpha}\cdot \bm{p} \, \phi - \left(m+\frac{\zeta}{r}
+\frac{1}{2}mR_{0\ell 0m}x^{\ell }x^m\right) \chi &=& E\chi .
\end{eqnarray}
From the second relation we can express $\chi$ in terms of
$\bm{\alpha}\cdot \bm{p} \, \phi$ and then we can insert this
expression into the first equation. This leads to
\begin{widetext}
\begin{eqnarray}
\left(E+m+\frac{\zeta}{r}+\frac{1}{2}mR_{0\ell 0m}x^{\ell }x^m\right)^{-1}
\left(\bm{\alpha}\cdot \bm{p}\right)^2\phi+\left(m-\frac{\zeta}{r}
+\frac{1}{2}mR_{0\ell 0m}x^{\ell }x^m\right) \phi&=& E\phi
\end{eqnarray}
or, using $\bm{\alpha}\cdot \bm{p} \, \phi=p^2\, \phi$ and
expanding the denominator of the first term in the above equation
\begin{eqnarray}
\left[1-\frac{1}{E+m}\frac{\zeta}{r}
-\frac{m}{2(E+m)}R_{0\ell 0m}x^{\ell }x^m\right]
\frac{p ^2}{E+m}\phi-\left(\frac{\zeta}{r}
-\frac{1}{2}mR_{0\ell 0m}x^{\ell }x^m\right) \phi&=& (E-m)\phi.
\end{eqnarray}
\end{widetext}
But, if we now perform an additional expansion assuming
non-relativistic velocities for the electron (which, in practice, is a
good approximation), one arrives at the following expression for the
term $p^2/(E+m)$
\begin{equation}
\frac{p^2}{E+m}=\frac{p}{m+m+p^2/(2m)+\cdots }
\simeq \frac{p^2}{2m}+\cdots 
\end{equation}
As a consequence, we conclude that the non relativistic limit is given
by
\begin{equation}
H_{_{\rm NR}}\phi=(E-m)\phi,
\end{equation}
where the non-relativistic Hamiltonian can be written
as~\cite{Parker:1980kw}
\begin{eqnarray}
H_{_{\rm NR}}&=& H_0+H_{_{\rm int}}=\frac{p^2}{2m}-\frac{\zeta}{r}
\nonumber \\ & &
+\frac{1}{2}mR_{0\ell 0m}x^{\ell }x^m\beta.
\end{eqnarray}
This is our final result. Endowed with this equation, one can
calculate, by means of perturbations theory, the modification of the
ground state due to the presence of a cosmological constant. This
leads to (the arguments of $E$ indicates the value of the quantum
numbers $n$ and $\ell$)
\begin{eqnarray}
\Delta E(1,0) &=& \frac{1}{2}mR_{0\ell 0m}\left \langle 1s\left \vert 
x^{\ell}x^m\right\vert 1s\right\rangle \\
&=& \frac{1}{2}mR_{0\ell 0m}\frac{1}{3}\delta ^{\ell m}
\left \langle 1s\left \vert 
r^2\right\vert 1s\right\rangle \\
&=& \frac{1}{6}mR_{00}\left \langle 1s\left \vert 
r^2\right\vert 1s\right\rangle
\end{eqnarray}
The ground state wave function is given by (at this order of
perturbations theory, there is no need to take into account the
modification on the wave-function; one can use the standard one)
\begin{equation}
\vert 1s\rangle =\frac{1}{\sqrt{\pi a_0^3}}{\rm e}^{-r/a_0},
\end{equation}
where we recall that $a_0$ is the Bohr radius. This implies that
\begin{equation}
\left \langle 1s\left \vert 
r^2\right\vert 1s\right\rangle=\frac{4\pi}{\pi a_0^3} \int _0^{\infty}
r^4{\rm e}^{-r/a_0}{\rm d}r=\frac{4\pi}{\pi a_0^3}\frac{3a_0^5}{4}
=3a_0^2.
\end{equation}
The vacuum Einstein equations are, see Eq.~(\ref{eq:Einsteinvac}),
$R_{\mu \nu}=\Lambda_{_{\rm eff}}g_{\mu \nu}$ and, therefore, at first
order in the cosmological constant one $R_{00}=-\Lambda_{_{\rm eff}}$
(here, of course, we only compute the corrections due to the
cosmological constant and not those originating from the Earth
gravitational field). This means that
\begin{eqnarray}
\label{eq:deltaElambda}
\Delta E(1,0) &=& -\frac{1}{2}mc^2\Lambda_{_{\rm eff}}a_0^2,
\end{eqnarray}
where we have re-established the speed of light for convenience. In
fact, this result could have been anticipated from the Newtonian
approach. Indeed, we have seen that the Newtonian potential, in
presence of a cosmological constant, is given by $-\Lambda_{_{\rm
    eff}}mc^2r^2/6$, see Eqs.~(\ref{eq:potlambda})
and~(\ref{eq:forcelambda}). Using the perturbation theory with this
potential exactly leads Eq.~(\ref{eq:deltaElambda}). Let us now
evaluate the correction. Using that the reduced mass of the atom is
essentially the electron mass, $m\simeq 9.1\times 10^{-31}\,
\mbox{kg}$ and that the Bohr radius is $a_0=0.52\times 10^{-10}\,
\mbox{m}$, one obtains
\begin{equation}
  \Delta E(1,0)=-6\times 10^{-69} h^2 
\left(\frac{\Lambda}{\Lambda _{_{\rm obs}}}\right)\, 
\mbox{eV},
\end{equation}
a quantity completely unobservable, probably for ever. For comparison
the Lamb shift is of the order $10^{-6}\, \mbox{eV}$. At this stage
this does not come as a surprise. As discussed in the case of planet
orbits, the size of the system is so small that the influence of the
cosmological constant is negligible. However, requiring that the shift
due to the cosmological constant be smaller than the Lamb shift leads
to
\begin{equation}
\rho_{_{\rm vac}}\lesssim 1.6\times 10^{62}\rho_{\rm cri}
\simeq 1.3\times 10^{16}
  \mbox{GeV}^4.
\end{equation}
This result is less good than the one obtained from planet motion by
about $50$ orders of magnitude. Moreover, the theoretical
expectation~(\ref{eq:rhovacnum}) satisfies this inequality. Let us
notice in passing that the result based on the wrong regularization
scheme, $\rho_{_{\rm vac}}\simeq M_{_{\rm Pl}}^4$ does not satisfy the
previous bound.

\par

We now have a clearer view of what the cosmological constant problem
is. On one hand, one can calculate the vacuum energy density by means
of, apparently, well known techniques of regularization. These
techniques usually lead very robust results, fully consistent with the
observations. For instance, the radiative corrections to the
calculation of cross-sections have proven in very good agreement with
various high energy physics experiments. One the other hand, various
observational results (and not only in cosmology) indicate that the
previous calculation is not correct. The theoretical expectation value
for the vacuum energy density is so large that it should have already
be seen by many experiences. Moreover, in cosmology, under a certain
number of hypothesis, one even measures $\rho_{_{\rm vac}}$ which is
found to be many orders of magnitude smaller than the number derived
from the theory. Since, as mentioned before, the theoretical framework
seems to be robust (let us recall again that it involves calculations
that can be found in the first pages of any textbook on quantum field
theory) and since the observations seem to be convincing, we are
facing a genuine mystery. However, before accepting this conclusion,
it is worth checking that no loophole is present in the previous
reasoning. This will be the goal of the next sections. A first
question that can be asked is whether the zero-point fluctuations that
are at the origin of the problem are a real physical phenomenon or
just an artifact of the formalism of quantum field theory. We turn to
this issue in the next section.

\section{Do the Vacuum Fluctuations Really Exist?}
\label{sec:existence}

We have seen in the previous sections that the zero-point fluctuations
manifest themselves as the vacuum energy. Since this leads to
conclusions that seem to be in contradiction with observations, it is
legitimate to ask whether these fluctuations really exist in
Nature. In particular, we want to investigate whether the zero-point
fluctuations could manifest themselves in other physical
phenomena. The answer to this question is usually positive and one
experiment which is considered as a proof that the vacuum
fluctuations are real is the measurement of the Lamb
shift~\cite{Welton:1948zz,Itzykson:1980rh}. In the next sub-section,
we discuss this phenomenon.

\subsection{The Lamb Shift}
\label{subsec:lambshift}

Let us consider the ``motion'' of an electron in an atom. Its
``location'' is described by the vector ${\bm r}$. The vector ${\bm
  r}$ is going to fluctuate because of the interaction between the
electron and the zero-point fluctuations of the electromagnetic
field. This interaction will slightly modify the position of the
atomic levels leading to the Lamb shift. Let us study in details how
this effect can be derived. Let $V({\bm r})$ be the Coulomb potential
which determines the properties of the atom. At the point ${\bm
  r}+\delta{\bm r}$, the potential can be Taylor expanded according to
\begin{equation}
  V({\bm r}+\delta{\bm r})=
  V({\bm r})+\frac{\partial V}{\partial x_i}\delta x_i+
  \frac12 \frac{\partial ^2V}{\partial x_i\partial x_j}
  \delta x_i\delta x_j+\cdots .
\end{equation}
Then, if we time average the previous expression, one obtains
\begin{equation}
\label{eq:perturbV}
\langle V({\bm r}+\delta{\bm r})\rangle \simeq 
V({\bm r})+\frac{\partial V}{\partial x_i}\langle \delta x_i\rangle +
\frac12 \frac{\partial ^2V}{\partial x_i\partial x_j}
\langle \delta x_i\delta x_j\rangle+\cdots ,
\end{equation}
and, because the problem is spherically symmetric, we have $\langle
\delta x_i\rangle =0$ and $\langle \delta x_i\delta x_j\rangle
=\langle \delta r^2\rangle \delta _{ij}/3$. The first term in
Eq.~(\ref{eq:perturbV}) is the unperturbed Coulomb potential and the
last one (since the second one vanishes) can be viewed as a small
perturbation of the Hamiltonian, namely
\begin{equation}
\label{eq:deltaHlamb}
\Delta H=\frac16 \frac{\partial ^2V}{\partial x_i\partial x_j}
\left \langle \delta r^2\right \rangle \delta _{ij}.
\end{equation}
In an atom with atomic number $Z$, the Coulomb potential is given by
\begin{equation}
V=-\frac{Z\alpha}{r},
\end{equation}
where $\alpha \simeq 1/137$ is the fine structure constant. This means
that $\delta _{ij}\partial ^2V/(\partial x_i\partial x_j) =4\pi
Z\alpha \delta \left({\bm r}\right)$.

\par

Let now us assume that the atom is placed in the quantum state $\vert
\ell,m\rangle $. As we have already discussed in
Sec.~\ref{subsec:atomlambda}, at first order, the time-independent
perturbations theory tells us that the correction to the energy of the
level $(\ell, m)$ is given by~\cite{tannoudji2006}
\begin{equation}
\Delta E(n,\ell)=\langle \ell, m\vert \Delta H\vert \ell, m\rangle.
\end{equation}
Using Eq.~(\ref{eq:deltaHlamb}), one obtains the following expression
for the energy displacement
\begin{equation}
\label{eq:DEenergyLS}
\Delta E(n,\ell)=\frac{2\pi}{3}Z\alpha 
\left\vert \Psi_{n\ell}(0)\right\vert ^2\left\langle 
\delta r^2\right \rangle , 
\end{equation}
where $\Psi_{n\ell}$ is the wave-function in the corresponding quantum
state. Since the wave-function at the origin can be expressed
as~\cite{tannoudji2006}
\begin{equation}
\left\vert \Psi_{n\ell}(0)\right\vert=\frac{1}{\pi^{1/2}}
\left(\frac{mZ\alpha}{n}\right)^{3/2}\delta _{\ell 0},
\end{equation}
where $m$ is the electron mass ($m\simeq 0.511\, \mbox{MeV}$) one
finally arrives at~\cite{Itzykson:1980rh}
\begin{equation}
\label{eq:deltaEinter}
\Delta E(n,\ell)=\frac{\left(2mZ\alpha\right)^3}{12}\frac{Z\alpha}{n^3} 
\left\langle \delta r^2\right 
\rangle \delta _{\ell 0}.
\end{equation}
Therefore, we have reduced the problem to the calculation of the
quantity $\left\langle \delta r^2\right \rangle$. In order to evaluate
this quantity, we must now evaluate the displacement of the
electron. The electron ``moves'' under the influence of the electric
field within the atom and, as a consequence, $\delta r_i$ obeys the
following equation
\begin{equation}
\label{eq:eomdeltar}
m\frac{{\rm d}^2}{{\rm d}t^2}\delta r_i=eE_i,
\end{equation}
where $E_{\mu}\equiv u^{\nu}F_{\mu \nu}=-\dot{A}_i$ is the electric
field (or, rather, its vacuum fluctuating component). Using the usual
expansion of the vector potential in terms of creation and
annihilation operators, see Eq.~(\ref{eq:expansionphoton}), one can
write
\begin{eqnarray}
E_{i}\left(t,{\bm x}\right)&=&\frac{-1}{\left(2\pi\right)^{3/2}}
\int \frac{{\rm d}{\bm k}}{\sqrt{2\omega (k)}}\sum _{\alpha=1}^2
\epsilon_{i}^{\alpha}({\bm k})\nonumber \\ &\times&
\biggl[-i\omega a_{\bm k}^{\alpha}
{\rm e}^{-i\omega t+i{\bm k}\cdot {\bm x}}
  +i\omega \left(a_{\bm k}^{\alpha}\right)^{\dagger}
  {\rm e}^{i\omega t-i{\bm k}\cdot {\bm x}}\biggr],\nonumber \\
\end{eqnarray}
In the same manner, one has to Fourier expand $\delta r_i$ which is
considered as an operator in this context. One obtains
\begin{eqnarray}
\delta r_{i}\left(t,{\bm x}\right)&=&\frac{1}{\left(2\pi\right)^{3/2}}
\int {\rm d}{\bm k}\sum _{\alpha=1}^2
\epsilon_{i}^{\alpha}({\bm k})
\biggl[\delta r({\bm k},t)a_{\bm k}^{\alpha}
{\rm e}^{i{\bm k}\cdot {\bm x}}\nonumber \\ 
  & &
  +\delta r^*({\bm k},t)\left(a_{\bm k}^{\alpha}\right)^{\dagger}
  {\rm e}^{-i{\bm k}\cdot {\bm x}}\biggr].
\end{eqnarray}
The Fourier amplitude of the displacement operator can be calculated
from the equation of motion~(\ref{eq:eomdeltar}). This leads to the
following expression
\begin{equation}
\delta r({\bm k},t)=\frac{e}{m}\frac{1}{\sqrt{2\omega ^3}}{\rm e}^{-i\omega t}.
\end{equation}
From this expression, one can now evaluate the quantity $\left\langle
  \delta r^2\right \rangle$. One finds
\begin{eqnarray}
\left\langle \delta r^2\right 
\rangle
&=&\left\langle 0\left \vert \delta ^{ij}\delta r_i\delta r_j\right \vert 0
\right \rangle 
\nonumber \\
&=&\frac{1}{(2\pi)^3}\int {\rm d}{\bm k}
\sum _{\alpha =1}^2
\delta ^{ij}\epsilon_i^{\alpha}\epsilon_j^{\alpha}\delta r({\bm k},t)
\delta r^*({\bm k},t)\nonumber \\
\\
\label{eq:meandr2}
&=&\frac{2\alpha }{m^2\pi}\int _0^{\infty}\frac{{\rm d}\omega}{\omega},
\end{eqnarray}
where we have used $\alpha =e^2/(4\pi)$. Therefore, inserting the
above formula into Eq.~(\ref{eq:deltaEinter}), one obtains the
following equation for the energy shift~\cite{Itzykson:1980rh}
\begin{equation}
\Delta E(n,\ell)=\frac{4 mZ^4\alpha ^5}{3\pi n^3}
\int _0^{\infty}\frac{{\rm d}\omega}{\omega}\delta _{\ell 0}.
\end{equation}
This equation is the main result of this section. It gives the energy
levels displacement due to the presence of the electromagnetic
zero-point fluctuations.

\par

As usual this expression is divergent and needs to be regularized. In
a simplified treatment, one simply assumes, see
Ref.~\cite{Itzykson:1980rh}, that the wavelengths in the above sum
must be larger than the Compton wavelength of the electron, which
implies $\omega <\omega_{\rm max}\simeq m$ and that $\omega >\omega
_{\rm min}\simeq 1/a_0 \simeq 1/[1/(\alpha m)]\simeq m \alpha$, where
$a_0$ is the Bohr radius. In this case, the previous results reduces to
\begin{equation}
\label{eq:LSfinal}
\Delta E(n,\ell)\simeq \frac{4 mZ^4\alpha ^5}{3\pi n^3}
\ln \left(\frac{1}{\alpha}\right)\delta _{\ell 0}.
\end{equation}
Let us now evaluate this quantity for the Hydrogen atom. Clearly,
there is no effect for $n=1$. But for $n=2$ and $\ell =0$, one has
\begin{equation}
\Delta E(2,0)\simeq \frac{4 m\alpha ^5}{3\pi 8}
\ln \left(\frac{1}{\alpha}\right)\delta _{\ell 0}\simeq 
668 \, \mbox{MHz},
\end{equation}
where we have used $m\simeq 0.511 \mbox{MeV}=h\nu$ with $h=4.135
\times 10^{-15}\mbox{eV}\times \mbox{s}$. 

\par

This shift has been experimentally observed and is usually taken as a
proof that the zero point fluctuations are real since they lead to an
observed physical phenomenon. Based on this result, it now seems
difficult to argue that the zero-point fluctuations are just an
artifact of the formalism of quantum field theory. Moreover, there
exists another experiment, the Casimir effect, where the effects of
the vacuum fluctuations can be observed. In the next section, we
investigate this case.

\subsection{The Casimir Effect}
\label{subsec:casimir}

\begin{figure*}
\begin{center}
\includegraphics[width=11cm]{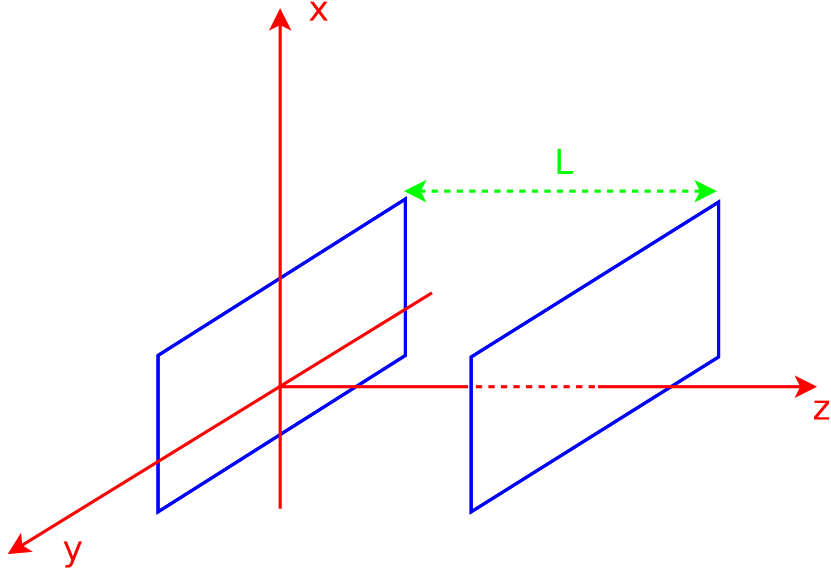}
\caption{Sketch of a Casimir cavity of width $L$. }
\label{fig:casimir}
\end{center}
\end{figure*}

In this section, we study the Casimir
effect~\cite{Plunien:1986ca,Milton:2004ya,Milton:2000av,Milton:2010qr,Lambrecht:2011qq,Milton:2002vm,Deutsch:1978sc}. Since
its discovery, this effect has been studied in great details and it is
clear that, here, one cannot review all the literature. We will give a
description of this phenomenon in a simplified context only and will
refer to Refs.~\cite{Plunien:1986ca,Milton:2004ya,Milton:2000av} for
readers interested in learning more about this interesting subject.

\subsubsection{The Casimir Force}
\label{subsubsec:casimirforce}

Let us consider an experimental situation where we have two conducting
plates in the $(x,y)$ plane separated by a distance $L$ long the $z$
direction, see Fig.~\ref{fig:casimir}. For simplicity, instead of
considering the electromagnetic field in the cavity (which is the case
in the real world, especially when measurements are performed), we
will just treat the case of a real massive scalar field $\Phi(t,{\bm
  x})$. This field obeys the Klein-Gordon equation already studied and
solved before, see Eq.~(\ref{eq:kgequation}), except that now the
boundary conditions are modified by the presence of the two
plates. Let us write $\Phi(t,{\bm x})=X(x)Y(y)Z(z)T(t)$. Then,
inserting this anzatz into the Klein-Gordon equation, one obtains
\begin{equation}
-\frac{\ddot{T}}{T}+\frac{1}{X}\frac{{\rm d}^2X}{{\rm d}x^2}
+\frac{1}{Y}\frac{{\rm d}^2Y}{{\rm d}y^2}
+\frac{1}{Z}\frac{{\rm d}^2Z}{{\rm d}z^2}-m^2=0,
\end{equation}
where $m$ is the mass of the scalar particle. Clearly, one can
separate the variables and, as a consequence, the above equation can
be split into four differential equations, namely
\begin{eqnarray}
-\frac{\ddot{T}}{T}-m^2 &=& C^2 \\
\frac{1}{X}\frac{{\rm d}^2X}{{\rm d}x^2}&=& k_x^2\\
\frac{1}{Y}\frac{{\rm d}^2Y}{{\rm d}y^2}&=& k_y^2\\
\frac{1}{Z}\frac{{\rm d}^2Z}{{\rm d}z^2}&=& k_z^2,
\end{eqnarray}
where the constants $C$, $k_x$, $k_y$ and $k_z$ are related by
\begin{equation}
\label{eq:relCk}
C-k_x^2-k_y^2-k_z^2=0.
\end{equation}
The solution in the $x$ and $y$ direction can be expressed in terms of
standard plane waves, $X(x)=A_x{\rm e}^{ik_xx}+B_x{\rm e}^{-ik_xx}$
and $Y(y)=A_y{\rm e}^{ik_yy}+B_y{\rm e}^{-ik_yy}$. The quantities
$A_{x,y}$ and $B_{x,y}$ are just integration constants. Of particular
interest is of course the solution along the $z$-axis. It can also be
written as
\begin{equation}
Z(z)=A_z{\rm e}^{ik_zz}+B_z{\rm e}^{-ik_zz},
\end{equation}
where $A_z$ and $B_z$ are two integration constants. But the
difference with the case of a free field is that the boundary
conditions ``feel'' the presence of the two plates. These boundary
conditions are given by $Z(0)=0$ and $Z(L)=0$. The first one implies
that $A_z=-B_z$ while the second one gives
\begin{equation}
2i\sin \left(k_zL\right)=0,
\end{equation}
or, equivalently,
\begin{equation}
k_z=n\frac{\pi}{L},
\end{equation}
where $n$ is an integer. Finally, the equation for the function $T(t)$
can also be easily solved and the solution reads
\begin{equation}
T(t)=A_t{\rm e}^{i\omega t}+B_t{\rm e}^{-i\omega t},
\end{equation}
where $\omega ^2=C^2+m^2$. Therefore, upon using Eq.~(\ref{eq:relCk}),
the frequency $\omega$ takes the form
\begin{equation}
\label{eq:defomegacasimir}
\omega =\sqrt{k_x^2+k_y^2+n^2\frac{\pi ^2}{L^2}+m^2}.
\end{equation}
In this expression, $k_x$ and $k_y$ take continuous values.

\par

Based on the previous considerations, one can now write the field
operator as an expansion in terms of creation and annihilation
operators. But this expansion encodes the fact that, along the
$z$-axis, the wave-number is discrete. Concretely, we have
\begin{widetext}
\begin{equation}
  \Phi (t,{\bm x})=\int \frac{{\rm d}k_x}{(2\pi)^{1/2}}
  \int \frac{{\rm d}k_y}{(2\pi)^{1/2}}
  \sum _{n=1}^{\infty}
  \sqrt{\frac{2}{L}}\frac{1}{\sqrt{2\omega}}\sin\left(n\frac{\pi}{L}z\right)
  \left(c_{k_x,k_y,n}{\rm e}^{-i\omega t+ik_xx+ik_yy} 
    +c_{k_x,k_y,n}^{\dagger}{\rm e}^{i\omega t-ik_xx-ik_yy}\right).
\end{equation} 
The operators $c_{k_x,k_y,n}$ and $c_{k_x,k_y,n}^{\dagger}$ obey the
usual commutation relations for bosons. It is interesting to check
that the above expansion is indeed consistent with the commutation
relations between the field and its conjugate momentum. For this
purpose, let us compute the following quantity
\begin{eqnarray}
\int {\rm d}x \int {\rm d}y \int _0^L{\rm d}z \, \Phi \, 
{\rm e}^{-ip_xx-ip_yy}
\sin \left(m\frac{\pi}{L}z\right)&=& 
\int \frac{{\rm d}k_x}{(2\pi)^{1/2}}
\int \frac{{\rm d}k_y}{(2\pi)^{1/2}}
\sum _{n=1}^{\infty}
\int {\rm d}x \int{\rm d}y
\sqrt{\frac{2}{L}}\frac{1}{\sqrt{2\omega}}
\biggl(c_{k_x,k_y,n}{\rm e}^{-i\omega t+ik_xx+ik_yy}
\nonumber \\ & & 
+c_{k_x,k_y,n}^{\dagger}{\rm e}^{i\omega t-ik_xx-ik_yy}\biggr)
{\rm e}^{-ip_xx-p_yy}
\int_0^L{\rm d}z
\sin\left(m\frac{\pi}{L}z\right)
\sin\left(n\frac{\pi}{L}z\right).
\end{eqnarray}
Using the fact that $\int_0^L{\rm d}z \sin(n\pi z/L)\sin(m\pi
z/L)=(L/2)\delta_{mn}$ (as can be checked directly) and $\int {\rm d}x
{\rm e}^{i(k_x-p_x)x}=(2\pi)\delta (k_x-p_x)$ one arrives at
\begin{eqnarray}
  \int {\rm d}x \int {\rm d}y \int _0^L{\rm d}z \, 
  \Phi(t,{\bm x})\, {\rm e}^{-ip_xx-ip_yy}
  \sin \left(m\frac{\pi}{L}z\right)&=& \frac{2\pi}{2}
  \sqrt{\frac{L}{\omega}}\biggl(c_{p_x,p_y,m}{\rm e}^{-i\omega t}
  +c_{-p_x,-p_y,m}^{\dagger}{\rm e}^{i\omega t}\biggr).
\end{eqnarray}
In the same manner, one can also evaluate the following integral
involving the conjugate momentum $\Pi(t,{\bm x})=\dot{\Phi}(t,{\bm
  x})$ (a dot means a derivative with respect to time)
\begin{eqnarray}
  \int {\rm d}x \int {\rm d}y \int _0^L{\rm d}z \, \Pi(t,{\bm x})\, 
  {\rm e}^{-ip_xx-ip_yy}
  \sin \left(m\frac{\pi}{L}z\right)&=& \frac{2\pi}{2}i\omega
  \sqrt{\frac{L}{\omega}}\biggl(-c_{p_x,p_y,m}{\rm e}^{-i\omega t}
  +c_{-p_x,-p_y,m}^{\dagger}{\rm e}^{i\omega t}\biggr)
\end{eqnarray}
From the two above expression, one can deduce a formula expressing the
creation and the annihilation operators in terms of the field operator
and its conjugate momentum. One obtains
\begin{eqnarray}
c_{p_x,p_y,m}&=&\sqrt{\frac{\omega }{L}}\frac{{\rm e}^{i\omega t}}{2\pi}
\left[\int {\rm d}x {\rm d}y 
\int _0^L{\rm d}z \, \Phi \, {\rm e}^{-ip_xx-ip_yy}
\sin \left(m\frac{\pi}{L}z\right)
+\frac{i}{\omega}\int {\rm d}x {\rm d}y 
\int _0^L{\rm d}z \, \Pi \, 
{\rm e}^{-ip_xx-ip_yy}
\sin \left(m\frac{\pi}{L}z\right)\right], \nonumber \\
c_{p_x,p_y,m}^{\dagger}&=&\sqrt{\frac{\omega }{L}}
\frac{{\rm e}^{-i\omega t}}{2\pi}
\left[\int {\rm d}x {\rm d}y 
\int _0^L{\rm d}z \, \Phi \, {\rm e}^{ip_xx+ip_yy}
\sin \left(m\frac{\pi}{L}z\right)
-\frac{i}{\omega}\int {\rm d}x {\rm d}y 
\int _0^L{\rm d}z \, \Pi \, 
{\rm e}^{ip_xx+ip_yy}
\sin \left(m\frac{\pi}{L}z\right)\right]\nonumber .
\end{eqnarray}
We are now in a position where we can compute the commutator of a
creation and an annihilation operators. It can be written as
\begin{eqnarray}
\left[c_{p_x,p_y,n},c_{q_x,q_y,m}^{\dagger}\right]&=& 
\frac{{\rm e}^{i(\omega_p-\omega_q)t}}{(2\pi)^2}
\frac{\sqrt{\omega_p\omega_q}}{L}
\int {\rm d}x{\rm d}y 
\int _0^L{\rm d}z \int {\rm d}\overline{x}{\rm d}\overline{y} 
\int _0^L{\rm d}\bar{z} 
\Biggl[\Phi (t,{\bm x})\, {\rm e}^{-ip_xx-ip_yy}
\sin \left(n\frac{\pi}{L}z\right)\nonumber \\ & & 
+\frac{i}{\omega_p}\Pi (t,{\bm x})\, 
{\rm e}^{-ip_xx-ip_yy}
\sin \left(n\frac{\pi}{L}z\right),
\Phi (t,\overline{{\bm x}})\, {\rm e}^{iq_x\overline{x}+iq_y\overline{y}}
\sin \left(m\frac{\pi}{L}\overline{z}\right)
\nonumber \\ & &
-\frac{i}{\omega_q}\Pi (t,\overline{{\bm x}})\, 
{\rm e}^{iq_x\overline{x}+iq_y\overline{y}}
\sin \left(m\frac{\pi}{L}\overline{z}\right)\Biggr].
\end{eqnarray}
Then, we make use of the canonical relation $[\Phi(t,{\bm
  x}),\Pi(t,{\bm y})]=i\delta ^3({\bm x}-{\bm y})$ in the previous
formula. This leads to the following complicated expression
\begin{eqnarray}
\left[c_{p_x,p_y,n},c_{q_x,q_y,m}^{\dagger}\right]&=& 
\frac{{\rm e}^{i(\omega_p-\omega_q)t}}{(2\pi)^2}
\frac{\sqrt{\omega_p\omega_q}}{L}
\int {\rm d}x{\rm d}y 
\int _0^L{\rm d}z \int {\rm d}\overline{x}{\rm d}\overline{y} 
\int _0^L{\rm d}\overline{z} 
\Biggl\{-\frac{i}{\omega _q}{\rm e}^{-ip_xx-ip_yy+iq_x\overline{x}+iq_y\overline{y}}
\sin \left(n\frac{\pi}{L}z\right)
\nonumber \\ & & \times
\sin \left(m\frac{\pi}{L}\overline{z}\right)
\left[\Phi(t,{\bm x}),\Pi(t,\bar{{\bm x}})\right]
+\frac{i}{\omega _p}{\rm e}^{-ip_xx-ip_yy+iq_x\overline{x}+iq_y\overline{y}}
\sin \left(n\frac{\pi}{L}z\right)
\sin \left(m\frac{\pi}{L}\overline{z}\right)
\nonumber \\ & & \times
\left[\Pi(t,{\bm x}),\Phi(t,\overline{{\bm x}})\right]\Biggr\}
\\
&=& \frac{{\rm e}^{i(\omega_p-\omega_q)t}}{(2\pi)^2}
\frac{\sqrt{\omega_p\omega_q}}{L}\left(\frac{1}{\omega _q}+\frac{1}{\omega _p}
\right)\int {\rm d}x{\rm d}y\int _0^L{\rm d}z
{\rm e}^{-i(p_x-q_x)x-i(p_y-q_y)y}
\sin \left(n\frac{\pi}{L}z\right)
\sin \left(m\frac{\pi}{L}z\right)\nonumber \\
&=& \frac{\omega}{L}\frac{2}{\omega}\delta(p_x-q_x)\delta(p_y-q_y)
\frac{L}{2}\delta _{mn}=\delta(p_x-q_x)\delta(p_y-q_y)\delta _{mn}. 
\end{eqnarray}
\end{widetext}
We have thus reached our goal, namely we have shown that the canonical
commutation relation between the field and its conjugate momentum
implies the usual commutation relation between the creation and the
annihilation operator.
 
We are now in a position where we can evaluate the energy density of
the field in the Casimir cavity. It is of course given by the same
expression as in the free case, see Eq.~(\ref{eq:hamiltonscalr}). This
means that one should first calculate the three terms that participate
in the expression of $T_{00}$, see Eqs.~(\ref{eq:phidot2}),
(\ref{eq:partialphi2}) and~(\ref{eq:phi2}). Let us start with $\langle
0\vert \dot{\Phi}^2\vert 0\rangle$. Using the Fourier expansion of the
field and the commutation relations established above, one obtains
\begin{eqnarray}
\langle 0\vert \dot{\Phi}^2\vert 0\rangle&=&
\int \frac{{\rm d}k_x}{2\pi}
\int \frac{{\rm d}k_y}{2\pi}
\sum _{n=1}^{\infty}
\frac{2}{L} \frac{\omega^2}{2\omega }\nonumber 
\\ & & \times \sin ^2\left(n\frac{\pi}{L}z\right),
\end{eqnarray}
and, if we integrate over the cavity volume
\begin{equation}
\int _{\rm Cavity}{\rm d}{\bm x}\, \langle 0\vert \dot{\Phi}^2\vert 0\rangle
=\frac{D^2}{2}\int \frac{{\rm d}k_x}{2\pi}
\int \frac{{\rm d}k_y}{2\pi}
\sum _{n=1}^{\infty}\omega,
\end{equation}
where $D^2$ is the area of the plates (of course, the two parallel
plates have the same surface). In the same manner, one finds
\begin{align}
\int _{\rm Cavity}{\rm d}{\bm x}\, &\langle 0\vert \delta ^{ij}
\partial _i\Phi\partial _j\Phi\vert 0\rangle
=\frac{D^2}{2}\int \frac{{\rm d}k_x}{2\pi}
\int \frac{{\rm d}k_y}{2\pi}\nonumber \\ &\times 
\sum _{n=1}^{\infty}\frac{1}{2\omega}
\left(k_x^2+k_y^2+n^2\frac{\pi^2}{L^2}\right).
\end{align}
Finally, the last relevant term reads
\begin{equation}
\int _{\rm Cavity}{\rm d}{\bm x}\, \langle 0\vert \Phi^2\vert 0\rangle
=\frac{D^2}{2}\int \frac{{\rm d}k_x}{2\pi}
\int \frac{{\rm d}k_y}{2\pi}
\sum _{n=1}^{\infty}\frac{1}{2\omega}.
\end{equation}
Putting everything together and using the
definition~(\ref{eq:hamiltonscalr}), the expression of the energy of
the field within the empty cavity takes the form
\begin{eqnarray}
& & \int _{\rm Cavity}{\rm d}{\bm x}\, \langle 0\vert T_{00}\vert 0\rangle
\\
&=& \left(\frac{D}{2\pi}\right)^2\int {\rm d}k_x
\int {\rm d}k_y\sum _{n=1}^{\infty}\frac{\omega}{2} \\ 
&=&
\left(\frac{D}{2\pi}\right)^2\int {\rm d}k_{x}
\int {\rm d}k_y\sum _{n=1}^{\infty}\frac{1}{2}
\sqrt{k_x^2+k_y^2+n^2\frac{\pi^2}{L^2}+m^2}.\nonumber \\
\end{eqnarray}
Of course, this expression has exactly the expected form. Roughly
speaking it can be written as ``$\sum \frac12\omega$'', see
Eq.~(\ref{eq:defomegacasimir}), the only difference being that, along
the $z$-axis, the sum takes into account the fact that the momentum is
discrete. Another common point with our previous considerations is
that the above number is actually divergent and must be
regularized. In the following, we assume $m=0$ for simplicity and use
dimensional regularization. Therefore, the expression of the energy
can now be written as
\begin{eqnarray}
  E &=& \left(\frac{D}{2\pi}\right)^d\int {\rm d}^{d}{\bm k}_{\perp}
  \sum _{n=1}^{\infty}\frac{1}{2}\sqrt{k_{\perp}^2+n^2\frac{\pi^2}{L^2}}
  \\
  &=& \left(\frac{D}{2\pi}\right)^d\int {\rm d}k_{\perp}k_{\perp}^{d-1}
  {\rm d}^{d-1}\Omega 
  \sum _{n=1}^{\infty}\frac{1}{2}\sqrt{k_{\perp}^2+n^2\frac{\pi^2}{L^2}}
  \nonumber \\ 
  &=& \left(\frac{D}{2\pi}\right)^d\frac{2\pi^{d/2}}{\Gamma (d/2)}
  \int {\rm d}k_{\perp}k_{\perp}^{d-1}
  \sum _{n=1}^{\infty}\frac{1}{2}\sqrt{k_{\perp}^2+n^2\frac{\pi^2}{L^2}}
  \nonumber \\
\end{eqnarray}
where $d$ is the ``dimension'' of the plates (\ie $d=2$ in the real
world) and $k_{\perp}$ denotes the wave-vector living in the
sub-manifold defined by the plates. Then using the change of variables
$y=Lk_{\perp}/(n\pi)$ and the following
definition~\cite{Abramovitz:1970aa,Gradshteyn:1965aa} of the Euler's
integral of first kind
\begin{equation}
B(x,y)=2\int_0^{\infty}\frac{t^{2x-1}}{(1+t^2)^{x+y}}{\rm d}t
=\frac{\Gamma(x)\Gamma(y)}{\Gamma(x+y)},
\end{equation}
the energy can be re-expressed as
\begin{equation}
\label{eq:sumecasimir}
E=\left(\frac{D}{2\pi}\right)^d\frac{\pi^{d/2}}{2}
\left(\frac{\pi}{L}\right)^{d+1}
\frac{\Gamma(-d/2-1/2)}{\Gamma(-1/2)}\sum _{n=1}^{\infty}n^{d+1}.
\end{equation}
The above formula can be expressed in term of the Riemann zeta
function defined by the following expression
\begin{equation}
\zeta(s)\equiv \sum_{n=1}^{\infty}n^{-s}.
\end{equation}
Moreover, upon using the two following equations 
\begin{eqnarray}
& & \zeta(-1-d)=2^{-1-d}\pi^{-2-d}\sin\left[\frac{\pi}{2}(-1-d)\right]
\nonumber \\ & & \times
\Gamma(2+d)\zeta(2+d), \\
& & \Gamma\left(-\frac{d}{2}-\frac12\right)\sin\left[\frac{\pi}{2}(-1-d)\right]
=\frac{\pi}{\Gamma(d/2+3/2)}, 
\nonumber \\
\end{eqnarray}
one can re-write the energy as
\begin{equation}
\frac{E}{D^d}=-\frac{1}{L^{d+1}}\frac{\Gamma(1+d/2)}{2^{d+2}\pi^{d/2+1}}
\zeta(2+d).
\end{equation}
The Casimir force arises because of a shift in the vacuum energy when
the plates are present compared to the situation where they are
absent. So the relevant quantity is in fact $\Delta E/D^d\equiv
(E-E_0)/D^d$ where $E_0/D^d$ is the above quantity in the limit
$L\rightarrow +\infty$. As a consequence, for $d=2$, one arrives at
\begin{equation}
\label{eq:ded2casimir}
\frac{\Delta E}{D^2}=-\frac{\pi^2}{1440L^{3}},
\end{equation}
where we have used $\zeta(4)=\pi^4/90$. Therefore, the corresponding
force is given
by~\cite{Plunien:1986ca,Milton:2004ya,Milton:2000av,Milton:2010qr}
\begin{equation}
F=-\frac{\partial}{\partial L}\left(\frac{\Delta E}{D^2}\right)
=-\frac{\pi^2}{480L^4},
\end{equation}
which is one half the result obtained in the case of the
electromagnetic field since we have only one state of polarization
instead of two. We notice that the force is attractive (this could
change in the case of more complicated geometries).

\par

It is also interesting to check that the final result is independent
from the regularization scheme used to obtain it. In order to test
this idea in the simplest situation (\ie where the calculations are
easy), let us consider again the energy of a Casimir cavity but, this
time, with $d=0$. In this case, our zeta-function regularized result
reads
\begin{equation}
\frac{E}{D^0}=-\frac{1}{L}\frac{\Gamma(1)\zeta(2)}{4\pi}=-\frac{\pi}{24L},
\end{equation}
since $\Gamma(1)=1$ and $\zeta(2)=\pi^2/6$. One the other hand, we can
return to the expression of $E$ in Eq.~(\ref{eq:sumecasimir}). In the
case $d=0$, the energy density $\rho(L)=E/L$ reads
\begin{equation}
\rho(L)=\frac{\pi}{2L^2}\sum _{n=1}^{\infty}n.
\end{equation}
In order to regularize this divergent quantity, we now proceed
differently and introduce an exponential cut-off such that
(previously, we argued that introducing a cut-off was not a good
method because this breaks Lorentz invariance. Here, of course, this
objection is no longer valid as the presence of the two plates along
the $z$-axis obviously breaks translation invariance; as a consequence
working with a cut-off seems legitimate in the present context)
\begin{equation}
\rho (L,\alpha)=\frac{\pi}{2L^2}\sum _{n=1}^{\infty}n\, {\rm e}^{-n\alpha/L}.
\end{equation}
The sum is easy to perform and one obtains
\begin{eqnarray}
\rho(L,\alpha) &=& \frac{\pi}{8L^2}\sinh ^{-2}\left(\frac{\alpha}{2L}\right)
\\
&\simeq & \frac{\pi}{2\alpha ^2}-\frac{\pi}{24L^2}+\cdots
\end{eqnarray}
Now, the regularized Casimir energy is the shift energy in the vacuum
after having sent the cut-off to zero; in other words
\begin{eqnarray}
E &=& L\lim_{\alpha \rightarrow 0}\left[\rho(L,\alpha)-\lim_{L\rightarrow \infty}
\rho(L,\alpha)\right]\\
&=& -\frac{\pi}{24L},
\end{eqnarray}
that is to say exactly the same expression obtained with another
regularization scheme. It is reassuring to notice that the final result
is independent of the regularization method used.

\subsubsection{The Casimir Stress-Energy Tensor}
\label{subsubsec:stresscasimir}

In this section, we aim at calculating the stress energy tensor of a
scalar field in a Casimir
cavity~\cite{Deutsch:1978sc,Plunien:1986ca,Milton:2004ya,Milton:2000av,Milton:2010qr}. Our
goal is of course to compare this tensor with the vacuum stress-energy
tensor $T_{\mu \nu}=-\rho_{_{\rm vac}}g_{\mu \nu}$. Here, we use a
method based on the calculations of the Green function. This is yet
another method of calculation of the Casimir effect, different from
the two ones exposed before and, therefore, it will be interesting to
compare this approach with the treatments used previously.

\par

Let us start with the calculation of the Green function. It obeys the
following equation
\begin{equation}
\label{eq:greencasimir}
  -\eta ^{\mu\nu}\partial_{\mu}\partial_{\nu}G\left(x,\overline{x}\right)
  =\delta^{(4)}\left(x^{\mu}-\overline{x}^{\mu}\right).
\end{equation}
It is interesting to compare the above formula with
Eq.~(\ref{eq:defgreen}). It is clear that this is the same equation
but with $g_{\mu \nu}=\eta_{\mu \nu}$ (flat space-time) and $m=0$
(mass-less field). In order to solve this equation, it is convenient to
Fourier transform $G(x,\overline{x})$ in frequency and transverse
momentum. Explicitly, one writes
\begin{eqnarray}
\label{eq:defgreencasimir}
G(x,\overline{x}) &=& \int \frac{{\rm d}\omega}{2\pi}
\frac{{\rm d}\bm{k}_{\perp}}{(2\pi)^2}
{\rm e}^{-i\omega\left(t-\overline{t}\right)
  +ik_x(x-\overline{x})+ik_y(y-\overline{y})}
\nonumber \\ & & \times
g(z,\overline{z};\omega,k_x,k_y).
\end{eqnarray}
Inserting this expression into the Green
equation~(\ref{eq:greencasimir}), one obtains
\begin{equation}
\label{eq:greenfourier}
-\left(\frac{{\rm d}^2}{{\rm d}z^2}+\lambda^2\right)g(z,\overline{z})
=\delta(z-\overline{z}),
\end{equation}
where we have defined
\begin{equation}
\label{eq:deflambda}
\lambda^2\equiv \omega^2-k_{\perp}^2=\omega^2-k_x^2-k_y^2.
\end{equation}
Our goal is now to determine explicitly the Green function. Let us
start with the solution within the plates, \ie
$0<z,\overline{z}<L$. When $z\neq \overline{z}$, the solution can be
written
\begin{equation}
\label{eq:solgreencos}
g(z,\overline{z})=A(\overline{z})\cos\left(\lambda z\right)
+B(\overline{z})\sin\left(\lambda z\right),
\end{equation}
where $A(\overline{z})$ and $B(\overline{z})$ are two unknown function
of $\overline{z}$. We must now take into account the boundary
conditions. If $z<\overline{z}$, the condition $g(z=0,\overline{z})=0$
implies that $A=0$ and therefore $g=B\sin(\lambda z)$ (for
$z<\overline{z}$). On the other hand, if $z>\overline{z}$, the
condition $g(z=L,\overline{z})=0$ means that
\begin{equation}
g=A\cos(\lambda z)-A\frac{\cos(\lambda L)}{\sin (\lambda L)}
\sin (\lambda z), \quad z>\overline{z}.
\end{equation} 
We have two unknown quantities and, therefore, we need two
equations. One is provided by writing the continuity of the Green
function when $z=\overline{z}$. The other is obtained by integrating
Eq.~(\ref{eq:greenfourier}) from $\overline{z}-\epsilon$ to
$\overline{z}+\epsilon$. This gives
\begin{equation}
  -\int_{\overline{z}-\epsilon}^{\overline{z}+\epsilon}
  \frac{{\rm d}^2g}{{\rm d}z^2}{\rm d}z-\lambda^2
  \int_{\overline{z}-\epsilon}^{\overline{z}+\epsilon}
  g{\rm d}z=\int_{\overline{z}-\epsilon}^{\overline{z}+\epsilon}
  \delta (z-\overline{z}){\rm d}z,
\end{equation}
which leads to
\begin{equation}
-\left[\frac{{\rm d}g}{{\rm d}z}(\overline{z}+\epsilon)
-\frac{{\rm d}g}{{\rm d}z}(\overline{z}-\epsilon)\right]=1.
\end{equation}
Therefore, taking the limit $\epsilon \rightarrow 0$, the solutions of
these two equations (\ie the one expressing the continuity of the
Green function and the one we have just derived) read
\begin{eqnarray}
A(\overline{z}) &=& \frac{1}{\lambda}\sin(\lambda \overline{z}), \\
B(\overline{z}) &=& \frac{1}{\lambda}\cos(\lambda \overline{z})
-\frac{\cos(\lambda L)}{\sin(\lambda L)}A(\overline{z}).
\end{eqnarray}
As a consequence, the Green function between the plates can be
expressed as~\cite{Milton:2010qr}
\begin{equation}
\label{eq:greeninside}
g(z,\overline{z})=
\begin{cases}
\displaystyle
\frac{\sin(\lambda z)\sin(\lambda L-\lambda \overline{z})}{\lambda 
\sin(\lambda L)}, & z<\overline{z}, \\ \\
\displaystyle
\frac{\sin(\lambda \overline{z})\sin(\lambda L-\lambda z)}{\lambda 
\sin(\lambda L)}, & z>\overline{z}.
\end{cases}
\end{equation}
The same method can be used to determine the Green function outside
the cavity. Let us apply it for $z$ and $\overline{z}$ larger than
$L$, \ie on the right hand side of the cavity. In this case, it is
more convenient to re-write the general
solution~(\ref{eq:solgreencos}) as
\begin{equation}
  g(z,\overline{z})=C(\overline{z}){\rm e}^{i\lambda z}
  +D(\overline{z}){\rm e}^{-i\lambda z}.
\end{equation}
When $z>\overline{z}$, we require only one branch to be present which
amounts to take $D=0$. For $z<\overline{z}$, the boundary condition is
still $g(L,\overline{z})=0$. Then, straightforward manipulations lead
to~\cite{Milton:2010qr}
\begin{equation}
\label{eq:greenoutside}
g(z,\overline{z})=
\begin{cases}
\displaystyle
\frac{1}{\lambda}{\rm e}^{i\lambda(\overline{z}-L)}
\sin(\lambda z-\lambda L), & z<\overline{z}, \\ \\
\displaystyle
\frac{1}{\lambda}{\rm e}^{i\lambda(z-L)}
\sin(\lambda \overline{z}-\lambda L), & z>\overline{z},
\end{cases}
\end{equation}
which completes our determination of the Green function. The
calculation of the Green function outside the cavity, but on the left
hand side, proceeds exactly in the same way. The corresponding
expression is obtained from the one above with $L=0$, reflecting the
fact that the boundary condition is now $g(0,\overline{z})=0$.

\par

Endowed with the Green function of the problem, one can now turn to
our main goal, namely the calculation of the stress-energy tensor. We
have already seen that, see also Eq.~(\ref{eq:propagator}), 
\begin{equation}
\label{eq:greencasimirT}
G(\bm{r},\overline{\bm{r}})=i\left \langle 0 \left 
\vert T\left[\Phi(\bm{r})\Phi(\overline{\bm{r}})\right]
\right \vert 0 \right \rangle.
\end{equation}
But we know that the $T_{00}$ component of the stress energy tensor
can be expressed as, see Eq.~(\ref{eq:tmunuscalarfield}),
\begin{equation}
T_{00}=\frac12\left[(\partial_0\Phi)^2+(\partial_x\Phi)^2
+(\partial_y\Phi)^2+(\partial_z\Phi)^2\right].
\end{equation}
Therefore, if we apply the following operator to the Green function in
Eq.~(\ref{eq:greencasimirT})
\begin{equation}
\frac12\left[\partial_0\partial_{\overline{0}}+\partial_x
\partial_{\overline{x}}
+\partial_y\partial_{\overline{y}}+\partial_z\partial_{\overline z}\right],
\end{equation}
where $\partial_{\overline{\mu}}$ means a derivative with respect to
$\overline{x}^{\mu}$, to $G(\bm{r},\overline{\bm{r}})$ and, then, take
the space-time points to be the same $\bm{r}=\overline{\bm{r}}$, one
should obtain an expression for $\langle T_{00}\rangle $. Using
Eqs.~(\ref{eq:defgreencasimir}) and~(\ref{eq:deflambda}), the result
reads
\begin{eqnarray}
\langle T_{00}\rangle
&=&\frac{1}{2i}\int \frac{{\rm d}\omega}{2\pi}
\frac{{\rm d}k_x{\rm d}k_y}{(2\pi)^2}
\left(\omega^2+k_x^2+k_y^2+\partial_z\partial_{\overline{z}}\right)
\nonumber \\ & & \times
g(z,\overline{z})\biggl \vert _{z=\overline{z}}.
\end{eqnarray}
Then, we use the expression of the Green function inside the
plates~(\ref{eq:greeninside}) and one arrives at
\begin{widetext}
\begin{eqnarray}
\langle T_{00}\rangle
&=&\frac{1}{2i}\int \frac{{\rm d}\omega}{2\pi}
\frac{{\rm d}k_x{\rm d}k_y}{(2\pi)^2}
\frac{1}{\lambda \sin\left(\lambda L\right)}
\biggl[(\omega^2+k_x^2+k_y^2)
\sin\left(\lambda z\right)
\sin \left(\lambda L-\lambda L\right)
-\lambda ^2\cos\left(\lambda z\right)\cos\left(\lambda L
-\lambda z\right)\biggr] \\
&=& 
-\frac{1}{2i}\int \frac{{\rm d}\omega}{2\pi}
\frac{{\rm d}k_x{\rm d}k_y}{(2\pi)^2}
\frac{1}{\lambda \sin\left(\lambda L\right)}
\left[\omega^2\cos\left(\lambda L\right)
-k^2\cos\left(2\lambda z
-\lambda L\right)\right]. 
\end{eqnarray}
In order to evaluate this integral, we perform the two Wick rotations
$\omega \rightarrow i\zeta $ and $\lambda \rightarrow i\kappa $. This leads to 
\begin{equation}
\langle T_{00}\rangle=
-\frac{1}{2}\int _0^{\infty}\frac{{\rm d}\zeta}{2\pi}
\int \frac{{\rm d}k_x{\rm d}k_y}{(2\pi)^2}
\frac{1}{\kappa \sinh\left(\kappa L\right)}
\left[\zeta^2\cosh\left(\kappa L\right)
+k^2\cosh\left(2\kappa z
-\kappa L\right)\right]. 
\end{equation}
Then, we introduce polar coordinates in the plane $(\zeta,k)$, \ie
$\zeta=\kappa \cos\theta $ and $k=\kappa \sin \theta $. This gives
\begin{eqnarray}
\langle T_{00}\rangle &=&
-\frac{1}{4\pi^2}\int _0^{\infty}{\rm d}\kappa \kappa
\int_0^{\pi/2}{\rm d}\theta \kappa^2
\frac{\sin \theta }{\sinh\left(\kappa L\right)}
\left[\cos^2\theta \cosh\left(\kappa L\right)
+\sin^2\theta \cosh\left(2\kappa z
-\kappa L\right)\right] \\
&=& -\frac{1}{12\pi^2}
\int _0^{\infty}{\rm d}\kappa \kappa^3
\left[\frac{\cosh(\kappa L)}{\sinh(\kappa L)}+
2\frac{\cosh(2\kappa z-\kappa L)}{\sinh(\kappa L)}\right].
\end{eqnarray}
\end{widetext}
Finally, writing that
\begin{equation}
\frac{\cosh(\kappa L)}{\sinh(\kappa L)}
=1+\frac{2}{{\rm e}^{2\kappa L}-1},
\end{equation}
the expression of $\langle T_{00}\rangle$ can be re-expressed as
\begin{eqnarray}
  \langle T_{00}\rangle &=&
  -\frac{1}{6\pi^2}
  \int _0^{\infty}{\rm d}\kappa \kappa^3
  \biggl[
  \frac{1}{{\rm e}^{2\kappa L}-1}
  +\frac12
  \nonumber \\ & & 
  +\frac{{\rm e}^{2\kappa z}+{\rm e}^{2\kappa (L-z)}}{{\rm e}^{2\kappa L}-1}
  \biggr].
\end{eqnarray}
The last step of the calculation consists in evaluating the last term
[we call it $g(z)$ in what follows; not to be confused with the Green
function] of the above equation. One has
\begin{eqnarray}
g(z)&=&-\frac{1}{6\pi^2}\frac{1}{16L^4}
\int _0^{\infty}{\rm d}y \, y^3
\frac{{\rm e}^{y/L}+{\rm e}^{y(1-z/L)}}{{\rm e}^{y}-1}
\\
&=& 
-\frac{1}{6\pi^2}\frac{1}{16L^4}
\sum_{n=1}^{\infty}
\int _0^{\infty}{\rm d}y \, y^3\biggl[
{\rm e}^{y(z/L-n)}
\nonumber \\ & &
+{\rm e}^{y(1-n-z/L)}\biggr],
\end{eqnarray}
where we have expanded the denominator in a geometric series. The two
above integrals can be expressed in terms of the Hurwitz zeta function
defined by~\cite{Abramovitz:1970aa,Gradshteyn:1965aa}
\begin{equation}
\zeta(s,a)\equiv \sum_{m=0}^{\infty}\frac{1}{\left(m+a\right)^s},
\end{equation}
and this leads to
\begin{equation}
g(z)=-\frac{1}{16\pi^2L^4}\left[\zeta\left(4,\frac{z}{L}\right)
+\zeta\left(4,1-\frac{z}{L}\right)\right].
\end{equation}
If we use the fact that $\int_0^{+\infty} \kappa^3/({\rm e}^{2\kappa
  L}-1){\rm d}\kappa=\pi^4/(240L^4)$, one arrives at our final
expression for the time-time component of the stress energy tensor,
namely
\begin{eqnarray}
\langle T_{00}\rangle &=& -\frac{\pi^2}{1440L^4}
-\frac{1}{6\pi^2}\int_0^{\infty}{\rm d}\kappa \frac{\kappa ^3}{2}
+g(z).
\end{eqnarray}
Of course we notice that the first term in the above equation is
similar to Eq.~(\ref{eq:ded2casimir}).

\par

Let us now evaluate $\langle T_{00}\rangle$ outside the cavity. The
calculation proceeds along the same lines but we now need to use the
Green function given by Eq.~(\ref{eq:greenoutside}). Straightforward
manipulations lead to
\begin{eqnarray}
\langle T_{00}\rangle &=& 
-\frac{1}{6\pi^2}\int_0^{\infty}{\rm d}\kappa \frac{\kappa ^3}{2}
-\frac{1}{16\pi^2}\frac{1}{\left(z-L\right)^4}.
\end{eqnarray}
We obtain the same structure as inside the cavity, namely a divergent
term and a surface divergence term, \ie a term which is divergent only
on the plate. As explained in Ref.~\cite{Milton:2010qr}, this type of
terms can be removed by restoring conformal invariance.

\par

Let us now calculate the quantity $\langle T_{zz}\rangle$ inside the
cavity. Clearly, following the previous considerations, it can be
written as
\begin{eqnarray}
  \langle T_{zz}\rangle
  &=&\frac{1}{2i}\int \frac{{\rm d}\omega}{2\pi}
  \frac{{\rm d}k_x{\rm d}k_y}{(2\pi)^2}
  \left(\partial_t\partial_{\overline{t}}-\partial_x\partial_{\overline{x}}
    -\partial_y\partial_{\overline{y}}+\partial_z\partial_{\overline{z}}\right)
  \nonumber \\ & & \times
  g(z,\overline{z})\biggl \vert _{z=\overline{z}}.
\end{eqnarray}
Following exactly the same procedure as before, we arrive at
\begin{eqnarray}
  \langle T_{zz}\rangle &=& -3\frac{\pi^2}{1440L^4}
  -\frac{3}{6\pi^2}\int_0^{\infty}{\rm d}\kappa \frac{\kappa ^3}{2}.
\end{eqnarray}
Let us now calculate $\langle T_{zz}\rangle$ outside the plates. As we
did for the time time component, we must now use the Green
function~(\ref{eq:greenoutside}). This gives
\begin{eqnarray}
\langle T_{zz}\rangle &=& \frac{1}{2i}\int \frac{{\rm d}\omega}{2\pi}
\frac{{\rm }k_x{\rm d}k_y}{(2\pi)^2}
\frac{{\rm e}^{i\lambda(z-L)}}{\lambda}\lambda^2
\biggl[\sin \left(\lambda z-\lambda L\right)
\nonumber \\ & &
+i\cos \left(\lambda z-\lambda L\right)\biggr]\\
&=& -\frac{3}{6\pi^2}\int _0^{\infty}\frac{\kappa ^3}{2}{\rm d}\kappa.
\end{eqnarray}
Of course, this quantity is divergent.

\par

Finally, in order to complete the determination of the stress-energy
tensor, one must calculate $\langle T_{xx}\rangle =\langle
T_{yy}\rangle $. It can be expressed as
\begin{eqnarray}
\label{eq:deftxx}
\langle T_{xx}\rangle
&=&\frac{1}{2i}\int \frac{{\rm d}\omega}{2\pi}
\frac{{\rm d}k_x{\rm d}k_y}{(2\pi)^2}
\left(\partial_t\partial_{\overline{t}}+\partial_x\partial_{\overline{x}}
-\partial_y\partial_{\overline{y}}-\partial_z\partial_{\overline{z}}\right)
\nonumber \\ & & \times
g(z,\overline{z})\biggl \vert _{z=\overline{z}}
\\ 
&=& \frac{1}{2i}\int \frac{{\rm d}\omega}{2\pi}
\frac{{\rm d}k_x{\rm d}k_y}{(2\pi)^2}\frac{1}{\lambda \sin(\lambda L)}
\biggl[\left(\omega^2+k_x^2-k_y^2\right)
\nonumber \\ & & \times
\sin(\lambda z)\sin(\lambda L-
\lambda z)
\nonumber \\ & & 
+\lambda^2\cos(\lambda z)
\cos(\lambda L-\lambda z)\biggr].
\end{eqnarray}
A first step consists in dealing with the terms proportional to
$k_x^2-k_y^2$. In fact these terms are of the form $\int {\rm
  d}\bm{k}_{\perp} (k_x^2-k_y^2){\cal F}(k_{\perp})$, where ${\cal
  F}(k_{\perp})$ represents the other terms in the integral which are
function of $k_x^2+k_y^2$ only. Therefore, one can evaluate them by
going to polar coordinates in the plane $(k_x,k_y)$. This leads to an
expression of the form $\int {\rm d}k_{\perp} {\rm d}\psi
k_{\perp}^2(\cos^2\psi-\sin ^2\psi){\cal F}(k_{\perp})=0$, which shows
that the corresponding contribution vanishes. Then, following the same
steps as before, one reduces the above expression to
\begin{eqnarray}
\langle T_{xx}\rangle&=&
\frac{1}{12\pi^2}
\int _0^{\infty}{\rm d}\kappa \frac{\kappa^3}{\sinh(\kappa L)}
\biggl[\cosh(2\kappa z-\kappa L)+
\nonumber \\ & & +
2\cosh(\kappa z)\cosh(\kappa L-\kappa z)\biggr]
=-\langle T_{00}\rangle,
\end{eqnarray}
where we have used that $2\cosh(\kappa z)\cosh(\kappa L-\kappa
z)=\cosh(\kappa L)+\cosh(2\kappa z-\kappa L)$. Of course, as already
mentioned, this also shows that $\langle T_{yy}\rangle=-\langle
T_{00}\rangle$. All the other components being zero, this completes
our calculation of the stress energy in the cavity. Only the
calculations of $\langle T_{xx}\rangle=\langle T_{yy}\rangle$ outside
the plates remains to be done and we now turn to this
question. Considering again Eq.~(\ref{eq:deftxx}), we have
\begin{eqnarray}
  \langle T_{xx}\rangle&=&
  \frac{1}{2i}
  \int \frac{{\rm d}\omega}{2\pi}
  \frac{{\rm d}k_x{\rm d}k_y}{(2\pi)^2}
  \frac{{\rm e}^{i\lambda(z-L)}}{\lambda}
  \biggl[\omega^2\sin (\lambda z-\lambda L)
  \nonumber \\ & & 
  -i\lambda^2\cos(\lambda z-\lambda L)
  \biggr]=-\langle T_{00}\rangle.
\end{eqnarray}
We are now in a position where we can write the full stress energy
tensor. Inside the cavity, it takes the form
\begin{eqnarray}
\langle T_{\mu \nu}\rangle _{_{\rm inside}}&=&
\left(\rho_{_{\rm vac}}+\rho_{_{\rm Casimir}}\right)
\begin{pmatrix}
1 & 0 & 0 & 0 \\
0 & -1 & 0 & 0 \\
0 & 0 & -1 & 0 \\
0 & 0 & 0 & 3 
\end{pmatrix}
\nonumber \\ & &
+g(z)
\begin{pmatrix}
1 & 0 & 0 & 0 \\
0 & -1 & 0 & 0 \\
0 & 0 & -1 & 0 \\
0 & 0 & 0 & 0
\end{pmatrix},
\end{eqnarray}
where we have defined the two quantities $\rho_{_{\rm vac}}$ and
$\rho_{_{\rm Casimir}}$ by
\begin{eqnarray}
\rho_{_{\rm vac}}=-\frac{1}{6\pi^2}\int _0^{\infty}{\rm d}\kappa 
\frac{\kappa ^3}{2}, \quad \rho_{_{\rm Casimir}}=-\frac{\pi^2}{1440 L^4}.
\end{eqnarray}
Outside the cavity (more precisely on the right hand side), the
results derived above imply that
\begin{align}
\langle T_{\mu \nu}\rangle _{_{\rm outside}}&=
\rho_{_{\rm vac}}
\begin{pmatrix}
1 & 0 & 0 & 0 \\
0 & -1 & 0 & 0 \\
0 & 0 & -1 & 0 \\
0 & 0 & 0 & 3 
\end{pmatrix}
\nonumber \\  
-& \frac{1}{16\pi^2(z-L)^4}
\begin{pmatrix}
1 & 0 & 0 & 0 \\
0 & -1 & 0 & 0 \\
0 & 0 & -1 & 0 \\
0 & 0 & 0 & 0
\end{pmatrix},
\end{align}
As discussed in Ref.~\cite{Milton:2010qr}, the terms proportional to
$g(z)$ and $(z-L)^{-4}$ are surface divergent terms. They can be
removed in the case where conformal invariance is re-established and,
for that reason, are not present in electromagnetism, \ie when one
considers an electric field inside the cavity rather than a scalar
field. In this case, we see that the difference between the
stress-energy tensors inside and outside the cavity reads
\begin{equation}
\label{eq:tmunucasimir}
\Delta\langle T_{\mu \nu}\rangle=
\rho_{_{\rm Casimir}}
\begin{pmatrix}
1 & 0 & 0 & 0 \\
0 & -1 & 0 & 0 \\
0 & 0 & -1 & 0 \\
0 & 0 & 0 & 3 
\end{pmatrix},
\end{equation}
which is a finite result. The infinite part has been removed by
subtraction. Finally, the stress energy tensor can be written in a
covariant form, namely~\cite{Deutsch:1978sc}
\begin{equation}
\label{eq:tmunucasimirfinal}
\Delta\langle T_{\mu \nu}\rangle=
\rho_{_{\rm Casimir}}\left(4\hat{z}^{\mu}\hat{z}^{\nu}-g^{\mu \nu}\right),
\end{equation}
where $\hat{z}^{\mu}$ in the unit vector in the $z$ direction.

\par

At this point, several comments are in order. Firstly, we notice in
Eqs.~(\ref{eq:tmunucasimir}) and~(\ref{eq:tmunucasimirfinal}) that the
energy density (\ie the time-time component) is similar to the
expressions already derived in Sec.~\ref{subsubsec:casimirforce} by
other methods. This confirms that this calculation is independent of
the regularization scheme used to obtain the result. Secondly, it is
clear that the stress-energy tensor~(\ref{eq:tmunucasimirfinal}) is
not similar to the cosmological constant stress-energy tensor $T_{\mu
  \nu}=-\rho_{_{\rm vac}}g_{\mu \nu}$. This illustrates the difference
between the Casimir case and the cosmological constant case. In the
latter case, one measures the absolute value of the vacuum energy
while, in the former case, one is only sensitive to the difference of
vacuum energy from one side of the plates to the other.

\par

Thirdly, the Casimir Force has been observed in the laboratory and
this is usually taken as another evidence in favor of the existence of
the zero-point fluctuations. After all, since we observe the Casimir
force and since this force is due to a change in the structure of the
vacuum that is perturbed by the presence of the plates, the fact is
that the vacuum must exist as a real physical phenomenon and is not an
artifact of the quantum field theory formalism. In addition, this also
validates the methods of regularization used to calculate $\rho_{_{\rm
    vac}}$. In the context of the Casimir effect, these methods seem
to work very well and to lead to predictions that are in agreement
with the experiments. Therefore, why would the very same techniques,
used in a similar context, fail to regularize the cosmological
constant?

\par

Fourthly, the claim that the calculation and the observation of the
Casimir force are evidences in favor of the reality of vacuum
fluctuations has been challenged in
Refs.~\cite{Milonni:1992zz,Jaffe:2005vp}. It is indeed possible to
derive the expression of the Casimir force without referring at all to
zero point fluctuations, by means of the ``source theory'' approach to
quantum field theory. However, to our knowledge, there is no attempt
to do the same for the Lamb shift (but this would certainly be an
interesting exercise). Therefore, it seems that zero-points
fluctuations are nevertheless necessary to quantum field theory even
if the above remark should lead us to tone down the claim that they
have been seen in the laboratory. Let us also notice that, very
recently, another situation where the vacuum fluctuations can lead to
an observable effect has been studied in
Ref.~\cite{Chernodub:2012ry}. This article investigates whether the
quantum zero point fluctuations could cause the rotation of a small
nano device. This ``rotational vacuum effect'' has in fact a similar
origin than the Casimir effect.

\par

Finally, it should be clear that, if the Lamb shift and the Casimir
effect seem to indicate that the vacuum fluctuations are real, they do
not say anything with regards to their gravitational properties. In
other words, they do not say how the vacuum fluctuations weigh (it is
worth noticing here that the gravitational properties of a Casimir
cavity have been studied in
Refs.~\cite{Calloni:2001hh,Bimonte:2007zt,Fulling:2007xa,Milton:2007hd,Milton:2008ks,Caldwell:2002im,Sorge:2005ed}). It
is clear that there exists the possibility that the zero-point
fluctuations do exist but have non-standard gravitational properties,
\ie that they abnormally weigh. It is therefore interesting to
investigate if there is an experiment which measures whether this is
true or not. In the next section, we turn to this question.

\section{Do the Quantum Fluctuations Gravitate?}
\label{sec:vacuumweight}

\begin{figure*}
\begin{center}
\includegraphics[width=12cm]{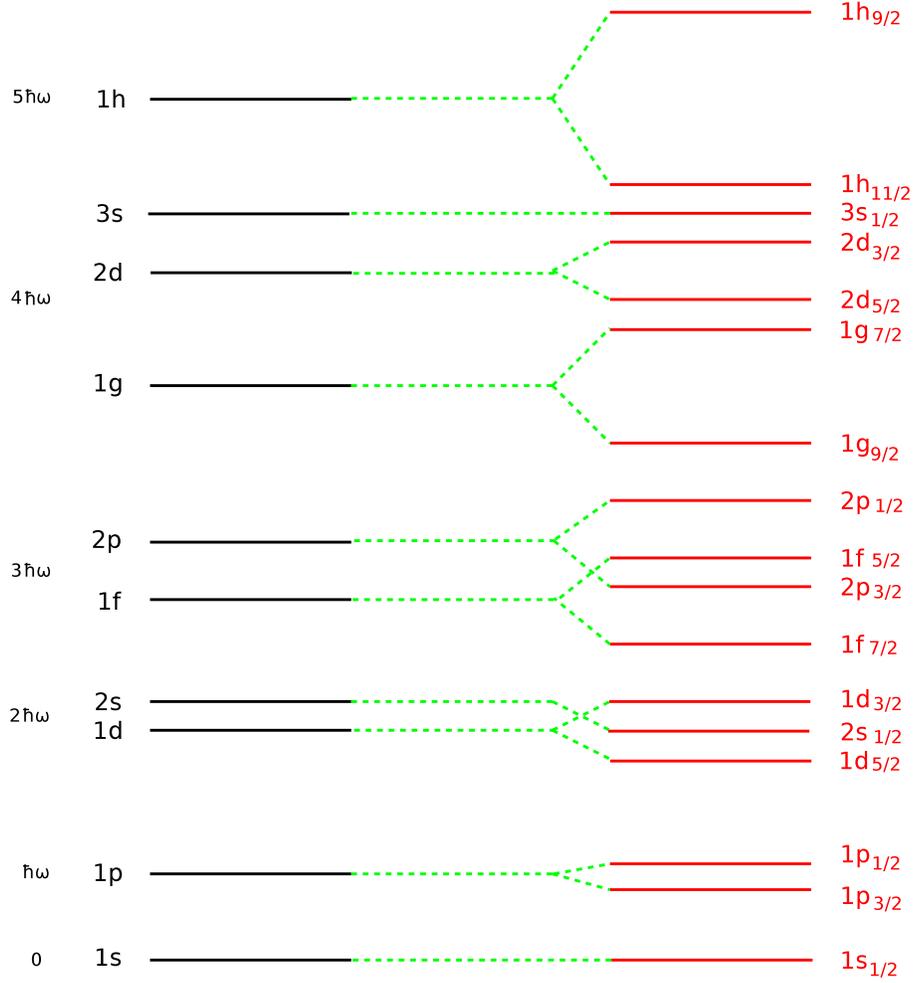}
\caption{Energy levels of the nucleus. On the left-hand side are
  represented the levels in the approximation where the Woods-Saxon
  potential is approximated by a parabola. In this case, each level
  (which includes a series of sub-levels) is separated by the energy
  $\hbar\omega$. On the right hand side are represented the levels
  where the spin orbit interaction~(\ref{eq:spinorbit}) is taken into
  account. This breaks the degeneracy of the sub-levels and reproduces
  the magic numbers.}
\label{fig:nucleus}
\end{center}
\end{figure*}

In this section, we study the gravitational property of the Lamb
shift~\cite{Polchinski:2006gy,Masso:2009zd}. We will follow the
treatment of Ref.~\cite{Masso:2009zd}. In Sec.~\ref{subsec:lambshift},
we showed that the interaction of the electron with the ``vacuum
electric field'' inside the atom leads to a shift of the atomic
levels. In some sense, this means that the zero-point fluctuations
affect the mass (energy) of the atom. Therefore, by studying how the
atom falls down, one can put constraints on the gravitational
properties of the vacuum fluctuations. This is the idea that we pursue
in this section. However, since the corresponding effect is small for
an atom and is larger for a nucleus (of course, the Lamb shift also
exists in this case), it is better to study the latter case. For this
reason, we start with recalling some basics facts about nuclear
models.

\par

The main idea is that the interaction between a given nucleon and the
other $A-1$ nucleons can be mimicked by a Woods-Saxon potential given
by
\begin{equation}
V_{_{\rm WS}}(r)=-V_0\left[1-\exp\left(\frac{r-R}{a}\right)\right],
\end{equation}
where $a$ describes how abrupt the potential is while $R$ is the
radius of the nucleus. As is well-known one can take $R=r_0A^{1/3}$
where $r_0\simeq 1.3\times 10^{-15}\, \mbox{m}=1.3 \, \mbox{fm}$. The
quantity $V_0$ represents the depth of the potential and can be taken
to be $V_0\simeq 45 \, \mbox{MeV}$. In fact, in order to have a
reasonable model, one must add to the Woods-Saxon potential a spin
orbit interaction term which can be expressed as
\begin{equation}
\label{eq:spinorbit}
V_{\rm L,s}=-f(r)\bm{L}\cdot \bm{s},
\end{equation}
where $\bm{L}$ is the angular momentum of the nucleon and $\bm{s}$ its
spin. With a good approximation, one can write $f(r)\simeq 24 A^{-2/3}
\, \mbox{MeV}$. Then, the Hamiltonian of the nucleus can be written as
\begin{equation}
H=\sum _{j=1}^A\left[\frac{p_j^2}{2m_j}+V_{_{\rm WS}}(r_j)
-f(r_j){\bm L}_j\cdot \bm{s}_j\right].
\end{equation}
The total wave-function can be taken as the product of the individual
wave functions (in fact, more precisely, one should consider a Slater
determinant in order to take into account the fact that the total wave
function must be anti-symmetrized), that is to say
\begin{equation}
\Psi\left(\bm{r}_1, \cdots,\bm{r}_A\right)
=\prod _{j=1}^A\Psi_j\left(\bm{r}_j\right).
\end{equation}
Each individual wave-function can be obtained from a Schr\"odinger
equation given by (for the moment, one ignores the spin orbit term
since it can be treated as a perturbation)
\begin{equation}
\left[-\frac{\hbar^2}{2\mu}\nabla ^2+V_{_{\rm WS}}(r_j)
\right]\Psi_j(r_j)=E_j\Psi_j(r_j),
\end{equation}
where $\mu$ is the reduced mass,
\begin{equation}
\frac{1}{\mu}\equiv \frac{1}{m_{\rm p}}+\frac{1}{(A-1)m_{\rm p}},
\end{equation}
or $\mu\simeq m_{\rm p}(A-1)/A\simeq m_{\rm p}\simeq 1.6 \times
10^{-27}\, \mbox{kg}$, $m_{\rm p}$ being the proton
mass. Unfortunately, the Schr\"odinger equation cannot be solved
analytically for the Woods-Saxon potential and, therefore, we are
forced to make approximations. The simplest one is just to describe
the Woods-Saxon potential with a parabola, \ie
\begin{equation}
V_{_{\rm WS}}(r)\simeq -V_0\left(1-\frac{r^2}{R^2}\right)
=-V_0+\frac12\mu\omega_0^2r^2,
\end{equation}
with $\omega_0\equiv \sqrt{2V_0/(\mu R^2)}$. In this case, the
question becomes solvable. We can use the spherical symmetry of the
problem and write the wave function as
\begin{equation}
\Psi_j(\bm{r_j})=\frac{u_{n,\ell}(r_j)}{r_j}Y_{\ell m}\left
(\theta_j,\varphi_j\right),
\end{equation}
where $Y_{\ell m}$ is a spherical
harmonics~\cite{Abramovitz:1970aa,Gradshteyn:1965aa}. The radial
function $u_{n\ell}$ is controlled by the equation
\begin{align}
\biggl[-\frac{\hbar^2}{2\mu}\frac{{\rm d}^2}{{\rm d}r_j^2}
+\frac{\ell (\ell+1)}{2\mu r_j^2}
&+\frac12\mu\omega _0^2r^2\biggr]u_{n,\ell}
\nonumber \\ & 
=(E_j+V_0)u_{n,\ell},
\end{align}
which can be solved in terms of Laguerre polynomials
$L_n^{\alpha}$~\cite{Abramovitz:1970aa,Gradshteyn:1965aa}. Explicitly,
one has
\begin{equation}
\Psi(r_j)= Cr_j^{\ell}{\rm e}^{-\mu\omega_0r_j^2/(2\hbar)}L_{n-1}^{\ell +1/2}
\left(\frac{\mu\omega_0}{\hbar}r^2\right)
Y_{\ell m}\left(\theta_j,\varphi_j\right),
\end{equation}
where $C$ is a normalization constant to be determined. In the above
expression, one has $n\ge 1$ and $\ell\ge 0$. The energy levels are
given by the expression
\begin{equation}
E(n,\ell)=-V_0+\hbar\omega_0\left[2(n-1)+\ell +\frac32\right].
\end{equation}
The factor $3/2=3\times 1/2$ originates from the fact that we have a
three-dimensional harmonic oscillator. The quantum numbers have been
chosen so that the ground state is $n=1$, $\ell=0$ which allows the
spectroscopic notation ``$1s$'', see Fig.~\ref{fig:nucleus}.

\par

The only thing which remains to be done is to determine the constant
$C$. Obviously, this is done by normalizing the wave-function, \ie
\begin{equation}
\vert C\vert^2 \int_0^{\infty}{\rm d}r_j r_j^2
\frac{\vert u_{n,\ell}\vert ^2}{r_j^2}
\int {\rm d}\Omega_j Y_{\ell m}=1, 
\end{equation}
Using the solution obtained above and performing the change of
variable $\rho=(\mu\omega_0/\hbar)r^2$, the previous condition reduces
to
\begin{align}
\frac{\vert C\vert^2}{2}&
\left(\frac{\mu \omega_0}{\hbar}\right)^{-\ell-3/2} \nonumber \\
& \times \int _0^{\infty}\rho^{\ell+1/2}{\rm e}^{-\rho}
L_{n-1}^{\ell +1/2}(\rho)L_{n-1}^{\ell +1/2}(\rho)=1.
\end{align}
Then, using Eq.~(7.414.3) of Ref.~\cite{Gradshteyn:1965aa}, one
obtains
\begin{equation}
\vert C\vert ^2=2(n-1)!\left(\frac{\mu \omega_0}{\hbar}\right)^{\ell+3/2}
\Gamma^{-1}\left(n+\ell+\frac12\right).
\end{equation}
We are now in a position where we can compute the value of the
wave-function at $r_j=0$ for $\ell=0$. This gives
\begin{equation}
\left\vert \Psi(0)\right\vert ^2=\vert C\vert^2\left[L_{n-1}^{1/2}(0)\right]^2
\frac{1}{4\pi}.
\end{equation}
Using the fact that $L_{n-1}^{1/2}(0)=\begin{pmatrix} n-1/2 \\ n-1 \end{pmatrix}
=\Gamma(n+1/2)/[\sqrt{\pi}(n-1)!/2]$, one obtains
\begin{equation}
\vert \Psi_{n,\ell=0}(0)\vert ^2=\frac{2}{\pi^2}
\frac{\Gamma(n+1/2)}{(n-1)!}
\left(\frac{\mu\omega_0}{\hbar}\right)^{3/2}.
\end{equation}
This completes the first part of the calculation. 

\par

We have established on very general grounds that the shift in energy
due to the interaction of a charged particle (in
Sec.~\ref{subsec:lambshift}, the electron in the atom, here the proton
in the nucleus) is given by Eq.~(\ref{eq:DEenergyLS}). The means value
of the square of the displacement has been calculated in
Eq.~(\ref{eq:meandr2}). Applying these formulas to the case of the
protons in the nucleus, one obtains~\cite{Masso:2009zd}
\begin{equation}
\Delta E(n,0)\simeq \frac{4\alpha^2}{3m_{\rm p}^2}
\sum_i Z^{(i)} \left\vert \Psi_{n,\ell=0}^{(i)}(0)
\right\vert ^2\int _0^{\infty}
\frac{{\rm d}\omega}{\omega}.
\end{equation}
As usual, this infinite integral must be regularized and usually
leads to logarithmic corrections, see for instance
Eq.~(\ref{eq:LSfinal}). We will ignore this factor in what follows
since this does not affect too much the final result. As explained in
Ref.~\cite{Masso:2009zd}, the sum over ``$(i)$'' runs only over
$s$-wave since this is the only way to get $\Psi_{n,\ell}(0)\neq
0$. Moreover, still following Ref.~\cite{Masso:2009zd}, it is
reasonable to approximate the field experienced by a proton by a field
produced by the other protons in inner shells. This is why $Z^{(i)}$
represents the total number of protons present in shells inner than
the proton ``$(i)$'' under consideration.

\par

Our goal is now to test how the vacuum fluctuations weigh. For this
purpose, we write the gravitational mass of a given nucleus
as~\cite{Masso:2009zd}
\begin{equation}
m_{\rm g}=m_{\rm i}-\eta \Delta E,
\end{equation}
where $m_{\rm i}$ is the inertial mass. The parameter $\eta$ is a
priori a free parameter which controls to which extent the weak
equivalence principle would be violated. If $\eta=0$, there is no
violation of the universality of free fall. A convenient measure of
potential violations of the weak equivalence principle is given by the
E\"otv\"os ratio
\begin{equation}
\eta (1,2)\equiv \frac{m_{\rm g}(1)}{m_{\rm i}(1)}
-\frac{m_{\rm g}(2)}{m_{\rm i}(2)},
\end{equation}
for two bodies $1$ and $2$. Using the previous calculations, this
ratio can be expressed as
\begin{eqnarray}
\eta(1,2)&=&\eta \frac{4\alpha^2}{3m_{\rm p}^3}
\biggl[\frac{1}{A_1}\sum_i Z^{(i)}_1 \left \vert 
\Psi_{n,\ell=0}^{(1,i)}(0)\right \vert ^2
\nonumber \\ & &
-\frac{1}{A_2}\sum_i Z^{(i)}_2 \left \vert 
\Psi_{n,\ell=0}^{(2,i)}(0)\right \vert ^2
\biggr],
\end{eqnarray}
where we have used $m_{\rm i}\sim Am_{\rm p}$, a relation valid since
we work at first order in the parameter $\eta$. Then, we use the fact
that
\begin{equation}
(\mu \omega_0)^{3/2}=\left(\frac{2m_{\rm p}V_0}{r_0^2}\right)^{3/4}
\frac{1}{A^{1/2}}
\end{equation}
to express the E\"otv\"os ratio as~\cite{Masso:2009zd}
\begin{eqnarray}
\eta (1,2)&=&\eta \frac{4\alpha^2}{3}
\left(\frac{2V_0}{m_{\rm p}^3r_0^2}\right)^{3/4}
\frac{2}{\pi^2}
\nonumber \\ & & \times
\biggl[\frac{1}{A_1^{3/2}}\sum _iZ_1^{(i)}
\frac{\Gamma(n+1/2)}{(n-1)!}
\nonumber \\ & &
-\frac{1}{A_2^{3/2}}\sum _iZ_2^{(i)}
\frac{\Gamma(n+1/2)}{(n-1)!}\biggr].
\end{eqnarray}
Let us evaluate this number for aluminum ($Z=13$, $A=27$) and
Platinum ($Z=78$, $A=195$). In the case of aluminum, since we only
have $13$ protons, the only $s$-shell protons are on the $1\, s_{1/2}$
level (in fact, we have two protons in $1\, s_{1/2}$, four protons in
$1\, p_{3/2}$, two protons in $1\, p_{1/2}$ and five protons in $1\,
d_{5/2}$, see Fig.~\ref{fig:nucleus}). As a consequence, there is no
inner protons for those protons (since, obviously, there is no inner
level than the ground state $1\, s_{1/2}$) and the corresponding
contribution in the E\"otv\"os ratio vanishes. In the case of
Platinum, all the levels are filled up until the level $1\, h_{11/2}$
which contains height protons, see Fig.~\ref{fig:nucleus}. Therefore,
we have two protons on $1\, s_{1/2}$, two protons on $2\, s_{1/2}$ and
two on $3\, s_{1/2}$, see again Fig.~\ref{fig:nucleus}. This means
that the $3\, s_{1/2}$ protons see four inner protons and the $2\,
s_{1/2}$ protons see two inner protons. As a consequence, one can
write
\begin{eqnarray}
\eta (\mbox{Al},\mbox{Pt})&\simeq & -\eta \frac{4\alpha^2}{3}
\left(\frac{2V_0}{m_{\rm p}^3r_0^2}\right)^{3/4}
\frac{2}{\pi^2}
\frac{1}{A_{\rm Pt}^{3/2}}
\nonumber \\ & \times &
\biggl[\frac{2}{(2-1)!}\Gamma \left(2+\frac12\right)
+\frac{4}{(3-1)!}\Gamma \left(3+\frac12\right)\biggr]
\nonumber \\
\end{eqnarray}
Using the values of $V_0$ and $r_0$ chosen previously this leads to 
\begin{equation}
\eta (\mbox{Al},\mbox{Pt})\simeq  -5.6\times 10^{-10} \, \eta 
\end{equation}
Given the fact that $\vert \eta (\mbox{Al},\mbox{Pt})\vert\lesssim
10^{-12}$~\cite{Su:1994gu,Damour:2009zy}, we reach the conclusion that
\begin{equation}
\eta \lesssim 1.7 \times 10^{-3}.
\end{equation}
Therefore, the vacuum fluctuations seem to have standard gravitational
properties to a very good accuracy. 

\par

The conclusion of this section is that it seems difficult to argue
that the zero-point fluctuations do not gravitate. Clearly, this makes
the cosmological constant problem more acute. However, the results
obtained here are by no means a final proof. For instance, it seems
reasonable to assume that the stress energy tensor of the Lamb shift
is not of the cosmological constant type. Therefore, the fact that a
nucleus falls down normally does not necessarily imply that the
cosmological constant weighs according to the Einstein
equations. Nevertheless, no sign of inconsistency appears in the above
situation and one is obviously tempted to assume that zero-point
fluctuations couple to gravity as any other type of matter. Again,
this makes the cosmological constant problem more mysterious.

\section{The Weak Equivalence Principle in Quantum Mechanics}
\label{sec:wepqm}

In the previous sections, we have argued that the zero-point
fluctuations seem to be real and seem to gravitate
normally. Therefore, these two reasons cannot be invoked to avoid the
cosmological constant problem. However, there is another (but related)
issue that it is interesting to investigate. When we consider the
vacuum stress energy tensor in the Einstein equations, we implicitly
assume that the zero-point fluctuations obey the weak equivalence
principle, \ie that they ``fall'' in a gravitational field as any
other type of matter. Clearly the zero point fluctuations are of
quantum-mechanical origin. The weak equivalence principle is well
established (and tested) in classical physics and general relativity
is based on this principle. But what is the status of this principle
in quantum mechanics? If it is not valid in quantum mechanics, maybe
we do not have the right to couple the quantum zero point fluctuations
in the standard way in the Einstein equations?  Could it be a way to
avoid the problem? In order to discuss this issue, we will consider
two situations, the Collela, Overhausser and Werner experiment (COW)
and the quantum Galileo experiment.

\subsection{The Collela, Overhausser and Werner (COW) Experiment}
\label{subsec:cow}

\begin{figure*}
\begin{center}
\includegraphics[width=16cm]{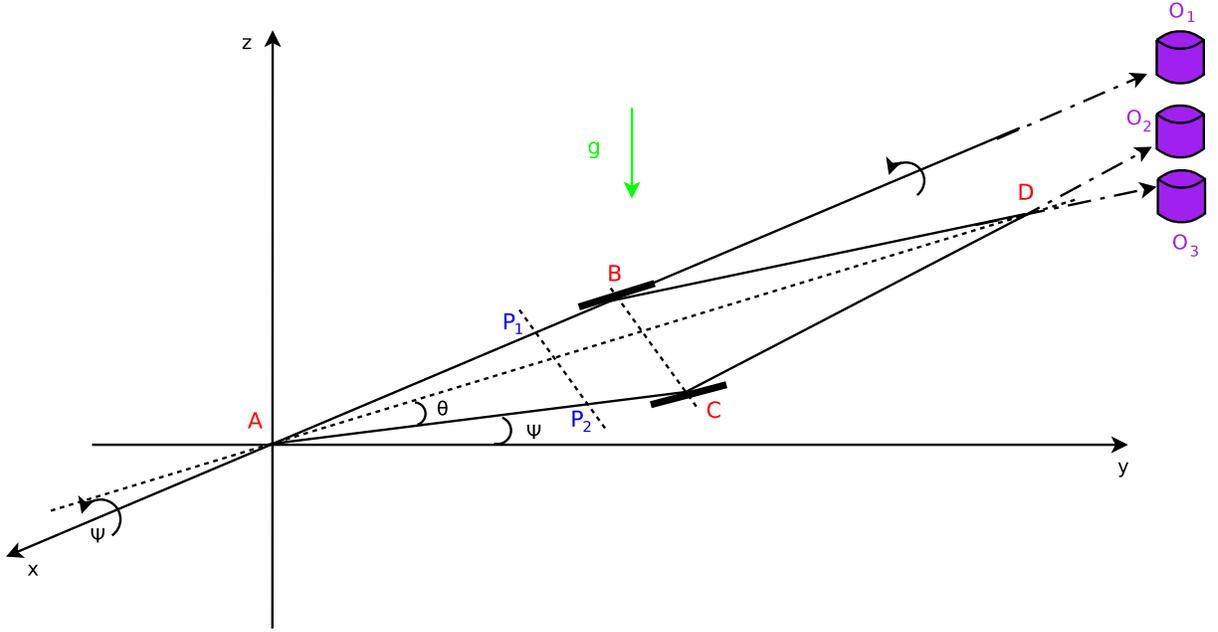}
\caption{Representation of the Collela, Overhausser and Werner (COW)
  interferometer. The interferometer is placed in a vertical
  gravitational field (represented by the green arrow) and is rotated
  around the axis $AB$.}
\label{fig:interfero}
\end{center}
\end{figure*}

The COW
experiment~\cite{Colella:1975dq,Greenberger:1979zz,Greenberger:1983zz}
consists in sending neutrons in an interferometer placed in a weak
gravitational field, see Fig.~\ref{fig:interfero}. The presence of the
gravitational field affects the wave-function of the neutrons and
changes the relative phases between the two beams. When the
interferometer is rotated, the gravitational phase shift is modified
because the height difference between the two arms is changed. This
effect can then be detected (and was detected) in the interference
pattern. The COW experiment is conceptually very important. This was
the first experiment to measure the effect of the gravitational field
on the wave-function. In fact, it shows that the gravitation is just
an ordinary force as far as the Schr\"odinger equation is concerned.

\begin{figure*}
\begin{center}
\includegraphics[width=16cm]{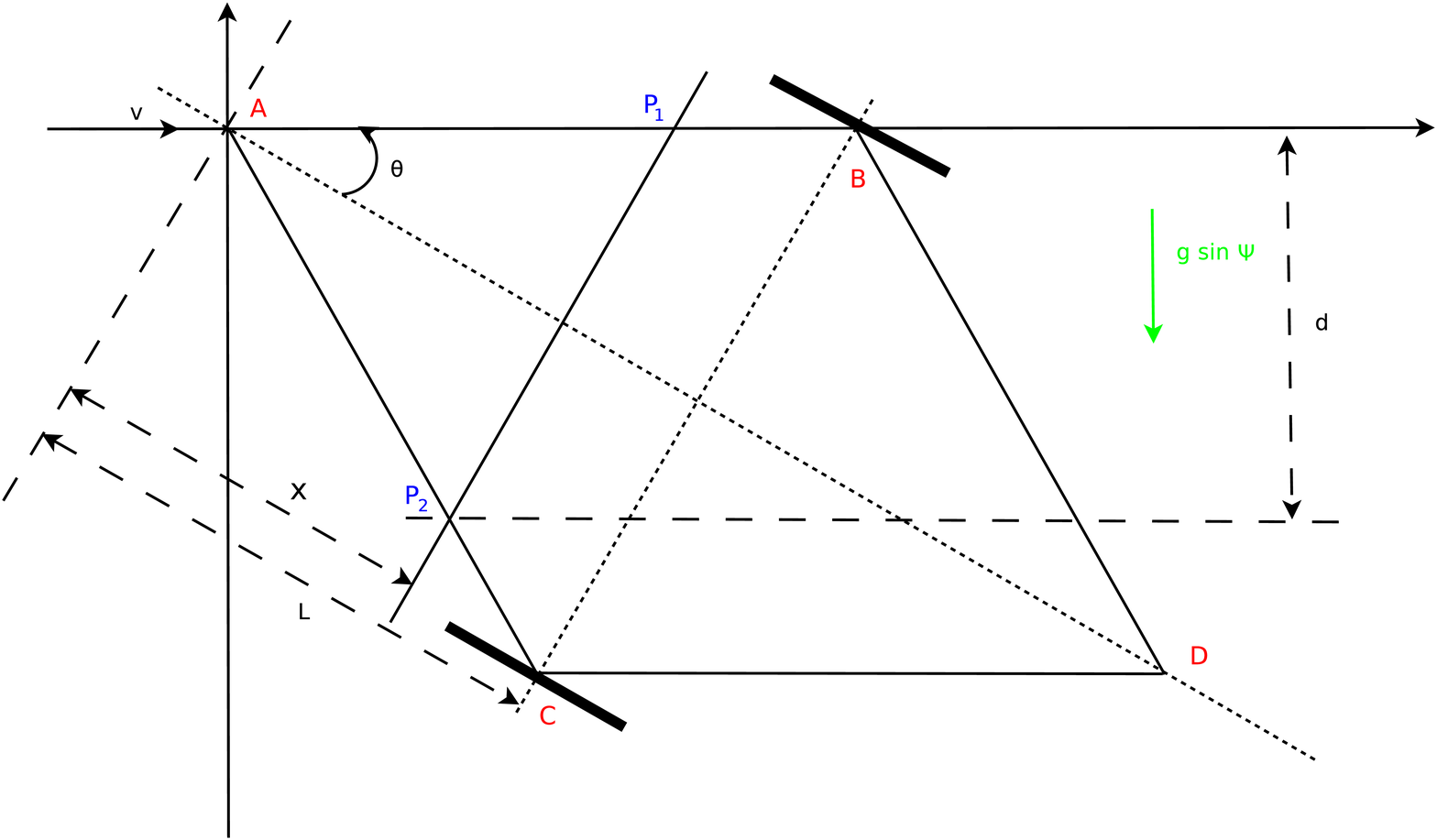}
\caption{Representation of the COW interferometer in the plane of the
  two beams.}
\label{fig:interfero2}
\end{center}
\end{figure*}

Let us now describe the experiment in more detail. The neutron
interferometer is represented in Fig.~\ref{fig:interfero}. Let us
assume that, along one of the two paths, a phase shift $\Delta
\varphi$ is introduced. Our first goal is to evaluate the intensity
seen by the detectors $O_2$ and $O_3$. After diffraction, the wave
function of the neutron takes the form
\begin{equation}
\Psi(\bm{r}) =\Phi T(\theta){\rm e}^{i \bm{k}_0\cdot \bm{r}}
+\Phi D(\theta){\rm e}^{-i \bm{k}_{\rm G}\cdot \bm{r}},
\end{equation}
where $\bm{k}_0$ is the incident wave-number and $\bm{k}_{\rm G}$ the
wave number after Bragg diffraction (which, therefore, must satisfy
the Bragg condition). The quantity $\Phi$ is the amplitude of the
incident wave and $T=T(\theta)$ (for ``transmitted'') and
$D=D(\theta)$ (for ``diffracted'') are two coefficients describing the
amplitude of the two branches of the wave-function after the neutron
has emerged from the crystal. Let us now consider the the path
$ABDO_3$. At point $B$, the wave function is given by (the origin of
the coordinates is chosen to be at point $A$)
\begin{equation}
\Psi(AB)=\Phi T(\theta){\rm e}^{i \bm{k}_0\cdot \bm{r}_{\rm B}},
\end{equation}
since the wave was just transmitted. Then, at point $B$, we have a
diffraction and, therefore, the wave-function at point $D$ reads
\begin{equation}
\Psi(ABD)=\Phi T(\theta){\rm e}^{i \bm{k}_0\cdot \bm{r}_{\rm B}}D(\theta)
{\rm e}^{i \bm{k}_{\rm G}\cdot (\bm{r}_{\rm D}-\bm{r}_{\rm B})}.
\end{equation}
Finally, at point $D$, we have another diffraction but this time with
opposite angle. The leads to 
\begin{eqnarray}
\Psi(ABDO_3) &=& \Phi T(\theta){\rm e}^{i \bm{k}_0\cdot \bm{r}_{\rm B}}D(\theta)
{\rm e}^{i \bm{k}_0\cdot (\bm{r}_{\rm D}-\bm{r}_{\rm B})}
\nonumber \\ & & \times
D(-\theta)
{\rm e}^{i \bm{k}_{\rm G}\cdot ({\bm r}-\bm{r}_{\rm D})}.
\end{eqnarray}
We also assume that, along that path, there is a phase shift $\Delta
\varphi$ that we do not specify for the moment. As a consequence, the
total wave function can be written as
\begin{eqnarray}
\Psi(ABDO_3) &=& \Phi T(\theta){\rm e}^{i \bm{k}_0\cdot \bm{r}_{\rm B}}D(\theta)
{\rm e}^{i \bm{k}_0\cdot (\bm{r}_{\rm D}-\bm{r}_{\rm B})}
\nonumber \\ & & \times
D(-\theta){\rm e}^{i\Delta \varphi}
{\rm e}^{i \bm{k}_0\cdot ({\bm r}-\bm{r}_{\rm D})},
\end{eqnarray}
where we have introduced the factor ${\rm e}^{i\Delta \varphi}$.

\par
  
It is straightforward to follow the same procedure and to establish the
expression for $\Psi(ACDO_3)$. This gives the expression of the wave
function at detector $O_3$, namely
$\Psi(O_3)=\Psi(ABDO_3)+\Psi(ACDO_3)$,
\begin{widetext}
\begin{eqnarray}
\Psi(O_3)=\Phi \left[
T(\theta){\rm e}^{i \bm{k}_0\cdot \bm{r}_{\rm B}}D(\theta)
{\rm e}^{i \bm{k}_0\cdot (\bm{r}_{\rm D}-\bm{r}_{\rm B})}
D(-\theta){\rm e}^{i\Delta \varphi}
+
D(\theta){\rm e}^{i \bm{k}_0\cdot \bm{r}_{\rm C}}D(-\theta)
{\rm e}^{i \bm{k}_0\cdot (\bm{r}_{\rm D}-\bm{r}_{\rm C})}
T(\theta)
\right]
{\rm e}^{i \bm{k}_0\cdot ({\bm r}-\bm{r}_{\rm D})}.
\end{eqnarray}
Using this expression, one can calculate the expected
intensity. Straightforward manipulations leads to the following expression
\begin{equation}
\label{eq:intensityd3}
I(O_3)=\vert \Psi(O_3)\vert ^2=\alpha \left(1+\cos \Delta \varphi\right),
\end{equation}
where $\alpha $ is a constant that can be determined from the above
expression and which depends on the incident flux and the crystal
structure. Then, one can repeat the same analysis and determine the
wave function at detector $O_2$. It reads
\begin{eqnarray}
\Psi(O_2)=\Phi \left[
T(\theta){\rm e}^{i \bm{k}_0\cdot \bm{r}_{\rm B}}D(\theta)
{\rm e}^{i \bm{k}_{\rm G}\cdot (\bm{r}_{\rm D}-\bm{r}_{\rm B})}
T(-\theta){\rm e}^{i\Delta \varphi}
+
D(\theta){\rm e}^{i \bm{k}_{\rm G}\cdot \bm{r}_{\rm C}}D(-\theta)
{\rm e}^{i \bm{k}_0\cdot (\bm{r}_{\rm D}-\bm{r}_{\rm C})}
D(\theta)
\right]
{\rm e}^{i \bm{k}_{\rm G}\cdot ({\bm r}-\bm{r}_{\rm D})}.
\end{eqnarray}
\end{widetext}
The intensity at detector $O_2$ can be computed along the same lines
and one is led to
\begin{equation}
I(O_2)=\beta -\alpha \cos \Delta \varphi,
\end{equation}
where $\alpha $ and $\beta $ are just constants that do not play an
important role in what follows [of course, the constant $\alpha $ is
the same as in Eq.~(\ref{eq:intensityd3})].

\par

Having estimated the form of the signal expected for any phase shift
$\Delta \varphi $, we now turn to its calculation in the case of the
COW experiment. For this purpose, we return to the Schr\"odinger
equation which can be written as
\begin{equation}
\nabla ^2\Psi +\omega^2\Psi=0,
\end{equation}
with 
\begin{equation}
\omega^2=\frac{2m}{\hbar^2}\left(E-V\right).
\end{equation}
Now, let us assume that the potential is given by 
\begin{equation}
V=V_0+\Delta V(\bm{r}),
\end{equation}
where $V_0$ is a constant and $\Delta V$ a small space-dependent small
perturbation, \ie $\Delta V\ll V_0$. If the perturbation does not
change abruptly, the Wentzel-Kramers-Brillouin (WKB) wave-function
\begin{equation}
\Psi \propto \exp \left[i\int \omega ({\bm r}){\rm d}{\bm r}\right],
\end{equation}
is a very good approximation to the actual solution. Introducing the
notation $k_0=2m(E-V_0)/\hbar^2$ and using that $p=mv=\hbar k_0$, an
expansion of the phase in $\Delta V/V_0$ leads to the following
expression
\begin{equation}
\Psi \propto \exp \left(i\bm{k}_0\cdot \bm{r}-\frac{1}{\hbar}
\int _{\cal P}\Delta V {\rm d}t\right),
\end{equation}
where the subscript ``${\cal P}$'' indicates that one integrates along
the non perturbed path. From the above equation, we conclude that
\begin{equation}
\Delta \varphi=-\frac{1}{\hbar}
\int _{\cal P}\Delta V {\rm d}t .
\end{equation}
The same result can also be expressed differently. Using energy
conservation, on can write
\begin{equation}
\frac{p^2}{2m}+V_0+\Delta V=E,
\end{equation}
and, since $\bm{p}=\bm{p}_0+\delta \bm{p}$, one obtains
\begin{equation}
\frac{p_0^2}{2m}+\frac{\bm{p}_0}{m}\cdot \delta \bm{p}+V_0+\Delta V=E,
\end{equation}
or,
\begin{equation}
\bm{v}_0\cdot \delta \bm{p}=-\Delta V.
\end{equation}
As a consequence, the phase shift can be re-written as the following
expression
\begin{equation}
\label{eq:phaseshiftmomentum}
\Delta \varphi=\frac{1}{\hbar}\int _{\cal P} \delta \bm{p}\cdot 
{\rm d}\bm {r},
\end{equation}
where the integral should again be calculated along the path of the
neutron.

\par

Let us now determine the phase shift for the COW experiment. Let us
consider two point $P_1$ and $P_2$ that correspond to one beam
transmitted and one beam diffracted in the interferometer, see
Fig.~\ref{fig:interfero2}. The potential difference between those two
points is given by
\begin{equation}
V(P_2)-V(P_1)=mgd\sin \psi ,
\end{equation}
where $d$ can be expressed as $d=AP_1 \cos(\pi/2-2\theta)$. Given that
the quantity $x$ in Fig.~\ref{fig:interfero2} can be written as
$x=AP_1 \cos \theta$, one deduces that $d=2x\sin \theta $. As a consequence, 
\begin{equation}
  V(P_2)-V(P_1)=2m_{\rm grav}gx\sin \theta \sin \psi ,
\end{equation}
where we have now carefully written that the mass which appeared in
the above expression is the gravitational mass (since this is the
``coupling constant'' to the gravitational field). This implies that
the phase shift can be written as
\begin{equation}
\Delta \varphi=-\frac{2}{\hbar}\int _0^L
2m_{\rm g}g_0x\sin \theta 
\sin \psi
\frac{{\rm d}x}{\cos \theta},
\end{equation}
where we have taken into account the fact that the line $P_1P_2$ is
``moving'' with a speed $v\cos \theta $. The factor two in front of
the whole expression comes from the fact that the phase shift
accumulates and is, along the total path, twice its value in the fist
part of the interferometer. The integration over the variable $x$ is
easily done and we obtain
\begin{equation}
\Delta \varphi=-2
\frac{m_{\rm g}g_0}{\hbar v}L^2\tan \theta \sin \psi.
\end{equation}
Finally, noticing that the total area of the interferometer is nothing
but ${\cal A}\equiv 2L^2\tan \theta $ and that the de Broglie relation
is $v=h/(m_{\rm in}\lambda )$, where $m_{\rm in}$ is the inertial mass
and $\lambda$ the wavelength of the incident beam, we arrive
at~\cite{Colella:1975dq,Greenberger:1979zz,Greenberger:1983zz}
\begin{equation}
\label{eq:gravityshift}
\Delta \varphi=-\frac{2\pi}{h^2}m_{\rm grav}m_{\rm in}g_0{\cal A} 
\lambda \sin \psi . 
\end{equation}
This expression is quite remarkable as it involves the gravitational
and the inertial mass together with the Planck constant. Moreover, we
deal with an observable quantity which does not depend on the ratio of
the gravitational mass to the inertial mass, as is usually the case,
but on the product of these quantities. We also see that the intensity
of the signal at detectors $O_2$ and/or $O_3$ depend on the
orientation of the interferometer $\psi$.

\par

However, in practice, one must deal with two other
effects~\cite{Colella:1975dq,Greenberger:1979zz,Greenberger:1983zz}. The
first one is the Sagnac effect due to the rotation of
Earth. Classically, the Hamiltonian of a neutron is given by
\begin{equation}
H=\frac{p^2}{2m_{\rm in}}+m_{\rm grav}\bm{g}\cdot \bm{r}
-\bm{\omega}\cdot {\bm L},
\end{equation}
where $\bm{L}=\bm{r}\times \bm{p}$ is the angular momentum of the
neutron and ${\bm \omega}$ the angular velocity of Earth. Upon using
the Hamilton equations, this implies that
\begin{equation}
\bm{p}=m_{\rm in}\, \dot{\bm{r}}+m_{\rm in}\, \bm{\omega}\times \bm{r}.
\end{equation}
Therefore, one obtains a Sagnac phase shift which can be expressed as
\begin{equation}
\Delta \varphi_{_{\rm Sagnac}}=\frac{m_{\rm in}}{\hbar}\oint 
\left( \bm{\omega}\times \bm{r}\right)\cdot {\rm d}\bm{r},
\end{equation}
where we have used Eq.~(\ref{eq:phaseshiftmomentum}). Using Stokes
theorem, this can also be written as
\begin{equation}
\Delta \varphi_{_{\rm Sagnac}}=\frac{4\pi m_{\rm in}}{h}\bm{\omega}\cdot 
\bm{{\cal A}},
\end{equation}
where $\bm{{\cal A}}$ is the vector associated to the interferometer
area. If $\Gamma $ is the incident neutron beam west of due south and
$\delta $ the co-latitude, the Sagnac phase shift takes the form
\begin{equation}
\Delta \varphi_{_{\rm Sagnac}}=\frac{4\pi m_{\rm in}}{h}
\omega {\cal A}
\left(\cos \psi \cos \delta +\sin \psi \sin \Gamma \sin \delta \right).
\end{equation}
The COW experiment is such that the incident beam is directed due
south which implies that $\Gamma =0$. In the case, one obtains
\begin{equation}
\label{eq:sagnacshift}
\Delta \varphi_{_{\rm Sagnac}}=\frac{4\pi m_{\rm i}}{h}
\omega {\cal A}\cos \delta \cos \psi ,
\end{equation}
a dependence in $\psi $ which is different from the gravitational
phase shift, see Eq.~(\ref{eq:gravityshift}).

\par

Finally, there is a shift due to the fact that the interferometer bends
and/or warps under its own weight. The COW experiment claim that this
effect can be described by the following equation
\begin{equation}
\Delta \varphi_{_{\rm bend}}=-q_{_{\rm bend}}\sin \psi.
\end{equation}
This expression can be justified by noticing that the bending effect
depends on the rotation angle. Clearly, this effect is difficult to
estimate from first principles.

\par

Therefore, the total phase shift expected in the COW experiment is the
sum of the three contributions discussed before, namely the gravity,
Sagnac and bending shifts. This leads to the following expression
\begin{equation}
\Delta \varphi =-q_{_{\rm grav}}\sin \psi +q_{_{\rm Sagnac}}\cos \psi 
-q_{_{\rm bend}}\sin \psi,
\end{equation}
where, in order to take into account Eqs.~(\ref{eq:gravityshift})
and~(\ref{eq:sagnacshift}), we have defined $q_{_{\rm grav}}\equiv
-2\pi m_{\rm grav}m_{\rm in}g_0{\cal A}\lambda/h ^2$ and $q_{_{\rm
    Sagnac}}=4\pi m_{_{\rm in}}\omega{\cal A}\cos\delta /h$. The above
expression can be re-written as
\begin{equation}
\Delta \varphi=q\sin \left(\psi -\psi_0\right),
\end{equation}
where $q^2=\left(q_{_{\rm grav}}+q_{_{\rm bend}}\right)^2+q_{_{\rm
    Sagnac}}^2$ and $\tan \psi_0=q_{_{\rm Sagnac}}/\left(q_{_{\rm
      grav}}+q_{_{\rm bend}}\right)$. In practice, the Sagnac effect
is only $2.5\%$ of the gravitational effect but we see it leads to a
global shift of the oscillatory pattern.

\par

Let us now describe the result of the experiment. According to the
previous considerations, the intensity observed at detector $O_3$ is 
\begin{equation}
I(O_3)=\alpha\left\{1+\cos\left[
q\sin \left(\psi -\psi_0\right)\right]\right\}.
\end{equation}
By Fourier transforming this signal, one can extract the frequency
$q_{_{\rm grav}}$ of this oscillations. If one repeat this procedure
for different wavelengths, one can fit the dependence of $q_{_{\rm
    grav}}$ with $\lambda$ which allows us to determine the product
$m_{\rm grav}m_{\rm in}$. Let us recall that the neutron mass is
usually obtained by mass spectroscopy on the deuteron which leads to
the following value
\begin{equation}
m_{\rm n}=m_{\rm D}-m_{\rm p}+\frac{E_{\gamma}}{c^2}
=1.6747\times 10^{-24}\mbox{g},
\end{equation}
where $E_{\gamma}$ is deuteron binding energy. On the other hand, the
COW experiment has found~\cite{Colella:1975dq}
\begin{equation}
\left(m_{\rm grav}m_{\rm in}\right)^{1/2}=(1.675\pm 0.003)
\times 10^{-24} \, \mbox{g}.
\end{equation}
This shows that the weak equivalence principle seems to be satisfied
even in a purely quantum mechanical situation. This also shows that
gravity can be ``coupled'' to the Schr\"odinger equation in a standard
way. The COW experiment is therefore very important since this was the
first time that this was demonstrated. Moreover, if we consider that
$m_{\rm n}=m_{\rm in}$ then the two previous equations shows that the
COW experiment proves that $m_{\rm in}=m_{\rm grav}$ even in the
quantum domain.

\subsection{The Quantum Galileo Experiment}

\begin{figure*}
\begin{center}
\includegraphics[width=12cm]{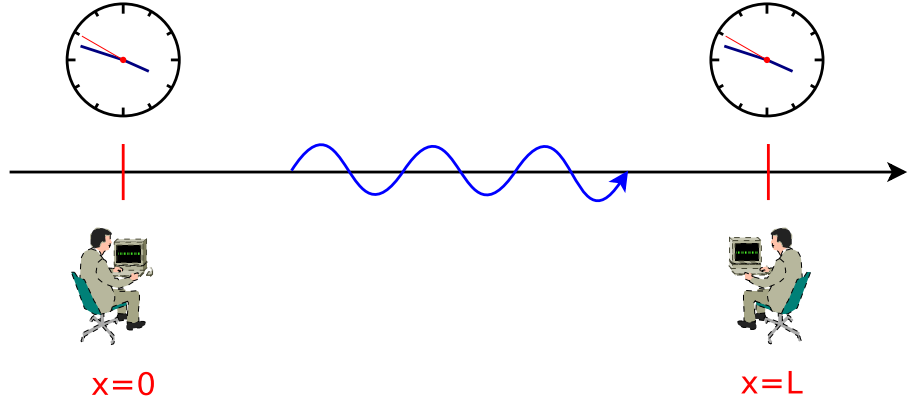}
\caption{Sketch of a quantum ``Peres'' clock. The clock is coupled to
  the particle traveling from the point $x=0$ to the point
  $x=L$. When the particle crosses $x=0$, the clock changes its state
  and starts running. When the particle reaches $x=L$, the clock stops
  and a comparison of the two clock states permits a measurement of
  the time of flight of the particle.}
\label{fig:peres}
\end{center}
\end{figure*}

It is well-known that, classically, the motion of a body in a constant
gravitational field only depends on the ratio $m_{\rm grav}/m_{\rm
  in}$ (which is measured to be unity). However, if one writes the
Schr\"odinger equation for the same situation, namely
\begin{equation}
i\hbar \frac{\partial \Psi(t,z)}{\partial t}
=-\frac{\hbar^2}{2m_{\rm in}}\frac{\partial ^2\Psi(t,z)}{\partial z^2}
+m_{\rm grav}gz\Psi(t,z),
\end{equation}
it is apparent that $m_{\rm in}$ and $m_{\rm grav}$ no longer cancel
out. Therefore, it is interesting to study how the universality of the
free fall is recovered in quantum mechanics. We investigate this point
in the following sections.

\subsubsection{The Salecker-Wigner-Peres Clock}
\label{subsub:peres}

In this section, we present a simple model of a quantum clock, the
so-called ``Salecker-Wigner-Peres
clock''~\cite{Salecker:1957be,Peres:1979nk,Leavens1993,Davies:2004gh,2004CQGra..21.2761D}. We
introduce this system because we want to study how a quantum particle
falls down in a constant gravitational field. One way to to do that is
to measure the time of flight of the particle and to compare it with
its classical analogue. Hence the question of how to define the time
of flight becomes important. As is well-known, the measurement of time
(or, to be more accurate, of time of arrival) is problematic in
quantum
mechanics~\cite{Grot:1996xu,Aharonov:1997md,Oppenheim:1998yy,Davies:2005uq,2011arXiv1112.4198M}. In
particular, there is no time operator because time is not a dynamical
variable. One way out is to introduce a quantum system, the ``clock'',
that is coupled to the particle and changes its state when the particle
crosses the starting and the ending points along its ``trajectory'',
see Fig.~\ref{fig:peres}. Reading these two states then gives
information about the time of flight. We now study how this general
idea works in practice.

\par

We assume that the clock has an odd number $N=2j+1$ of states. The
Hamiltonian of the clock is given by
\begin{equation}
  H_{\rm clock}=\omega J=-i\hbar \omega \frac{\partial}{\partial \theta},
\end{equation}
where the quantity $\theta$ can be viewed as describing the position
of the clock's hand. The corresponding states of the clock are
represented by the following wave-function
\begin{equation}
u_n(\theta)=\frac{1}{\sqrt{2\pi}}{\rm e}^{in\theta},
\end{equation}
that are eigenfunction of $H_{\rm clock}$ since $H_{\rm
  clock}u_n(\theta) =n\hbar \omega u_n(\theta)$. It is also easy 
to show that
\begin{equation}
\label{eq:evolu}
{\rm e}^{-iH_{\rm clock}t/\hbar}u_n(\theta)=u_n(\theta-\omega t),
\end{equation}
which gives the time evolution of the system.

\par

It is also convenient to introduce another basis $v_s(\theta)$ defined
by ($s=0, \cdots ,N-1$)
\begin{eqnarray}
\label{eq:defv}
v_s(\theta) &=& \frac{1}{\sqrt{N}}\sum _{n=-j}^{n=j} 
{\rm e}^{-\frac{2i\pi sn}{N}}u_n(\theta) \\
&=& \frac{1}{\sqrt{2\pi N}}
\frac{\sin \left[N\left(\theta-2\pi s/N\right)/2\right]}
{\sin \left[\left(\theta-2\pi s/N\right)/2\right]}.
\end{eqnarray}
If $N$ is large enough, these functions peak at $\theta =2\pi s/N$ and
this corresponds to the hand of the clock pointing to the $s$-th hour
with an angle uncertainty of $\pm \pi/N$ (hence, as expected, if $N$
is large, the accuracy is good). Moreover, using Eq.~(\ref{eq:evolu}),
we have
\begin{equation}
\label{eq:evolv}
{\rm e}^{-iH_{\rm clock}t/\hbar}v_s(\theta)=v_s(\theta-\omega t).
\end{equation}
This means that if we evaluate the above equation for a time $\tau$
such that $\omega\tau =2\pi/N$, then one obtains
\begin{equation}
{\rm e}^{-iH_{\rm clock}\tau/\hbar}v_s(\theta)
=v_{s+1\, (\mbox{mod}\, N)}(\theta),
\end{equation}
which confirms the interpretation given before.

\par

Let us now consider a free particle traveling along the $x$-axis. We
want to measure the time of flight between $x=0$ and $x=L$, see
Fig.~\ref{fig:peres}. In order for the clock to record this time, we
must of course couple it to the particle. Following
Ref.~\cite{Peres:1979nk}, we assume that the Hamiltonian of the total
system ``particle+clock'' can be written as
\begin{equation}
H=\frac{p^2}{2m}+{\cal P}(x)H_{\rm clock},
\end{equation}
where ${\cal P}(x)$ is the projector operator which is one if $0<x<L$
and zero otherwise and $m$ is the mass of the particle. We write the
eigen wave-functions of the system as
\begin{eqnarray}
\label{eq:defpsi}
\Psi(t,x,\theta)&=&\psi(x,\theta)\, {\rm e}^{-iEt/\hbar}
\nonumber \\ 
&=&\frac{1}{\sqrt{N}}
\sum_{n=-j}^{n=j}\psi_n(x)u_n(\theta){\rm e}^{-iEt/\hbar},
\end{eqnarray}
and inserting this form into the Schr\"odinger equation, one obtains
\begin{align}
\label{eq:schroperes}
\sum _{n=-j}^{n=j}\biggl[-\frac{\hbar^2}{2m}
\frac{{\rm d} ^2\psi_n}{{\rm d}x^2} &+n\hbar \omega{\cal P}(x)\psi_n(x)
\nonumber \\ &
-E\psi_n(x)\biggr]u_n(\theta)=0.
\end{align}
Therefore, we just have to solve the time-independent Schr\"odinger
equation in presence of a rectangular barrier. Of course, this
exercise is solved in any textbooks on quantum mechanics, see for
instance Ref.~\cite{tannoudji2006}. It is straightforward to show that
\begin{equation}
\label{eq:solpsim}
\psi_n(x)=
\begin{cases}
\displaystyle
A{\rm e}^{ikx}+B{\rm e}^{-ikx}, & x<0, \\ 
\displaystyle
C{\rm e}^{ipx}+D{\rm e}^{-ipx}, & 0<x<L, \\
\displaystyle
E {\rm e}^{ikx}, & x>L,
\end{cases}
\end{equation}
where we have defined 
\begin{eqnarray}
\label{eq:defk}
k &=& \frac{\sqrt{2mE}}{\hbar}, \\
p &=& \frac{\sqrt{2m(E-n\hbar \omega)}}{\hbar}=k\sqrt{1-\epsilon},
\end{eqnarray}
with $\epsilon\equiv n\hbar\omega/E$, a quantity which measures how
the clock perturbs the particle. The quantities $A$, $B$, $C$, $D$ and
$E$ are integration constants that are fixed by requiring the
wave-function and its derivative to be continuous at the
barrier. Notice that, after the barrier there is only one branch; this
is of course because we assume that the particle is coming from the
left, see Fig.~\ref{fig:peres}. Matching $\psi_n$ and ${\rm
  d}\psi_n/{\rm d}x$ at $x=0$ and $x=L$, one obtains
\begin{widetext}
\begin{eqnarray}
A&=& \frac{1}{4p}\biggl[(k+p){\rm e}^{i(k-p)L}-(k-p){\rm e}^{i(k+p)L}\biggr]E
+\frac{1}{4k}\biggl[(k+p){\rm e}^{i(k-p)L}+(k-p){\rm e}^{i(k+p)L}\biggr]E, \\
B&=& \frac{1}{4p}\biggl[(k+p){\rm e}^{i(k-p)L}-(k-p){\rm e}^{i(k+p)L}\biggr]E
-\frac{1}{4k}\biggl[(k+p){\rm e}^{i(k-p)L}+(k-p){\rm e}^{i(k+p)L}\biggr]E .
\end{eqnarray}
\end{widetext}
Then, we assume that the clock does not perturb the particle too much
and, as a consequence, that $\epsilon\ll 1$. In this case, $B\simeq 0$
and $A\simeq {\rm e}^{i(k-p)L}E$. Indeed, we have
$p=k\sqrt{1-\epsilon}$ which, at leading order, simply gives $k\simeq
p$ (but, of course, one should use the next to leading order when one
estimates the argument of the exponentials, see below). Then,
initially, this means that the wave-function can be written as, see
Eq.~(\ref{eq:defpsi})
\begin{equation}
\psi(x,\theta)=
\frac{1}{\sqrt{N}}\sum_{n=-j}^{n=j}A{\rm e}^{ikx}u_n(\theta)
=
A{\rm e}^{ikx}v_0(\theta), \quad x\ll 0,
\end{equation}
because $k$ does not depend on $n$, see Eq.~(\ref{eq:defk}). We see
that, initially, the clock points to zeroth hour, as expected. After
the barrier, using the previous considerations, the expression of the
wave-function can be written as, see Eq.~(\ref{eq:defpsi})
\begin{equation}
\psi(x,\theta)\simeq \frac{A}{\sqrt{N}}
{\rm e}^{ikx}\sum_{n=-j}^{n=j}{\rm e}^{-i(k-p)L}u_n(\theta), \quad x\gg L,
\end{equation}
where we have used Eq.~(\ref{eq:defv}). But we have
\begin{equation}
p\simeq k\left(1-\frac{\epsilon}{2}\right)\simeq k
-\frac{n\omega}{(2E/m)^{1/2}},
\end{equation}
which implies that (for $x\gg L$)
\begin{eqnarray}
\psi(x,\theta) &=& 
\frac{A}{\sqrt{2\pi N}}
{\rm e}^{ikx}\sum_{n=-j}^{n=j}{\rm e}^{in\theta- in\omega L/(2E/m)^{1/2}} 
\\
&=& A{\rm e}^{ikx}v_0\left(\theta-\frac{\omega L}{\sqrt{2E/m}}\right).
\end{eqnarray}
This means, see Eq.~(\ref{eq:evolv}), that the clock now indicates 
the time
\begin{equation}
t=\frac{L}{(2E/m)^{1/2}}=\frac{L}{v_{\rm clas}},
\end{equation}
where $v_{\rm clas}\equiv (2E/m)^{1/2}$ is the classical
velocity. Therefore, the clock has measured a time of flight which is
nothing but the classical time of flight $L/v_{\rm
  clas}$~\cite{Peres:1979nk}.

\begin{figure*}
\begin{center}
\includegraphics[width=12cm]{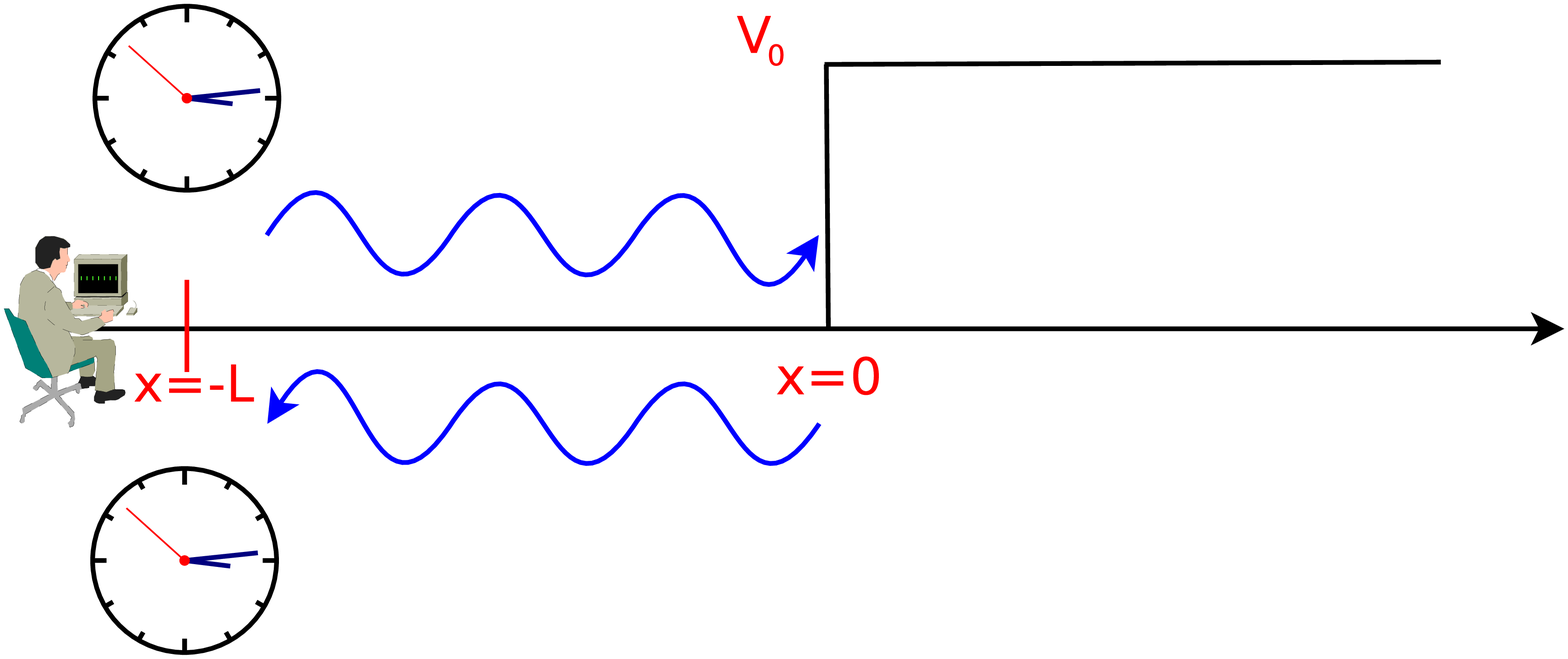}
\caption{Measurement of an out-and-back time of flight. An observer,
  located at $x=-L$ sends a particle towards a square potential step
  located at $x=0$. The physicist turns on the clock when the particle
  leaves and turn it off when the reflected wave (\ie the reflected
  particle) comes back and reaches $x=-L$.}
\label{fig:tunnel}
\end{center}
\end{figure*}

We now consider the situation represented in Fig.~\ref{fig:tunnel},
see Ref.~\cite{2004CQGra..21.2761D}. An observer, located at $x=-L$,
sends a particle towards a reflecting barrier of height $V_0$ located
at $x=0$. When the particle leaves the Peres clock is turned on. Then
the particle bounces back and when it comes back at the observer
position, the Peres clock is turned off. With this experimental set
up, one can measure the out-and-back time of flight. Following the
same approach as before, the wave-function can be written as
\begin{equation}
\label{eq:solpsitunnel}
\psi_n(x)=
\begin{cases}
\displaystyle
A{\rm e}^{ikx}+B{\rm e}^{-ikx}, & x<0, \\ 
\displaystyle
C{\rm e}^{-px}, & x>0 ,
\end{cases}
\end{equation}
where we now have
\begin{eqnarray}
k &=& \frac{\sqrt{2m(E-n\hbar \omega)}}{\hbar}=\frac{\sqrt{2mE}}{\hbar}
\sqrt{1-\epsilon}\, \\
\label{eq:newdefp}
p &=& \frac{\sqrt{2m(V_0+n\hbar \omega -E)}}{\hbar} \nonumber \\
&=&\frac{\sqrt{2m(V_0-E)}}{\hbar}\left(1+\frac{E}{V_0-E}\epsilon\right)^{1/2}
.
\end{eqnarray}
Then, we have to match the wave-function and its derivative at
$x=0$. This gives
\begin{eqnarray}
B &=& \frac{k-ip}{k+ip}A\equiv R A, \\
C &=& \frac{2k}{k+ip} A.
\end{eqnarray}
As before, we assume that, initially, the clock was is a given state,
here $v_0(\theta)$. Therefore, the wave-function in the region $x<0$
can be expressed as, see Eq.~(\ref{eq:defpsi})
\begin{equation}
\psi(x,\theta)=\frac{A}{\sqrt{N}}
\sum_{n=-j}^{n=j}\left({\rm e}^{ikx}+R{\rm e}^{-ikx}
\right)u_n(\theta).
\end{equation}
But $\vert R\vert =1$ and its phase is given by $-2\arctan(p/k)$. As a
consequence, one has
\begin{eqnarray}
\psi(x=-L,\theta)&=&\frac{A}{\sqrt{N}}\sum_{n=-j}^{n=j}\biggl[{\rm e}^{-ikL}
\nonumber \\
&+& {\rm e}^{-2i\arctan(p/k)}{\rm e}^{ikL}\biggr]u_n(\theta).
\end{eqnarray}
The next step is similar to the case of a free particle, namely we
expand the phases of the two wave-function branches in terms of the
parameter $\epsilon$. This gives
\begin{widetext}
\begin{eqnarray}
\psi(x=-L,\theta)&=&\frac{A}{\sqrt{2\pi N}}\sum_{n=-j}^{n=j}
\exp\Biggl[-\frac{iL}{\hbar}\sqrt{2mE}+in\left(\theta 
+\frac{\omega L}{v_{\rm clas}}\right)\Biggr]
+\exp\Biggl[-2i \arctan\left(\sqrt{\frac{V_0-E}{E}}\right)
+\frac{iL}{\hbar}\sqrt{2mE}
\nonumber \\
&+&in\left(\theta -\frac{\omega L}{v_{\rm clas}}
-\frac{\hbar \omega }{\sqrt{E(V_0-E)}}\right)\Biggr]\\
&=& \frac{A}{\sqrt{N}}
\exp\left(-\frac{iL}{\hbar}\sqrt{2mE}\right)
v_0\left(\theta 
+\frac{\omega L}{v_{\rm clas}}\right)
+\frac{A}{\sqrt{N}}
\exp\Biggl[-2i \arctan\left(\sqrt{\frac{V_0-E}{E}}\right)
+\frac{iL}{\hbar}\sqrt{2mE}\Biggr]
\nonumber \\ & & \times
v_0\left(\theta -\frac{\omega L}{v_{\rm clas}}
-\frac{\hbar \omega }{\sqrt{E(V_0-E)}}\right),
\end{eqnarray}
\end{widetext}
and this allows us to directly read the time indicated by the
clock. For the first branch, it is $-L/v_{\rm clas}$ while for the
second one, it is given by $L/v_{\rm clas}+\hbar/\sqrt{E(V_0-E)}$. As
a consequence, the time of flight of the particle can be expressed as
\begin{equation}
\Delta t=\frac{2L}{v_{\rm clas}}+\frac{\hbar}{\sqrt{E(V_0-E)}}.
\end{equation}
Therefore, the time of flight is the classical one, $2L/v_{\rm clas}$,
but, this time, there is an additional contribution. This one can be
understood as follows. The quantity $\Delta t$ can be re-written
as~\cite{2004CQGra..21.2761D}
\begin{equation}
\label{eq:tofsquare}
\Delta t=\frac{2(L+d)}{v_{\rm clas}},
\end{equation}
where $d=1/p$, see Eq.~(\ref{eq:newdefp}) (evaluated at $\epsilon=0$)
is the penetration depth into the potential step. This means that
there is an additional delay due to the tunnel effect and the fact
that the particle has a non vanishing probability to be below the
barrier~\cite{2004CQGra..21.2761D}. In this regime, the ``velocity''
is nevertheless given by its classical value $v_{\rm
  clas}$~\cite{2004CQGra..21.2761D}.

\par

This concludes this section on the Peres clock. We have seen that this
is a useful device to measure times of flight. We now use it to study
the motion of a quantum particle in an uniform gravitational field.

\subsubsection{The Gravitational Case}
\label{subsub:gravity}

\begin{figure*}
\begin{center}
\includegraphics[width=12cm]{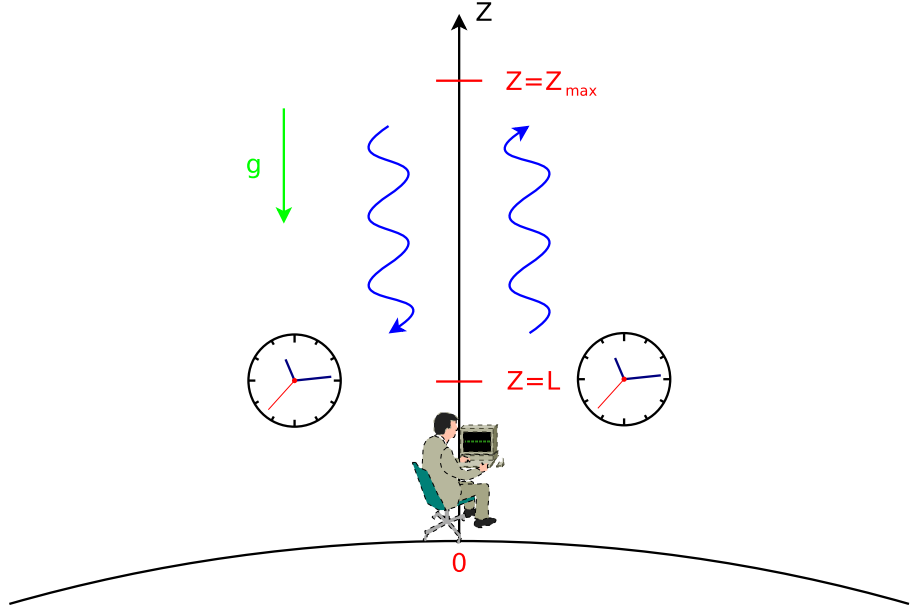}
\caption{Sketch of a quantum particle in an uniform gravitational
  field. The particle is sent upward, the initial time being recorded
  by a quantum Peres clock. The particle reaches a maximum altitude
  and then falls back to the detector at $z=L$ where the Peres clock
  measures this return time.}
\label{fig:galileo}
\end{center}
\end{figure*}

In this section, we study the status of the weak equivalence principle
and of the universality of the free fall in quantum
mechanics~\cite{Viola:1996de,2004CQGra..21.2761D,Davies:2004gh,Ali:2006ub,Chowdhury:2012zz,Onofrio:2010zza,Giulini:2011xf}.

\par

Following Ref.~\cite{2004CQGra..21.2761D}, let us consider the
situation where a particle is sent upwards in a uniform gravitational
field, see Fig.~\ref{fig:galileo}. If $z$ denotes the vertical
coordinate, the corresponding classical situation is described by the
following Lagrangian
\begin{equation}
L(\dot{z},z)=\frac{m_{\rm in}}{2}\dot{z}^2-m_{\rm grav}gz,
\end{equation}
where $g$ is the gravitational field and where we have made the
difference between the inertial mass $m_{\rm in}$ and the
gravitational mass (the ``gravitational charge'') $m_{\rm g}$. The 
solution to the classical equation of motion reads
\begin{equation}
\label{eq:classicaltrajec}
z(t)=-\frac12 \frac{m_{\rm grav}}{m_{\rm in}}gt^2+v_{\rm ini}t+L,
\end{equation}
where $z=L$ is the initial altitude (at $t=0$) and $v_{\rm ini}$ is the 
initial velocity, related to the energy $E$ of the particle by the 
following expression
\begin{equation}
v_{\rm ini}=\sqrt{2g\frac{m_{\rm grav}}{m_{\rm in}}}
\left(\frac{E}{m_{\rm grav}g}-L\right)^{1/2}.
\end{equation} 
Of course, the motion of the particle only depends on the ratio
$m_{\rm grav}/m_{\rm in}$. The maximum altitude, $z_{\rm max}$,
reached by the particle, see Fig.~\ref{fig:galileo}, is given by
\begin{equation}
\label{eq:zmax}
  z_{\rm max}=\frac{E}{m_{\rm grav}g},
\end{equation}
which means that, from its initial position to $z_{\rm max}$, the
particle can rise a distance of $E/(m_{\rm grav}g)-L$. After having
reached $z_{\rm max}$, the particle falls back and the up-to-down
classical time of flight reads
\begin{equation}
\label{eq:tofgravity}
\Delta t=2\sqrt{\frac{2}{g}}\left(\frac{m_{\rm in}}{m_{\rm g}}\right)^{1/2}
\left(\frac{E}{m_{\rm grav}g}-L\right)^{1/2}.
\end{equation}
If $m_{\rm grav}=m_{\rm in}$, this equation can also be written as
$\Delta t=2v_{\rm ini}/g$.

\par

Then, we would like to consider the same situation but from a
quantum-mechanical point of view~\cite{2004CQGra..21.2761D}. In
particular, we would like to measure the quantum time of flight. Given
the discussion of the previous section, it is clear that we must
couple the quantum particle to a Peres clock located at $z=L$. Then,
the next step is, of course, to solve the Schr\"odinger
equation. Following the previous considerations, we just have to
consider the equivalent of Eq.~(\ref{eq:schroperes}), but for the case
of an uniform gravitational field. It reads
\begin{equation}
\frac{{\rm d}^2\psi_n}{{\rm d}z^2}
+\Biggl[\frac{2m_{\rm in}\left(E-n\hbar \omega\right)}{\hbar^2}
-\frac{2m_{\rm in}m_{\rm grav}g}{\hbar^2}z\Biggr]\psi_n=0.
\end{equation}
It is straightforward to obtain that the solution can be expressed in
terms of Airy functions~\cite{Abramovitz:1970aa,Gradshteyn:1965aa}
$\mbox{Ai}(z)$ and $\mbox{Bi}(z)$. If we retain only the branch that
leads to a bounded wave-function, the result is given by
\begin{equation}
\label{eq:solpsigrav}
\psi(z,\theta)=\frac{A}{\sqrt{N}}\sum_{n=-j}^{n=j}\mbox{Ai}\left(
\frac{z-b}{a}\right)u_n(\theta),
\end{equation}
where $A$ is constant fixed by the normalization of the wave-function
and $a$ and $b$ are defined by
\begin{eqnarray}
a&=& \left(\frac{\hbar^2}{2m_{\rm in}m_{\rm grav}g}\right)^{1/3}, \\
b&=& \frac{E-n\hbar\omega}{m_{\rm grav}g}.
\end{eqnarray}
The next step consists in evaluating the wave-function at $x=L$, with
the assumption that the return point is located far from the maximum
altitude. This means that $L-b$ is a negative quantity and that $\vert
L-b\vert $ is large [more precisely, one has $ L-b=L-E/(m_{\rm
  grav}g)+E\epsilon/(m_{\rm grav}g)$ and the quantity $L-E/(m_{\rm
  grav}g)$ is minus the distance the particle can rise, see the remark
after Eq.~(\ref{eq:zmax}). Therefore, $L-b<0$ and, if this distance is
large, $\vert L-b\vert $ is indeed a large quantity]. In this limit,
one has
\begin{eqnarray}
\label{eq:limitpsigrav}
\psi(L,\theta)&\simeq &\frac{A}{2i\sqrt{\pi N}}\sum_{n=-j}^{n=j}
\left(
\frac{z-b}{a}\right)^{-1/4} \nonumber \\ & & \times 
\left({\rm e}^{i\xi+i\pi/4}-{\rm e}^{-i\xi-i\pi/4}
\right)u_n(\theta),
\end{eqnarray}
where $\xi \equiv (2/3)[(b-L)/a]^{3/2}$. Explicitly, this quantity can 
be expressed as
\begin{eqnarray}
\label{eq:wavegravity}
\xi &=& \frac{2}{3a^{3/2}}\left[\frac{E}{m_{\rm grav}g}\left(1-\epsilon\right)
-L\right]^{3/2} \\
&\simeq & \frac{2}{3a^{3/2}}
\left(\frac{E}{m_{\rm grav}g}-L\right)^{3/2}
\nonumber \\ & & 
-\frac{1}{a^{3/2}}\left(\frac{E}{m_{\rm grav}g}-L\right)^{1/2}
\frac{n\hbar \omega}{m_{\rm grav}g}. \nonumber \\
\end{eqnarray}
In order to calculate the difference between the two times indicated
by the clock and corresponding to the two branches of the
wave-function, it is in fact sufficient to estimate the part of the
phase shift that depends on $n$. Therefore, it is clear that the phase
shift $\pm \pi/4$ in the expression~(\ref{eq:limitpsigrav}) will not
contribute because it is $n$-independent. Moreover, the term
$[(z-b)/a]^{-1/4}$ is $n$-dependent but is common to the two branches
of the wave-functions. As a consequence, it will cancel out when the
phase shift is determined (but would participate to the ``absolute''
time indicated by the clock when the particle leaves $z=L$ and returns
to $z=L$). Summarizing, the up-and down time of flight is given by
\begin{eqnarray}
\Delta t&=&\frac{2}{a^{3/2}}\left(\frac{E}{m_{\rm grav}g}-L\right)^{1/2}
\frac{\hbar}{m_{\rm grav}g} \\
&=& 2\sqrt{\frac{2}{g}}\left(\frac{m_{\rm in}}{m_{\rm grav}}\right)^{1/2}
\left(\frac{E}{m_{\rm grav}g}-L\right)^{1/2},
\end{eqnarray}
that is to say exactly Eq.~(\ref{eq:tofgravity}), see
Ref.~\cite{2004CQGra..21.2761D}. Several remarks are in order
here. Firstly, although it is of quantum-mechanical origin, the above
equation does not contain $\hbar$ that cancels out. Secondly, this
result seems to indicate that the weak equivalence principle holds at
the quantum level. Thirdly, it may appear surprising that the quantum
$\Delta t$ is not corrected by a term corresponding to the penetration
depth, see Sec.~\ref{subsub:peres}, in particular
Eq.~(\ref{eq:tofsquare}), as it is the case for a square
potential. The interpretation given in Ref.~\cite{2004CQGra..21.2761D}
is that there is a finite probability that the particle tunnels below
the barrier but there is also a finite probability that the particle
is scattered back before the classical turning point. If these two
probabilities cancel then one obtains the classical
prediction. Fourthly, the above result is valid only if the return
time is measured far from the maximum altitude. What happens if, on
the contrary, the particle starts from a position which is close to
the turning point? To investigate this case, we re-start from
Eq.~(\ref{eq:solpsigrav}) and write the solution as
\begin{widetext}
\begin{equation}
\psi(z,\theta)=\frac{A}{\sqrt{N}}\sum_{n=-j}^{n=j}
\frac{\sqrt{y}}{3}\left\{\left[{\rm e}^{i\pi/3}J_{1/3}\left(\zeta\right)
+{\rm e}^{-i\pi/3}J_{-1/3}\left(\zeta\right)\right]
+
\left[\left(1-{\rm e}^{i\pi/3}\right)J_{1/3}\left(\zeta\right)
+\left(1-{\rm e}^{-i\pi/3}\right)J_{-1/3}\left(\zeta\right)\right]\right\}
u_n(\theta),
\end{equation}
where $y\equiv (b-z)/a>0$ and $\zeta\equiv 2y^{3/2}/3$. In this way,
we have identified the incident (first term in the squared bracket)
and the reflected wave (second term in the square bracket). Indeed,
using the asymptotic behavior of the Bessel functions at infinity, one
can check that each branches give precisely ${\rm
  e}^{\pm(i\xi+i\pi/4)}$ if $z=L$, see Eq.~(\ref{eq:limitpsigrav}). If
we now consider the limit $y\ll 1$, then one arrives at
\begin{eqnarray}
\psi(z,\theta)&=&\frac{A}{\sqrt{N}}\sum_{n=-j}^{n=j}
\Biggl\{\left[\frac{3^{-2/3}}{2\Gamma(2/3)}+\frac{3^{-4/3}}{2\Gamma(4/3)}y
\right]+i\sqrt{3}\left[-\frac{3^{-2/3}}{2\Gamma(2/3)}
+\frac{3^{-4/3}}{2\Gamma(4/3)}y\right]
+
\left[\frac{3^{-2/3}}{2\Gamma(2/3)}+\frac{3^{-4/3}}{2\Gamma(4/3)}y
\right]
\nonumber \\ & &
-i\sqrt{3}\left[-\frac{3^{-2/3}}{2\Gamma(2/3)}
+\frac{3^{-4/3}}{2\Gamma(4/3)}y\right]+\cdots 
\Biggr\}u_n(\theta),
\end{eqnarray}
\end{widetext}
the two first terms corresponding to the incident wave and the two
last ones to the reflected wave. The two branches are of course
complex conjugate to each other. Therefore, if we denotes by $\aleph$
the complex number corresponding to the two first terms in the above
equation, then the wave-function takes the form
\begin{equation}
\psi(z,\theta)\simeq \frac{A}{\sqrt{N}}\sum_{n=-j}^{n=j}
\vert \aleph \vert \left({\rm e}^{i\arg \aleph}
+{\rm e}^{-i\arg \aleph}\right)u_n(\theta),
\end{equation}
where 
\begin{equation}
\arg \aleph =\arctan \left[
\sqrt{3}
\frac{3^{2/3}\Gamma(2/3)y-3^{4/3}\Gamma(4/3)}
{3^{2/3}\Gamma(2/3)y+3^{4/3}\Gamma(4/3)}\right].
\end{equation}
The rest of the calculation proceeds as before. We first expand the
phase in $\epsilon$ and, then, take the limit $E/(m_{\rm g}g)-L
\rightarrow 0$. This leads to
\begin{equation}
\arg \aleph \simeq -\frac{\pi}{3}
-\frac{3^{5/6}}{6}\frac{\Gamma(2/3)}{\Gamma(4/3)}
\frac{\hbar}{m_{\rm grav}ga}n\omega.
\end{equation}
As a consequence, one finds that the time of flight is given by
\begin{eqnarray}
\Delta t&\simeq& 2 \frac{3^{5/6}}{6}\frac{\Gamma(2/3)}{\Gamma(4/3)}
2^{1/3}\left(\frac{m_{\rm in}}{m_{\rm grav}}\right)^{1/3}
\left(\frac{\hbar}{m_{\rm grav}g^2}\right)^{1/3}\\
\label{eq:qtofgravi}
&\simeq & 1.59 \times \left(\frac{m_{\rm in}}{m_{\rm grav}}\right)^{1/3}
\left(\frac{\hbar}{m_{\rm grav}g^2}\right)^{1/3},
\end{eqnarray}
in rough agreement with Ref.~\cite{2004CQGra..21.2761D}. This time, we
notice that $\hbar$ has not canceled in the final expression and that
the time of flight is different from its classical
counterpart~(\ref{eq:tofgravity}). This also means, as already
mentioned, that the cancellation which leads to a quantum time of
flight equals to the classical one is in fact valid only if the
particle starts far from the turning point.

\par

We conclude that the weak equivalence principle (or the universality
of the free fall) can be extended to the quantum regime. The linear
potential seems to possess precisely the shape which leads to
identical times of flight. It is important to realize that this
property depends on the shape of the potential and would not be
obtained with another shape, see
Ref.~\cite{2004CQGra..21.2761D}. Moreover, this seems to be true only
far from the turning point. If one identifies the wave-vector to
$k=(2m_{\rm in}m_{\rm grav}g/\hbar^2)^{1/3}$, see
Ref.~\cite{2010ApPhB.100...43K}, the de Broglie relation $v\simeq
\hbar k/m$ allows us to estimate a velocity. Then, from the quantum
time of flight, see Eq.~(\ref{eq:qtofgravi}), we can construct a
length $\simeq [\hbar^2/(m^2g)]^{1/3}$ which typically controls the
validity of the approximation just mentioned.

\subsubsection{Atom Trampoline}
\label{subsubsec:trampoline}

In this section, we want to briefly mention the case of the ``atom
trampoline''~\cite{Giulini:2011xf,2010ApPhB.100...43K}. Let us
consider again the experimental set up of the previous section but let
us now assume that a reflecting wall has been installed on the ground,
at $z=0$. Then, the boundary condition $\psi_n(0)=0$ leads to
$\mbox{Ai}(-b/a)=0$. If $z_n$ denotes the zero of the Airy function,
this means that we now have a discrete spectrum of energy levels given
by
\begin{equation}
E_n=-z_n\left(\frac{\hbar^2m_{\rm grav}^2g^2}{2m_{\rm in}}\right)^{1/3}.
\end{equation}
It is interesting to notice that this result depends on the ratio
$m_{\rm grav}^2/m_{\rm in}$ and not on $m_{\rm grav}/m_{\rm
  in}$. Moreover, this spectrum has been observed (with ultra-cold
neutrons which leads to $E_n$ of the order of $\sim
10^{-12}\, \mbox{eV}$) in the gravitational field of the
Earth~\cite{Nesvizhevsky:2003ww,Abele:2006xd}. Therefore, this result
confirms the discussion about the COW experiment described in
Sec.~\ref{subsec:cow}. It experimentally establishes that the
gravitational force ``couples'' to the Schr\"odinger equation in a
standard way.

\subsubsection{Schr\"odinger Equation in an Accelerated Frame}
\label{subsubsec:accelerated}

It is also frequent to refer to the weak equivalence principle as the
property stating that, locally, the effect of a constant gravitational
field can be mimicked by an accelerating frame. Therefore, it is
interesting to study whether this claim holds in quantum
mechanics. This question has been studied in
Refs.~\cite{Viola:1996de,Giulini:2011xf,2010ApPhB.100...43K}. Here, we
follow the treatment of Ref.~\cite{Viola:1996de}.

\par

Let us consider the free Schr\"ondinger equation (in one dimension to
simplify the problem). It reads
\begin{equation}
i\hbar \frac{\partial \Psi(t,z)}{\partial t}=-\frac{\hbar^2}{2m}
\frac{\partial ^2\Psi(t,z)}{\partial z^2}.
\end{equation}
Then, let us consider the following coordinate transformation
\begin{eqnarray}
z' &=& z-vt-\frac12gt^2 \\
t' &=& t.
\end{eqnarray}
It is straightforward to show that the function $\varphi(t',z')$
defined according to
\begin{eqnarray}
\Psi(t,z) &=& \varphi\left[t'(t),z'(t,z)\right]
\nonumber \\ & & \times
\exp\Biggl[\frac{imv}{\hbar}\left(z'+\frac{vt'}{2}\right)
\nonumber \\ & &
+\frac{imgt'}{\hbar}
\left(z'+\frac{vt'}{2}
+\frac{gt'^2}{6}\right)\Biggr],
\end{eqnarray}
satisfies the equation~\cite{Viola:1996de}
\begin{equation}
i\hbar \frac{\partial \varphi(t',z')}{\partial t'}=-\frac{\hbar^2}{2m}
\frac{\partial ^2\varphi(t',z')}{\partial z'^2}+mgz'\varphi(t',z').
\end{equation}
The potential in the above equation is precisely of the form
describing a constant gravitational field. In this sense, we find that
the equivalence between a constant gravitational field and an
accelerated frame propagates to quantum mechanics.

\subsubsection{Falling Composite Quantum Objects}
\label{subsubsec:composite}

In the previous sections, we have studied how a quantum particle falls
in a uniform gravitational field. For instance, we have investigated
under which circumstances one can say that the weak equivalence
principle holds in quantum mechanics. Here, we would like to study the
same question but in the case of a composite quantum object. In
particular, it is interesting to ask whether the motion of the
composite system (the atom) can depend on its internal state.

\par

Let us start by recalling the Ehrenfest theorem~\cite{tannoudji2006}
for a quantum particle. It states that the time derivatives of the
mean position and momentum are given by (for simplicity, we consider
the case of a one-dimensional system since we have in mind a particle
falling down along the $z$-axis)
\begin{eqnarray}
\frac{{\rm d}\langle z\rangle}{{\rm d}t} &=& 
\frac{\langle p_z\rangle}{m_{\rm in}},\\
\frac{{\rm d}\langle p_z\rangle}{{\rm d}t} &=& 
-\left\langle \frac{\partial V}{\partial z}\right 
\rangle .
\end{eqnarray}
In the case of a linear potential, $V(z)=m_{\rm grav}gz$, the two
previous equations expressing the Ehrenfest theorem can be combined in
such a way that we arrive at
\begin{equation}
\label{eq:d2zgravi}
\frac{{\rm d}^2\left \langle z\right\rangle}{{\rm d}t^2}
=-\frac{m_{\rm grav}}{m_{\rm in}}g.
\end{equation}
This means that the average position $\langle z\rangle(t)$ evolves
exactly as the classical $z$ given in
Eq.~(\ref{eq:classicaltrajec}). However, let us also notice that the
variance of the position is not a function of the ratio $m_{\rm
  grav}/m_{\rm in}$ only, see in particular
Refs.~\cite{Viola:1996de,Ali:2006ub,Chowdhury:2012zz}.

\par

After this short reminder, let us consider a bound system of two
particles interacting through a potential $V$ which only depends on
the distance between them. We also assume that this system is embedded
in a region where there is an exterior field described by the
potential $U({\bm r})$. The corresponding Lagrangian reads
\begin{eqnarray}
\label{eq:lagrangeatom}
  L\left(\dot{\bm r}_1,\dot{\bm r}_2,{\bm r}_1,{\bm r}_2\right)
&=&\frac{m_{\rm in,1}}{2}\dot{\bm r}_1^2+\frac{m_{\rm in,2}}{2}\dot{\bm r}_2^2
  -V\left({\bm r}_1-{\bm r}_2\right)
\nonumber \\ & &
-U\left({\bm r}_1\right)
  -U\left({\bm r}_2\right),
\end{eqnarray}
where ${\bm r}_1$ and ${\bm r}_2$ are the positions of the two
particles and $m_{\rm in,1}$, $m_{\rm in,2}$ their mass. If we
now introduce the center of mass position vector,
\begin{equation}
{\bm r}_{_{\rm G}}=\frac{m_{\rm in,1}}{m_{\rm in,1}+m_{\rm in,2}}{\bm r}_1
+\frac{m_{\rm in,2}}{m_{\rm in,1}+m_{\rm in,2}}{\bm r}_2,
\end{equation}
and the relative position vector,
\begin{equation}
\label{eq:defrelavector}
{\bm r}={\bm r}_1-{\bm r}_2,
\end{equation}
then Eq.~(\ref{eq:lagrangeatom}) takes the form
\begin{eqnarray}
L\left(\dot{\bm r}_{_{\rm G}},\dot{\bm r},{\bm r}_{_{\rm G}},{\bm r}\right)
&=&\frac{M}{2}\dot{\bm r}_{_{\rm G}}^2+\frac{\mu}{2}{\bm r}^2
-V({\bm r})
\nonumber \\ & &
-U({\bm r}_1)-U({\bm r}_2),
\end{eqnarray}
where $M\equiv m_{\rm in,1}+m_{\rm in,2}$ is the total mass and
$\mu\equiv m_{\rm in,1}m_{\rm in,2}/(m_{\rm in,1}+m_{\rm in,2})$ is
the reduced one. Of course, in the above expression ${\bm r}_1$ and
${\bm r}_2$ must be viewed as function of ${\bm r}_{_{\rm G}}$ and
${\bm r}$\footnote{The relation between these quantities is given by
\begin{eqnarray}
{\bm r}_1 &=& {\bm r}_{_{\rm G}}+\frac{m_{\rm in,2}}{M}{\bm r}, \\
{\bm r}_2 &=& {\bm r}_{_{\rm G}}
-\frac{m_{\rm in,1}}{M}{\bm r}.
\end{eqnarray}}. From the above considerations, it is straightforward to 
obtain the Hamiltonian of the system. We arrive at the following expression
\begin{equation}
\label{eq:hamiltonatom}
H=\frac{{\bm p}_{_{\rm G}}^2}{2M}+\frac{{\bm p}^2}{2\mu}
+V({\bm r})+U({\bm r}_1)+U({\bm r}_2),
\end{equation}
where ${\bm p}_{_{\rm G}}\equiv {\bm p}_1+{\bm p}_2$ and ${\bm
  p}=(\mu/m_{\rm in,1}){\bm p}_1-(\mu/m_{\rm in,2}){\bm p}_2$. If we
now assume that the external field is a uniform gravitational field
(directed along the $z$-axis), then it is easy to show that
\begin{align}
  U({\bm r}_1)+ U({\bm r}_2)&=\left(m_{\rm grav,1}+m_{\rm grav,2}\right)
gz_{_{\rm G}}
\nonumber \\ &+
\frac{1}{M}\left(m_{\rm grav,1}m_{\rm in,2}-m_{\rm grav,2}m_{\rm in,1}\right)
gz \\
&\equiv \overline{U}(z_{_{\rm G}})+\tilde{U}(z),
\end{align}
where, according to Eq.~(\ref{eq:defrelavector}), $z\equiv
z_1-z_2$. As a consequence, the Hamiltonian~(\ref{eq:hamiltonatom})
takes the form
\begin{equation}
H=H_{_{\rm G}}+H_{_{\rm rel}},
\end{equation}
where the center of mass and relative Hamiltonians can be expressed as
\begin{eqnarray}
H_{_{\rm G}}&=&\frac{{\bm p}_{_{\rm G}}^2}{2M}+\overline{U}(z_{_{\rm G}}) \\
H_{_{\rm rel}} &=& \frac{{\bm p}^2}{2\mu}+V({\bm r})+\tilde{U}(z).
\end{eqnarray}
The physical interpretation of these equations is clear. The center of
mass behaves as a free particle embedded in a uniform gravitational
field. On the other hand, the internal states of the composite system
are determined by the potential $V({\bm r})$ but are also affected by
the gravitational field through the term $\tilde{U}(z)$. In practice,
one expects that $\tilde{U}(z)\ll V({\bm r})$ and, therefore, the
perturbations of the energy levels will be either very small or even
totally negligible. Moreover, because $\left[z,p_{_{\rm G},z}\right]
=\left[z_{_{\rm G}},p_z\right]=0$, the two above Hamiltonians commute
\begin{equation}
\left[H_{_{\rm G}},H_{_{\rm rel}}\right]=0,
\end{equation}
which implies that the total state space is in fact the tensorial
product of the center of mass state space and of the relative state
space.

\par

Let us now study the behavior of $\langle z_{_{\rm G}}\rangle$. One has
\begin{eqnarray}
\label{eq:dzg}
\frac{{\rm d}\left \langle z_{_{\rm G}} \right \rangle}{{\rm d}t}
&=&\frac{1}{i\hbar}\left \langle \left[z_{_{\rm G}},H_{_{\rm G}}+H_{_{\rm rel}}
\right]\right \rangle
=\frac{1}{i\hbar}\left \langle \left[z_{_{\rm G}},H_{_{\rm G}}
\right]\right \rangle \nonumber \\
&=&\frac{\left \langle p_{_{\rm G},z}\right \rangle}{M}.
\end{eqnarray}
In the same manner, one can now calculate the evolution of $\langle
p_{_{\rm G},z}\rangle$. One obtains
\begin{eqnarray}
\frac{{\rm d}\left \langle p_{_{\rm G},z} \right \rangle}{{\rm d}t}
&=&\frac{1}{i\hbar}\left \langle \left[p_{_{\rm G},z},H_{_{\rm G}}+H_{_{\rm rel}}
\right]\right \rangle
=\frac{1}{i\hbar}\left \langle \left[p_{_{\rm G},z},H_{_{\rm G}}
\right]\right \rangle \nonumber \\
&=&\frac{1}{i\hbar}\left \langle \left[p_{_{\rm G},z},\overline{U}
(z_{_{\rm G}})\right]\right \rangle \nonumber \\
\label{eq:dpg}
&=&-\left(m_{\rm grav,1}+m_{\rm grav,2}\right)g.
\end{eqnarray}
Therefore, combining Eqs.~(\ref{eq:dzg}) and~(\ref{eq:dpg}), we arrive
at the following expression
\begin{equation}
\label{eq:d2zGfinal}
\frac{{\rm d}^2\left \langle z_{_{\rm G}}\right\rangle}{{\rm d}t^2}
=-\frac{m_{\rm grav,1}+m_{\rm grav,2}}{m_{\rm in,1}+m_{\rm in,2}}g.
\end{equation}
Of course, this equation is similar to Eq.~(\ref{eq:d2zgravi}) and
this implies that the center of mass mean value of the quantum
composite system will follow the classical trajectory. In particular,
this means that $\left \langle z_{_{\rm G}}\right\rangle$ does not
depend in which internal energy state the system is placed. Therefore,
if, for instance, we consider a falling Hydrogen atom, the behavior of
the position mean value of the atom will not depend of whether the
atom is in the state, say, $1s$ or $2p$. We also notice that the
coefficient in front of $g$ in Eq.~(\ref{eq:d2zGfinal}) is the ratio
of the total gravitational mass to the total inertial mass. In this
sense, the weak equivalence principle is also satisfied by the
composite quantum systems.

\par

Finally, it is also interesting to repeat the gedanken experiment in
Fig.~\ref{fig:galileo} but with a composite system. It seems
reasonable to couple the Peres clock to the center of mass of the
system which means that the Hamiltonian of the system is still given
by
\begin{equation}
H=H_{_{\rm G}}+H_{_{\rm rel}},
\end{equation}
where, now, the center of mass and relative Hamiltonians can be
expressed as
\begin{eqnarray}
H_{_{\rm G}}&=&\frac{{\bm p}_{_{\rm G}}^2}{2M}+\overline{U}(z_{_{\rm G}})
+{\cal P}(z_{_{\rm G}})H_{\rm clock} \\
H_{_{\rm rel}} &=& \frac{{\bm p}^2}{2\mu}+V({\bm r})+\tilde{U}(z).
\end{eqnarray}
Then, following Eq.~(\ref{eq:defpsi}), we write the wave-function of
the system as
\begin{equation}
\Psi\left(t,{\bm r}_{_{\rm G}},{\bm r},\theta\right)
=\frac{1}{\sqrt{N}}\sum_{n=-j}^{n=j}
\psi_n({\bm r}_{_{\rm G}})\chi_n({\bm r})u_n(\theta){\rm e}^{-iEt/\hbar}.
\end{equation}
The next step consists in inserting this wave-function in the
Schr\"odinger equation in order to obtain the equivalent of
Eq.~(\ref{eq:schroperes}). The result can be written as
\begin{widetext}
\begin{align}
\sum_{n=-j}^{n=j}\Biggl\{\chi_n({\bm r})&
\left[-\frac{\hbar^2}{2M}\Delta _{_{\rm G}}\psi_n({\bm r}_{_{\rm G}})
+\overline{U}(z_{_{\rm G}})\psi_n({\bm r}_{_{\rm G}})
+n\hbar \omega {\cal P}(z_{_{\rm G}})\psi_n({\bm r}_{_{\rm G}})
-E_{_{\rm G}}\psi_n({\bm r}_{_{\rm G}})\right]
\nonumber \\ & 
+\psi_n({\bm r}_{_{\rm G}})
\left[-\frac{\hbar^2}{2m}\Delta \chi_n({\bm r})
+\tilde{U}(z)\chi_n({\bm r})-E_{_{\rm rel}}\chi_n({\bm r})\right]\Biggr\}
u_n(\theta)=0,
\end{align}
\end{widetext}
where $E=E_{_{\rm G}}+E_{_{\rm rel}}$. This implies that only the
center of mass part, $\psi_n({\bm r}_{_{\rm G}})$, of the total
wave-function will depend on $n$. As a consequence, the internal part
should in fact be written as $\chi({\bm r})$ instead of $\chi_n({\bm
  r})$. This means that $\chi({\bm r})$ cannot cause a rotation of the
Peres clock. Therefore, the time indicated by the clock will not
depend on the internal state of the composite system. We conclude that
the time of flight of an atom, defined and calculated in terms of the
Peres clock, does not depend in which quantum state this atom is
placed.

\par

This concludes the section on the weak equivalence principle in
quantum mechanics. The result is mixed. We have shown that, very
often, it is satisfied in the framework of quantum mechanics. But, on
the other hand, it seems sometimes to be modified by quantum effects,
see in particular Eq.~(\ref{eq:qtofgravi}). Maybe this indicates that
its application to the more complicated case of quantum vacuum
fluctuations of a field is rather suspicious?

\section{Conclusions}
\label{sec:conclusions}

We are now in a position where we can really formulate the
cosmological constant problem. Vacuum fluctuations seem to exist in
Nature and to have normal gravitational properties. Therefore, it
seems natural to postulate that they participate in the value of the
cosmological constant. However, when one tries to use well controlled
techniques of quantum field theory to compute their energy density,
one obtains a number which seems to be in contradiction with the
measurement of $\Lambda$ that one obtains in cosmology and with the
constraints that one deduces from (for instance) the motion of the
planets in our solar system. The difficulty of the problem stems from
the fact that, in order to remove the above contradiction, one
necessarily has to abandon something that is considered as robust, \ie
renormalization in quantum field theory, weak equivalence principle,
high accuracy measurements of the expansion of the universe, etc \dots
.

\par

Even if this is not the purpose of this article, let us conclude this
review with a few words about the solutions that have been proposed to
solve the cosmological constant problem, for a complete overview of
the subject, see for instance Ref.~\cite{Nobbenhuis:2006yf}. Of
course, the most obvious solution is
super-symmetry~\cite{Bailin:1994qt,Aitchison:2005cf,Binetruy:2006ad}. However,
as we have discussed in Sec.~\ref{subsec:susy}, super-symmetry has to
be broken and this destroys the ``miraculous'' cancellation of the
various terms participating the vacuum energy. In fact, the only
common point between these various contributions is
gravity. Therefore, this suggests that the mechanism that cancels the
cosmological constant makes use of gravity. This idea has been
explored in the so-called adjustment
mechanisms~\cite{Dolgov:1982tj,Ford:1987de,Dolgov:2008rf}. These
models work reasonably well but, unfortunately, they suffer from a
severe disease. Indeed, in these scenarios, the Newton constant is
time-dependent and it turns out that the variation of the
gravitational coupling is too strong to be compatible with the known
constraints on $\vert \dot{G}/G\vert$. Therefore, although the idea is
very attractive at first sight, it seems that it does not yet exist a
realistic realization of it. It is also worth signaling that
back-reaction mechanisms have also been studied in
Refs.~\cite{Abramo:1997hu,Mukhanov:1996ak}. Here, the idea is that the
long wavelength (super-Hubble) cosmological perturbations are
described by a stress-energy tensor that exactly cancels the
cosmological constant.

\par

Another class of solution is based on quantum gravity and quantum
cosmology~\cite{Hartle:1983ai,Baum:1984mc,Coleman:1988tj}. It is
argued that the no-boundary Hartle-Hawking wave-function is such that
it peaks at $\Lambda =0$. However, this approach suffers from various
limitations among which is the fact that the path integral is not
properly defined and that probabilities are not positive definite
(this comes from the fact that the Wheeler-De Witt equation is in fact
similar to a Klein-Gordon equation. The positivity of probabilities in
quantum cosmology can be only be restored in the WKB approximation).

\par

Recently, following the pioneered work of Ref.~\cite{Rubakov:1983bz},
there have been many attempts to solve the cosmological constant
problem in the framework of extra dimensions, see chapter seven of
Ref.~\cite{Nobbenhuis:2006yf} or Ref.~\cite{Burgess:2004kd}. Another
popular solution, also based on high-energy physics considerations, is
the landscape approach~\cite{Polchinski:2006gy} in string
theory~\cite{Zwiebach:2004tj}. This approach assumes that all the
pocket universes of the landscape are populated during eternal
inflation. Combining this fact with the anthropic
principle~\cite{Weinberg:1987dv}, it is argued that the most probable
value of $\Lambda $ is approximately the value observed today. This
solution suffers from the difficulty of defining a measure on the
landscape~\cite{Linde:2010xz} and the use of the anthropic principle
in this context has been criticized in
Refs.~\cite{Starkman:2006at,Trotta:2006uj}.

\par

Finally, another possibility that has recently been investigated is to
modify the Einstein equations such that they become blind to a stress
energy tensor of the form $T_{\mu \nu}\propto g_{\mu \nu}$, see
Refs.~\cite{Afshordi:2008xu,Ellis:2010uc,Aslanbeigi:2011si}.

\par

In order to be complete, let us mention that models of analogue
gravity have also been developed to study the vacuum energy
problem~\cite{Volovik:2000ua,Volovik:2004gi,Sindoni:2011ej}. In these
models, one can calculate the equivalent of the cosmological constant
and study precisely its origin since the physics controlling the
microscopic constituents is explicitly known. These models are based
on many different types of analogue physical systems such as condensed
matter systems~\cite{Volovik:2001fm}, super-fluid
Helium~\cite{Volovik:2003fe}, Fermi liquid~\cite{Alexander:2008yg},
Bose-Einstein condensates~\cite{Finazzi:2011zw,Finazzi:2012wz} etc
... Then, one can show that a simple calculation of the ground state
may lead to an incorrect result, the only possibility in order to
reach the correct answer being to fully take into account the
microscopic theory. This would indicate that a correct calculation of
the cosmological constant must necessarily be based on the fine
structure of space-time, \ie probably on a consistent theory of
quantum gravity.

\par

As the variety of the subjects studied is this review shows, the
cosmological constant problem is a very rich one. Given its
difficulty, it is probable that it will have to wait for a new theory
beyond the current standard model to find its resolution. This theory
will have to describe the gravitational properties of the vacuum
fluctuations and to regulate the ultra-violet infinities that appear
in the calculation of vacuum energy density. This list of requirements
almost constitutes an identity card for a theory of quantum
gravity. But the most interesting aspect maybe goes the other way
round. Given the fact that cosmology has enabled us to grasp
observational signatures related to the vacuum energy, maybe these
experimental results will help us to deduce and to establish a
convincing theory of quantum gravity?  This is the reason why the
cosmological constant problem appears to be so important and so
interesting.

\acknowledgements

I would like to thank P.~Brax, C.~Ringeval and V.~Vennin for careful
reading of the manuscript and enjoyable discussions. Some parts of
this review article were presented as a lecture on the cosmological
constant problem at the Research Center for the Early Universe
(RESCEU)/Dark Energy NETwork (DENET) Summer school, in Kochi, Shikoku
island, Japan, in $2010$. I would like to thank the organizers,
especially Y.~Suto and J.~Yokoyama, and all the participants for very
helpful conversations.

\bibliography{bibliocc}

\end{document}